\documentclass[letterpaper ]{cernrep}

\usepackage[utf8]{inputenc}
\usepackage{subcaption}
\usepackage{multirow}
\usepackage{pgfplots}
\pgfplotsset{compat=1.12}
\usepackage{tikz}
\usetikzlibrary{patterns}
\usepackage{graphicx}
\usepackage{amssymb}
\usepackage{xspace}
\usepgfplotslibrary{fillbetween}

\usepackage[margin=1in]{geometry}
\usepackage{url}
\usepackage[hidelinks]{hyperref}
\usepackage{graphicx} %
\usepackage{color}
\usepackage{xcolor}
\usepackage{amsmath}

\newcommand{\ml}{\mathcal}

\usepackage{enumitem}
\usepackage{lineno}
\usepackage{authblk}

\hypersetup{%
  linkbordercolor=blue,%
  pdfborderstyle={/W 1}%
}

\newcommand{\sqrts}{\sqrt{s_\textrm{NN}}}

\renewcommand{\AA}{A+A}
\newcommand{\AB}{A+B}
\newcommand{\pA}{p+A}
\newcommand{\dA}{d+A}
\newcommand{\HeA}{$^3$He+A}
\newcommand{\PbPb}{Pb+Pb}
\newcommand{\AuAu}{Au+Au}
\newcommand{\RuRu}{Ru+Ru}
\newcommand{\ZrZr}{Zr+Zr}
\newcommand{\OO}{O+O}
\newcommand{\ArAr}{Ar+Ar}
\newcommand{\UU}{U+U}
\newcommand{\pp}{p+p}
\newcommand{\ep}{e+p}
\newcommand{\pPb}{p+Pb}
\newcommand{\pAu}{p+Au}
\newcommand{\pO}{p+O}
\newcommand{\dAu}{d+Au}

\newcommand{\SiW}{Si+W}
\newcommand{\pAr}{p+Ar}
\newcommand{\PbAr}{Pb+Ar}

\newcommand{\LRP}{Long Range Plan}
\newcommand{\pt}{$p_{\rm T}$\xspace}
\newcommand{\rpa}{$R_{\rm pA}$\xspace}
\newcommand{\raa}{$R_{\rm AA}$\xspace}
\newcommand{\rcp}{$R_{\rm CP}$\xspace}
\newcommand{\jpsi}{J/$\psi$\xspace}

\newcommand{\nch}{N_{\mathrm{ch}}}
\newcommand{\snn}{\sqrts}

\newcommand{\invnb}{\mathrm{nb}^{-1}}

\usepackage{tocloft}
\title{Hot QCD White Paper}
\date{October 2022}

\newcommand{\YaleUniversityNewHavenConnecticutUSA}{1}
\newcommand{\DukeUniversityDurhamNorthCarolinaUSA}{2}
\newcommand{\RiceUniversityHoustonTexasUSA}{3}
\newcommand{\BenemritaUniversidadAutnomadePueblaPueblaPueblaMexico}{4}
\newcommand{\UniversityofFlorenceFirenzeFirenzeItaly}{5}
\newcommand{\UniversityofHoustonHoustonTexasUSA}{6}
\newcommand{\PetertheGreatStPetersburgPolytechnicUniversityStPetersburgStPetersburgRussia}{7}
\newcommand{\CzechTechnicalUniversityinPraguePragueCzechRepublic}{8}
\newcommand{\UniversityofTexasatAustinAustinTexasUSA}{9}
\newcommand{\OakRidgeNationalLaboratoryOakRidgeTennesseeUSA}{10}
\newcommand{\NationalCentreforNuclearResearchWarsawMasovianVoivodeshipPoland}{11}
\newcommand{\MassachusettsInstituteofTechnologyCambridgeMassachusettsUSA}{12}
\newcommand{\GeorgiaStateUniversityAtlantaGeorgiaUSA}{13}
\newcommand{\BrookhavenNationalLaboratoryUptonNewYorkUSA}{14}
\newcommand{\EotvosLorandUniversityBudapestHungary}{15}
\newcommand{\LosAlamosNationalLaboratoryLosAlamosNewMexicoUSA}{16}
\newcommand{\UniversityofCaliforniaLosAngelesLosAngelesCaliforniaUSA}{17}
\newcommand{\StonyBrookUniversityStonyBrookNewYorkUSA}{18}
\newcommand{\KentStateUniversityKentOhioUSA}{19}
\newcommand{\LawrenceBerkeleyNationalLaboratoryBerkeleyCaliforniaUSA}{20}
\newcommand{\McGillUniversityMontrealQubecCanada}{21}
\newcommand{\UniversityofCaliforniaBerkeleyBerkeleyCaliforniaUSA}{22}
\newcommand{\GSIHelmholtzzentrumfurSchwerionenforschungDarmstadtHesseGermany}{23}
\newcommand{\InstituteforTheoreticalPhysicscharCGoetheUniversityFrankfurtamMainHesseGermany}{24}
\newcommand{\FrankfurtInstituteforAdvancedStudiesFrankfurtamMainHesseGermany}{25}
\newcommand{\UniversityofIllinoisChicagoChicagoIllinoisUSA}{26}
\newcommand{\CharlesUniversityPragueCzechRepublic}{27}
\newcommand{\OhioUniversityAthensOhioUSA}{28}
\newcommand{\FloridaStateUniversityTallahasseeFloridaUSA}{29}
\newcommand{\WayneStateUniversityDetroitMichiganUSA}{30}
\newcommand{\AugustanaUniversitySiouxFallsSouthDakotaUSA}{31}
\newcommand{\VanderbiltUniversityNashvilleTennesseeUSA}{32}
\newcommand{\CityUniversityofNewYorkNewYorkNewYorkUSA}{33}
\newcommand{\NaraWomensUniversityNaraNaraJapan}{34}
\newcommand{\TheOhioStateUniversityColumbusOhioUSA}{35}
\newcommand{\KoreaUniversitySeoulSeoulRepublicofKorea}{36}
\newcommand{\NRCKurchatovInstitutePNPIGatchinaLeningradskayaoblastRussia}{37}
\newcommand{\UniversityofMinnesotaMinneapolisMinnesotaUSA}{38}
\newcommand{\IndianaUniversityBloomingtonIndianaUSA}{39}
\newcommand{\CaliforniaPolytechnicStateUniversitySanLuisObispoCaliforniaUSA}{40}
\newcommand{\LehighUniversityBethlehemPennsylvaniaUSA}{41}
\newcommand{\UniversityofKansasLawrenceKansasUSA}{42}
\newcommand{\NationalCentralUniversityChungliTaoyuanCityTaiwan}{43}
\newcommand{\IowaStateUniversityAmesIowaUSA}{44}
\newcommand{\PusanNationalUniversityBusanRepublicofKorea}{45}
\newcommand{\UniversityofIllinoisUrbanaChampaignUrbanaIllinoisUSA}{46}
\newcommand{\CentralChinaNormalUniversityWuhanHubeiChina}{47}
\newcommand{\UniversityofSaoPauloSaoPauloSaoPauloBrazil}{48}
\newcommand{\UniversityofMarylandCollegeParkMarylandUSA}{49}
\newcommand{\UniversityofColoradoBoulderBoulderColoradoUSA}{50}
\newcommand{\UniversityofTennesseacharCKnoxvilleKnoxvilleTennesseeUSA}{51}
\newcommand{\UniversityofTsukubaTsukubaIbarakiJapan}{52}
\newcommand{\NationalResearchNuclearUniversityMEPhIMoscowMoscowRussia}{53}
\newcommand{\MississippiStateUniversityStarkvilleMississippiUSA}{54}
\newcommand{\PennsylvaniaStateUniversityUniversityParkPennsylvaniaUSA}{55}
\newcommand{\JagiellonianUniversityKrakowMalopolskaPoland}{56}
\newcommand{\CEAParisSaclayledeFranceFrance}{57}
\newcommand{\TexasAMUniversityCollegeStationTexasUSA}{58}
\newcommand{\CentroBrasileirodePesquisasFisicasRiodeJaneiroRiodeJaneiroBrazil}{59}
\newcommand{\RutgersUniversityPiscatawayNewJerseyUSA}{60}
\newcommand{\NorthCarolinaStateUniversityRaleighNorthCarolinaUSA}{61}
\newcommand{\CreightonUniversityOmahaNebraskaUSA}{62}
\newcommand{\UniversityofCaliforniaRiversideRiversideCaliforniaUSA}{63}
\newcommand{\BanarasHinduUniversityVaranasiUttarPradeshIndia}{64}
\newcommand{\PekingUniversityBeijingBeijingChina}{65}
\newcommand{\CERNcharCEuropeanOrganizationforNuclearResearchGenevaGenevaSwitzerland}{66}
\newcommand{\NuclearphysicisinstituteCASeCzechRepublic}{67}
\newcommand{\AdiyamanUniversityAdiyamanAdiyamanTurkey}{68}
\newcommand{\AkitaInternationalUniversityYuwaAkitacityJapan}{69}
\newcommand{\TheUniversityofTexasatAustinAustinTexasUSA}{70}
\newcommand{\UniversityofReginaReginaSaskatchewanCanada}{71}
\newcommand{\PurdueUniversityWestLafayetteIndianaUSA}{72}
\newcommand{\NationalChengKungUniversityTainanTainanTaiwan}{73}
\newcommand{\UniversityofWashingtonSeattleWashingtonUSA}{74}
\newcommand{\ColumbiaUniversityNewYorkNewYorkUSA}{75}

\author[\YaleUniversityNewHavenConnecticutUSA]{M. Arslandok}
\affil[\YaleUniversityNewHavenConnecticutUSA]{Yale University, New Haven, Connecticut, USA}
\author[\DukeUniversityDurhamNorthCarolinaUSA]{S. A. Bass}
\affil[\DukeUniversityDurhamNorthCarolinaUSA]{Duke University, Durham, North Carolina, USA}
\author[\RiceUniversityHoustonTexasUSA]{A. A. Baty}
\affil[\RiceUniversityHoustonTexasUSA]{Rice University, Houston, Texas, USA}
\author[\BenemritaUniversidadAutnomadePueblaPueblaPueblaMexico]{I. Bautista}
\affil[\BenemritaUniversidadAutnomadePueblaPueblaPueblaMexico]{Benemérita Universidad Autónoma de Puebla, Puebla, Mexico}
\author[\YaleUniversityNewHavenConnecticutUSA]{C. Beattie}
\author[\UniversityofFlorenceFirenzeFirenzeItaly]{F. Becattini}
\affil[\UniversityofFlorenceFirenzeFirenzeItaly]{University of Florence, Firenze, Italy}
\author[\UniversityofHoustonHoustonTexasUSA]{R. Bellwied}
\affil[\UniversityofHoustonHoustonTexasUSA]{University of Houston, Houston, Texas, USA}
\author[\PetertheGreatStPetersburgPolytechnicUniversityStPetersburgStPetersburgRussia]{Y. Berdnikov}
\affil[\PetertheGreatStPetersburgPolytechnicUniversityStPetersburgStPetersburgRussia]{Peter the Great St.Petersburg Polytechnic University, St.Petersburg, Russia}
\author[\PetertheGreatStPetersburgPolytechnicUniversityStPetersburgStPetersburgRussia]{A. Berdnikov}
\author[\CzechTechnicalUniversityinPraguePragueCzechRepublic]{J. Bielcik}
\affil[\CzechTechnicalUniversityinPraguePragueCzechRepublic]{Czech Technical University in Prague, Prague, Czech Republic}
\author[\UniversityofTexasatAustinAustinTexasUSA]{J. T. Blair}
\affil[\UniversityofTexasatAustinAustinTexasUSA]{University of Texas at Austin, Austin, Texas, USA}
\author[\OakRidgeNationalLaboratoryOakRidgeTennesseeUSA]{F. Bock}
\affil[\OakRidgeNationalLaboratoryOakRidgeTennesseeUSA]{Oak Ridge National Laboratory, Oak Ridge, Tennessee, USA}
\author[\NationalCentreforNuclearResearchWarsawMasovianVoivodeshipPoland]{B. Boimska}
\affil[\NationalCentreforNuclearResearchWarsawMasovianVoivodeshipPoland]{National Centre for Nuclear Research, Warsaw, Poland}
\author[\YaleUniversityNewHavenConnecticutUSA]{H. Bossi}
\author[\YaleUniversityNewHavenConnecticutUSA]{H. Caines}
\author[\MassachusettsInstituteofTechnologyCambridgeMassachusettsUSA]{Y. Chen$^\mathsection$}
\affil[\MassachusettsInstituteofTechnologyCambridgeMassachusettsUSA]{Massachusetts Institute of Technology, Cambridge, Massachusetts, USA}
\author[\GeorgiaStateUniversityAtlantaGeorgiaUSA]{Y.-T. Chien$^\mathsection$}
\affil[\GeorgiaStateUniversityAtlantaGeorgiaUSA]{Georgia State University, Atlanta, Georgia, USA}
\author[\BrookhavenNationalLaboratoryUptonNewYorkUSA]{M. Chiu}
\affil[\BrookhavenNationalLaboratoryUptonNewYorkUSA]{Brookhaven National Laboratory, Upton, New York, USA}
\author[\GeorgiaStateUniversityAtlantaGeorgiaUSA]{M. E. Connors$^*$\thanks{$^\mathsection$ Writing committee}\thanks{$^*$ Writing committee co-chairs: \href{mailto:mconnors@gsu.edu}{mconnors@gsu.edu}, \href{mailto:jean-francois.paquet@vanderbilt.edu}{jean-francois.paquet@vanderbilt.edu}}$^\mathsection$}
\author[\EotvosLorandUniversityBudapestHungary]{M. Csan\'ad}
\affil[\EotvosLorandUniversityBudapestHungary]{E\"otv\"os Lor\'and University, Budapest, Hungary}
\author[\LosAlamosNationalLaboratoryLosAlamosNewMexicoUSA]{C.L. da Silva$^\mathsection$}
\affil[\LosAlamosNationalLaboratoryLosAlamosNewMexicoUSA]{Los Alamos National Laboratory, Los Alamos, New Mexico, USA}
\author[\UniversityofCaliforniaLosAngelesLosAngelesCaliforniaUSA]{A. P. Dash}
\affil[\UniversityofCaliforniaLosAngelesLosAngelesCaliforniaUSA]{University of California Los Angeles, Los Angeles, California, USA}
\author[\StonyBrookUniversityStonyBrookNewYorkUSA,\BrookhavenNationalLaboratoryUptonNewYorkUSA]{G. David}
\affil[\StonyBrookUniversityStonyBrookNewYorkUSA]{Stony Brook University, Stony Brook, New York, USA}
\author[\StonyBrookUniversityStonyBrookNewYorkUSA]{K. Dehmelt}
\author[\KentStateUniversityKentOhioUSA]{V. Dexheimer$^\mathsection$}
\affil[\KentStateUniversityKentOhioUSA]{Kent State University, Kent, Ohio, USA}
\author[\LawrenceBerkeleyNationalLaboratoryBerkeleyCaliforniaUSA]{X. Dong$^\mathsection$}
\affil[\LawrenceBerkeleyNationalLaboratoryBerkeleyCaliforniaUSA]{Lawrence Berkeley National Laboratory, Berkeley, California, USA}
\author[\StonyBrookUniversityStonyBrookNewYorkUSA]{A. Drees}
\author[\McGillUniversityMontrealQubecCanada]{L. Du}
\affil[\McGillUniversityMontrealQubecCanada]{McGill University, Montreal, Québec, Canada}
\author[\LosAlamosNationalLaboratoryLosAlamosNewMexicoUSA]{J. M. Durham$^\mathsection$}
\author[\LawrenceBerkeleyNationalLaboratoryBerkeleyCaliforniaUSA,\UniversityofCaliforniaBerkeleyBerkeleyCaliforniaUSA]{R.J. Ehlers}
\affil[\UniversityofCaliforniaBerkeleyBerkeleyCaliforniaUSA]{University of California Berkeley, Berkeley, California, USA}
\author[\GSIHelmholtzzentrumfurSchwerionenforschungDarmstadtHesseGermany,\InstituteforTheoreticalPhysicscharCGoetheUniversityFrankfurtamMainHesseGermany,\FrankfurtInstituteforAdvancedStudiesFrankfurtamMainHesseGermany]{H. Elfner}
\affil[\GSIHelmholtzzentrumfurSchwerionenforschungDarmstadtHesseGermany]{GSI Helmholtzzentrum f\"ur Schwerionenforschung, Darmstadt, Germany}
\affil[\InstituteforTheoreticalPhysicscharCGoetheUniversityFrankfurtamMainHesseGermany]{Institute for Theoretical Physics\char"2C \, Goethe University, Frankfurt am Main, Germany}
\affil[\FrankfurtInstituteforAdvancedStudiesFrankfurtamMainHesseGermany]{Frankfurt Institute for Advanced Studies, Frankfurt am Main, Germany}
\author[\UniversityofIllinoisChicagoChicagoIllinoisUSA]{O. Evdokimov}
\affil[\UniversityofIllinoisChicagoChicagoIllinoisUSA]{University of Illinois Chicago, Chicago, Illinois, USA}
\author[\CharlesUniversityPragueCzechRepublic]{M. Finger}
\affil[\CharlesUniversityPragueCzechRepublic]{Charles University, Prague, Czech Republic}
\author[\CharlesUniversityPragueCzechRepublic]{M. Finger Jr.}
\author[\OhioUniversityAthensOhioUSA]{J. Frantz}
\affil[\OhioUniversityAthensOhioUSA]{Ohio University, Athens, Ohio, USA}
\author[\FloridaStateUniversityTallahasseeFloridaUSA]{A. D. Frawley}
\affil[\FloridaStateUniversityTallahasseeFloridaUSA]{Florida State University, Tallahassee, Florida, USA}
\author[\McGillUniversityMontrealQubecCanada]{C. Gale}
\author[\RiceUniversityHoustonTexasUSA]{F. Geurts}
\author[\WayneStateUniversityDetroitMichiganUSA]{V. Gonzalez}
\affil[\WayneStateUniversityDetroitMichiganUSA]{Wayne State University, Detroit, Michigan, USA}
\author[\AugustanaUniversitySiouxFallsSouthDakotaUSA]{N. Grau}
\affil[\AugustanaUniversitySiouxFallsSouthDakotaUSA]{Augustana University, Sioux Falls, South Dakota, USA}
\author[\VanderbiltUniversityNashvilleTennesseeUSA]{S. V. Greene}
\affil[\VanderbiltUniversityNashvilleTennesseeUSA]{Vanderbilt University, Nashville, Tennessee, USA}
% ...

\begin{abstract}
Hot QCD physics studies the nuclear strong force under extreme temperature and densities. Experimentally these conditions are achieved via high-energy collisions of heavy ions at the Relativistic Heavy Ion Collider (RHIC) and the Large Hadron Collider (LHC). In the past decade, a unique and substantial suite of data was collected at RHIC and the LHC, probing hydrodynamics at the nucleon scale, the temperature dependence of the transport properties of quark-gluon plasma, the phase diagram of nuclear matter, the interaction of quarks and gluons at different scales and much more.
This document, as part of the 2023 nuclear science long range planning process, was written to review the progress in hot QCD since the 2015 Long Range Plan for Nuclear Science, as well as highlight the realization of previous recommendations, and present opportunities for the next decade, building on the accomplishments and investments made in theoretical developments and the construction of new detectors.
Furthermore, this document provides additional context to support the recommendations voted on at the Joint Hot and Cold QCD Town Hall Meeting, which are reported in a separate document.
\end{abstract}

\begin{document}    

{ \hypersetup{hidelinks} 
	\urlstyle{sf}

\maketitle

}

\newpage

{ \hypersetup{hidelinks}  \tableofcontents }

\newpage
\section{Executive Summary}

Collisions of large nuclei at the Relativistic Heavy Ion Collider (RHIC) and the Large Hadron Collider (LHC) create a plasma of quarks and gluons with the properties of the early Universe, including an extraordinarily high temperature and near symmetry between matter and antimatter. 
This ``hot'' regime of Quantum Chromodynamics (QCD) lies at the intersection of nuclear physics with many-body quantum field theory, relativistic fluid dynamics, and condensed matter, probing the dynamical properties of quarks and gluons --- the fundamental degrees of freedom of nuclear matter --- at extreme densities and temperature. 
The newly built and mission-critical sPHENIX detector, and complementary STAR upgrades at RHIC, together with increased luminosity at the LHC and upgraded detectors at ALICE, ATLAS, CMS and LHCb,   
will enable a multimessenger era for hot QCD based on the combined constraining power of low-energy hadrons, jets, thermal electromagnetic radiation, heavy quarks, and exotic bound states.
Successful theory collaborations~\cite{An:2021wof,Putschke:2019yrg,Phillips:2020dmw} %
have paved the way for significant developments and breakthroughs in the next decade; state-of-the-art numerical simulations, assisted by machine learning techniques, will provide quantified uncertainties on the viscosities, jet transport coefficients, and other properties of the plasma, including detailed information on these quantities' dependence on the plasma's temperature.
Meanwhile, experimental measurements exploring the formation of quark-gluon plasma at the proton scale will provide an unparalleled opportunity to understand the regime of applicability of relativistic fluid dynamics, which will transform our understanding of this topic.
The phase diagram of nuclear matter will be redrawn by the analysis results of the completed RHIC systematic scan of collision energies, which probes the high baryon density regime of hot QCD, in synergy with astrophysical constraints from electromagnetic and gravitational observations of neutron stars and their mergers.

The purpose of this document is to serve as input to the U.S. Long Range Plan for Nuclear Physics by providing additional context and support for the goals of the hot QCD community described in the white paper from the Joint Hot/Cold QCD Town Hall meeting.  The recommendations and initiatives established by the QCD community during the Town Hall Meeting at MIT in September 2022 are described in a separate white paper. %
The present document focuses on hot QCD physics by highlighting the scientific progress since the 2015 Long Range Plan, describing the recent successes and plans for the experimental facilities, and future prospects for addressing remaining open questions.  Following the immensely successful implementation of the previous Long Range Plan recommendations, we recommended the following, based on the requirements to advance the understanding of hot QCD physics in the next decade and beyond.

{\bf Capitalizing on past investments at RHIC, the LHC, and theory }

Building upon previous investments, we recommend support for full operation of the RHIC facilities, maintaining U.S. leadership within the LHC heavy-ion program, and a healthy base theory program, as well as for the scientists involved in all these efforts.

\begin{itemize}
    \item Our understanding of QCD was transformed by the discoveries at RHIC in previous decades.
    To successfully complete the previously established RHIC scientific mission, it is crucial to support the RHIC running necessary for the sPHENIX program \cite{sPHENIX:2022BUR} as well as the resources to fully analyze the data collected from all RHIC experiments, including the highly successful Beam Energy Scan-II. The planned RHIC running will also enable STAR to extend measurements in the rapidity with its new forward upgrades \cite{STAR:2022BUR}. As RHIC is a unique and versatile facility capable of addressing several open questions, we note the additional opportunities described in the sPHENIX and STAR Beam Use Proposals if an opportunity for additional RHIC running occurs \cite{sPHENIX:2022BUR,STAR:2022BUR}.
    \item The LHC, the world's premier facility at the energy frontier, enables detailed studies of QCD at the highest temperatures through its ultrarelativistic heavy-ion program. It is essential for the U.S. to maintain its established leadership within the heavy-ion LHC program through data analysis as well as important detector upgrades across ALICE, ATLAS, CMS and LHCb, including the proposed ALICE 3 detector\cite{ALICE:2022wwr}.
    \item Advancements of QCD and interpretation of measurements rely on a strong theory community pushing to advance the field, working in close collaboration with the experimental community. Collaborations such as 
    BEST and JETSCAPE,
    focused on collectively addressing specific outstanding challenges, have proven extremely successful. Robust support for the base theory program, including opportunities for theorists to learn and exchange ideas, are also crucial for advancing our understanding of QCD and developing the direction of future experimental pursuits.  
\end{itemize}

{\bf \noindent Increasing U.S. participation in the LHC Heavy-Ion Program}

As RHIC is transformed into the Electron-Ion Collider (EIC), the LHC heavy-ion program will become the world's leading hot QCD facility. To maintain U.S. leadership in experimental hot QCD physics, supporting increased participation in the LHC programs is crucial, including detector upgrades required to capitalize on the planned luminosity increase. This luminosity will enable more precise measurements of rare probes and more differential measurements, strengthening our understanding of the underlying physics as the field evolves from discoveries to precision science. 

Supporting targeted detector R\&D and U.S.-led upgrades at the LHC experiments will enable opportunities for new and unique measurements, train the next generation of nuclear physicists and enhance the synergy between collider programs at LHC and the future EIC. 

{\bf \noindent  Increasing resources for high-performance computing and the development of machine-learning applications and open-source software}

With the increased luminosity and massive amount of data to be collected, high-performance computing is essential for the acquisition, storage, and timely analysis of experimental data. Likewise, the growth of theoretical physics depends on the use of cutting-edge computing facilities and algorithms. 
Machine learning techniques are now being applied to address theoretical, numerical, and experimental challenges and require increasing investment in cutting-edge computing resources, including hardware, and in the development of a high-quality software base accessible to the community. 
Past successes, including the dramatic advances made in finite temperature lattice QCD calculations, would not have been possible without access to leading class computing resources, USQCD computing facilities as well as the SciDAC program. Therefore access to these state-of-the-art computing resources will be vital for continued success and additional breakthroughs~\cite{ai_ml_workshop}.

{\bf \noindent  Exploring the high baryon density frontier}

Intermediate-energy nuclear collisions at RHIC and in international facilities provide a unique opportunity to search for the QCD critical point and for evidence of a first-order phase transition at high baryon density. The field is in an unequaled position to contribute to the study of neutron stars and their mergers, combining collider constraints on the nuclear equation of state with the invaluable information now provided by gravitational wave measurements. Following the highly successful BES-II at RHIC, U.S. leadership in mapping the QCD phase diagram should continue through funding theoretical developments~\cite{Lovato:2022vgq,Sorensen:2023zkk} and collaborations such as BEST\cite{An:2021wof}, as well as exploring opportunities for participation at international experimental facilities \cite{Almaalol:2022xwv}. 

{\bf \noindent  Enhanced investment in the growth and development of a diverse, equitable workforce }

 With the suite of opportunities 
in the field of hot QCD 
in the coming decades, a robust workforce is vital for continued success. Pursuing the above recommendations successfully will critically depend on increased attention to workforce development. Therefore, it is crucial to have an enforced community code of conduct, as well as programs to support underrepresented minority scientists and to increase diversity at every level, including faculty and leadership positions to ensure a thriving and productive community.

\newpage

\section{Introduction}

Ultra-relativistic heavy ion collisions seek to understand the most extreme form of nuclear matter, the quark-gluon plasma, which formed the early universe in the first few microseconds after the Big Bang and which may be formed in the extraordinary pressures present in the merger of neutron stars. 
The strength of interactions of quarks and gluons allows the plasma to maintain cohesion even under its explosive relativistic expansion.
This quark-gluon collectivity leads to a clear correlation between the initial geometry of colliding nuclei and the momentum anisotropy of the final particles. 

Evidence for collectivity in the collision of smaller systems was confirmed by a series of experiments at RHIC with proton-gold, deuteron-gold, and $^3$He-gold collisions, which showed a distinct correlation between small nuclei's variable geometry and the final hadron’s anisotropy measure  $v_n$ (Section~\ref{sec:progress:mesoscopic:small_size_limit_of_qgp}). This observation of fluid behavior in ever-smaller collision systems provided a renewed impetus to understand the limits of applicability of relativistic fluid dynamics, and how the system can approach local equilibrium shortly after the impact of the nuclei. Insights from weakly-coupled finite-temperature field theory, holography, and kinetic theory are being combined to better understand the approach to equilibrium of strong-coupled quark-gluon plasma (Section~\ref{sec:progress:mesoscopic}).

The limits of the formation of quark-gluon plasma have also been probed in two extensive Beam Energy Scan experiments at RHIC. The larger net-baryon density found at lower collision energies gives access to a wider region of the phase diagram of nuclear matter, enabling the search for a critical point (Section~\ref{sec:progress:macroscopic:finite_muB}). The search for a clear experimental signal of a critical point has advanced side-by-side with crucial new lattice calculations of the equation of state at finite baryon chemical potential; the latter has not found signs of a critical point at $\mu_B/T \lesssim 2\textrm{-}3$, driving the search for a critical point in the lower range of collision energies. The increasing complexity of modeling nuclear collisions at intermediate energies has been met with considerable theoretical progress, with new models of the initial condition, generalized hydrodynamic simulations with conserved charges, and more; these advances benefited considerably from the effort of the dedicated BEST topical collaboration (Section~\ref{sec:progress:macroscopic:hydro}). Connections with the astrophysics community have been strengthened by a joint interest in constraining the phase diagram of nuclear matter in a broad range of conditions, including those found in neutron stars and in neutron-star mergers, which can now be studied by a combination of electromagnetic signals and gravitational waves.

The diverse set of measurements provided by soft hadrons, jets, soft and hard electromagnetic probes, and heavy flavors also make possible multimessenger studies in heavy-ion collisions. Aided by machine learning, multiple systematic comparisons of heavy-ion simulations with large sets of measurements have now been performed. This has provided increasingly methodical constraints on the properties of strongly-coupled quark-gluon plasma, in particular, its shear and bulk viscosity  (Section~\ref{sec:progress:macroscopic:hydro}) and the interaction rate of energetic light and heavy partons with the plasma (Section~\ref{sec:progress:microscopic}). A combination of new lattice calculations and experimental measurements, along with theoretical advances, have advanced our understanding  of the formation of open heavy hadrons and quarkonium, and the insight they can provide about the properties of quark-gluon plasma (Sections~\ref{sec:progress:microscopic:theory_heavy_flavor}-\ref{sec:progress:microscopic:experiment_open_heavy_flavor}).

The large experimental effort to study the modification of jets in heavy-ion collisions can now benefit from the large increase in statistical precision due to increased luminosity and the upgraded detectors which enable refined measurements of correlations between electroweak bosons and jets (Section~\ref{sec:progress:microscopic:jets_and_leading_hadrons}), and increasingly fine-grained understanding of the internal structure of jets (Section~\ref{sec:progress:microscopic:jet_substructure}). Open software modeling environments such as the JETSCAPE framework can now provide a systematic comparison of jet calculations with this ensemble of new measurements. Modeling advances now make it possible to study both the interaction of hard partons with the quark-gluon plasma and the resulting wake left in the plasma by the parton (Section~\ref{sec:progress:mesoscopic:medium_response}). This ability to study the plasma's response to these perturbations provides an important connection to the plasma's approach to equilibrium early in the collision (Section~\ref{sec:progress:mesoscopic:onset_of_hydro}).

The isobar program at RHIC made possible a careful search for signals of the chiral magnetic effect in nuclear collisions (Section~\ref{sec:progress:macroscopic:chirality_and_vorticity}). The observed sensitivity of the results to small differences in the nuclear structure of the two isobars exemplifies the challenge of this undertaking, with only upper limits established at this point. The large magnetic field behind the chiral magnetic effect may also have a measurable effect on the polarization of hyperons. While not yet showing sensitivity to the magnetic field, hyperon polarization from the plasma's large vorticity has been successfully measured (Section~\ref{sec:progress:macroscopic:chirality_and_vorticity}).

The observed sensitivity of heavy-ion collisions to the nuclear structure is one of many examples that highlight the interdisciplinary nature of heavy-ion collisions (Section~\ref{sec:progress:interdisciplinary}). Studies of ultra-peripheral collisions have made a number of connections with other fields, including with the physics of the future Electron-Ion Collider. A joint interest in dense nuclear matter and relativistic fluids is also fostering close collaborations between the heavy-ion and astrophysics communities. Additional interdisciplinary collaborations include the community's involvement with machine learning, quantum computing, holographic studies of strongly-coupled systems, and more.

The progress accomplished since the last \LRP{} is organized below into three broad categories. The ``macroscopic'' section covers work related to collectivity, flow, and the thermal and near-thermal properties of strongly-coupled nuclear matter. This covers hydrodynamic simulations of heavy-ion collisions, including the initial conditions used for these simulations, along with studies of the nuclear equation of state, the production of soft thermal electromagnetic probes by the plasma, and studies of chirality and vorticity. The ``mesoscopic section'' focuses on the limits of collectivity in nuclear collisions: evidence of collectivity in collisions of small nuclei and the evolving understanding of the applicability of hydrodynamics in lower multiplicity systems. The thermalization of systems in the early stage of heavy-ion collisions is also discussed, and the related topic of jet wake thermalization closes this subsection. The final ``microscopic'' section transitions from a macroscopic description of strongly-coupled quark-gluon plasma to a microscopic description of particles interacting with the plasma and other particle production. Jet physics represents a major part of this section, including the theory of parton-plasma interactions, jet substructure as well as light and heavy quark jet observables.   Heavy quarks and their dynamical evolution in the plasma are further discussed. Also included in this section are the production of electroweak bosons, the study of the exotic hadronic bound states, and the rich physics of ultra-peripheral collisions. This same breakdown of topics is used in the future prospects section. The facilities section describes how the facilities have contributed to the progress to date as well as how the current and future upgrades including the newly constructed sPHENIX experiment recommended in the previous Long Range Plan will contribute to the next decade of progress and beyond.

While not a complete list, some of the key objectives of the hot QCD community in the next decade include:
\begin{itemize}

    \item Continue the community's pioneering work on understanding the foundation of relativistic fluid dynamics, the limits of its applicability in low multiplicity or far-from-equilibrium systems, and the connections of these topics with attractors;
    \item Further constrain the location of a QCD critical point and study the phase diagram at high baryon chemical potential;
    \item Explore innovative ways to experimentally observe the chiral magnetic effect and chiral symmetry restoration;
    \item Use thermal photons and dileptons to probe the early stages of large and small system collisions;
    \item Determine the origin of collectivity in small systems;    
    \item Utilize measurements of the jet-induced medium response to further our understanding and theoretical description of the jet-medium interface;
    \item Advance our understanding of the jet fragmentation process and its modification in heavy-ion collisions through detailed jet substructure measurements.
    \item Utilize new heavy-quark measurements to determine the heavy-quark diffusion coefficient and its temperature dependence.
    \item Improve the description of the three-dimensional structure of the initial state and its connection to the evolution of the produced QGP in heavy-ion collisions.
\end{itemize}

The summary, Sec. \ref{sec:summary}, describes the resources needed to address these questions in the next several years. These needs align well with the recommendations resulting from the Joint Hot/Cold QCD Town Hall meeting.

\clearpage

\section{Progress since the last \LRP}
\label{sec:progress}
\subsection{Macroscopic: collectivity, flow and thermal properties}
\label{sec:progress:macroscopic}

\subsubsection{The hydrodynamic limit of QCD:  collectivity, flow and transport coefficients of quark-gluon plasma}
\label{sec:progress:macroscopic:hydro}

Through long-standing efforts from both theory and experiment, the field of heavy-ion collisions is at the state of the art in terms of developing, testing, and simulating relativistic viscous fluid dynamics \cite{Luzum:2013yya,Heinz:2013th,DerradideSouza:2015kpt}.  
Since the last long-range plan, significant progress has been made in understanding the initial state immediately after two heavy ions collide, extracting transport coefficients at vanishing baryon densities, precisely reproducing experimental data, and the nature of collective flow in small systems, the development of relativistic spin hydrodynamics, the theoretical description of a multi-component relativistic fluid, and coupling hydrodynamics to jets and hadronic interactions.  Because of these advances in relativistic viscous hydrodynamics, it has paved the way for new discoveries such as  interdisciplinary connections through nuclear structure \cite{Lim:2018huo,Summerfield:2021oex,Bally:2021qys,Giacalone:2021udy,Bally:2022vgo}, out-of-equilibrium effects in neutron star mergers \cite{Most:2021zvc,Most:2022yhe}, cold atom quantum analogs of the quark-gluon plasma (QGP) \cite{Bluhm:2017rnf,Floerchinger:2022qem}, and statistical analysis techniques \cite{Bernhard:2019bmu}. Below we will outline the developments in the field over the last 7 years and then we will explore open questions and new opportunities.  

{\bf Standard Model of Heavy-ion collisions}

\begin{figure}[h]
    \centering
    \includegraphics[width=\textwidth]{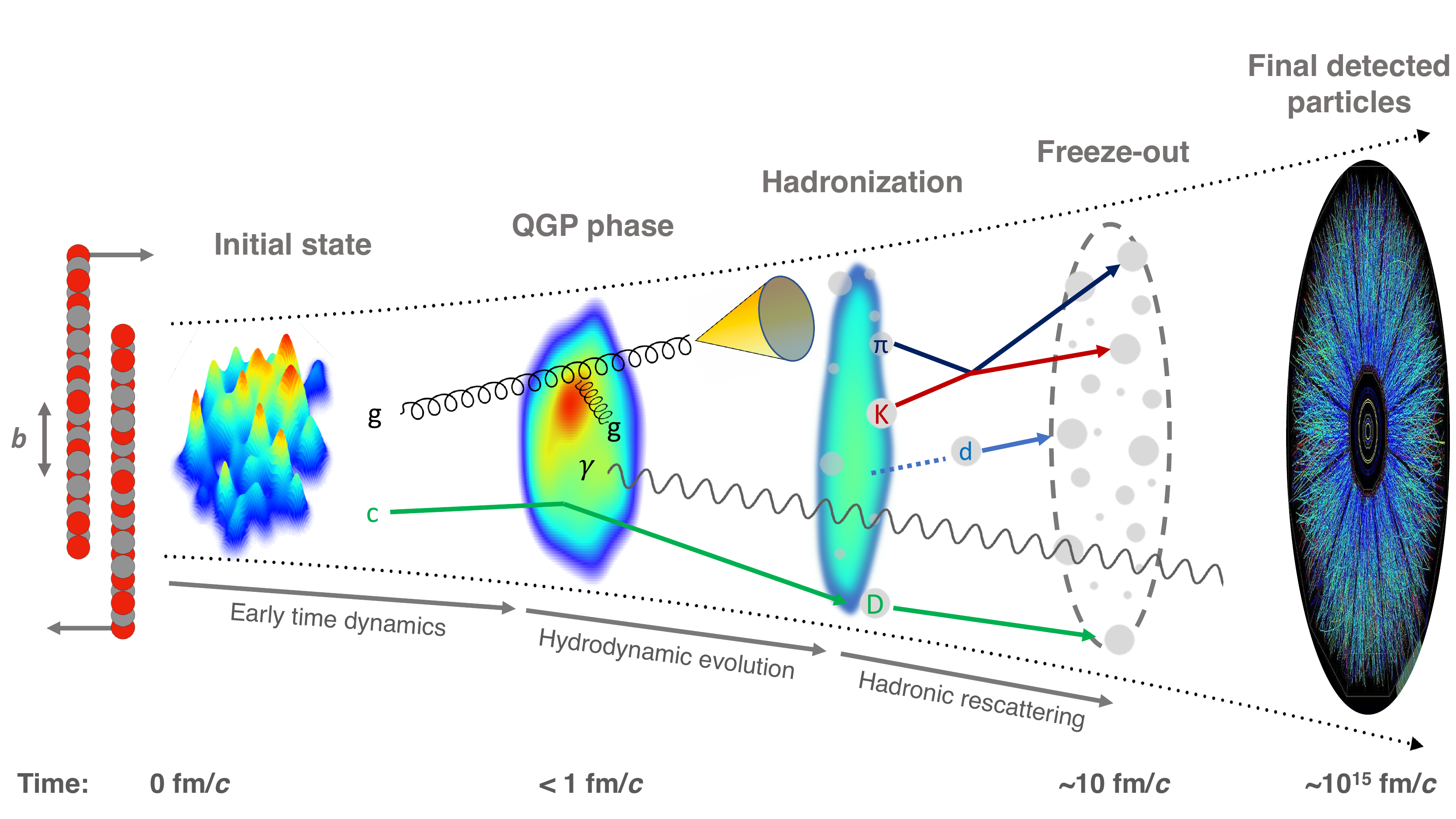}
    \caption{An illustration of different stages in a heavy-ion collision. The event display at the final stage is an iconic picture from the \href{https://www.bnl.gov/newsroom/news.php?a=217221}{STAR experiment at RHIC.}}
    \label{fig:standard_model_hic}
\end{figure}

Figure~\ref{fig:standard_model_hic} provides a schematic demonstration of the ``standard model" of heavy-ion collisions, that has been developed over a number of decades\footnote{This figure was inspired by illustrations from ~\href{https://u.osu.edu/vishnu/2014/08/06/sketch-of-relativistic-heavy-ion-collisions}{Chun Shen}.}. Radial flow occurs due to a greater pressure at the centre of the QGP compared to the outskirts, and this leads to a common velocity field outwards. The rate of the hydrodynamic expansion is influenced by the QGP's bulk viscosity ($\zeta/s$), which is its resistance to volume growth. Anisotropic flow is the result of a directional dependence to these pressure gradients. This occurs due to spatial anisotropies in the initial state. These arise if the collision zone is almond shaped (at $b>0$), or due to the lumpiness of the initial state. Such spatial anisotropies are converted to momentum anisotropies via the hydrodynamic response. This is influenced by the QGP's shear viscosity ($\eta/s$), which quantifies the resistance to fluid deformation. 

A key development during recent times was the democratization of initial state models that can reproduce experimental data.  The geometrical shape of the impact region of two colliding heavy ions can be quantified through a series of eccentricities 
\begin{equation}
    \mathcal{E}_{n,m}\equiv -\frac{\int r^m e^{i n\phi}\rho(r,\phi) rdrd\phi}{\int r^m \rho(r,\phi) rdrd\phi}
\end{equation}
wherein the indices $n,m$ correlate various harmonics such that $n=m=2$ is the elliptical eccentricity and $n=m=3$ is the triangular eccentricity. Large pressure gradients (that appear due to the geometrical shape of the initial state) lead to flow such that over time an elliptical shape in the initial state is transformed into an elliptical shape in momentum space, which is possible to measure experimentally using azimuthal anisotropies or otherwise known as collective flow harmonics $V_n$.  Careful studies using Pearson coefficients \cite{Gardim:2014tya} have determined that linear response ($V_n\sim \kappa_n \mathcal{E}_{n,n}$ where $\kappa_n$ is some constant that encodes medium effects) is the predominant contribution to $V_n$, especially in central collisions. Because of linear response in certain regimes, one can directly compare ratios of eccentricities e.g. $\frac{v_n\left\{4\right\}}{v_n\left\{2\right\}}\sim \frac{\varepsilon_n\left\{4\right\}}{\varepsilon_n\left\{2\right\}}$ to experimental data excluding a number of initial state models \cite{Yan:2013laa,Giacalone:2017uqx} and providing a clear benchmark for models to pass.  However, in peripheral collisions \cite{Noronha-Hostler:2015dbi}, low beam energies \cite{Rao:2019vgy}, and in small systems, the non-linear response of the final flow harmonics to the initial state \cite{Sievert:2019zjr} as well as out-of-equilibrium effects \cite{Schenke:2019pmk} begin to dominate.   Now open-source codes that pass benchmarks of multi-particle cumulant ratios of eccentricities (e.g. T$_{\rm{R}}$ENTo \cite{Moreland:2014oya} and IP-Glasma \cite{Schenke:2012wb}) exist,  such that the entire community has access to initial conditions that can reproduce experimental data. This has significantly reduced uncertainty in the extraction of transport coefficients and allows physicists to study questions that require greater precision than was previously possible. 

Due to the far-from-equilibrium nature of the initial state in heavy-ion collisions, it is possible that the energy-momentum tensor from the initial state, $T^{\mu\nu}_0$, 
cannot be used
directly input into hydrodynamics.  At finite densities this is further complicated by out-of-equilibrium currents for initial baryon $B^\mu$, strangeness $S^\mu$, and electric charge $Q^\mu$.
Hydrodynamics is applicable in the limit when the system has a large separation of scales. For example, the small scale, $l$, of a pond relates to the interactions of individual $H_2O$ molecules whereas the large scale, $L$, of a pond is simply the size of the pond.  Hydrodynamics is applicable when $l\ll L$. In heavy-ion collisions (especially for the initial state), it is not as clear that $l\ll L$ and, therefore, it is generally believed that a pre-equilibrium stage exists in heavy ions where some dynamics occur until time $\tau_{hyd}$ such that the initial energy momentum tensor, $T^{\mu\nu}_0$, is evolved in time until an energy momentum tensor is reached that is more compatible with hydrodynamics, $T^{\mu\nu}_{hyd}$.  There are a number of strong theoretical arguments which highlight the need for a pre-equilibrium stage such as the violation of causality and stability if $T^{\mu\nu}_0$ is applied directly to hydrodynamics \cite{Chiu:2021muk,Plumberg:2021bme}, the relation between $\tau_{hyd}$ to the $R_{AA}\times v_2$ puzzles in jets, and in order to handle  large momentum-space anisotropies in the local rest frame \cite{Strickland:2013uga}. These physics topics are covered in more detail in other sections, discussing the many open questions and future opportunities to explore, especially as one begins to explore jet-medium interactions and their influence on hydrodynamics (Sections~\ref{sec:progress:microscopic} and \ref{sec:progress:mesoscopic:medium_response})
and far-from-equilibrium systems that are relevant to ultra-small systems (Section~\ref{sec:progress:mesoscopic:small_systems}), such as those presumably created in collisions involving very small nuclei such as protons or deuterons.

The backbone of heavy-ions simulations is relativistic viscous hydrodynamics wherein a variety of 2+1D and 3+1D codes exist that incorporate shear and bulk viscosity \cite{Bozek:2009dw,Karpenko:2013wva,Noronha-Hostler:2014dqa,Shen:2014vra,Ryu:2015vwa,Weller:2017tsr,Okamoto:2017ukz} and are beginning to incorporate finite baryon densities \cite{Denicol:2018wdp,Du:2019obx} and even strangeness and electric charge \cite{Monnai:2012jc,Fotakis:2019nbq,Schafer:2021csj,Almaalol:2022pjc}.
Then relativistic viscous  hydrodynamic equations of motion have the general form of \cite{Denicol:2012cn}
\begin{eqnarray}
\tau_X \dot{X} + X& =& -(\omega + \frac{\tau_X}{2}X)\theta +\texttt{2nd order terms}+\texttt{coupling terms}
\end{eqnarray}
where $X=\pi^{\mu\nu},\; \Pi,\; n^{\mu}, \dots$ for the shear stress tensor, bulk pressure, and  the diffusion currents associated with baryon number, strangeness, and electric charge, which have the associated first order transport coefficients of shear viscosity $\eta$, bulk viscosity $\zeta$, and charge diffusion matrix $\kappa_{ij}$ where $ij$ can be a combination of multiple conserved charges. These first-order transport coefficients have been studied quite extensively in a variety of theoretical frameworks. Due to the fermion sign problem \cite{Troyer:2004ge}, one cannot currently calculate these transport coefficients from lattice QCD but it is possible using perturbative QCD.   Using these perturbative QCD calculations, $\eta/s$ has been calculated for vanishing baryon densities \cite{Ghiglieri:2018dib} at next-to-leading order and at finite baryon densities $\eta T/(\varepsilon+p)$ has been calculated at leading log  \cite{Danhoni:2022xmt} (note that $\varepsilon+p=sT$  for vanishing densities such that $\eta T/(\varepsilon+p)$ simplifies to $\eta/s$ at $\mu_B=0$).  In the future, it may be possible to calculate $\eta/s$ directly from quantum computing.  In the meantime, there is interest in developing quantum algorithms in order to calculate $\eta/s$ directly from Quantum Chromodynamics (QCD) \cite{Cohen:2021imf}.  Unlike $\eta/s$, bulk viscosity $\zeta/s$ is not possible to calculate within a conformal field theory and is precisely most relevant in the extremely non-perturbative regime around the QCD phase transition.  Thus, it is not possible to calculate $\zeta/s$ in the cross-over region from perturbative QCD, and other approaches are required. Recent calculations of bulk and shear viscosity both at $\mu_B=0$ and at finite $\mu_B$ from other theoretical approaches include  non-conformal holography \cite{Grefa:2022sav}, hadron resonance gas (HRG) model \cite{Kadam:2015xsa,Kadam:2015fza,Kadam:2018hdo,Mohapatra:2019mcl,McLaughlin:2021dph}, transport theory \cite{Rose:2017bjz,Rais:2019chb}, Color String Percolation Model \cite{Sahu:2020mzo}, linear sigma model \cite{Heffernan:2020zcf}, quasi-particle models \cite{Soloveva:2019xph,Mykhaylova:2019wci}, or QCD-motivated alternatives \cite{Haas:2013hpa,Christiansen:2014ypa,Dubla:2018czx,Ghiglieri:2018dib}.  In the cross-over phase transition, relevant at vanishing baryon densities, then one expects a continuous matching between these transport coefficients in the hadron gas phase to the QGP phase \cite{Noronha-Hostler:2008kkf}. Thus, calculations of $\eta$ and $\zeta$ within a hadron resonance gas also are of extreme interest and it appears that it is only possible to reproduce the extremely small minimum of $\eta/s$ (from phenomenological approaches) or a peak in $\zeta/s$ if sufficiently many hadrons exist and they are also very strongly interacting \cite{Karsch:2007jc}. Transport coefficients associated with diffusion, $\kappa_{ij}$, have many other subtleties that will be explored further later on and also in Section~\ref{sec:future:macroscopic}.  However, we note that even at the LHC and top RHIC energy, it may be possible to constrain baryons, strangeness, and electric charge diffusion using charge fluctuations due to gluons splitting into quark anti-quark pairs \cite{Carzon:2019qja}.

After the fluid has expanded and cooled sufficiently to fall below the QCD (crossover) phase transition, one must switch from a fluid into a gas of interacting hadrons. Generally, simulations rely on the Cooper-Frye approach~\cite{Cooper:1974mv} wherein one uses a Fermi-Dirac distribution to account for equilibrium $f_{eq}$ and out-of-equilibrium corrections from shear viscosity $\delta f_\eta$, bulk viscosity $\delta f_\zeta$,  and diffusion $\delta f_{\kappa_{ij}}$ such that
\begin{equation}
    f=f_{eq}+\delta f_\eta +\delta f_\zeta+\delta f_{\kappa_{ij}}
\end{equation}
which can be used to calculate the particle spectra for each hadron species. Typically groups sample over this spectrum (at finite densities conservation of baryons, strangeness, and electric charges is important to consider)
to produce particles that are then fed into hadronic afterburners like SMASH~\cite{Weil:2016zrk}
where the particles first undergo chemical freeze-out, followed by kinetic freeze-out. From the final particle spectra after kinetic freeze-out one can then obtain collective flow observables that can be compared directly to experimental data. 

This standard model of heavy-ion collision simulation setup is not just for soft observables, though.  These hydrodynamic backgrounds are then used to produce jets, heavy flavor, or electromagnetic probes as well.  In fact, it has been shown that subtle details in hydrodynamics as in the shear viscosity, using event-by-event simulations, or varying $\tau_{hydr}$ can affect jets and heavy flavor observables. Recently, groups are taking this a step further 
and fully incorporating jets into the hydrodynamic simulations using source terms that dump energy into the medium (see Section~\ref{sec:progress:mesoscopic:medium_response}).  Such approaches provide a fantastic opportunity to study far-from-equilibrium physics in the dense medium as well as multiscale problems across the entire range of momentum. Hydrodynamic predictions have been successful at reproducing collective flow in small systems and even making predictions \cite{Habich:2014jna,Shen:2016zpp} for the PHENIX small system scan \cite{PHENIX:2018lia}.

Over the past decade, one intriguing puzzle has arisen in the field of relativistic viscous fluid dynamics.  What is the small droplet of the QGP and at what point does the fluid dynamic picture break down? In systems as small as proton-ion, proton-proton, and even $\gamma^*$-ion signatures of the QGP have appeared for collective flow, strangeness enhancement, and heavy flavor flow.  However, the energy loss of high $p_T$ particles has puzzled theorists and experimentalists alike (see Sec.\ \ref{sec:progress:microscopic}). On the theoretical side, progress has been made in understanding these tiny droplets of fluid. It has been found that the pre-equilibrium phase can play a crucial role \cite{Liu:2015nwa,Weller:2017tsr,Kurkela:2018wud,Schenke:2019pmk,daSilva:2022xwu}, the new for sub-nucleonic fluctuations have been necessary to describe certain flow observables \cite{Mantysaari:2016ykx,Moreland:2018gsh,Nijs:2021clz}, and a new core-corona approach wherein the interior of QGP is treated as a fluid but the exterior is treated either as jet-like or via hadronic interactions \cite{Kanakubo:2019ogh,Kanakubo:2021qcw} (especially important for strange particles).

Another direction that relativistic viscous hydrodynamics has taken is to explore large baryon densities that are relevant for the Beam Energy Scan (and more specifically at low beam energies).  At low beam energies, the initial conditions become full 3-dimensional and may not reach hydrodynamization at a single initial time. Thus, groups have developed 3D initial conditions with baryon stopping with strings \cite{Shen:2017bsr} or used hadron transport models \cite{Schafer:2021csj}.  Then, hydrodynamics must include the conservation of baryon number that also leads to a finite baryon diffusion, which has been recently implemented in a number of models \cite{Denicol:2018wdp,Du:2019obx}. Meanwhile, groups have derived the theoretical framework for understanding critical fluctuations with hydrodynamics \cite{Nahrgang:2018afz,Rajagopal:2019xwg,An:2019osr,Nahrgang:2020yxm,Bluhm:2020mpc,An:2020vri,Martinez:2019bsn,Du:2020bxp,Pradeep:2022mkf,Pihan:2022xcl,Schaefer:2022bfm,An:2022jgc}, termed ``Hydro+" \cite{Stephanov:2017ghc}, and have applied this framework to low-dimension hydrodynamic toy models \cite{Du:2021zqz}. Studies in these low-dimensional hydrodynamic toys models have found interesting connections to out-of-equilibrium effects that may lead to potential signatures of the QCD critical point \cite{Monnai:2016kud,Dore:2020jye,Dore:2022qyz}, but they must be tested first in more realistic simulations.  However, a number of further steps are required in order to search for the QCD critical point or other interesting phase structures, which will be discussed in Section~\ref{sec:future:macroscopic}.

{\bf Collective flow Observables}

\begin{figure}[h]
    \centering
    \includegraphics[width=0.7\textwidth]{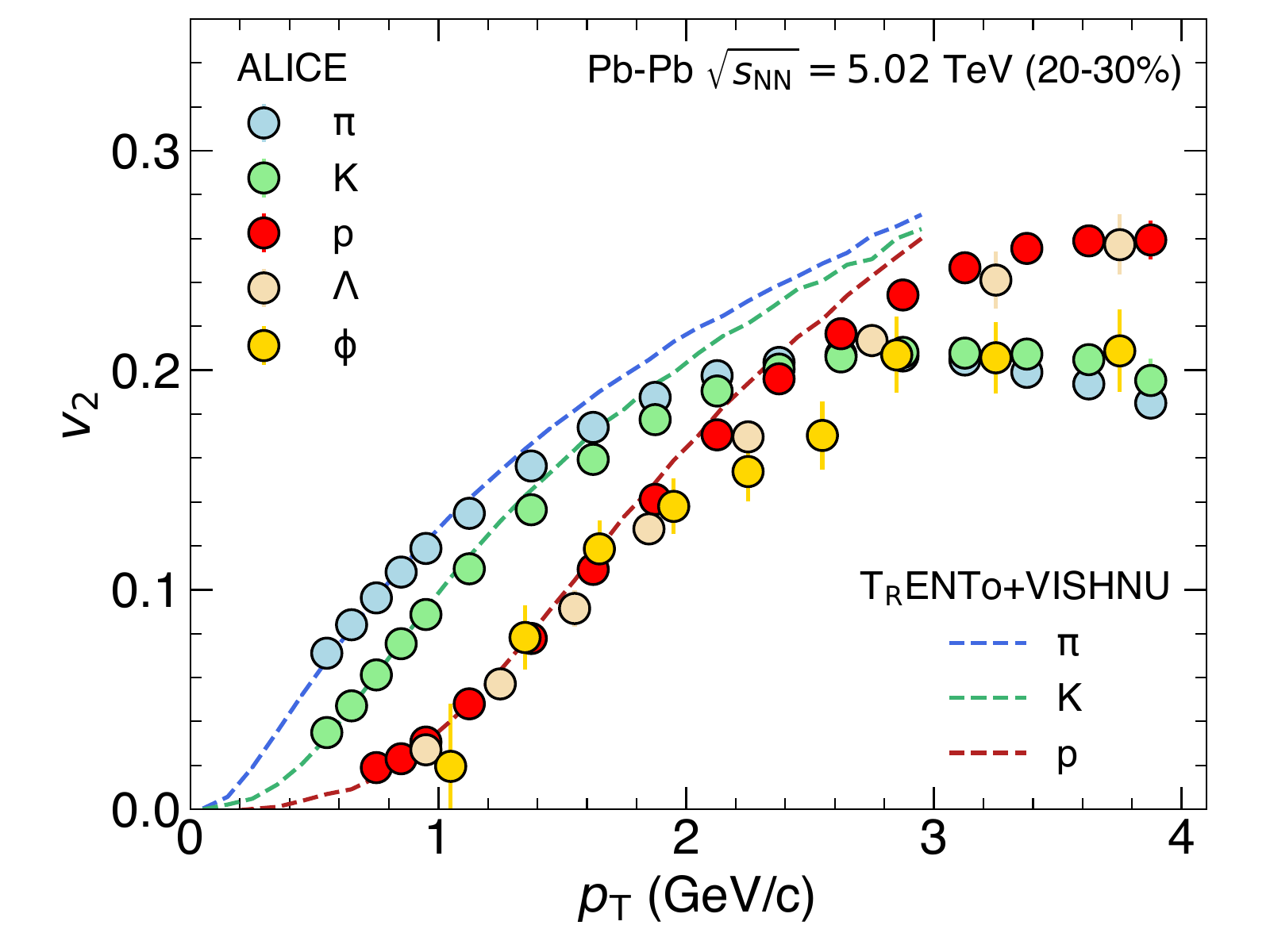}
    \caption{The \pt\ dependence of elliptic flow in \PbPb\ collisions at the LHC for various particle species. The data are from the ALICE experiment~\cite{Acharya:2018zuq}. The curves represent estimations extracted from T$_{\rm{R}}$ENTo+VISHNU based on a Bayesian analysis that determines the optimal hydrodynamic parameters from other observables~\cite{Bernhard:2019bmu}.    
    }
    \label{fig:pidVn}
\end{figure}

Since the previous \LRP, a number of measurements sensitive to the hydrodynamic response have entered the precision era. The uncertainties associated with some of these bulk observables are typically on the order of a percent in heavy-ion collisions - for a recent review of such observables at the LHC, see here~\cite{ALICE:2022wpn}. These measurements are therefore now used to provide strong constraints to the medium properties relevant for the studies of heavy-flavor and jet interactions in the QGP. A hallmark of the hydrodynamic response is the mass ordering observed for the \pt\ dependence of elliptic flow of various hadron species in the light flavor sector. Figure~\ref{fig:pidVn} shows the most recent measurements in \PbPb\ collisions at the LHC. The corresponding measurements are simultaneously sensitive to elliptic and radial flow, with elliptic flow leading to finite $v_{2}$ values, and radial flow inducing a mass splitting at a fixed \pT. Therefore, these measurements constrain both $\eta/s$ and $\zeta/s$ respectively. The hydrodynamic predictions shown from T$_{\rm{R}}$ENTo+VISHNU are based on a Bayesian analysis of separate observables in the soft sector. They describe the mass ordering extremely well at low \pt. A divergence is observed for \pt $> 1.5$ GeV/c, where non-equilibrium effects start to contribute. The clearest manifestation of these effects is observed via the baryon/meson grouping of $v_2$ at higher \pt, which can be explained by quark coalescence processes. Identified light hadron flow measurements have also been performed for a variety of collision systems recently. At RHIC, these include collisions of Cu+Au at $\sqrt{s_{\mathrm{NN}}} = 200$ GeV~\cite{PHENIX:2015zbc} and U+U at $\sqrt{s_{\mathrm{NN}}} = 193$ GeV~\cite{STAR:2021twy}, while at the LHC these include Xe+Xe at $\sqrt{s_{\mathrm{NN}}} = 5.44$ TeV~\cite{ALICE:2021ibz}. They have also been explored for higher orders of anisotropic flow, with examples from RHIC for Au+Au collisions at $\sqrt{s_{\mathrm{NN}}} = 200$ GeV~\cite{PHENIX:2014uik}, and Pb+Pb collisions at the top LHC energy~\cite{Acharya:2018zuq,CMS:2022bmk}.

A key example of more differential endeavors regarding the investigations into hydrodynamic response, pursued since the last \LRP, is given by measurements of Symmetric Cumulants (SC$(k,l)$). These extend the study of the individual flow amplitudes $v_n$ at low \pt to correlations between event-by-event fluctuations of flow coefficients~\cite{Niemi:2012aj,Bilandzic:2013kga,Aad:2015lwa,Qian:2016pau,Zhu:2016puf}. Hydrodynamic calculations show that while $v_2$ and $v_3$ exhibit an approximately linear dependence on the corresponding eccentricities $\epsilon_2$ and $\epsilon_3$, respectively, the higher order $v_n$ coefficients (i.e., for $n > 3$) have also non-linear contributions from $\epsilon_2$ and $\epsilon_3$ in addition to the linear ones from $\epsilon_n$~\cite{Qin:2010pf,Qiu:2011iv,Yan:2015jma,Kolb:2003zi}. These observations lead to non-trivial correlations between different flow coefficients which result in new and independent constraints on the initial conditions and $\eta/s$. 
\begin{figure}[h]
    \begin{center}
    \includegraphics[width = 0.99\textwidth]{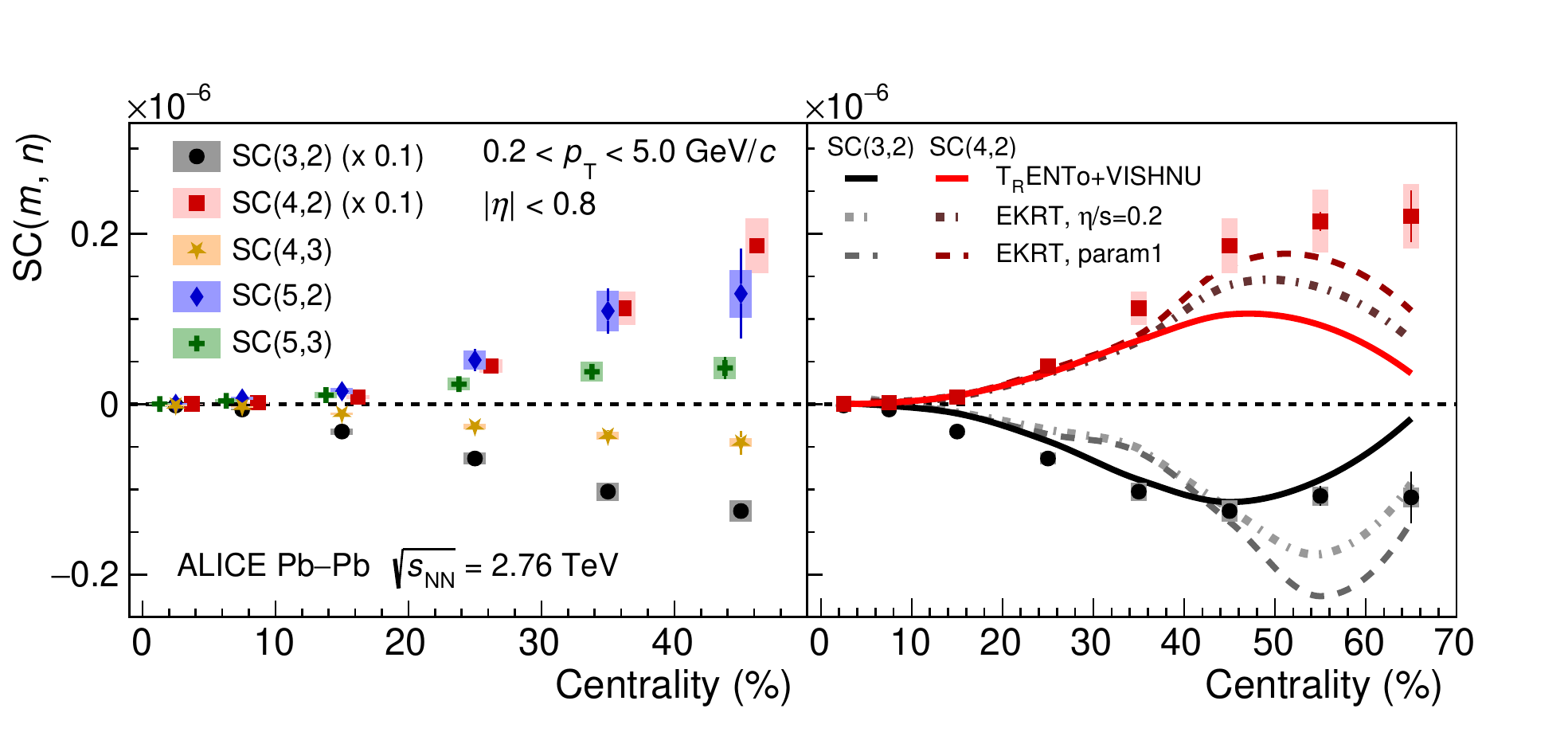}
    \end{center}
    \caption{Centrality dependence of event-by-event flow harmonic correlations in \PbPb{} collisions at $\sqrt{s_{\mathrm{NN}}} = 2.76$ TeV~\cite{ALICE:2016kpq,Acharya:2017gsw} compared with various hydrodynamic calculations~\cite{Niemi:2015qia, Bernhard:2019bmu}.} 
    \label{fig:SC}
\end{figure}
The left panel of Figure~\ref{fig:SC} presents the centrality dependence of correlations between $v_n$ coefficients (up to the $5^{\mathrm{th}}$ order) from Symmetric Cumulants, which are defined as:
\begin{equation}
{\rm SC}(k,l) \equiv \langle v_k^2 v_l^2\rangle - \langle v_k^2\rangle\langle v_l^2\rangle 
\end{equation}
The results shown are for \PbPb{} collisions at $\sqrt{s_{\mathrm{NN}}} = 2.76$ TeV~\cite{ALICE:2016kpq, Acharya:2017gsw}. The correlations among different flow coefficients depend on harmonic, as well as collision centrality~\cite{Acharya:2017gsw}. Positive values of SC(4,2), SC(5,2), and SC(5,3) and negative values of SC(3,2) and SC(4,3) are observed for all centralities. These indicate that event-by-event fluctuations of $v_2$ and $v_4$, $v_2$ and $v_5$, and $v_3$ and $v_5$ are correlated, while $v_2$ and $v_3$, and $v_3$ and $v_4$ are anti-correlated. Furthermore, the lower-order harmonic correlations are much larger than the higher-order ones.  Measurements from ATLAS and CMS for \PbPb{} collisions at the LHC~\cite{ATLAS:2015qwl,CMS:2017kcs}, and STAR for variety of \AuAu{} collision energies~\cite{STAR:2018fpo}, show similar trends. 
The SC observables are compared with EKRT~\cite{Niemi:2015qia} and T$_{\rm{R}}$ENTo+VISHNU~\cite{Bernhard:2019bmu} predictions in the right panel of Figure ~\ref{fig:SC}. The EKRT calculations are shown for two temperature-dependent $\eta/s$ parameterizations that provide the best description of RHIC and LHC $v_n$ data: the constant $\eta/s=0.2$, and ``param1"~\cite{Niemi:2015qia}. The ``param1" parameterization is characterized by a moderate slope in the temperature dependence of $\eta/s$ which decreases (increases) in the hadronic (QGP) phase and the phase transition occurs around 150 MeV. The measurements of SC(3,2) and SC(4,2) are not described simultaneously in each centrality interval by the EKRT calculations. However, such a comparison does demonstrate the sensitivity of SC$(m,n)$ to the temperature dependence of $\eta/s$ - the EKRT predictions are clearly different for the two parameterizations. Measurements of SC(4,2) are better described by the T$_{\rm{R}}$ENTo+VISHNU predictions for \PbPb{} collisions at $\sqrt{s_{\mathrm{NN}}} = 2.76$ TeV, however this is not the case for SC(3,2). These predictions utilize the same hydrodynamic parameters as those shown in Figure ~\ref{fig:pidVn}.
Hydrodynamic calculations describe a wide variety of additional experimental results, which reveal many facets of the QGP dynamical evolution. Such results include the energy dependence of $\Lambda$ global polarization values~\cite{STAR:2017ckg,STAR:2018gyt,ALICE:2019onw}, and $\Lambda$ longitudinal polarization measurements at RHIC and the LHC~\cite{STAR:2019erd,ALICE:2021pzu}. These will be discussed further in Section~\ref{sec:progress:macroscopic:chirality_and_vorticity}. They can also describe measurements of higher-order flow coefficients (up to $v_8$)~\cite{ATLAS:2018ezv,ALICE:2020sup}, non-linear  contributions to higher-order flow coefficients~\cite{ALICE:2017fcd,CMS:2019nct}, and symmetry plane correlations~\cite{ATLAS:2014ndd,CMS:2015xmx,ALICE:2017fcd}. All of these results are new with respect to the last \LRP. On the other hand, for reasons subject to much theoretical attention, hydrodynamic predictions cannot describe measured anisotropic flow coefficients at LHC energies in ultra central \PbPb{} collisions to the same degree of accuracy as mid-central \PbPb{} collisions. This is referred to as the ``ultra central flow puzzle"~\cite{Shen:2015qta}, and persists even with the most state of the art hydrodynamic models~\cite{Giannini:2022lbj}. In addition, hydrodynamic calculations cannot describe balance-function widths~\cite{ALICE:2021hjb,ALICE:2022wpn}, which might be explained by charge diffusion effects not implemented in these calculations. The differences with respect to the data can therefore provide a crude measure of charge diffusion, another key QGP transport parameter.

Nonetheless, the success of hydrodynamic models in describing a wide variety of experimental observables implies that the system is strongly coupled at momentum scales corresponding to the QGP temperature, as hydrodynamics describes the evolution of the QGP in terms of a liquid. The question then becomes quantitative -- how strong is the coupling -- and values of the shear ($\eta/s$) and bulk ($\zeta/s$) viscosities per entropy density play a central role in addressing this question. In the strong-coupling picture, both transport properties $\eta/s$ and $\zeta/s$ are proportional to the shear and bulk relaxation times. As the initial state is highly non-uniform, this leads to large non-equilibrium shear and bulk pressures at early times. The shear and bulk relaxation times determine how quickly the system can ease these pressures and develop flow -- if the coupling is large, these times will be small, which corresponds to small values of $\eta/s$ and $\zeta/s$, and vice versa. 
\begin{figure}[h]
    \centering
    \includegraphics[width=0.8\textwidth]{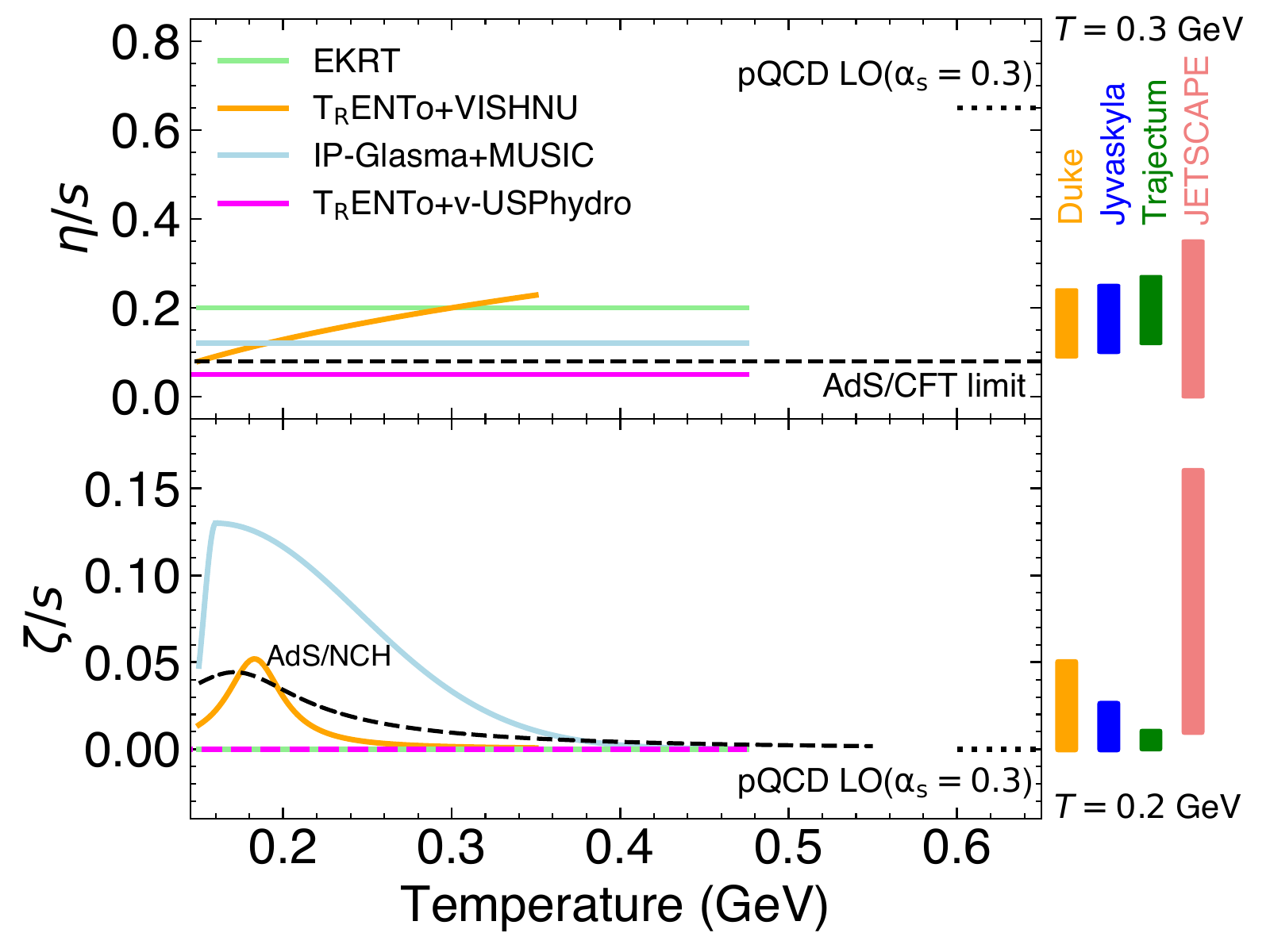}
    \caption{The temperature dependence of the shear (top panel) and bulk (bottom) viscosities over entropy density in the QGP phase constrained by RHIC and LHC measurements from various hydrodynamic models described in the text. Limits from pQCD~\cite{Arnold:2006fz}, AdS/CFT~\cite{Kovtun:2004de}, and AdS/Non-Conformal Holographic~\cite{Finazzo:2014cna} approaches are also shown. The ranges on the right of the plot represent 90\% posterior intervals from the Bayesian analyses. The figure is taken from elsewhere~\cite{ALICE:2022wpn}.}
    \label{fig:eta_zetas}
\end{figure}
Measurements from a suite of soft sector observables at RHIC (e.g.~\cite{STAR:2008med,STAR:2004jwm,STAR:2013qio}) and the LHC (e.g.~\cite{ALICE:2010mlf,ALICE:2016igk,ALICE:2013mez,ALICE:2011ab,ALICE:2014gvd}) have allowed for an exploration of the temperature dependence of $\eta/s$ and $\zeta/s$ in heavy-ion collisions. 

Figure~\ref{fig:eta_zetas} compares these dependencies for four hydrodynamic model chain calculations constrained by RHIC and LHC data. Regarding $\eta/s$, many other fluids exhibit a temperature dependence for $\eta/s$, with a minimum occurring at the phase transition temperature between the gas and liquid phases. Whether such a temperature dependence exists for the QGP phase is an open question for the temperatures probed by heavy-ion collisions, as demonstrated in the top panel of Figure~\ref{fig:eta_zetas}. The models utilize either an independence (IP-Glasma+MUSIC~\cite{Schenke:2020mbo} and EKRT~\cite{Niemi:2015qia}), or a weak dependence (T$_{\rm{R}}$ENTo+VISHNU~\cite{Bernhard:2019bmu}). All provide a reasonable description of anisotropic flow measurements. The corresponding shear relaxation time range is $\tau_{\pi}=$ 0.15--0.40~fm/$c$ at $T=0.5$~GeV, which implies that spatial anisotropies from the initial state are rather quickly developed into momentum anisotropies in the QGP phase. Two coupling limits are shown: an AdS/CFT limit which is calculated for infinite coupling, and a pQCD limit, which is determined in leading order for $\alpha_{\rm S}=0.3$. In the weak-coupling picture, $\eta/s$ is inversely related to the coupling strength -- therefore the $\alpha_{\rm S}$ chosen corresponds to the strongest possible coupling (therefore the lowest $\eta/s$) in that scheme. It is clear that the extracted $\eta/s$ values are closer to the infinite-coupling limit. Nonetheless, it should be pointed out that next-to-leading-order corrections for the pQCD determination of $\eta/s$ are very large~\cite{Ghiglieri:2018dib}. The values of $\eta/s$ from the QGP are roughly four times smaller than for Helium after it transitions to a superfluid~\cite{lemmon2010thermophysical}.

The constraints on $\zeta/s$ from RHIC and LHC measurements are shown in the bottom panel of Figure~\ref{fig:eta_zetas}. Predictions from an infinitely-coupled AdS Non-Conformal Holographic approach (AdS/NCH) are also shown for comparison. As the applicability of Conformal Symmetry regarding the strong potential assumed in the AdS/CFT scheme at temperatures close to the $T_{\rm pc}$ is expected to break down, an alternative approach is needed to determine $\zeta/s$. The breaking of conformal symmetry leads to $\zeta/s$ rising near $T_{\rm pc}$ (it is zero otherwise). This approach was also used to reevaluate the $\eta/s$ in the limit of infinite coupling at all temperatures and was found to also give $1/4\pi$, which suggests this limit is universal. Its prediction that $\zeta/s$ should depend strongly on the temperature in this region is utilized for the T$_{\rm{R}}$ENTo+VISHNU and IP-Glasma+MUSIC models.  The high-temperature pQCD limit for $\zeta/s$ is close to 0, which appears to apply for all the models shown at temperatures above 0.4 GeV. This then implies that bulk pressures in the initial state are washed out in the QGP phase even more quickly than the shear pressures e.g $\tau_{\pi} < 0.1$~fm/$c$ for IP-Glasma+MUSIC at $T=0.4$ GeV. 

Finally, the posterior distributions for $\eta/s$ and $\zeta/s$ have been evaluated using Bayesian parameter estimation techniques on RHIC and LHC data. They have been carried out by the Duke~\cite{Bernhard:2019bmu}, JETSCAPE~\cite{JETSCAPE:2020mzn}, Trajectum~\cite{Nijs:2020ors}, and Jyv\"{a}skyl\"{a}~\cite{Parkkila:2021tqq} groups. These are shown on the right of Figure~\ref{fig:eta_zetas}, at $T=0.3$~GeV for $\eta/s$ and $T=0.2$~GeV for $\zeta/s$. The size of these posterior ranges is influenced by the prior ranges and data sets included. For example, the JETSCAPE prior ranges were larger than those by the Duke and Trajectum groups and yielded a larger upper limit for $\eta/s$. The Jyv\"{a}skyl\"{a} group used an even larger prior range, but included measurements of Symmetric Cumulants. As discussed, these highly constrain the temperature dependence of $\eta/s$, and therefore reduce the upper limit. Generally, the $\zeta/s$ ranges differ more than the $\eta/s$ ranges for all the Bayesian analyses. The JETSCAPE group also found that the duration of the pre-equilibrium phase has a strong impact on the extracted viscosity transport parameters. This is clearly demonstrated by the v-USPhydro chain~\cite{Noronha-Hostler:2013gga}; the hydrodynamic evolution starts without any pre-equilibrium phase, and therefore requires low values of $\eta/s=0.05$ and $\zeta/s=0$ in order to develop enough flow to describe LHC data~\cite{Giacalone:2017dud}. Those values are also shown in Figure~\ref{fig:eta_zetas} in both panels.

\subsubsection{Mapping the QCD phase diagram}
\label{sec:progress:macroscopic:finite_muB}

\begin{figure}[h]
  \begin{center}
    \includegraphics[width=0.9\textwidth]{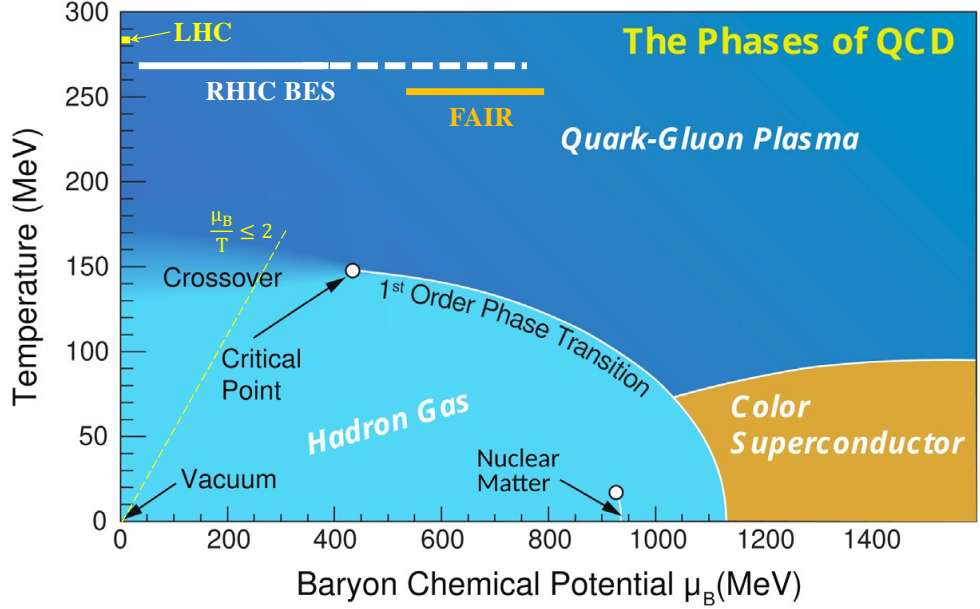}
  \end{center}
  \vspace{-1cm}
\linespread{1.0}\selectfont{} 
    \caption{Sketch of the QCD phase diagram, incorporating a conjectured critical endpoint and first-order transition regime. The yellow line indicates the smooth crossover region up to $\mu_B/T \leq 2$. The coverage of the LHC, the RHIC-BES including the fixed target program (dashed line), and the future CBM Experiment at FAIR, are indicated at the top of the figure.
    }
\label{fig:3.1.2:phasestructure} 
\end{figure}

At high temperatures and vanishing net-baryon density, the QGP-hadron transition is understood to be a smooth crossover based on lattice QCD calculations~\cite{Aoki:2006we}. The most recent result for the transition temperature is $T_C=158.0\pm0.6$ MeV \cite{Borsanyi:2020fev}, in agreement with the previously quoted $T_C = 156\pm1.5$ MeV \cite{HotQCD:2018pds}. Due to the fermionic sign problem, it is currently not possible to perform lattice simulations of QCD thermodynamics at finite baryon density. For this reason, the knowledge of the QCD phase diagram and equation of state from first principles is limited to small densities. The transition temperature as a function of baryon chemical potential $\mu_B$ can be parameterized as a Taylor series, of which we currently know the 0-th, 2-nd and 4-th order coefficients with high precision \cite{Borsanyi:2020fev,HotQCD:2018pds}.
Model studies indicate that a first-order phase boundary is expected at large net-baryon density ~\cite{Fukushima:2013rx,Bzdak:2019pkr}.
If this is the case, then there is a QCD critical point, %
separating the first-order phase boundary and the smooth crossover. %
The state-of-the-art lattice calculations on the QCD transition line predict that the chiral crossover region extends to a finite chemical potential $\mu_B/T \le 2$~\cite{Bazavov:2020bjn,Borsanyi:2020fev}, see Figure~\ref{fig:3.1.2:phasestructure}. Recent results on the Equation of State hint that this region might extend up to $\mu_B/T\sim$3--3.5 \cite{Bollweg:2022fqq,Borsanyi:2021sxv}. Due to the current lattice QCD limitations, experimental measurements are essential to answer the question on the existence of the QCD critical point.  

\begin{figure}[htb]
\centering
\includegraphics[width=0.9\columnwidth]{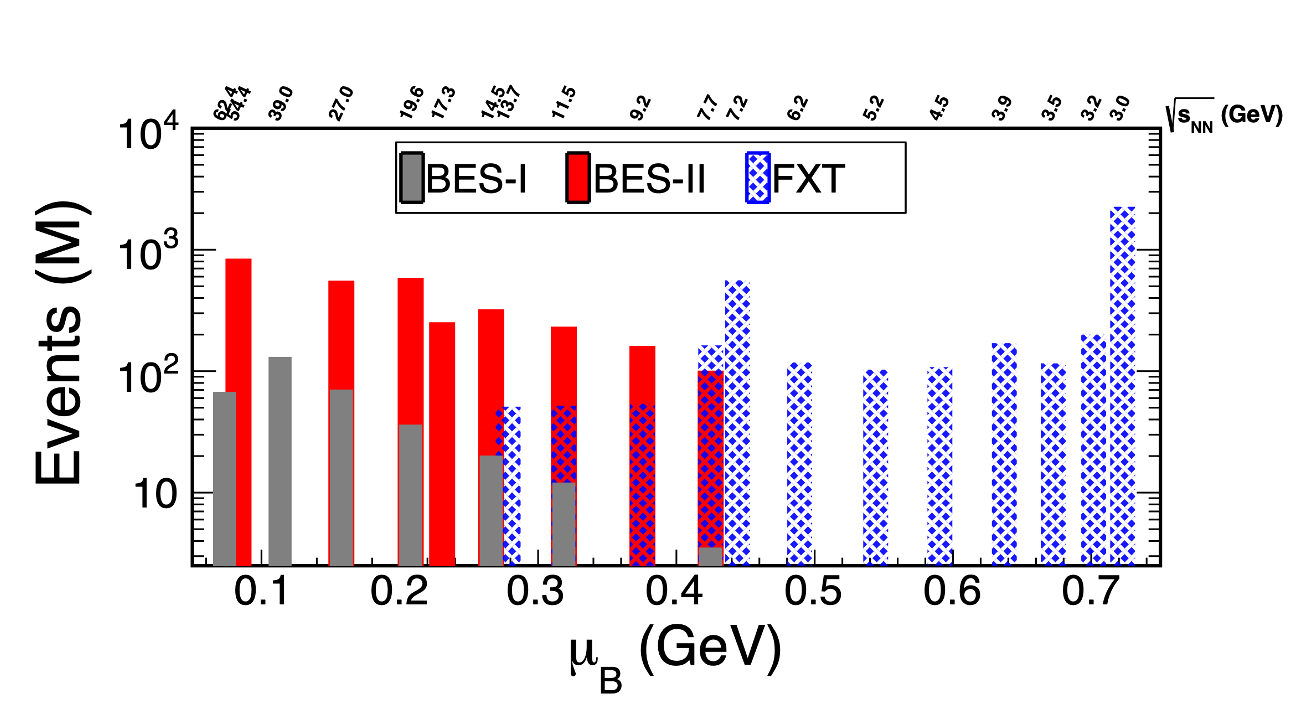}
\linespread{1.0}\selectfont{} 
\caption{Recorded BES-I collider (grey bars), BES-II collider (red bars) and fixed target (blue hatched bars) datasets at different collision energies and their corresponding $\mu_B$ values. }
\label{fig:3.1.2:dataset}
\end{figure}

The Beam Energy Scan (BES) program at RHIC, colliding heavy nuclei at the center-of-mass energy range $\sqrt{s_{NN}}$\,=\,3\,--\,200\,GeV, was initiated in 2008 in order to search for the QCD critical point and study the nuclear matter equation of state at high baryon density~\cite{STAR:BESII2014,US:NPLRP2015}. The BES phase-I (BES-I) program was conducted during 2010--2014 covering collision energies between 200 and 7.7 GeV (solid white line in Figure~\ref{fig:3.1.2:phasestructure}). The second phase of BES (BES-II) took place during 2019--2021 focusing on the center-of-mass energy range $\sqrt{s_{NN}}$\,=\,3\,--\,19.6\,GeV. While data in the energy range 19.6 -- 7.7 GeV were collected in the collider mode, the fixed target mode was also used to collect data over the energy range 3 -- 13.1 GeV (dashed white line in Figure~\ref{fig:3.1.2:phasestructure}). In the overlap energy range, the event statistics from BES-II are improved by a factor of 20 to 40 compared to that of BES-I. Figure~\ref{fig:3.1.2:dataset} summarizes the datasets collected at BES-I and BES-II including both collider and fixed target modes in different collision center-of-mass energies and their corresponding $\mu_B$ values.

Experimental measurements have been focusing on searching for the evidence for the dominance of the QGP phase and the hadronic phase at different collision energies. All of the BES-I data have been analyzed and most of the results are published. While the first set of BES-II data, particularly at lowest energy at RHIC (3 GeV), was recently published and started to shed new insights into the understanding of the QCD phase structure at high density region. These observations include:
(i) Jet Energy Loss: High momentum hadrons and fully reconstructed jets are unique probes for early-time QGP dynamics in heavy-ion collisions. The strong suppression in the leading hadron $R_{AA}$ at $p_T \ge$ 5 GeV/c, a signature of the formation of the QGP in central \AuAu{} collisions at $\sqrt{s_{NN}}$ = 200 GeV, was found to be gradually increased and to even become larger than unity in \AuAu{} central collisions when energy is lower than 19.6 GeV~\cite{STAR:2017ieb}.  (ii) Partonic Collectivity: Collective observables such as $v_1$ and $v_2$ are widely used for studying properties of the hot and dense medium created in high-energy nuclear collisions. The number of quark scaling found in the $v_2$ for all hadrons, the fingerprint of the QGP, has been found to persist down to the 7.7 GeV \AuAu{} collisions~\cite{STAR:2013cow}. This implies that the partonic activities remain to be dominant at these high-energy collisions. (iii) Critical Fluctuation: Higher order non-Gaussian cumulants of particle multiplicities and their ratios are expected to be sensitive to the existence of the critical point and phase boundary~\cite{Stephanov:2008qz,Athanasiou:2010kw}. The cumulants of net-proton protons (a proxy for net-baryons~\cite{Hatta:2003wn}) from top 5\% central 200 GeV \AuAu{} collisions, $C_6/C_2$, $C_5/C_1$, and $C_4/C_2$, are all found to be consistent with lattice QCD predictions of a smooth crossover chiral transition~\cite{STAR:2020tga,STAR:2021rls,Borsanyi:2018grb,Bazavov:2020bjn,Bellwied:2021nrt}. In \AuAu{} collisions at 3 GeV, on the other hand, hadronic interactions are evident from the measurements of proton distributions, collective flow, and strangeness productions~\cite{STAR:2021fge,STAR:2021yiu,STAR:2021hyx}. These results imply that the QCD critical point if it exists, can only be experimentally detected in heavy-ion collisions at energies $\sqrt{s_{NN}}$ between 3 and 20 GeV.

\begin{figure}[htb]
\centering
\includegraphics[width=0.85\columnwidth]{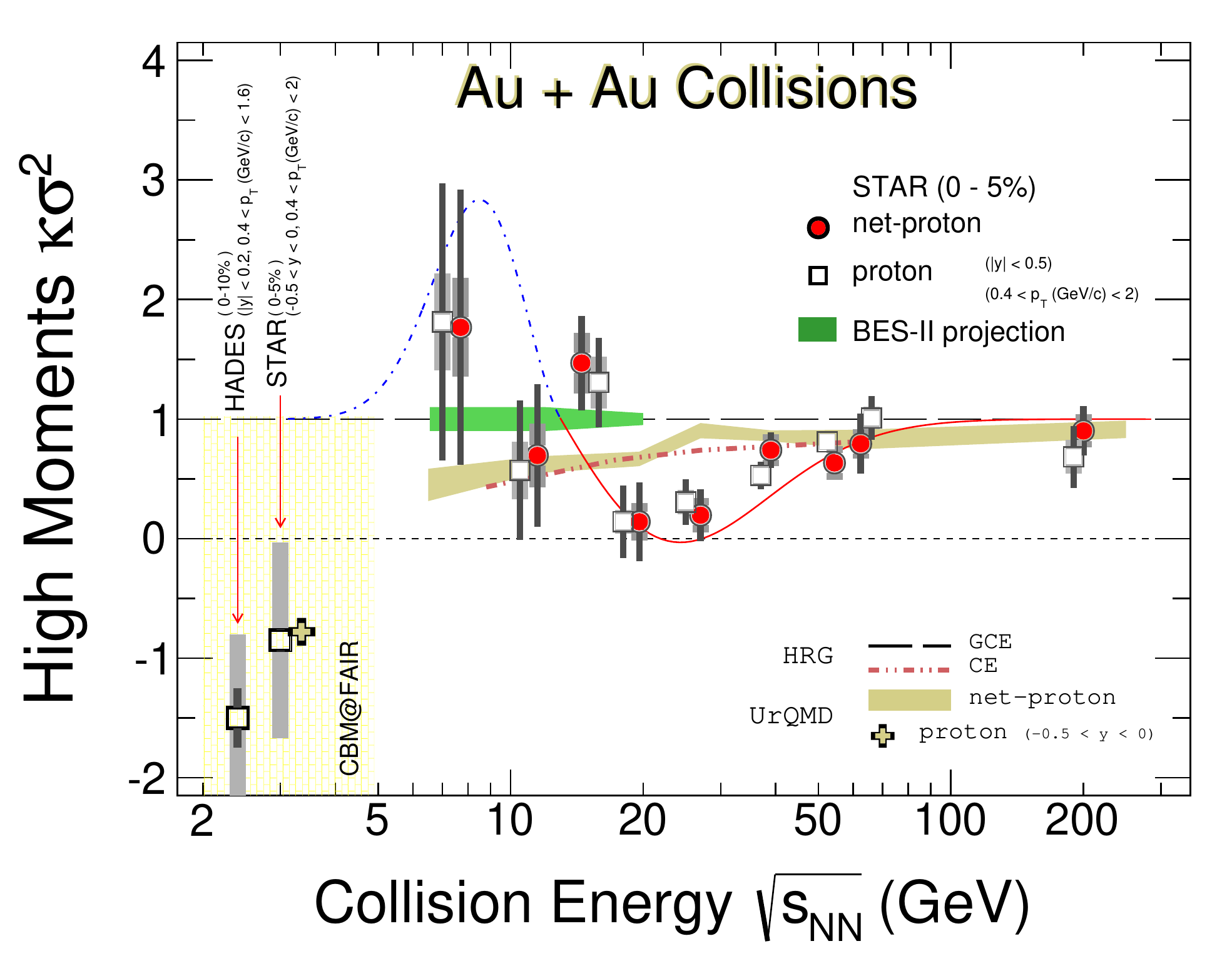}
\linespread{1.0}\selectfont{} 
\caption{Energy dependence of the net-proton (filled circles) and proton (open squares) high moments from central \AuAu{} collisions at RHIC BES~\cite{STAR:2021iop,Adamczewski-Musch:2020slf,STAR:2021fge}. For comparison, model results from the HRG model based on both Canonical Ensemble (CE) and Grand-Canonical Ensemble (GCE)~\cite{Braun-Munzinger:2020jbk}, and transport model UrQMD~\cite{Bleicher:1999xi,Bass:1998ca} are also shown. The energy range covered by the CBM experiment is shown as the yellow hatched area. The non-monotonic curve indicates the qualitative shape of the contribution of critical fluctuations \cite{Stephanov:2011pb}.}
\label{fig:3.1.2:net-p}
\end{figure}

Figure~\ref{fig:3.1.2:net-p} shows recent results on the fourth-order net-proton (filled red circles) and proton (open squares) moments in central \AuAu{} collisions from the RHIC BES~\cite{STAR:2020tga,STAR:2021fge,HADES:2020wpc}, compared to several model predictions. The thin red and blue dot-dashed lines are expected from a qualitative prediction 
due to critical behavior~\cite{Stephanov:2011pb}, even though recent quantitative results pointed out that the possibility of observing the dip depends on the distance of the freeze-out line from the critical point \cite{Mroczek:2020rpm}. The hadronic transport model UrQMD~\cite{Bass:1998ca,Bleicher:1999xi} (gold band)
and a thermal model with canonical ensemble~\cite{Braun-Munzinger:2020jbk} (red dot-dashed line) represents a non-critical dynamic baseline. The RHIC BES-I data is inconsistent with the monotonic predictions of non-critical models at the 3.1$\sigma$ level~\cite{STAR:2020tga,STAR:2021fge}. RHIC BES-II results will provide significantly improved statistical precision (and likely reduced systematic uncertainties) measurements, as indicated by the green band in the figure. The extended acceptance and particle identification to a larger rapidity region (from $|y|<$0.5 to $|y|<$0.8) will allow a more systematic investigation of the nature of these fluctuations. %

The equation of state of QCD from first principles is currently known up to $\mu_B/T=3$ from a resummed Taylor expansion \cite{Bollweg:2022fqq} and up to $\mu_B/T=3.5$ from an alternative expansion scheme which was recently proposed in Refs. \cite{Borsanyi:2021sxv,Borsanyi:2022qlh}. These results are obtained at either $\mu_Q=\mu_S=0$, or for $\mu_Q$ and $\mu_S$ as functions of $\mu_B$ and $T$ such that they satisfy the phenomenological conditions $\langle n_S\rangle=0$ and $\langle n_Q\rangle=0.4\langle n_B\rangle$. A full, four-dimensional Taylor-expanded equation of state in the $T,~\mu_B,~\mu_S,~\mu_Q$ parameter space is available in the range $\mu_B/T<2$ \cite{Noronha-Hostler:2019ayj,Monnai:2019hkn}. Extensions beyond these $\mu_B$ ranges are works in progress and currently limited due to the fermionic sign problem.

Although neither the strength nor $\mu_B$ value for the deconfinement and chiral phase transitions have been pinpointed at low temperatures, since the last long-range plan was written, significant progress has been made in understanding the equation of the state of dense matter. This advance came as a result of low-energy laboratory experiments and astrophysics observations, combined with a large theoretical effort from the nuclear physics and astrophysics communities. As a result, it was understood that dense matter is neither entirely soft nor stiff (as previously discussed) but presents nuances that can be translated into the structure in the {(effectively) zero-temperature} speed of sound \cite{Bedaque:2014sqa,Tews:2018kmu,Baym:2019iky,Kojo:2009ha}.

First, from nuclear physics, there are indications that the symmetry energy of matter around saturation density is not large \cite{SRIT:2021gcy,Fan:2014rha,Li:2019xxz} (although controversy still exists \cite{Reed:2021nqk,Reinhard:2021utv,Miyatsu:2022bhh}). This means that isospin-asymmetric matter, although stiffer than isospin-symmetric matter, is still relatively soft at low densities. This is corroborated by electromagnetic and gravitational-wave observations of intermediate-mass neutron stars, that point to small stars \cite{Miller:2019cac,Riley:2019yda,Miller:2021qha,Riley:2021pdl} with lower tidal deformability \cite{LIGOScientific:2018cki}. On the other hand, electromagnetic \cite{Fonseca:2021wxt,Romani:2022jhd} and possible gravitational-wave observations of massive stars \cite{LIGOScientific:2020zkf} point towards a stiff equation of state at large densities. Finally, at asymptotically large densities, perturbative QCD predicts that the conformal limit is recovered from below, pointing once more to a soft equation of state \cite{Annala:2019puf}. Perturbative QCD results can be extrapolated to lower densities, providing a limit for the stiffness of the equation of state all the way down to a few times saturation density \cite{Komoltsev:2021jzg}.

\begin{figure}[t]
\centering
\includegraphics[width=0.7\columnwidth]{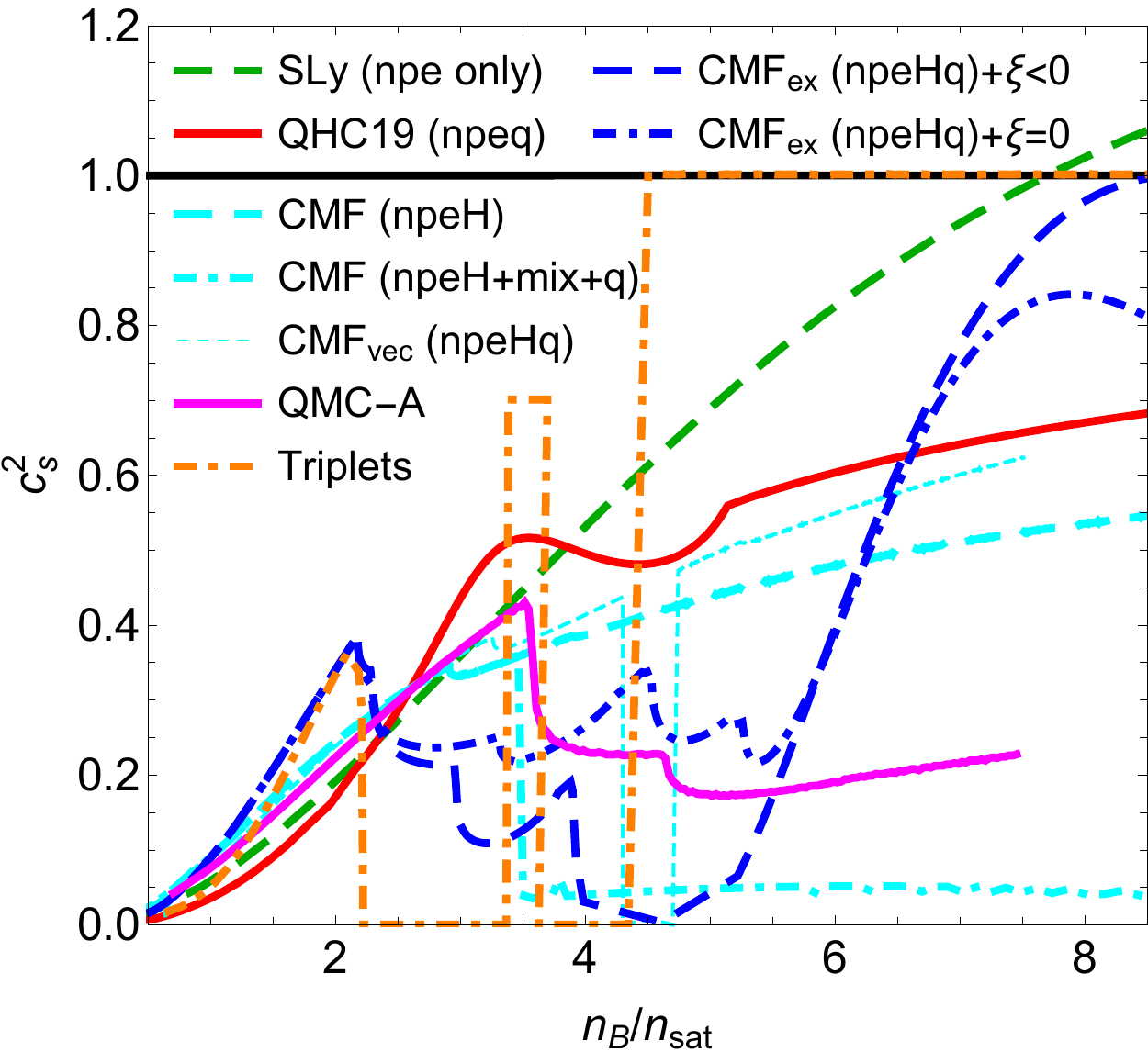}
\caption{Comparison of the speed of sound
squared calculated from state-of-the-art dense-matter models including different compositions and reproducing different-order phase transitions. Figure from Ref.~\cite{Tan:2021ahl}}
\label{cs2}
\end{figure}

Together, these results show that either the degrees of freedom \cite{Stone:2019blq,Annala:2019puf,Motornenko:2019arp,Baym:2019iky,McLerran:2018hbz,Aloy:2018jov,Xie:2020rwg,Han:2022rug}, symmetries \cite{Hippert:2021gfs,Fujimoto:2022ohj}, or interactions \cite{Dexheimer:2018dhb,Dexheimer:2020rlp,Lopes:2020dvs,Pisarski:2021aoz,Most:2021ktk,Pfaff:2021kse} are changing within the relevant portion of the QCD phase diagram, implying a phase transition of some order at zero temperature, as shown in Figure~\ref{cs2}. Besides the purely nucleonic Sly equation of state, all the others shown in the figure present a phase transition, either of the first order, with the speed of sound going all the way to zero, or higher order, including a mixed phase with very low speed of sound. These state-of-the-art dense-matter models present different compositions (including hyperons, negative parity states, and quarks) and symmetries (including chiral symmetry and different pairing schemes).

At finite temperature, meaning tens of MeV's in order to modify significantly the properties of hadrons, dense matter beyond a couple of times saturation density has not been yet experimentally probed. This is because such conditions can only be found in supernova explosions (and within the first minute of life of proto-neutron stars) and in neutron star-mergers. So far, although much theoretical work has been laid \cite{Sagert:2008ka,Ouyed:2009dr,Nakazato:2013iba,Fischer:2017lag,Kuroda:2021eiv,Jakobus:2022ucs,Most:2018eaw,Weih:2019xvw,Blacker:2020nlq}, no neutrinos have been detected from supernovae in our neighborhood, and the post-merger part of gravitational-wave signals from neutron star mergers has not yet been detected by LIGO and Virgo detectors.

\subsubsection{Thermal photons/dileptons}

\label{sec:progress:macroscopic:thermal_em_probes}

Heavy-ion collision experiments conducted at the Relativistic Heavy Ion Collider (RHIC) and the Large Hadron Collider (LHC) create an environment at a temperature of a trillion degrees, to study the property of dense nuclear matter. 
Electromagnetic (EM) probes, such as direct photons and dileptons ($e^+ e^-$ and $\mu^+ \mu^-$ pairs) are recognized as valuable messengers in relativistic nuclear collisions~\cite{David:2019wpt, Geurts:2022xmk}.
Because photons and dileptons interact only electromagnetically, they can penetrate the medium, carry almost undistorted dynamical information, and report on conditions existing at their production point.
Such probes are sensitive to the early stages of the collision system, to thermal and transport properties of the quark-gluon plasma (QGP), and the dynamical evolution proceeding from the cross-over regions to the hadronic phase~\cite{Strickland:1994rf,Schenke:2006yp,Martinez:2008di,Shen:2015nto, Shen:2016odt, Bhattacharya:2015ada,Ryblewski:2015hea,Paquet:2017wji,Kasmaei:2018oag,Kasmaei:2019ofu,Coquet:2021lca,Coquet:2021gms,Vujanovic:2022itq,Monnai:2022hfs}. In addition, dilepton invariant mass spectrum is the only observable that can measure the in-medium spectral function of hadrons, and its relation to chiral symmetry restoration \cite{Rapp:2009yu, Hohler:2013eba}.

\medskip

\noindent \textit{Electromagnetic radiation in and out-of-equilibrium}

Electromagnetic emission rates from the QGP are directly related to the electromagnetic current-current correlation function \cite{Bellac:2011kqa, Kapusta:2006pm},
\begin{equation}
    d\Gamma_k = \frac{d^3 k}{(2\pi)^3 2 k} \sum_{a = 1, 2} \epsilon_{(a)}^\mu (k) \epsilon_{(a)}^\nu (k) \int d^4 x e^{-i k \cdot x} \langle J_{em, \mu}(0) J_{em, \nu}(x) \rangle,
    \label{eq:EMcorr}
\end{equation}
where $\epsilon_{(a)}^\mu(k)$ is the polarization vector. The current-current correlation function in Eq.~\eqref{eq:EMcorr} encodes the response of QGP to electromagnetic fields. It is directly related to quarks, the electric charge carriers in the QGP, and indirectly related to gluons which interact with quarks during photon emission processes.
In thermal equilibrium, the EM current-current correlator can be related to the vector spectral function and the photon emission rate can be written as
\begin{equation}
    k \frac{d\Gamma_k}{d^3 k} \propto \frac{\alpha_{\rm EM}}{\pi^2} n_B(k) \rho_v(k, \vec{k}). 
\end{equation}
With the same vector spectral function, the equilibrated dilepton production rates are
\begin{equation}
    \frac{d \Gamma_{l^+ l^-}(\omega, \vec{k})}{d \omega d^3 \vec{k}} \propto \frac{\alpha_{\rm EM}^2}{\pi^3 M^2} n_B(\omega) \rho_v(\omega, \vec{k}),
\end{equation}
with $M^2 \equiv \omega^2 - k^2$.
This EM vector spectral function $\rho_v(\omega, \vec{k})$ in the QGP phase has been calculated using both perturbative QCD (pQCD) and non-perturbative lattice methods. Theoretical progress on the pQCD calculations has yielded full next-to-leading (NLO) EM spectral functions for both real $(M = 0)$ photons \cite{Ghiglieri:2013gia} and virtual photons, which subsequently decay into dileptons \cite{Laine:2013vma, Ghisoiu:2014mha, Ghiglieri:2014kma}. The NLO rates allow for comparisons and connections with non-perturbative approaches \cite{Ghiglieri:2016tvj, Jackson:2019yao}. Ref.~\cite{Ghiglieri:2016tvj} showed the remarkable agreement between NLO pQCD calculations of EM spectral functions and lattice QCD results~\cite{Brandt:2019shg, Ce:2022fot}. These advancements in computing the QGP EM spectral function provide carrying out reliable phenomenological studies with the RHIC and LHC experiments.

In the hot hadronic phase, the in-medium modifications of the vector spectral function in the hadronic phase, which has a direct connection to chiral symmetry restoration during the phase transition, have been studied extensively with hadronic many-body theories \cite{Rapp:1999us, Hohler:2015iba}.

The dynamically evolving QGP does not stay in thermal equilibrium, and the local momentum anisotropy of the plasma generates viscous corrections to photon emission rates. These out-of-equilibrium corrections to the EM emission rates are important for phenomenological studies. 
Calculations at the leading order of strong coupling $\alpha_s$ were carried out for two-to-two scattering channels~\cite{Shen:2014nfa, Bhattacharya:2015ada, Hauksson:2016nnm, Czajka:2017wdo, Kasmaei:2018yrr, Kasmaei:2019ofu} and inelastic channels with bremsstrahlung and pair annihilation processes with Landau-Pomeranchuk-Migdal (LPM) effects~\cite{Hauksson:2017udm,  Hauksson:2020wsm}. The viscous corrections to the dilepton emission rates were computed~\cite{Vujanovic:2015nwv, Bandyopadhyay:2015wua, Vujanovic:2016anq, Vujanovic:2017psb, Vujanovic:2019yih}.

\medskip
\noindent \textit{Thermal EM probes -- the QGP multi-messenger}

Because the momentum distributions of photons and dileptons can carry dynamic information about their production points, they are excellent diagnostic probes to infer the dynamical evolution of QGP in relativistic nuclear collisions.

\medskip
\noindent \textbullet \textit{Thermometer and Chronometer}

Real and virtual photons are useful tools for experimentally accessing the temperature of the QGP created in heavy-ion collisions~\cite{Shuryak:1978ij,Hwa:1985xg}. The yields of photons are directly proportional to the space-time volume and the average temperature of the hot QGP matter \cite{Shen:2016mmv, Gale:2019abf}. The dilepton yields in the low invariant-mass region showed a strong correlation with the fireball lifetime \cite{Rapp:2011is, Rapp:2014hha}.

\begin{figure}[t]
\centering
\includegraphics[width=0.44\columnwidth]{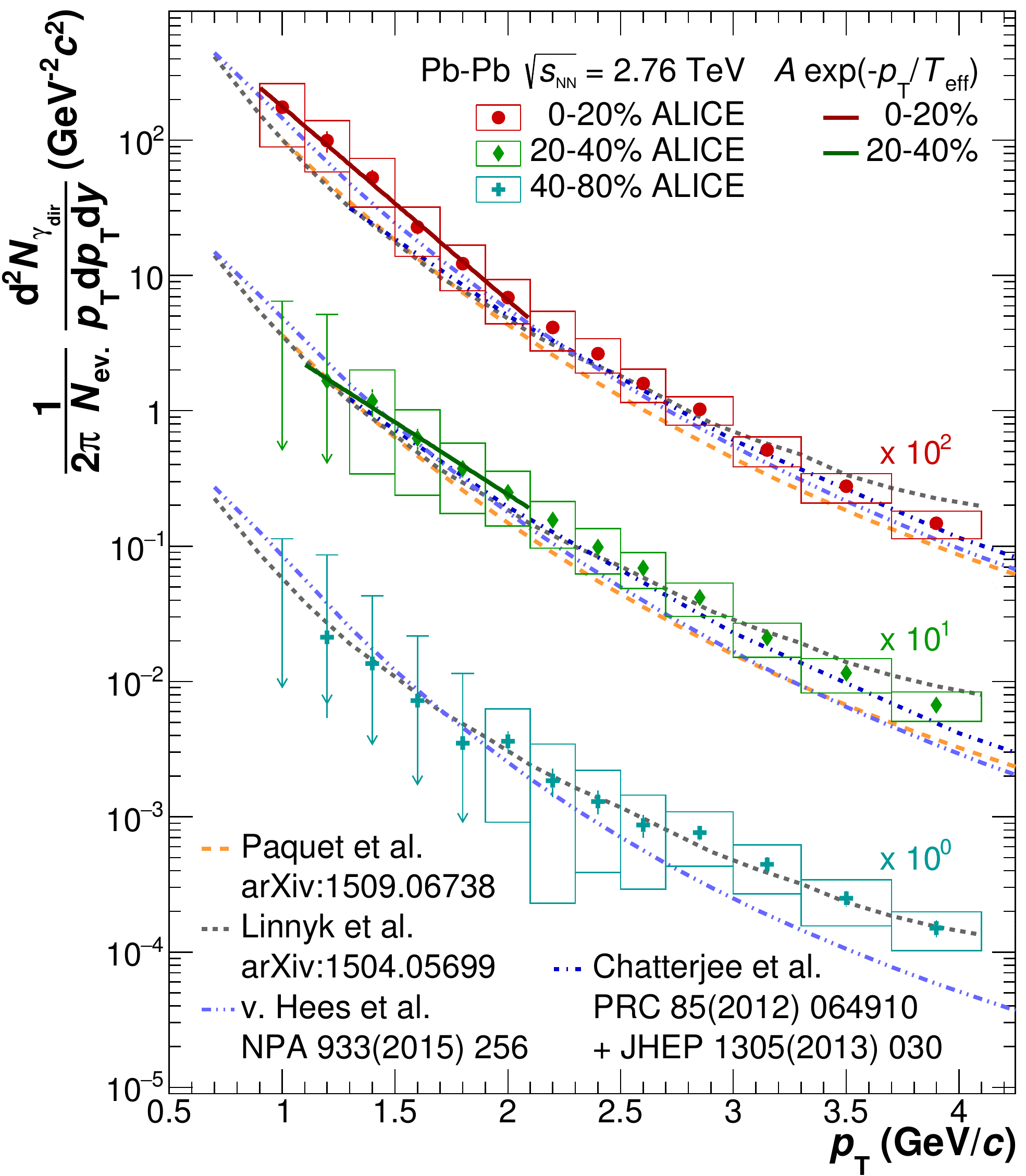}
\includegraphics[width=0.48\columnwidth]{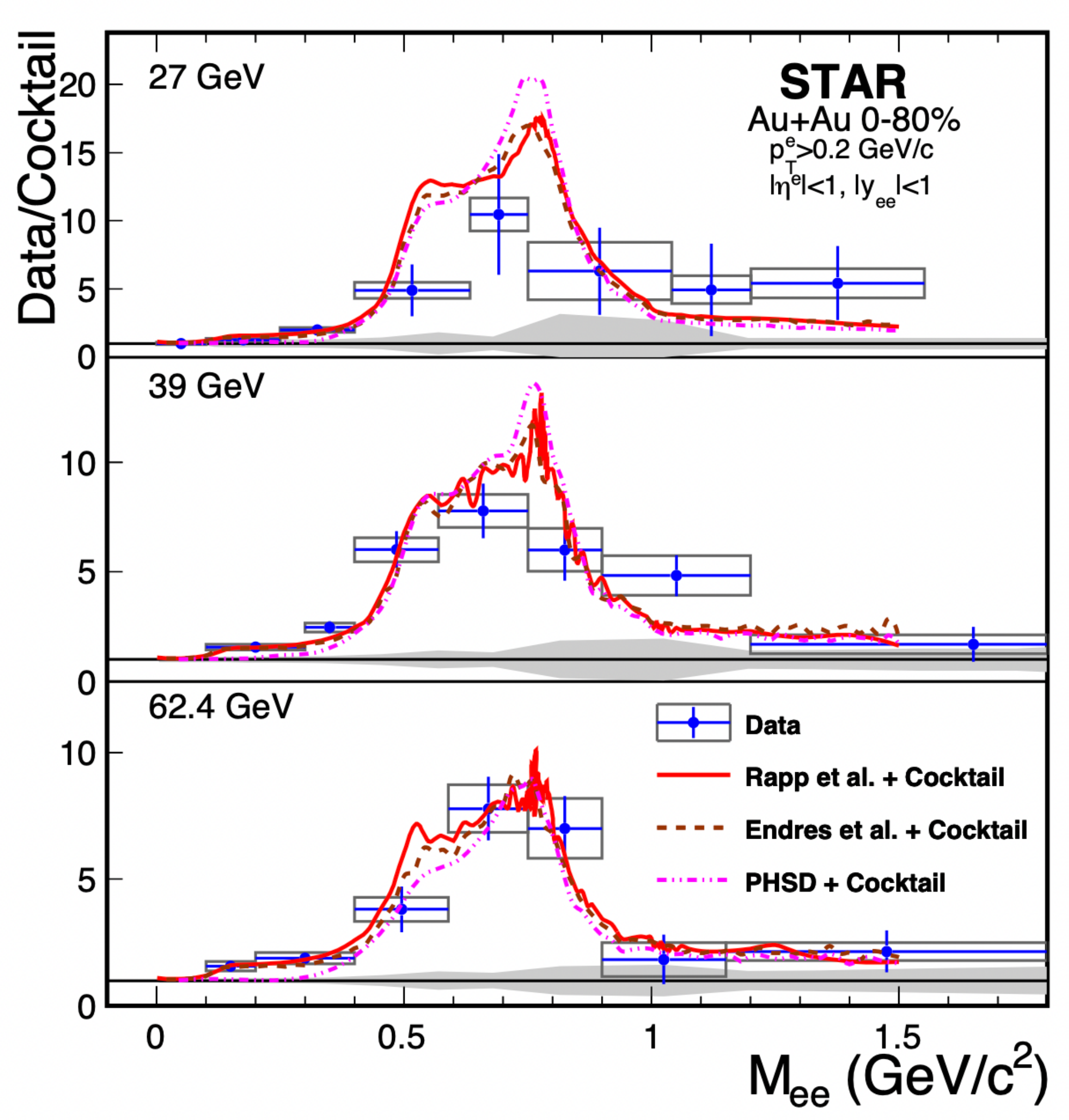}
\caption{Left panel: The direct photon spectra from the ALICE experiment in \PbPb{} collisions in three different centrality classes compared with three different model calculations. Figure from Ref.~\cite{ALICE:2015xmh}. Right panel: STAR result on the dilepton invariant mass divided by the expected background from the cocktail simulation at three different collision energies. The cocktail contributions from $\omega$ and $\phi$ mesons have actually been removed. Figure from Ref.~\cite{STAR:2018xaj}.}
\label{fig:thermometer}
\end{figure}

The slopes of the photon and dilepton spectra encode temperature information of the collision system.
Direct photon $p_T$-spectra have been measured by the PHENIX, STAR, and ALICE Collaborations in heavy-ion collisions at the top RHIC and LHC energies\,\cite{PHENIX:2014nkk, ALICE:2015xmh, STAR:2016use}. The low $p_T$ part of the spectra can be well characterized by their inverse logarithmic slope $T_\mathrm{eff}$. The PHENIX Collaboration reported $T_\mathrm{eff} = (239 \pm 25^\mathrm{stat} \pm 7^\mathrm{sys})$\,MeV for 0-20\% \AuAu{} collisions at $\sqrt{s_\mathrm{NN}} = 200$\,GeV \cite{PHENIX:2014nkk} and the ALICE Collaboration found $T_\mathrm{eff} = (304 \pm 11^\mathrm{stat} \pm 40^\mathrm{sys})$\,MeV in 0-20\% \PbPb{} collisions at $\sqrt{s_\mathrm{NN}} = 2.76$\,TeV \cite{ALICE:2015xmh}. The ALICE measurement of direct photon production at $\sqrt{s_\mathrm{NN}} = 2.76$\,TeV constraint to the $p_{\rm T}< 4~{\rm GeV}/c$ region is shown on the left panel in Figure~\ref{fig:thermometer} with comparison to the current theoretical models. Quantitative studies \cite{vanHees:2011vb, Shen:2013vja} have shown that thermal photons emitted from $T > 300$\,MeV during early stages of the evolution were reflecting their local temperatures while those from $T < 250$\,MeV received significant blue-shift from the hydrodynamic flow. Therefore, the measured effective temperatures set strong constraints on hydrodynamic evolution. 
A solid extraction of the initial temperature of the collision system requires detailed comparisons between experimental data and dynamical model simulations. 

On the other hand, thermal dilepton invariant-mass spectra are free from blue-shift flow contamination. The authors in Ref.\,\cite{Rapp:2014hha} demonstrated that the slope of the dilepton spectrum in the intermediate mass region (IMR), $1.5$\,GeV\,$< M < 2.5$\,GeV, could provide temperature information about the collision system. However, Ref.~\cite{Ryblewski:2015hea} shows that the local momentum anisotropy of the quarks can lead to small but non-negligible corrections to the slope of the dilepton spectrum. 
In addition, Refs.\,\cite{Rapp:2014hha,Rapp:2013ema,Bratkovskaya:2011wp,Vujanovic:2015gba} showed that dilepton invariant-mass spectra were valuable tools to probe the properties of the baryon-rich fireball in the RHIC Beam Energy Scan (BES) program. The right panel in Figure~\ref{fig:thermometer} shows the recent STAR measurement of the low mass region enhancement when compared to the cocktail sum of different decay production in three different collision energies. The comparison to the model calculations are all consistent with the scenario of in-medium broadening of the $\rho$ meson, together with previous measurements providing an order of magnitude range in collision energy. Recently, phenomenological studies of dilepton invariant-mass spectra have been carried out in heavy-ion collision at HADES energies~\cite{Endres:2015fna, Galatyuk:2015pkq, Staudenmaier:2017vtq}. In such a baryon-rich environment, the interactions with the excess of baryons play a crucial role in EM emissions. The average temperature information has been extracted by HADES Collaboration~\cite{HADES:2019auv}.

\medskip
\noindent \textbullet \textit{Constraining out-of-equilibrium QGP dynamics}

The large expansion rates in relativistic heavy-ion collisions drive the system away from thermal equilibrium during its evolution. Out-of-equilibrium dynamics leave their fingerprints on the electromagnetic observables. 

Effects on thermal photon emission owing to locally anisotropic particle distributions were investigated in Ref.\,\cite{Schenke:2006yp,Dusling:2009bc,Dion:2011pp,Shen:2014nfa,Shen:2014lye}. The related phenomenological impacts were studied in Ref.\,\cite{Dion:2011pp,Shen:2013cca,Shen:2014cga,Bhattacharya:2015ada,Shen:2016zpp, Kasmaei:2018oag, Kasmaei:2019ofu}. Effects from bulk viscosity were recently studied  Ref.\,\cite{Ryu:2015vwa,Paquet:2015lta,Hauksson:2016nnm,Gale:2021emg}. The inclusion of a non-vanishing bulk viscosity near the phase transition generates extra entropy production and increases the space-time volume in the late hadronic phase by about 50\%, which allows more thermal photon radiation \cite{Paquet:2015lta}. Another consequence of including bulk viscosity is reducing the hydrodynamical radial flow at the late stage of the evolution, which weakens the blue shift of the thermal photon spectrum compared to the simulations without bulk viscosity. Both effects together increase the thermal photon yields in the low $p_T$ regions and shift the peak of the direct photon $v_2(p_T)$ towards the low $p_T$ regions\,\cite{Paquet:2015lta}.

Dileptons have also been shown to be a clean and sensitive probe of the out-of-equilibrium dynamics of the system~\cite{Ryblewski:2015hea, Vujanovic:2014vwa, Coquet:2021gms}. Recent studies have demonstrated that, compared to hadronic observables, the thermal dilepton spectrum and its flow anisotropy show a larger sensitivity to the early time dynamics, to the system's shear stress tensor $\pi^{\mu\nu}$ and bulk viscous pressure $\Pi$, to the temperature dependence of shear viscosity $\eta/s(T)$, and even to the choice of the second order transport coefficient $\tau_\pi$ \cite{Vujanovic:2016anq, Vujanovic:2017psb, Vujanovic:2019yih}.

\medskip
\noindent \textbullet \textit{Diagnosing early-stage chemical equilibrating QGP}

The hot QGP emits photons and dileptons during its entire evolution. The early stage of heavy-ion collisions is overpopulated by gluons. The chemical evolution of QGP leaves its signatures at high $p_T$ photon spectra and $M > 1 $\,GeV in dilepton invariant-mass spectra. Because the EM emission rates are directly related to the quarks in QGP, the yields of high $p_T$ photons and large invariant-mass dileptons carry information about chemical equilibration dynamics~\cite{Gale:2021emg, Coquet:2021lca}. 

\begin{figure}[t]
\centering
\includegraphics[width=0.44\columnwidth]{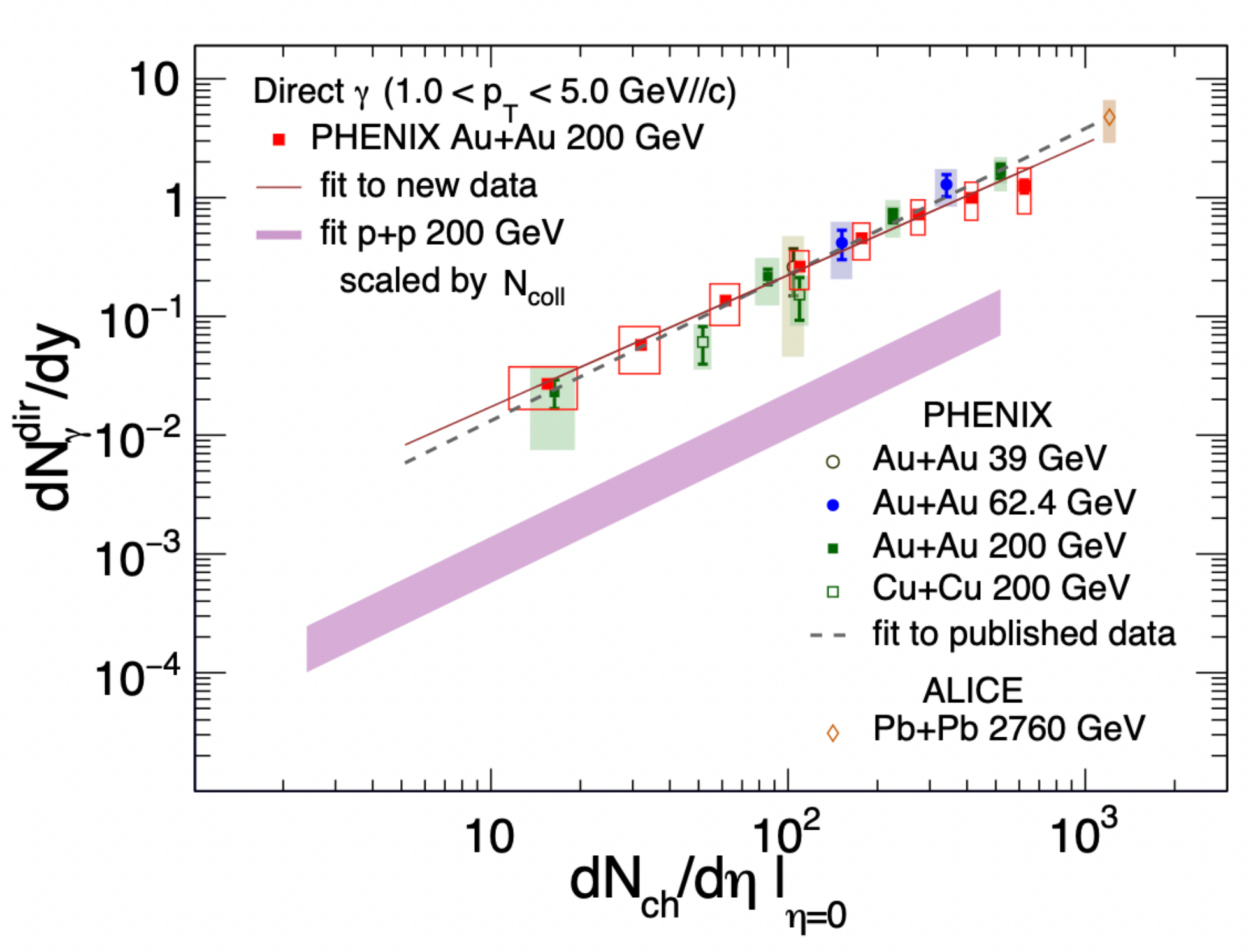}
\caption{The direct photon yield compilation from the large collisions systems from PHENIX and ALICE as a function of the event multiplicity, figure from~\cite{PHENIX:2022rsx}.}
\label{fig:dirphotonscaling}
\end{figure}

One interesting way of studying the direct photon production is discussed in ~\cite{PHENIX:2018for, PHENIX:2022rsx}, where the charged hadron multiplicity of the system is proportional to the temperature and size of the system. At the same energy, the centrality closely correlates with the size, while at the same centrality, the collision energy would correlate with the temperature and lifetime of the system. Figure~\ref{fig:dirphotonscaling} shows the recent results from PHENIX, where the important point is that A+A data from different centralities and a wide range of collision energies can be empirically described in terms of $dN_{ch}/d\eta$ with just two parameters, suggesting some fundamental commonality in the underlying physics. There seems to be no qualitative change in the photon sources and their relative contributions for different collision centrality or collision energy.
This scaling behavior shows that the direct photons emit from the (3+1)D space-time volume of the collisions while hadrons only emit from the final freeze-out surface.
The measurement includes all non-prompt photons which would include all electromagnetic radiation from non-perturbative QCD processes.

\medskip
\noindent \textbullet \textit{Thermal emission signatures from small QGP droplets}

Unraveling the collective origin of small collision systems has been a hot topic in our field. If there were a hot and nearly thermal-equilibrated QGP medium produced in these collisions, the QGP droplet would radiate thermal photons and dileptons and enhance the measured spectra for these EM probes~\cite {Shen:2015qba}. Systematic calculations have been carried out as a function of collision system size to study the hadronic observables and direct photon production as the same time~\cite{Shen:2016egw, Shen:2016zpp, Iatrakis:2016ugz, Shen:2016mmv, Gale:2021emg}. The theory predicted thermal enhancement of direct photon at low $p_T$ was consistent with the PHENIX measurements.

\medskip
\noindent \textbullet \textit{Photon emission in the dilute hadronic phase}

The photon emissions from the hadronic bremsstrahlung processes in the dilute phase were found to be important to explain the large excess of direct photon spectra measured by the PHENIX Collaboration at the RHIC~\cite{Linnyk:2015tha}. Such out-of-equilibrium photon emission in the dilute hadronic phase was recently studied via a systematic comparison between photon emission from the hydrodynamic medium at low temperature and microscopic production from hadronic transport model SMASH~\cite{Schafer:2021slz}. One finds consistent results from photon $p_T$ spectra from the two approaches. The photon elliptic flow is larger from the transport model than that from the near-equilibrium hydrodynamic description.
A more comprehensive description requires the dynamical models to include additional photon emission processes from baryonic channels \cite{Holt:2020mwf}. These processes are expected to play an important role in heavy-ion collisions in a baryon-rich environment at Beam Energy Scan energies.

\medskip
\noindent \textbullet \textit{Probing the nature of hadron-QGP phase transition at high baryon density}

Unlike the yields of hadronic particles, photons and dileptons production are correlated to the total space-time volume of heavy-ion collisions. 
Recent studies show that dilepton production can serve as a signature for the first-order phase transition~
\cite{Seck:2020qbx, Savchuk:2022aev}. The fireball space-time volume will increase if the system goes through a first-order phase transition compared to that of a smooth cross-over situation. A factor of 2 enhancement in dilepton yields was predicted from model calculations for \AuAu{} collisions at the $E_{\rm lab} = 1 - 10$\,GeV.

\medskip

\subsubsection{Initial condition}

\newcommand{\lr}[1]{\left\langle #1\right\rangle}
\newcommand{\lrp}[1]{\left( #1\right)}

\label{sec:progress:macroscopic:initial_cond}

\paragraph{Challenges towards understanding the heavy-ion initial condition}
The study of the initial stages of heavy-ion collisions is of paramount interest to the community of high-energy nuclear physics. This interest ranges from a better understanding of the wave functions of the colliding nuclei to the description of the early pre-equilibrium stage that is formed immediately after the collisions. The initial condition is followed by the QGP that is describable by effective kinetic theory or fluid dynamics. Therefore, any limitations in our knowledge of the initial condition contribute to uncertainties in the fluid-dynamic description of the matter formed in heavy-ion collisions and the interpretation of final state observables~\cite{Schenke:2020mbo}.

There are two dominant sources of uncertainties related to the initial condition. First, we have limited knowledge of the wave functions of the colliding ions in terms of 1) the nuclear structure, deformation parameters, and neutron skin and 2) the spatial, momentum, and collision energy dependence of the parton distributions of the colliding ions. Some of these quantities can be constrained through independent measurements. For example, constraints on nuclear structure come from measurements performed in low-energy nuclear physics experiments. The distribution of the partons inside a colliding proton or ion as a function of Bjorken-x and momentum transfer ($Q^2$) can be constrained using electron-proton/ion scattering at HERA~\cite{Rezaeian:2012ji}, recently using ultra-peripheral collisions and, at the future EIC (see Section~\ref{sec:progress:microscopic:ultraperipheral_collisions}). However, uncertainties arise due to limited available data on nuclear parton distribution and we lack control over the kinematics of the heavy-ion initial state. We normally resort to theoretical frameworks such as QCD evolution equations such as BFKL, DGLAP, or JIMWLK that can extrapolate the parton distributions to the desired regions of the $x-Q^2$ landscape~\cite{Balitsky:1995ub,Jalilian-Marian:1996mkd,Iancu:2000hn,Ferreiro:2001qy,Kovchegov:1999ua}. 

The second source of uncertainty of the initial condition is associated with incorporating sources of initial-state fluctuations at different lengths scales and the description of the pre-equilibrium phase of matter formed immediately after the collisions. For theoretical modeling of heavy-ion collisions, initial-state models are expected to predict the initial full energy-momentum tensor that is input to fluid-dynamic simulations. No direct measurement can provide insights about this early phase of collisions as final-state effects always complicate the interpretation. Modeling of small collision systems have indicated that incorporating sub-nucleonic hot spots constrained by the HERA data~\cite{Mantysaari:2016ykx} as well as pre-equilibrium flow play an important role in the description of the experimental data~\cite{Weller:2017tsr,Mantysaari:2017cni}. However, discerning the relative contribution of the sub-nucleonic hot spots and flow in pre-hydrodynamic phase is experimentally challenging. 

Over the years, attempts have been made to address these issues and to come up with the best possible solution description of the heavy-ion initial state. The commonly-used MC-Glauber model distributes the initial nucleons in coordinate space and uses various choices of energy or entropy deposition schemes to provide the necessary initial conditions. One example of such a scheme is the wounded nucleon model that involves free parameters that can be constrained by data-model comparison or Bayesian analysis with global data. In turns MC-Glauber description already provides the necessary lumpiness of the initial state to describe the most commonly used experimental measurements of flow harmonics ($v_n$). If the kinematics of collisions are such that the colliding ions can be described as saturated sheets of gluonic matter one can use more sophisticated models of the initial state such as IP-Glasma. With inputs from HERA data on the gluonic profile of the proton, IP-Glasma can construct initial color-charge distribution inside the colliding nuclei. It then simulates the collisions of the color fields due to such color charges and also the evolution of the pre-equilibrium matter after the collisions. The spatial inhomogeneities at various length scales (down to inverse of nuclear saturation scale $1/Q_S^2(x)$, the intrinsic momentum space correlations of the gluons), and the flow of the pre-equilibrium matter are dynamically generated in this framework. The output of IP-Glasma, such as the full stress-energy tensor corresponding to a given value of the number density of gluons can be coupled directly to a fluid-dynamic simulation~\cite{Schenke:2020mbo}. The IP-Glasma model has been quite successful in describing bulk observables over a wide range of collision systems~\cite{Gale:2012rq}. Such success of the IP-Glasma model can be attributed to its predictive power of the right correlation between transverse initial-state spatial anisotropy and gluon multiplicity. The IP-Glasma model combined with the RHIC data on deformed \UU{} collisions provided the stepping stones for the development of the TRENTO model which parametrizes the initial energy/entropy distribution in terms of the nuclear thickness function~\cite{Moreland:2014oya}. Due to its intrinsic simplicity and flexibility, TRENTO is one of the most widely used initial-state models and a common tool for Bayesian analyses. 

\paragraph{Understanding the three-dimensional structure of the initial state}

\begin{figure}
    \centering
    \includegraphics[width=0.8\textwidth]{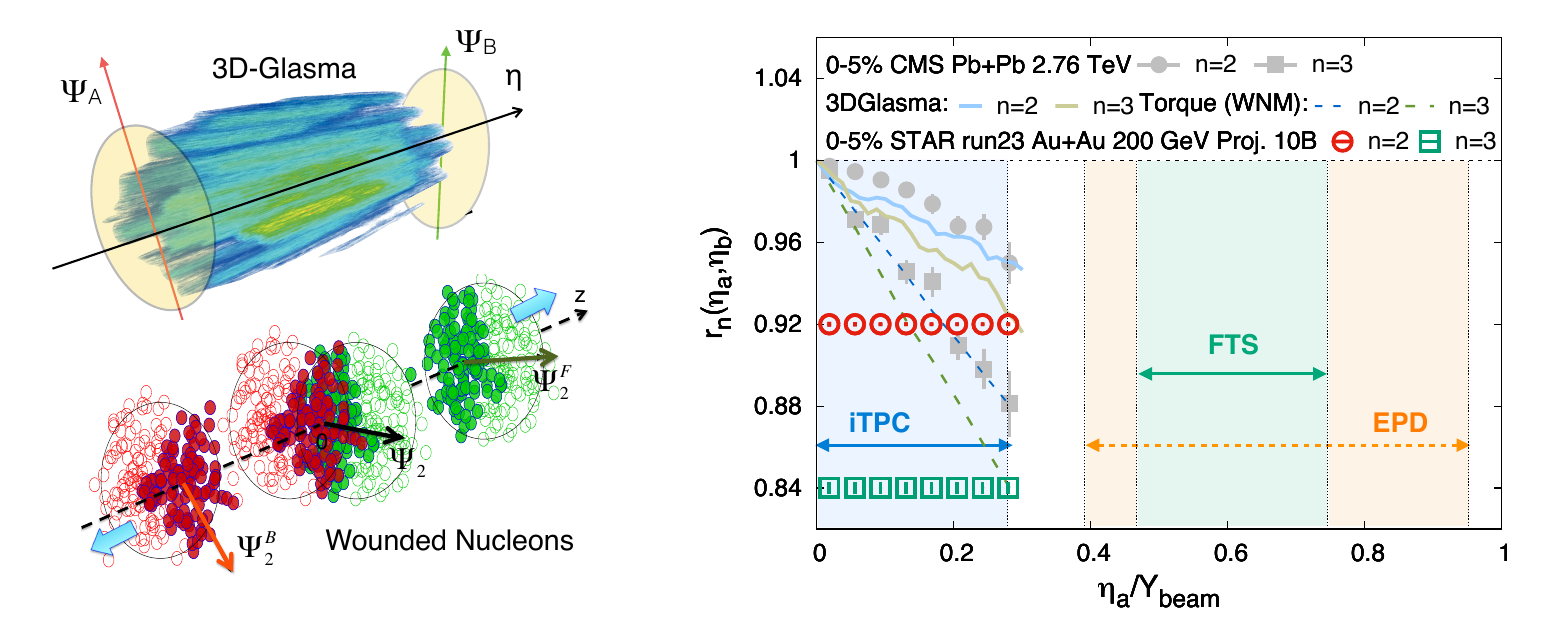}
    \caption{(Left) A cartoon depicting the de-correlation of event planes in the longitudinal direction of collision, based on the gluon saturation model 3D-Glasma~\protect\cite{Schenke:2016ksl} and the wounded nucleon model~\protect\cite{Li:2017qvf,Bozek:2010vz}. Both models predict the breaking of boost invariance, but at different scales. (Right) The longitudinal de-correlation of the elliptic anisotropy plane as a function of pseudorapidity, in units of beam rapidity. The colored regions show that current and future capabilities at STAR (with iTPC+EPD+FST) can extend such measurements with good precision by covering up to $60\%$ of the beam rapidity at 200 GeV. As $Y_{\rm beam}$ decreases at lower (BES-II) energies, a wider longitudinal extent can be measured.
    }
    \vspace{-0.25cm}
    \label{ep_decorrelation}
\end{figure}

The success of the IP-Glasma or TRENTO is associated with the right description of the transverse profile of the initial energy density distributions. Since the last long-range plan, the development of the initial state has been focused on a deeper understanding of the three-dimensional structure of the initial state that remained largely unexplored. It turns out the measurements of pseudorapidity decorrelation of event planes and flow harmonics are ideal probes to constrain the three-dimensional initial state of heavy-ion collisions. The cartoon in Figure~\ref{ep_decorrelation} (left) demonstrates how the initial-state longitudinal fluctuations and fluid dynamical response of the medium formed in HICs can lead to decorrelations of the harmonic anisotropies planes $\Psi_n$ at different pseudorapidity. Such effects are often referred to as torque or twist of the event shape~~\cite{Bozek:2010vz,Jia:2014ysa,Pang:2015zrq} that eventually leads to a breaking of longitudinal/boost/rapidity invariance. Depending on whether the initial state is determined by gluon saturation or the wounded nucleon model the effect of such de-correlation will be different. 
If the initial state is described by gluon saturation, as simulated by the 3D-Glasma model, the breaking of boost invariance is determined by the QCD equations (BK, JIMWLK) which predict the evolution of gluons in the saturation regime. The typical rapidity scale is over which boost invariance is broken depending on the strong coupling constant and the gluon saturation scales  $\Delta y \sim 1/\alpha_S(Q_s^2)$. On the other hand if one expects the initial state is described by the wounded nucleon model the trivial scale controlling the breaking of boost invariance will be different. There is a trivial scale of decorrelation determined by the beam rapidity $\Delta y \sim 2 Y_{\rm beam}$ -- simply because the orientation of the harmonic anisotropy planes in the target and projectile fragmentation regions are different. With the decrease of collision energy, one expects $Y_{\rm beam}$ to change and therefore a stronger breaking of boost invariance is expected. In order to explore this effect, one needs to go beyond the conventional two-particle azimuthal correlations over a narrow window around mid-rapidity. Several promising observables have been proposed to study this effect, 
Figure~\ref{ep_decorrelation} shows one which can be expressed as $r_n(\eta_a, \eta_b) = V_{n\Delta} (-\eta_a,\eta_b)/V_{n\Delta} (\eta_a,\eta_b)$ that measures the de-correlation of flow vectors while going from $-\eta_a$ to $\eta_a$ using reference particles taken from $\eta_b$~\cite{Khachatryan:2015oea}. Here, $V_{n\Delta} (\eta_a,\eta_b)$ is the Fourier coefficient calculated with pairs of particles taken from different pseudorapidity regions $\pm\eta_a$ and $\eta_b$.  In order to remove the trivial effect of de-correlation due to beam rapidity the x-axis of the plot is scaled by $Y_{\rm beam}$.    
Data from CMS collaboration~\cite{Khachatryan:2015oea} are compared to 3D-Glasma models of saturation~\cite{Schenke:2016ksl} and wounded nucleon model~\cite{Bozek:2015bna}. At RHIC predictions from both gluon saturation~\cite{Schlichting:2020wrv} as well as 3D-Glauber calculations with valence quarks and rapidity loss fluctuations~\cite{Shen:2017bsr} are available that remain to be tested. 
Further studies indicate that when scaled by the beam rapidity the ATLAS data~\cite{Aaboud:2017tql} for $\sqrt{s_{_{NN}}}$=2.76 and 5 TeV fall on top of each other but due to large uncertainty it is not possible to see if RHIC data deviates from such scaling. 
Hydrodynamics models with 3D initial state from AMPT initial conditions ~\cite{Pang:2015zrq} have been compared to data on Figure~\ref{ep_decorrelation} that predict stronger decorrelations for RHIC measurements at 200~GeV. In summary, it is essential to incorporate the rpaidity dependence of initial energy density to perform a full 3+1 dimensional hydrodynamic simulations. Other than the aforementioned models, the same has been performed by initializing the hydrodynamic simulations with inputs from UrQMD~\cite{Petersen:2008dd}, AMPT~\cite{Pang:2014pxa}  and EPOS~\cite{Werner:2010aa} initial conditions. The other models that include the rapidity dependent initial conditions are the strong coupling based Holographic initial conditions that employ AdS/CFT to match with hydrodynamic simulations~\cite{vanderSchee:2013pia,vanderSchee:2015rta}. 

The BES-II and forward upgrades of the STAR experiment include the Even Plane Detector and the Forward Tracking Systems improve these measurements with greater precision and over a wider range of $\eta/Y_{\rm beam}$ extend these measurements up to about $60\%$ of the beam rapidity at the top RHIC energy where $Y_{\rm beam}=5.36$ as shown by the yellow and green bands in Figure~\ref{ep_decorrelation} (right). 
With the decrease of collision energy the value of $Y_{\rm beam}$ decreases thereby covering an even wider longitudinal extent of the initial state. This way it will be possible to map the 3D structure of the initial state.

The de-correlations of the event plane can also be probed by measurements of mixed harmonic correlations of reaction planes. Such correlations have been measured by the STAR collaboration in Ref~\cite{STAR:2017idk} using the relative pseudorapidity dependence of the three-particle correlator $$C_{m,n,m+n}(\eta_a,\eta_b, \eta_c)=\left<\cos(m\phi_1(\eta_a)+n\phi_2(\eta_b)-(m+n)\phi_3 (\eta_c)\right>$$ with BES-I data. In the limited (currently available) acceptance at STAR, a small but significant decorrelation of the event planes was already observed~\cite{STAR:2017idk} and a hint of stronger decorrelations at lower energy was also observed. The origin of such decorrelations by analyzing the BES-II data that includes the wider acceptance of the iTPC detector. The future measurements of rapidity correlations at RHIC and LHC will extend our knowledge beyond the conventional two-dimensional (rapidity-invariant) picture of the initial state and help to constrain the rapidity evolution of the gluon densities inside colliding hadrons or nuclei. 

Another observable studied by the ATLAS collaboration to study the three-dimensional structure of the initial state is the two-particle correlation function in pseudorapidity $C(\eta_1,\eta_2)$~\cite{ATLAS:2016rbh}. The coefficients $a_{n,m}$ of the correlation function decomposed in terms of Legendre polynomials can be used to constrain the scale over which longitudinal invariance is  broken~\cite{Jia:2015jga,Bzdak:2012tp,Shen:2017bsr}. The coefficients $a_{1,1}$ measured in \pp{} collisions have been argued to constrain the intrinsic fluctuations of the saturation scale inside a proton~\cite{Bzdak:2016aii}. 

One of the direct consequences of the breaking of longitudinal invariance is the creation of vortical structure in the initial stages of collisions. Such a vorticity of the medium is transferred to the spin polarization of hadrons and has been measured in experiment. The polarization of hyperons is a novel probe of 3D initial state as they can help extract information such as the gradient of the fluid velocity that were not accessible before. Measurements of the global polarization of hyperons if different masses such as $\Lambda$, $\Xi$ and $\Omega$ that have different formation times can provide the dynamics evolution of the velocity fields of the medium created in heavy-ion collisions. More discussion on this can be found in the later sections of this document. 

\paragraph{Initial state of conserved charge distributions and baryon stopping}

At lower energy collisions, the initial state is intimately related to understanding the nucleon wave function in the nucleus, in particular, the momentum distribution and short-range correlations of the nucleons confined in a nucleus~\cite{Voloshin:2016ppr,Alvioli:2010yk}. In addition, knowledge of baryon transport and beam fragmentation is also needed for a complete picture of the initial state at lower energy. The three-dimensional structure of the initial state becomes more important at lower collision energies.%
Due to limited constraints from experimental data, the initial state of heavy-ion collisions at lower energy is poorly understood. This limits the modeling of heavy-ion collisions at lower energy and the interpretation of measurements from the RHIC BES program, which is dedicated to the search for the onset of the first-order phase transition and QCD critical point.%

\begin{figure}[htb]
  \begin{center}
    \includegraphics[width=0.65\textwidth]{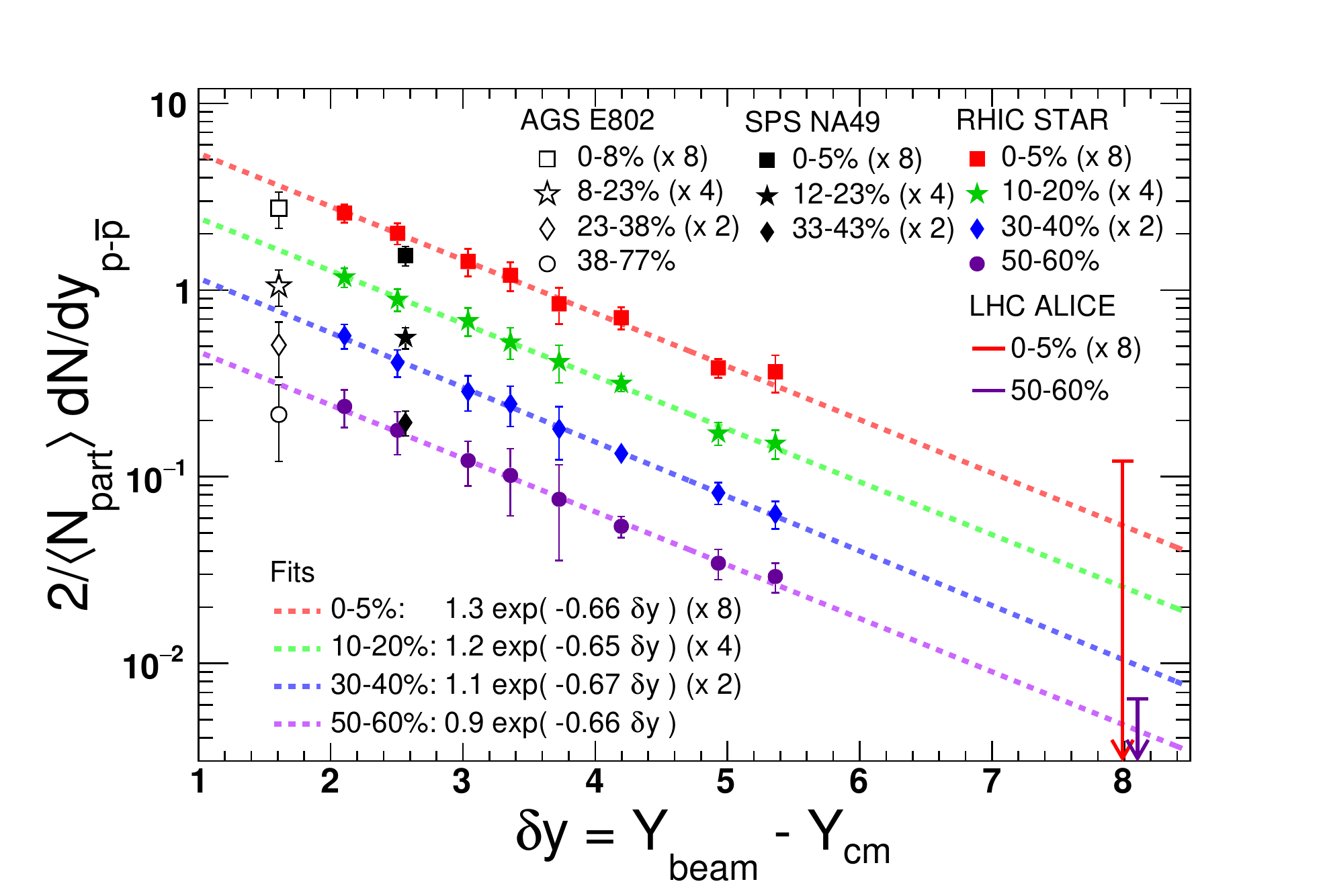}
  \end{center}
  \caption{\label{fig_baryon} Exponential dependence of midrapidity ($y\approx0$) baryon density per participant pair in heavy-ion collisions with $Y_{\rm beam}$ which is equal to the rapidity difference between beam and detector midrapidity ($\delta y$)~\cite{STAR:2008med,STAR:2017sal,Bearden:2003hx,E802:1998cxv,NA49:1998gaz,ALICE:2013mez}. An exponential fit function of $A\times\exp{(-\alpha_B\delta y)}$ is also included.}
\end{figure}

The promise of the BES program is predicated upon the ability to create a Quark Gluon Plasma that can be doped with baryons in heavy-ion collisions. This enables us to walk along the chemical potential axis of the QCD phase diagram. For experiments, this boils down to the measurements of finite baryon asymmetry in the central rapidity region of the collisions where observables sensitive to the first-order phase transition and QCD critical point are also measured (see the section on the QCD critical point). In addition, we must have the ability to achieve a lever arm to vary the baryon asymmetry by changing the collision energy so that a wide range of the chemical potential axis can be scanned.
The question is what is the dynamics that controls the initial density-distribution of the net-baryons in heavy-ion collisions. Since the net-baryon number cannot be created in the system and must come from the colliding target and projectile, a question is raised related to the understanding of the microscopic mechanism that makes baryons shift from the target and projectile rapidity to the midrapidity~\cite{Brandenburg:2022hrp}.
The observation of substantial baryon asymmetry in the central rapidity (mid-rapidity) region both at RHIC~\cite{Bearden:2003hx,STAR:2008med,STAR:2017sal} and at LHC energies ($\sqrt{s_{_{NN}}}=$900 GeV)~\cite{ALICE:2010hjm, ALICE:2013yba} is indeed a puzzling feature of heavy-ion collisions. Thus far, no approach from first principles has been able to explain this feature of the data. About three decades ago, it was argued that the question of how a baryon is stopped may be related to what is the true carrier of the baryon quantum numbers.
In a conventional picture, valence quarks carry baryon quantum number in a nucleus. At sufficiently high energies, it is expected that these valence quarks will pass through each other and end up far from mid-rapidity in the fragmentation regions~\cite{Kharzeev:1996sq,Andersson:1983ia}. However, the global data on net-proton yields at midrapidity from the Alternating Gradient Synchrotron (AGS)\cite{E802:1998cxv}, the Super Proton Synchrotron (SPS)\cite{NA49:1998gaz}, RHIC~\cite{STAR:2008med,STAR:2017sal,Bearden:2003hx}, and LHC~\cite{ALICE:2013mez}, as shown in Figure~\ref{fig_baryon}, indicate non-zero baryon stopping. The data also show that, for all centralities in heavy-ion collisions, the mid-rapidity net-baryon density follows an exponential distribution $A\exp{(-\alpha_B\delta y)}$ with the variable $\delta y=Y_{\rm beam}-Y_{\rm cm}$ and an exponent $\alpha_B$ ranging from 0.65-0.67. Other ways of characterizing the substantial baryon asymmetry observed in collisions include studying the average rapidity loss~\cite{BRAHMS:2009wlg} or the $\bar{p}/p$ ratio~\cite{STAR:2001rbj,ALICE:2010hjm, ALICE:2013yba}. Over the years, due to a lack of fundamental understanding, initial-state models of heavy-ion collisions have parametrized baryon stopping to reproduce experimental data.

A recent modeling of heavy-ion collisions indicates that the inclusion of the baryon junction is essential for describing net-proton density at RHIC~\cite{Shen:2022oyg}. Clearly some of the earlier implementations of baryon junctions (for example, the HIJING/B~\cite{Vance:1998vh,ALICE:2010hjm}, HIJING/B$\bar{B}$~\cite{ToporPop:2007df}), which attempted to match the earlier experimental data with certain parameter tunes, do not reproduce the experimental results presented in Figure~\ref{fig_baryon}. The idea that the flow of the baryon number can be traced by the flow of a non-perturbative Y-shaped configurations of gluon fields, called  baryon-junctions, rather than valence quarks was proposed in Ref~\cite{Kharzeev:1996sq}. Such junctions are the only possible gauge invariant structure of the baryon wave function, and have been widely studied in Lattice QCD~\cite{Suganuma:2004zx,Takahashi:2000te}. The gluon junctions as baryon carriers can lead to
significantly larger baryon-stopping at mid-rapidity compared to that of
quarks and could resolve the puzzle. Recently, a few experimental observables have been proposed to test such ideas~\cite{Brandenburg:2022hrp}. In the next half a decade we might be  able to have a much better understanding of the initial-state physics that describes the phenomenon of baryon stopping from first principles (see more in the ultra-peripheral collisions section).

\paragraph{Imprints of nuclear structure on the collective effects:}

The 2015 NSAC long-range plan points out that there is an important advantage of colliding nuclei of different shapes, such as uranium, copper or gold. In such collisions, one can measure observables that are related to collectivity and see the imprints of the initial-state geometry and fluctuations. 
Over the past seven years, measurements at RHIC and the LHC have revealed that bulk observables in relativistic heavy-ion collisions are very sensitive to nuclear shape and nuclear radial profile. Leveraging the knowledge of nuclear structure from low-energy experiments, we then have a new tool to independently probe and constrain the initial condition of heavy-ion collisions.

\begin{figure}[!h]
\includegraphics[width=1\linewidth]{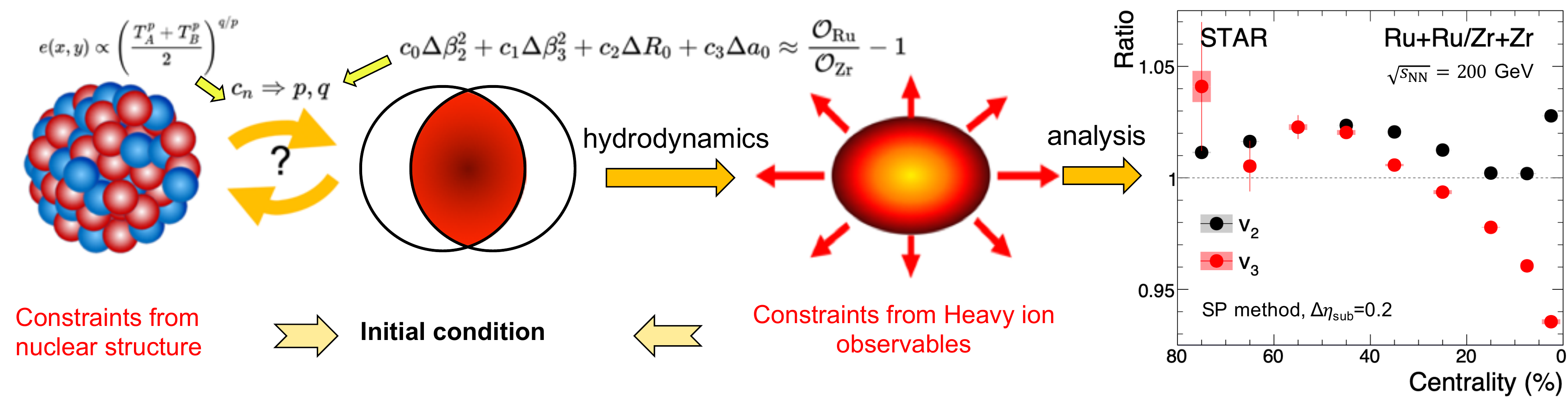}
\caption{\label{fig:ns1} 
Impact of isobar-like collisions on the initial condition of QGP. Better control on the initial condition can be achieved by exploiting the constraints from both the ratios of final-state observables ($v_2$ and $v_3$ on the right side)~\cite{STAR:2021mii} and the nuclear structure knowledge (left side). The left formula is a generic Ansatz for the initial energy of the system~\cite{Nijs:2022rme,Giacalone:2022hnz}, parametrized from the thickness functions of two nuclei, $T_{\rm A}$ and $T_{\rm B}$. The parameters $p,q$ can be constrained from $c_{\mathcal{O},n}$ in Eq.~\ref{eq:ns1}.}
\end{figure}

The complications of the final-state effects can be largely eliminated if one considers collisions of systems with similar mass but different structural properties, which are well-known for many stable isotopes. These properties can be characterized, for instance, by parameters of a deformed Woods-Saxon distribution such as quadrupole deformation, $\beta_2$, octupole deformation, $\beta_3$, triaxiality $\gamma$, half-width radius, $R_0$, and surface diffuseness, $a_0$. In particular, we focus on the ratio of a given observable $\mathcal{O}$ in collisions of isobars $X$ and $Y$, and ask:
$\mathcal{O}_{\rm X+X}/\mathcal{O}_{\rm Y+Y } \stackrel{?}{=} 1$. Any significant departure from unity must originate from the structural differences. Based on a Taylor expansion, numerical simulations show that isobar ratios can be approximated by \cite{Giacalone:2021uhj,Jia:2021tzt,Jia:2021oyt}:
\small{\begin{equation}
\label{eq:ns1}
\frac{\mathcal{O}^{\rm ini}_{{\mathrm X+X}} }{\mathcal{O}^{\rm ini}_{\mathrm Y+Y} } \approx \frac{\mathcal{O}_{{\mathrm X+X}} }{\mathcal{O}_{\mathrm Y+Y} }= 1+c_{\mathcal{O},1} (\beta_{2,\rm X}^2-\beta_{2,\rm Y}^2)+c_{\mathcal{O},2} (\beta_{3,\rm X}^2-\beta_{3,\rm Y}^2)+ c_{\mathcal{O},3} (R_{0,\rm X}-R_{0,\rm Y})+ c_{\mathcal{O},4} (a_{0,\rm X}-a_{0,\rm Y})\;,
\end{equation}}\normalsize
where $\mathcal{O}\equiv p(\nch), v_n, \mbox{or}\; \lr{p_{T}}$ are related to  corresponding initial-state estimators, $\mathcal{O}^{ini}$, such as $\varepsilon_n$ for the $v_n$.  The parameters $c_{\mathcal{O},1}$ to $c_{\mathcal{O},4}$ are found to be insensitive to final state effects~\cite{Xu:2021uar,Zhang:2022fou}, and hence reflect directly the response of the initial condition to changes in the nuclear structure parameters (see Figure~\ref{fig:ns1}). From the experimentally measured isobar ratios on the left-hand side and known nuclear structure differences on the right-hand side, the $c_{\mathcal{O},n}$ can be determined directly. Relations similar to Eq.~\eqref{eq:ns1} are also established for higher-order observables, such as $\lr{v_2^2\delta p_{T}}\sim a-b \beta_2^3\cos(3\gamma)$~\cite{Jia:2021qyu}. Therefore, isobar ratios offer a new tool to constrain the initial condition by exploiting nuclear structure information. 

The impact of nuclear shapes on flow observables has been observed in isobar-like comparisons between $^{238}$U+$^{238}$U and $^{197}$Au+$^{197}$Au at RHIC~\cite{STAR:2015mki,Giacalone:2021udy} and between $^{129}$Xe+$^{129}$Xe and $^{208}$Pb+$^{208}$Pb at the LHC~\cite{ALICE:2018lao,Bally:2021qys,ATLAS:2022dov}.  The most striking evidence, however, was recently obtained from $^{96}$Ru+$^{96}$Ru and $^{96}$Zr+$^{96}$Zr collisions at RHIC~\cite{STAR:2021mii}. Ratios of more than ten observables have been measured, all displaying distinct and centrality-dependent deviations of up to 30\% from unity, two of which are reported in the right panel of Figure~\ref{fig:ns1}~\cite{chunjianhaojie}. The ratios in central collisions are mostly impacted by deformation, while in mid-central collisions they are impacted by $R_0$ and $a_0$~\cite{Jia:2021oyt,Jia:2022qgl,Xu:2021vpn,Xu:2021uar}. The behavior of $v_3$ and $v_2$ suggests a large $\beta_{3,\rm Zr}$, not predicted by mean field structure models~\cite{Cao:2020rgr}. As argued above, such rich and versatile information provides a new type of constraint on the initial condition.

\subsubsection{Chirality and vorticity in QCD}

\label{sec:progress:macroscopic:chirality_and_vorticity}

Gauge fields describe the fundamental interactions in the Standard Model of particle physics. Gauge field configurations with nontrivial topology, such as instantons and sphalerons, are known to play crucial roles in many important phenomena, from matter-anti-matter asymmetry of today's universe to nonperturbative structures of the QCD vacuum. Their presence is however elusive for experimental detection.  
A novel approach for accessing such topological structures is to look for the Chiral Magnetic Effect (CME) in heavy-ion collisions. The CME predicts an electric charge separation along the large magnetic fields created at early times in these collisions, given the presence of nonzero quark chirality which in turn is a direct consequence of gluon topological fluctuations. Probing the CME signatures provides a unique way to explore the phenomenological roles of the quantum anomaly that connects quark chirality with gluon topology. In addition, the observation of the CME could provide important evidence for the
chiral symmetry restoration at high temperature --- a fundamental feature of QCD theory.     

Since the initial measurements published by STAR in 2009 with the first hint of possible CME signal, extensive theoretical studies as well as enthusiastic experimental search at both RHIC and LHC have been conducted. The key issue for an unambiguous detection of CME  is a rather  small signal embedded in a very large contribution from bulk background correlations. As noted in the 2015 Hot QCD White Paper: {\em ``While there have been hints of the CME in experiments, conventional explanations of
these data exist as well.''} 
Since then, substantial progress has been achieved in addressing this challenge. 

On the theoretical front, the critical need was clearly identified in the 2015 Hot QCD White Paper: 
{\em ``Quantifying the
predictions regarding signatures of quantum anomalies is crucial. This requires inclusion of
the anomalies into the standard hydrodynamic framework.''}  Such a goal has been achieved through the development of EBE-AVFD (event-by-event anomalous-viscous fluid dynamics)~\cite{Shi:2019wzi,Shi:2017cpu} as part of the Beam Energy Scan Theory (BEST) Collaboration project~\cite{An:2021wof}. This comprehensive framework 
provides the quantitative simulations of CME transport on top of realistic bulk hydrodynamic evolution and implements relevant background correlations. It has now been widely used for studying anomalous transport, and in particular, adopted as an essential tool for developing,  calibrating and interpreting various experimental observables~\cite{Choudhury:2021jwd,Milton:2021wku,Magdy:2017yje,Tang:2019pbl,Christakoglou:2021nhe}.

\begin{figure}[!hbt]
	\begin{center}
		\includegraphics[width=4in]{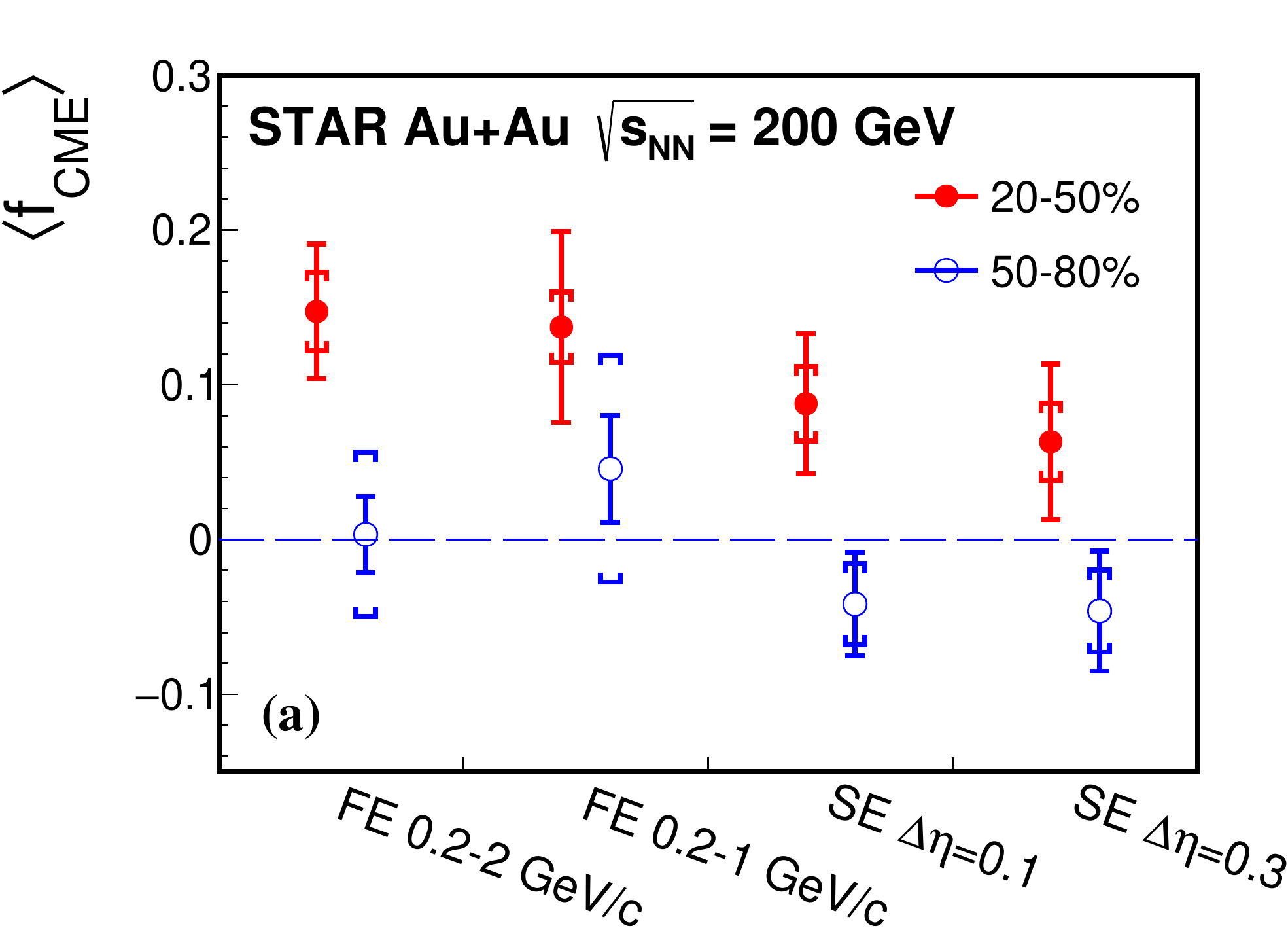}
		\caption{The CME signal fraction $\langle \rm f_{CME} \rangle$ after flow-background removal signal in 50–80\% (open markers) and 20–50\% (solid markers) centrality \AuAu{} collisions at $\rm \sqrt{s_{NN}}= 200 GeV$, extracted by various analysis methods (FE: full-event, SE: sub-event) and kinematic cuts. Error bars show statistical uncertainties; the caps indicate the systematic uncertainties. Figure from  ~\cite{STAR:2021pwb}.  }
		\vspace{-0.5cm}
		\label{fig_cme01}
	\end{center}
\end{figure}

On the experimental side, a number of new methods and observables have been developed, aiming at the extraction of possible signals and separation of background correlations. By utilizing event-shape engineering approach as well as comparison with small colliding systems, the CMS and ALICE Collaborations have put stringent limits on the existence of CME at LHC energies~\cite{CMS:2016wfo,CMS:2017lrw,ALICE:2017sss,ALICE:2020siw}. On the other hand, the latest measurements from the STAR Collaboration using a variety of analysis methods~\cite{STAR:2021pwb,STAR:2020gky,Lin:2020jcp} indicate a nonzero CME signal in 200~GeV \AuAu{} collisions at RHIC, albeit with still limited statistical significance. 
Fig.~\ref{fig_cme01} shows an example of the finite signal fraction $\langle \rm f_{CME} \rangle$ at  $1\sim3 \sigma$ level  for $20-50\%$ centrality~\cite{STAR:2021pwb}, extracted from the overall charge-dependent azimuthal correlations by exploiting the azimuthal fluctuations of the magnetic fields~\cite{Bloczynski:2012en} and comparing measurements relative to spectator versus participant planes~\cite{Xu:2017qfs,Voloshin:2018qsm}.  Another significant step for the CME search was taken by completing the isobar collision experiment at RHIC, where data for several billions of events were recorded for each of the \RuRu{} and \ZrZr{} colliding systems. Collisions of such isobar pairs were expected to produce identical bulk flow backgrounds while generating different CME signals due to differing magnetic field strength arising from their respective nuclear charges. A rigorous blind analysis procedure was developed and completed by the STAR Collaboration~\cite{STAR:2019bjg}, with the results announced in 2021~\cite{STAR:2021mii}. 
The unprecedented high precision achieved in the isobar data analysis was able to decisively reveal the few-percent level of difference in the bulk properties (such as multiplicity and elliptic flow in the same centrality class) of \RuRu{} and \ZrZr{} systems, thus pointing to a non-negligible variation of background correlations between them and complicating their comparative analysis. A recent attempt to account for such observed bulk differences and identify appropriate baseline (zero-signal scenario) for isobar comparison has revealed a potential signal fraction at a few percent level~\cite{Kharzeev:2022hqz}, which however may be vulnerable to various uncertainties such as nonflow effect~\cite{Feng:2021pgf}.   
Clearly, a final conclusion on the CME search from isobar collisions will require further experimental analyses as well as theoretical studies.

While the CME pertains to novel spin transport under magnetic fields, the spin degrees of freedom could also respond nontrivially to macroscopic fluid motions with large vorticity fields. One of the major discoveries since the last \LRP~is the observation of hyperon global spin polarization by STAR Collaboration~\cite{STAR:2017ckg}, openning a new direction of investigation in the field. See recent reviews in e.g. \cite{Becattini:2020ngo,Becattini:2021lfq,Becattini:2022zvf}. When two nuclei collide with each other, especially in non head-on collisions, the system carries a large orbital angular momentum, of the order of $L\sim10^5\hbar$, which is partially transferred to the created medium. Particles produced in the collisions are ``globally'' polarized on average along the initial orbital angular momentum direction~\cite{Liang:2004ph,Voloshin:2004ha,Becattini:2007sr}.
Under the assumption of local thermal equilibrium, the polarization $\boldsymbol{P}$ can be determined by the local vorticity of the fluid $\boldsymbol{\omega}$~\cite{Becattini:2007sr}. In a non-relativistic limit, the polarization is simplified as~\cite{Becattini:2016gvu}
\begin{equation}
    \boldsymbol{P} \approx \frac{(s+1)}{3}\frac{(\boldsymbol{\omega}+\mu \boldsymbol{B}/s)}{T},\label{eq:PH}
\end{equation}
where $T$ is the temperature, $s$ and $\mu$ are the spin and magnetic moment of the particle, and $\boldsymbol{B}$ is the magnetic field whose direction coincides with the direction of $L$, i.e., perpendicular to the reaction plane. A more general expression of the above equation that is relativistically  covariant and properly in terms of thermal vorticity can be found in e.g.   \cite{Becattini:2013fla}. 
More recently it has been found that the spin polarization is also
sensitive to the so-called thermal shear tensor contribution arising from the linear and  symmetric part of the hydrodynamic field gradients~\cite{Fu:2021pok,Becattini:2021iol}, which induces nontrivial local spin polarization patterns while leaves the global spin polarization unchanged.

\begin{figure}[htb]
\begin{minipage}[c]{0.55\hsize}    
    \centering
    \includegraphics[width=\linewidth]{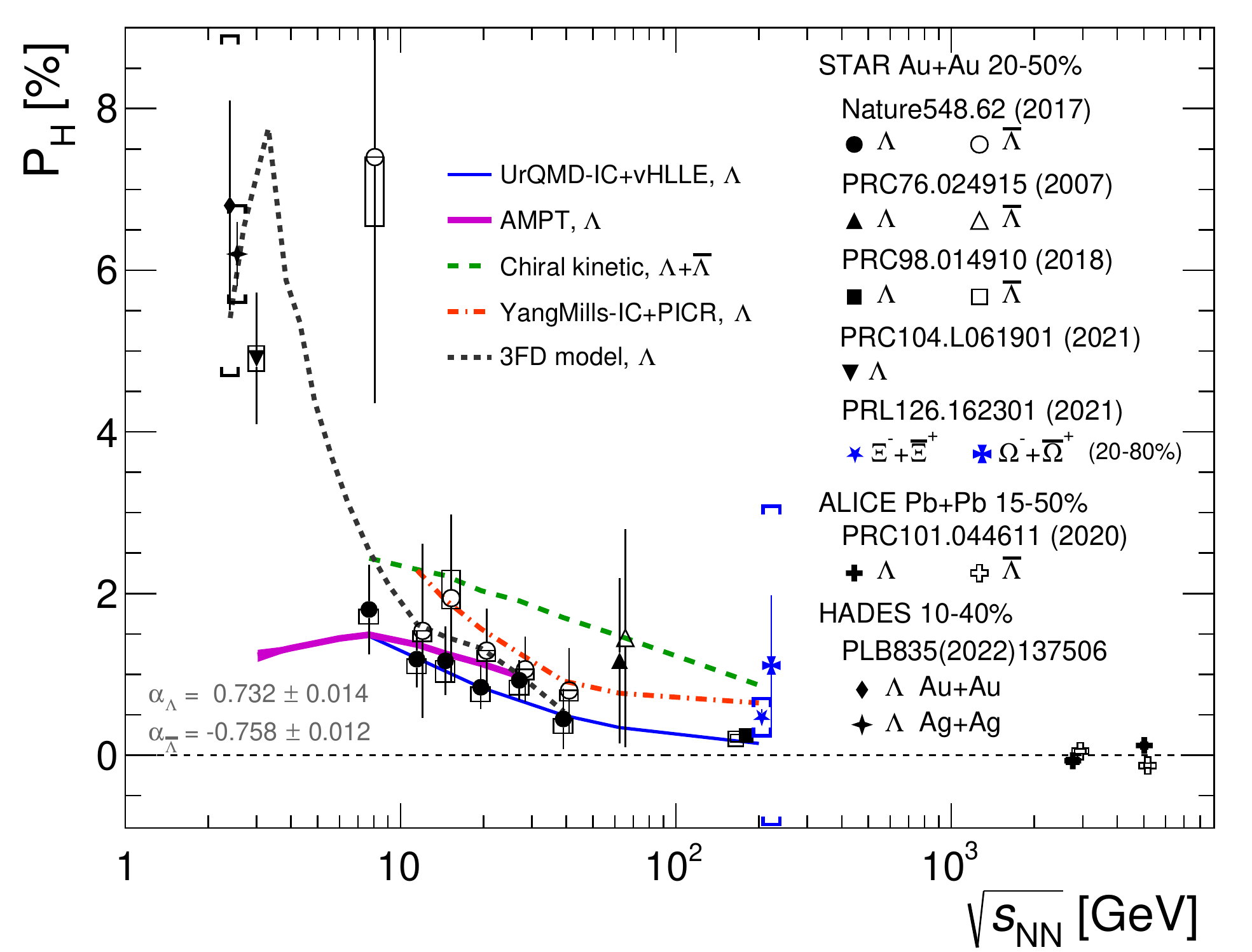}
\end{minipage}
\begin{minipage}[c]{0.44\hsize}
    \centering
    \includegraphics[width=\linewidth]{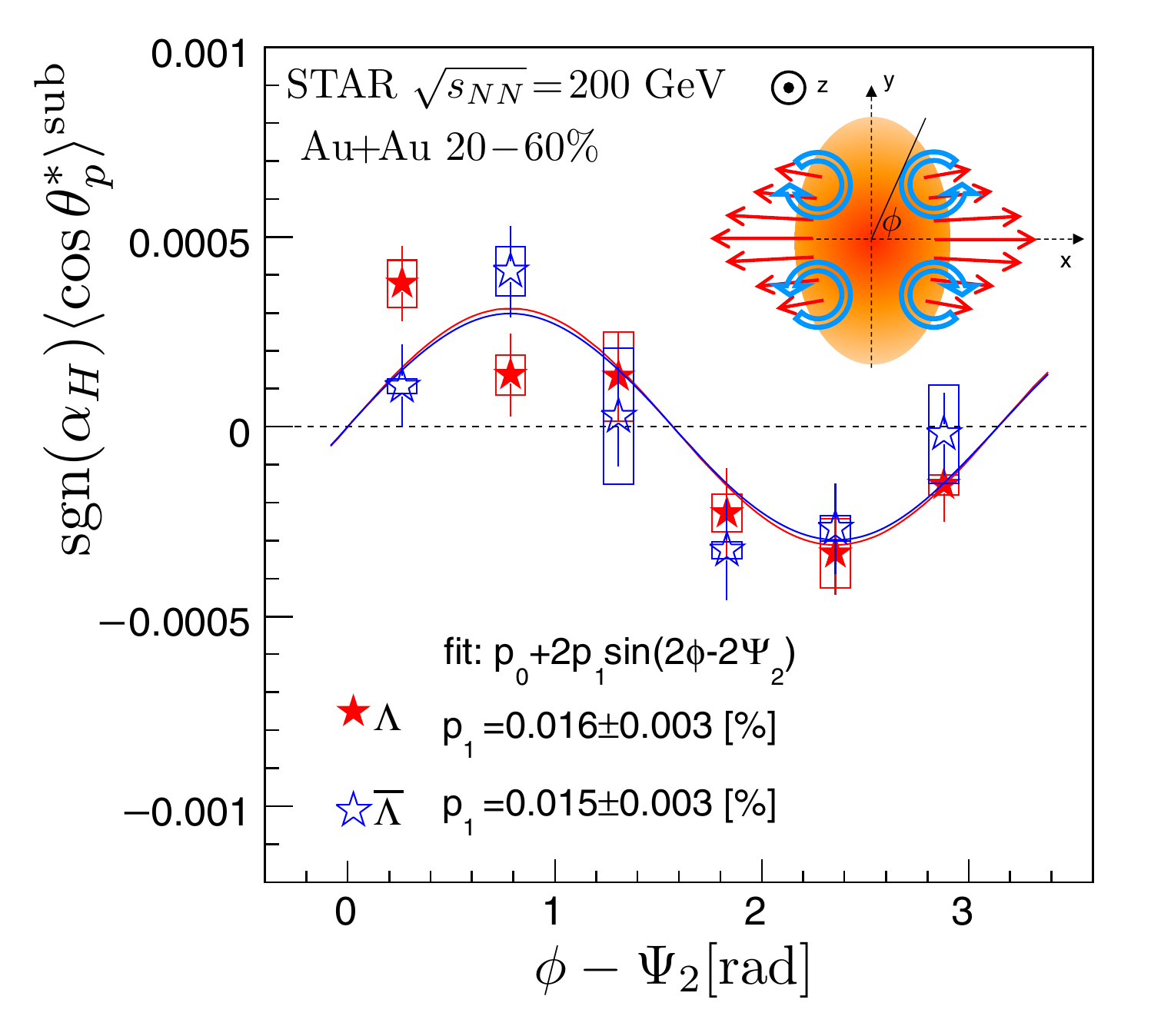}
\end{minipage}
\caption{(Left) Hyperon global polarization as a function of collision energy, compared to various theoretical calculations. (Right) Polarization along the beam direction, $\langle\cos\theta_p^\ast\rangle \approx \alpha_H P_z/3$, as a function of azimuthal angle for $\Lambda$ and $\bar{\Lambda}$ hyperons in \AuAu{} collisions at $\sqrt{s_{NN}}=200$ GeV.}
\label{fig:Pol}
\end{figure}

Global polarization of $\Lambda$ hyperons was first observed in the beam energy scan of \AuAu{} collisions~\cite{STAR:2017ckg} and later confirmed at $\sqrt{s_{NN}}=$ 200 GeV with more differential measurements~\cite{STAR:2018gyt}. 
As shown in Fig.~\ref{fig:Pol} (left), $\Lambda$ global polarization shows the collision energy dependence where a larger signal is observed when decreasing the collision energy --- a nontrivial trend that could be understood from the predicted beam energy dependence of average vorticity in these collisions~\cite{Jiang:2016woz}. The data are in good agreement with theoretical calculations, such as hydrodynamics, transport model, and chiral kinetic theory. This reveals that the QGP is the most vortical fluid ever observed, of the order of $\omega\sim10^{21}~s^{-1}$.  Recent measurements at STAR fixed-target and HADES experiments~\cite{STAR:2021beb,HADES:2022enx} show the largest polarization at a few GeV, which is close to the $\Lambda$ production threshold and in which lower energies the system would be dominated by hadronic interactions rather than partonic interactions. At LHC energies, the signal is expected to be small and current results are consistent with zero~\cite{ALICE:2019onw}.

As indicated in Eq.~(\ref{eq:PH}), there could be a contribution from the magnetic field which leads to larger polarization of $\bar{\Lambda}$ than that of $\Lambda$ due to the sign of $\mu$. In other words, one could probe the later-stage magnetic field with this measurement~\cite{Becattini:2016gvu,Muller:2018ibh,Guo:2019joy} which will be an important input for theoretical predictions of the CME signal, although there are other possible mechanisms to make such a difference~\cite{Fang:2016vpj,Csernai:2018yok,Vitiuk:2019rfv,Guo:2019mgh}. Current measurements show a hint of the difference at lower energies but the difference is not significant. The possible difference between $\Lambda$ and $\bar{\Lambda}$ and the energy dependence in lower energies are being investigated with high statistics data collected during  BES-$\mathrm{I}\hspace{-1.2pt}\mathrm{I}$ program at RHIC. 

Recently, STAR also measured global polarization of multistrangeness such as $\Xi$ and $\Omega$ hyperons at $\sqrt{s_{NN}}=200$ GeV~\cite{STAR:2020xbm}. It is of great importance to confirm the global polarization and vorticity picture using different particles with different spins and/or magnetic moments to better understand the polarization mechanism in heavy-ion collisions. The uncertainties need to be reduced with future measurements but there seems to be a hint of a hierarchy in global polarization, i.e., $P_{\Lambda}<P_{\Xi}<P_{\Omega}$, qualitatively consistent with a predicted mass-ordering effect~\cite{Wei:2018zfb}. Measurement of non-zero $P_{\Omega}$ also helps to determine the sign of unmeasured decay parameter $\gamma_{\Omega}$~\cite{ParticleDataGroup:2022pth}. %

Richer vortical structures due to density fluctuations coupled with the system collective expansion have been predicted~\cite{Pang:2016igs,Becattini:2017gcx,Voloshin:2017kqp,Xia:2018tes}. In heavy-ion collisions, elliptic flow has been extensively studied and known as a consequence of stronger expansion into the reaction plane angle, which was predicted to lead the polarization along the beam direction $P_z$ changing the sign depending on azimuthal angle. Such a local polarization was observed by STAR at RHIC~\cite{STAR:2019erd} and later by ALICE at the LHC~\cite{ALICE:2021pzu}. Figure~\ref{fig:Pol} (right) shows $P_z$ ($\approx 3\langle\cos\theta_p^\ast\rangle/\alpha_H$) as a function of azimuthal angle relative to elliptic flow plane angle ($\Psi_2$). A quadrupole pattern of the polarization was observed which is consistent with the expectation from elliptic flow (see a cartoon inside the figure) and the blast-wave model with parameters determined by the fits to $p_T$ spectra and HBT measurements~\cite{Voloshin:2017kqp,STAR:2019erd}. However unlike the case for the average global polarization, many theoretical models fail to describe the data of $P_z$, implying a possible lack of understanding the origin of the polarization and importance of non-equilibrium dynamics of spin in heavy-ion collisions. The latest studies in \cite{Fu:2021pok,Becattini:2021iol} suggest that the inclusion of the thermal shear  contribution to spin polarization could be  important for explaining the data. 
A more quantitative investigation in \cite{Alzhrani:2022dpi} finds that the resulting longitudinal polarization results differ between two proposed implementations of the thermal shear terms and also depend upon the hydrodynamic initial conditions.  

The polarization measurement of strongly decaying particles is more difficult but one can measure the $00^{\rm th}$ element of the spin density matrix $\rho_{00}$ of vector mesons~\cite{Liang:2004xn}, which represents the probability to have spin projection onto the polarization axis to be zero. The deviation of $\rho_{00}$ from $1/3$ indicates spin alignment of vector mesons. Global spin alignment of $\phi$ and $K^{\ast0}$ mesons has been measured both at RHIC~\cite{STAR:2022fan} and the LHC~\cite{ALICE:2019aid} and significant deviation from $1/3$ was observed; $\rho_{00}\sim 0.2$--$0.3~ (<1/3)$ for $K^{\ast0}$ at the LHC and $\rho_{00}\sim 0.36~(>1/3)$ for $\phi$ in lower energies at RHIC. These large deviations cannot be explained by the vorticity picture and possible explanation might be a strong force field as proposed in Refs.~\cite{Sheng:2019kmk,Sheng:2022wsy,Sheng:2022ffb} which needs further investigation.

\clearpage

\subsection{Mesoscopic: emergence of the quark-gluon plasma and approach to equilibrium}
\label{sec:progress:mesoscopic}

\subsubsection{Small Collision Systems}
\label{sec:progress:mesoscopic:small_systems}

Systems of different sizes and shapes,  such as \pp{}, \pA{}, \dA{}, and \HeA{}
present challenges and opportunities for the study of mesoscopic QCD matter. In
such systems the thermalization process has a significant impact on flow
observables, offering a  unique experimental window into the evolution of the
QCD system from a quantum wave function to a hydrodynamically expanding quark-gluon plasma \cite{Schenke:2021mxx}.  It
is this transition and its rich non-Abelian many-body dynamics that we wish to briefly
review and explore further both theoretically and experimentally.  

The initial wave function of the system is characterized by its mean and fluctuations,  and
these initial-state fluctuations can yield measurable correlations in the
final state \cite{Luzum:2013yya}. Characterizing the fluctuation of the incoming particles (such as
the fluctuation in the proton radius) is interesting in its own right, and is
essential if the dynamics of the initial state are to be disentangled from the
flow-like correlations built up by the many-body response of the system in its approach to equilibrium. Addressing this concretely is possible due to advances in experimental
measurements and probes, and theory and simulation.

This section is structured as follows. First, in Section~\ref{sec:progress:mesoscopic:small_size_limit_of_qgp} we  will review the measurements,
and discuss a variety of new probes which can concretely characterize initial
state fluctuations, both in rapidity and in the transverse plane.  Ultimately,
these experimental  inputs will be the basis for numerical simulations that
describe the system's thermalization process in detail. Since the last \LRP,
there has been enormous progress in understanding the transition to
hydrodynamics both at weak and strong coupling from simulations. Indeed, there are
features of the transition to hydrodynamics that are seemingly universal, and
independent of the coupling. This is reviewed in Section~\ref{sec:progress:mesoscopic:onset_of_hydro}. The challenge is to use the detailed experimental
probes as a function of system size, together with advances in theory and
simulation that incorporate these universal features, to provide a comprehensive picture of thermalization and the onset of hydrodynamics in small systems. Section~\ref{sec:progress:mesoscopic:medium_response} describes the connection between the interplay between the medium and hard probes.

\subsubsection{Small systems \& small size limit of the QGP}
\label{sec:progress:mesoscopic:small_size_limit_of_qgp}

In the last decade, the experimental exploration of small collision systems has brought a paradigm shift in the understanding of hot QCD. Previously, proton-nucleus or deuteron-nucleus collisions were considered to be control experiments with no QGP formation that provide opportunities to study how cold nuclear matter in the initial state affects final-state observables. Proton-proton collisions are usually used as a reference baseline to quantify the nuclear effects. Most notably, the absence of suppression in the production of high-p$_T$ hadrons and in back-to-back di-hadron correlations in \dAu{} collisions at RHIC solidified the discovery of jet quenching in heavy-ion collisions~\cite{PHENIX:2003qdw,STAR:2003pjh,BRAHMS:2003sns,PHOBOS:2003uzz}. The subsequent discovery of long-range correlations ("ridge") in particle productions in high-multiplicity \pp{}~\cite{CMS:2010ifv} and pPb~\cite{CMS:2012qk, ALICE:2012eyl,ATLAS:2012cix,LHCb:2015coe} collisions at the LHC came as a surprise. At RHIC, reanalysis of previously recorded \dAu{} data~\cite{PHENIX:2013ktj,PHENIX:2014fnc,STAR:2015kak} revealed that long-range angular correlations are also present in \dAu{} collisions at 200 GeV. In nucleus-nucleus (\AA{}) collisions the ridge is associated with the anisotropic collective expansion of QGP resulting from anisotropies in the initial collision geometry, which are subsequently transferred to the azimuthal distributions of the produced particles. If the ridge in \pp{} and p/d+A collisions is of the same origin as in \AA{} collisions, then the formation of small QGP droplets cannot be excluded in small systems and their role as control experiments has to be re-examined.   

Following the initial discoveries, a vigorous experimental exploration ensued testing most every aspect of the bulk system dynamics associated with QGP formation in large systems (reviews can be found in Refs.~\cite{Dusling:2015gta,Loizides:2016tew,Nagle:2018nvi}). The LHC experiments collected new pPb data at 8.16 TeV to complement the 2013 and 2016 data sets at 5.02 TeV. The higher collision energy gave access to higher multiplicities than previously experimentally accessible. At LHC, high-multiplicity \pp{} collisions have been studied at several center-of-mass energies ranging from 2.76 TeV to 13 TeV. In 2014-2016 RHIC conducted a geometry scan with p/d/$^3$He+Au collisions~\cite{PHENIX:2016cfs, PHENIX:2017nae,PHENIX:2015idk, PHENIX:2018lia} at 200 GeV providing means to test if the initial geometry is reflected in the final-state particle distributions. Additionally, a beam energy scan (200, 62.4, 39, and 19.6 GeV) with \dAu{} collisions~\cite{PHENIX:2017nae,PHENIX:2017xrm} was performed to investigate if the apparent collectivity turns off at lower collision energy. The geometry and beam energy scan with small systems comprised a major part of the last three years of the PHENIX experimental program. Since then, additional \dAu{} and \OO{} collision data at 200 GeV were collected by the STAR experiment in 2021 with the addition of a dedicated event-plane detector covering the pseudorapidity region of $2.2<|\eta|<5.2$ and extended coverage in the TPC to $|\eta|<1.5$. The LHC experiments utilized ultra-peripheral collisions (UPC) in \PbPb{}~\cite{ATLAS:2021jhn} and \pPb{}~\cite{CMS:2022doq} to examine if there are signs of collectivity in photon-nucleus and photon-proton collisions. Two-particle correlations were also studied in archived data from hadronic Z-boson decays in  $e^+e^-$ collisions measured by ALEPH at LEP~\cite{Badea:2019vey}, and in hadronic $e^+e^-$ annihilation events at 10.52 GeV off-resonance from the $\Upsilon(4S)$ state with the Belle experiment at KEKB~\cite{Belle:2022fvl}, as well as in archived deep inelastic scattering events from \ep{} collisions with Zeus at HERA~\cite{ZEUS:2019jya}.   

\begin{figure}[h!]
\centering
\includegraphics[width=0.85\textwidth]{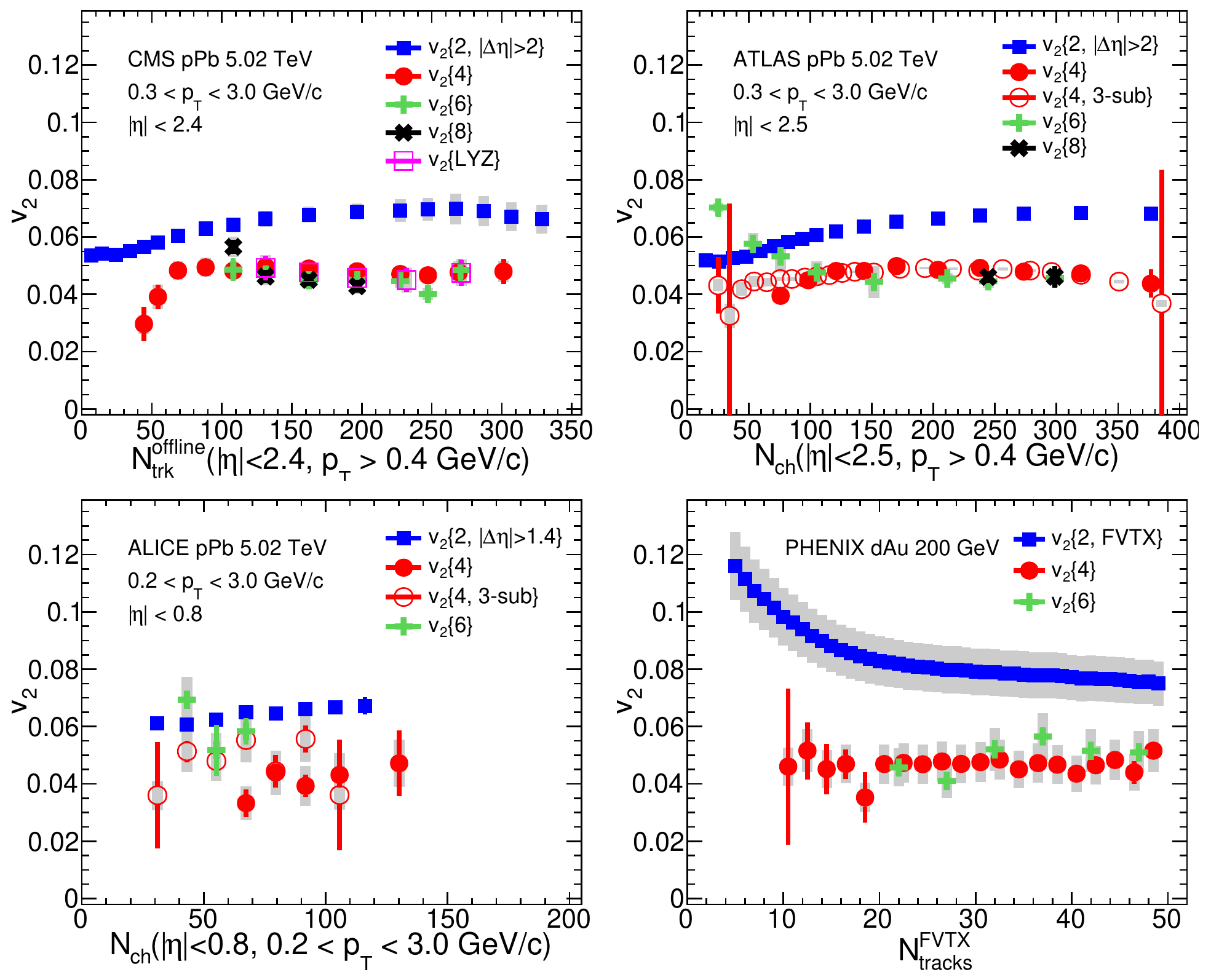}
\caption{The measured $v_{2}$ using two- and multi-particle correlations from  CMS~\cite{CMS:2016fnw}, ATLAS~\cite{ATLAS:2017hap}, and ALICE collaborations~\cite{ALICE:2019zfl} in pPb collisions at 5.02 TeV, and PHENIX collaboration~\cite{PHENIX:2017xrm} in \dAu{} collisions at 200 GeV.}
\label{fig:multiparticlev2}
\end{figure}

The collective behavior is characterized with single-particle azimuthal anisotropies that are quantified by $n^{th}-$order Fourier coefficients, or flow harmonics, $v_{n}$. There are different methods of measuring the flow harmonics that have different sensitivity to event-by-event flow fluctuations and to nonflow correlations that arise from resonance decays, Bose–Einstein correlations, and jet production. To reduce short-range nonflow correlations in the two-particle correlation analyses the particles are separated by rapidity gaps. By construction, the multi-particle cumulant measurements suppress nonflow by subtracting off lower order correlations. Measurements of $v_{2}$ with multi-particle correlations in pPb~\cite{CMS:2013jlh,ATLAS:2013jmi,CMS:2015yux,ALICE:2014dwt} and in \pp{}~\cite{CMS:2016fnw, ATLAS:2017rtr, ALICE:2019zfl} collisions at the LHC established that the observed long-range correlations are a collective phenomenon. At RHIC, up to 6-particle cumulants were studied in \dAu{} collisions at 200 GeV~\cite{PHENIX:2017xrm}, and also found to agree with expectations for hydrodynamic behavior both in the measured values with 2-, 4-, 6- particles and the expected relation between the cumulants and the event-by-event flow fluctuations. Going down in energy the multi-particle measurements are more challenging as the multiplicity is reduced and nonflow correlations are more difficult to suppress. Nevertheless, real-valued $v_{2}\{4\}$ were observed down to the lowest center-of-mass energy measured in \dAu{} collisions, indicating that collective effects may persist. Figure~\ref{fig:multiparticlev2} shows the measurements of $v_{2}$ using two- and multi-particle correlations from ALICE, ATLAS, and CMS collaborations from \pPb{} collisions at 5.02 TeV, and the PHENIX collaboration in \dAu{} collisions at 200 GeV. 

Another feature of the collectively expanding QGP in large systems is the mass dependence of the particle spectral shapes and the $v_n$ coefficients. These features have also been seen in small systems and studied extensively with light-flavor hadrons~\cite{CMS:2013pdl, ALICE:2013snk, CMS:2014und, ALICE:2015mpp, ALICE:2016fzo, CMS:2016zzh, PHENIX:2017djs, CMS:2019isl}. Surprisingly, open-charm hadrons and even $J/\psi$ also exhibit collective behavior in both pPb and in \pp{} collisions, while there is no clear evidence of bottom hadron flow in these systems~\cite{ATLAS:2019xqc, ALICE:2017smo, CMS:2018loe, CMS:2020qul, CMS-PAS-HIN-21-001}.

\begin{figure}[h!]
\centering
\includegraphics[width=1\textwidth]{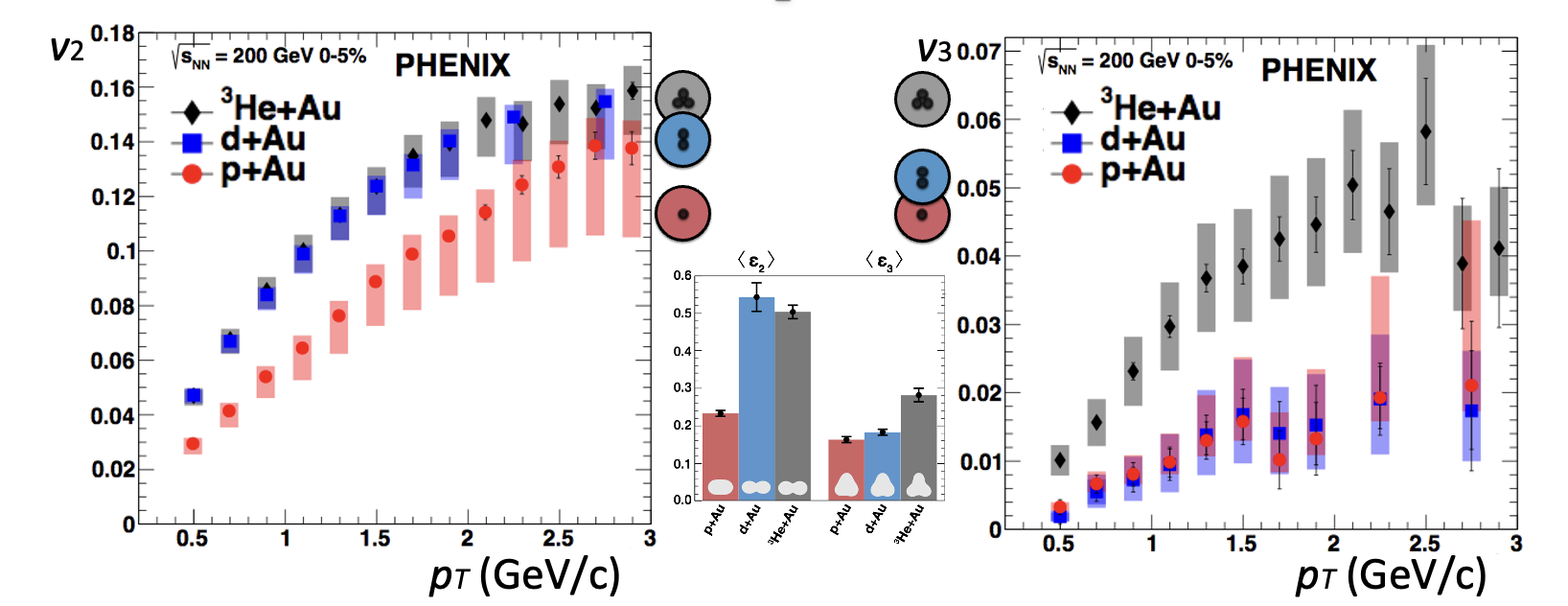}
\caption{PHENIX measurements~\cite{PHENIX:2018lia} of $v_{2} (p_T)$  (left) $v_{3} (p_T)$ (right) in p/d/$^3$He+Au collisions are compared to the initial spatial eccentricities (middle)  calculated with the Glauber model.}
\label{fig:PHENIX_small_systems}
\end{figure}

Testing the response of the final-state particle production to the initial geometry motivated the geometry scan with p/d/$^3$He+Au collisions at RHIC~\cite{Nagle:2013lja}. The three different projectiles provide initial geometries that have different elliptic and triangular eccentricities that can be probed in the final state with measurements of the second and third harmonics $v_2$ and $v_3$. The PHENIX measurements~\cite{PHENIX:2016cfs, PHENIX:2017nae,PHENIX:2015idk, PHENIX:2018lia} shown in Figure~\ref{fig:PHENIX_small_systems} indicate that the final-state particle distributions retain memory of the initial-state geometry as expected in viscous hydrodynamics~\cite{Romatschke:2015gxa, Shen:2016zpp}. Recent STAR measurements~\cite{STAR:2022pfn}, on the other hand, find $v_{3}(p_{T})$ values that are system-independent within the uncertainties suggesting that sub-nucleonic fluctuations may influence the initial eccentricities. Kinematic dependencies in the flow and nonflow correlations, as well as longitudinal decorrelations may contribute to the observed differences~\cite{PHENIX:2021ubk}. In the PHENIX analysis the nonflow correlations are suppressed by imposing a large separation in pseudorapidity between the particles used to analyze flow and the event-plane detectors, the nonflow is estimated based on measurements in \pp{} collisions and included in the systematic uncertainties instead of being subtracted. The STAR measurements are performed in the range $|\eta|< 0.9$ where nonflow correlations are large and are therefore subtracted. Nonflow correlations depend strongly on the particle separation in rapidity and nonflow estimates are challenging since there are significant model dependencies~\cite{Nagle:2021rep}. Large kinematic dependence in $v_3$ values has also been observed by STAR in peripheral \RuRu{} and \ZrZr{}~\cite{STAR:2021mii} collisions. Another challenge is the fluctuations in rapidity. Because of this complication, which is becoming large in small systems, flow measurements in different rapidity windows do not have to agree. From a theoretical point of view characterizing these three-dimensional fluctuations is particularly important. Recent calculations~\cite{Zhao:2022ugy} with $(3+1)$D framework combining a dynamic initial state with hydrodynamics and hadronic transport give a better description of the data. 
Further discussion has been given in Section~\ref{sec:progress:macroscopic:hydro}. In the future, measurements of flow decorrelations at different rapidities, such as e.g.~\cite{CMS:2015xmx}, with the upgraded STAR detector and in sPHENIX will be essential to a complete understanding of the dynamics. 

Overall, the hydrodynamic description of the system evolution gives the most complete and quantitative description of the vast amount of experimental data collected with a variety of small collision systems spanning center-of-mass energies from 20 GeV to 13 TeV. However, certain features observed in the data are not unique to the hydrodynamic response to the initial-state pressure gradients. For example, momentum correlations in the initial state of the collision could persist into the final state as well. The relative contribution to the final-state anisotropy is expected to become more important at low multiplicity and lower collision energy~\cite{Giacalone:2020byk}. The correlations between $v_n$ and mean $p_T$ may provide means to distinguish  initial-state momentum correlation effects~\cite{Giacalone:2020byk} and hydrodynamic response~\cite{Bozek:2016yoj}. The $v_n$ and mean $p_T$ correlations have been measured in  pPb collisions by the ATLAS Collaboration~\cite{ATLAS:2019pvn}, but the studied multiplicities may not be small enough to reveal any initial-state momentum correlations. Additionally, recent studies from the CMS collaboration~\cite{Tuo:2022qm} indicate that nonflow has to be carefully removed, since the theoretical models do not include these effects. 

The observations of ubiquitous collective phenomena in large and small hadronic systems prompted searches for collectivity in photon-nucleus, photon-proton, and $e^+e^-$ collisions. 
Significant nonzero flow coefficients are observed in \PbPb{} UPCs~\cite{ATLAS:2021jhn}. Calculations based on Color Glass Condensate~\cite{Shi:2020djm} and full $(3+1)$D dynamical framework with 3D-Glauber+MUSIC+UrQMD~\cite{Zhao:2022ayk} are both able to describe the data. In these models, the virtual photon in each UPC event is treated as a vector meson with a lifetime longer than the time of the interactions, therefore it is not a surprise to have flow in these calculations using meson-nucleus collisions. With the current multiplicity reach, no collective flow is observed in pPb UPC events~\cite{CMS:2022doq}. If the photon-proton collisions are modeled as meson-proton interactions, it is possible that collective flow is revealed in pPb UPCs with future data sets that have larger statistical reach.

To account for collective effects in small systems string-based models, such as PYTHIA with color reconnections~\cite{Christiansen:2015yqa}, the DIPSY rope hadronization~\cite{Bierlich:2015rha}, and the string-shoving~\cite{Bierlich:2020naj}, have been introduced and have demonstrated qualitative agreement with certain aspects of the data. The AMPT framework~\cite{Lin:2004en} based on parton transport also describes most of the collective observables well. A recent calculation in $e^+e^-$ collisions with AMPT including string melting, parton scattering, hadronization and hadron re-scattering shows that a strong $v_2$ can be created with two initial strings, while there is no observable signal of collectivity with one string~\cite{Nagle:2017sjv}. The two-particle correlation measurements in $e^+e^-$ collisions and deep inelastic \ep{} scattering do not indicate collective behavior~\cite{Badea:2019vey, Belle:2022fvl, ZEUS:2019jya}. The upcoming Electron-Ion Collider will provide new opportunities to explore collective phenomena in small systems.

\subsubsection{Onset of hydrodynamics}
\label{sec:progress:mesoscopic:onset_of_hydro}
There has been extraordinary progress in understanding the evolution of the system in the far from equilibrium situations relevant to heavy-ion collisions \cite{Florkowski:2017olj,Romatschke:2017ejr,Schlichting:2019abc,Berges:2020fwq}. Significant developments have been made towards understanding the emergence of hydrodynamic behavior in relativistic systems both in strong coupling \cite{Chesler:2010bi,Heller:2011ju,Heller:2012km,vanderSchee:2013pia,Heller:2013oxa,Casalderrey-Solana:2013aba,Chesler:2013lia,Chesler:2015wra,Keegan:2015avk,Spalinski:2017mel} and also in weak coupling/kinetic theory \cite{Denicol:2014xca,Denicol:2014tha,Kurkela:2015qoa,Keegan:2015avk,Bazow:2015dha,Denicol:2016bjh,Heller:2016rtz,Romatschke:2017vte,Strickland:2017kux,Strickland:2018ayk,Kurkela:2018wud,Kurkela:2018vqr,Kurkela:2019set,Strickland:2019hff,Denicol:2019lio,Almaalol:2020rnu,Du:2020dvp,Du:2020zqg,Ambrus:2021fej, Jaiswal:2022udf,Alalawi:2022pmg,Mullins:2022fbx,Alalawi:2022pmg,Ambrus:2022qya,Ambrus:2022koq,Rocha:2022ind,Du:2022bel}.
Indeed, with the advent of hydrodynamic attractors \cite{Heller:2015dha}, there is an emerging picture with which one can estimate the value of $\tau_{\rm hydro}$ where some form of hydrodynamics begins to apply at a given multiplicity see, for instance, Refs.\ \cite{Keegan:2015avk,Strickland:2018ayk}. To illustrate and explain the main arguments behind these developments, here we consider a simple toy model of the QGP in terms of conformal kinetic theory \cite{Baier:2007ix,Denicol:2014xca,Denicol:2014tha} in the relaxation time approximation \cite{Baym:1984np,Florkowski:2013lza,Florkowski:2013lya,Florkowski:2014sfa}. This type of approximation has provided powerful insights in the determination of the onset of hydrodynamic behavior in rapidly expanding systems \cite{Florkowski:2013lza,Florkowski:2013lya,Florkowski:2014sfa,Denicol:2014xca,Denicol:2014tha,Strickland:2015utc,Alalawi:2022pmg}.
In fact, although the relaxation time approach may seem  primitive, and the QGP is not a conformal system, many of the conclusions are expected to have
a universal character as they reflect the competition between free streaming and
dissipative dynamics of an approximately conformal system expanding longitudinally \cite{Brewer:2019oha}. As we discuss below, the simple conformal kinetics captures the scaling behavior expected from QCD at weak coupling~\cite{Baier:2000sb}, and the scaling properties displayed by strongly coupled conformal theories based on the AdS/CFT correspondence~\cite{Heller:2011ju}.  Although below we focus on conformal systems for simplicity, we note that there has been recent evidence of the existence of hydrodynamic attractors for certain moments of the one-particle distribution function in non-conformal systems \cite{Chattopadhyay:2021ive,Jaiswal:2021uvv,Alalawi:2022gul}.

\noindent {\bf Longitudinal Expansion}. 
Let us first assume the system undergoes homogeneous Bjorken expansion \cite{Bjorken:1982qr}. We consider a conformal medium, which essentially implies that the energy density, $e$, and the isotropic pressure, $P$, of the system are related via $e=3P$. Furthermore, at finite temperature $T$, any dimensionful quantity scales with appropriate powers of $T$ (for example, $e\sim T^4$).  
For a Bjorken expanding system, conformality together with energy-momentum conservation, $\partial_{\mu}T^{\mu\nu}=0$, gives
\begin{equation}
    \frac{de}{d\tau} = \frac{e + P_L}{\tau} \, .
    \end{equation}
    where $P_{L}$ is the system's longitudinal pressure and $\tau$ is the proper time.
In first-order Navier-Stokes viscous hydrodynamics the longitudinal pressure takes the form  $P_{L}/e = \tfrac{1}{3}  - \frac{16 \eta/s}{9 \tau T}$, while second-order viscous corrections are of order $(\eta/sT \tau)^2$ \cite{Romatschke:2017ejr}, where $s$ is the entropy density and $\eta$ is the shear viscosity. The hydrodynamic attractor concept \cite{Heller:2015dha} anticipates an all-order resummation of these terms into a constitutive relation of the form $P_L/e=f(w)$,  where  $w(\tau)=\tau T_{\rm eff}/(4\pi \eta/s)$ \cite{Florkowski:2017olj}. The   
effective temperature $T_{\rm eff}$ is defined through the non-equilibrium energy density and the equation of state,  $ s T = \tfrac{4}{3} e =  b_{\rm qgp} T^4$,  where $b_{\rm qgp} \simeq 17.6$ is estimated from the lattice equation of state at high temperatures \cite{Borsanyi:2010cj,HotQCD:2014kol}. Traditionally, hydrodynamics is expected to emerge when the Knudsen number, determined by the ratio between microscopic scales $\ell$ (e.g., the mean free path of a gas) and macroscopic scales $L$ associated with the spatial gradients of conserved quantities, $\mathrm{Kn} = \ell/L$, becomes much smaller than unity. Here, $w\sim 1/\mathrm{Kn}$ so the hydrodynamic limit is valid at late times  when $w \gg 1$,  and corrections are organized in powers of $w^{-1}$ (or, equivalently, in powers of $\mathrm{Kn}$).  
If the attractor constitutive relation is appropriate,
the energy density takes the form~\cite{Giacalone:2019ldn}
\begin{equation}
   \label{attractor}
  \frac{e(\tau) \tau^{4/3}}{ (e \tau^{4/3})_{\rm \infty} } =  \mathcal  E (w).
\end{equation}
Here $(e\tau^{4/3})_{\infty} \propto (\tau s)_{\infty}^{4/3}$  normalizes the entropy in the system and $\mathcal E \rightarrow 1$ at late times. 

Numerous simulations in the relaxation time approximation~\cite{Romatschke:2017vte,Strickland:2017kux,Strickland:2018ayk,Heller:2018qvh,Strickland:2019hff,Jaiswal:2022udf,Alalawi:2022pmg}, in complete QCD kinetic theory~\cite{Kurkela:2015qoa,Keegan:2015avk,Kurkela:2018wud,Kurkela:2018vqr,Kurkela:2019set,Almaalol:2020rnu}, and the AdS/CFT correspondence~\cite{Heller:2011ju,Heller:2012km,vanderSchee:2013pia,Heller:2013oxa,Casalderrey-Solana:2013aba,Chesler:2013lia,Chesler:2015wra,Keegan:2015avk,Spalinski:2017mel}, show that the simple interpolating form of Eq.~\eqref{attractor} captures the overall dynamics.
The form of $\mathcal E(w)$ is strongly constrained by its behavior at early and late times.  
At late times the entropy per rapidity $(\tau s)_{\infty} $ is constant,
while at  early times $\tau\rightarrow 0$ the energy per rapidity $(\tau \epsilon(\tau))_0$ is constant, fixing the behavior $\mathcal E \simeq C_{\infty}^{-1} w^{4/9}$~\cite{Giacalone:2019ldn}.
The constant $C_{\infty}$ depends only very weakly on the theory and is approximately unity. Simulations with QCD kinetic theory give $C_{\infty} \simeq 0.87$, while the AdS/CFT correspondence gives $C_{\infty}=1.06$ \cite{Heller:2011ju,Heller:2015dha}.

\begin{figure}
\centering
\includegraphics[width=0.8\textwidth]{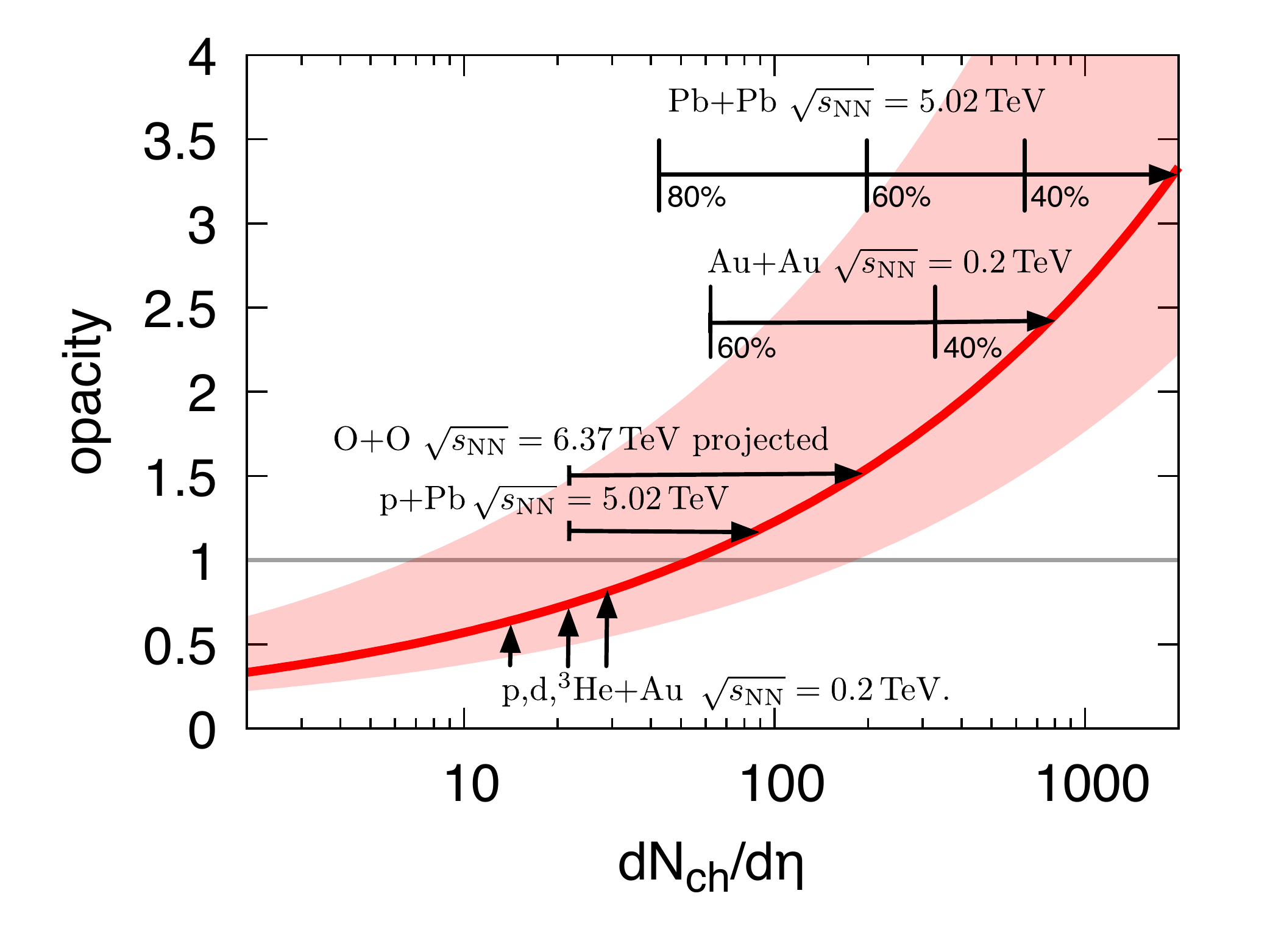}
\caption{The opacity, $\chi$, as a function of multiplicity defined  in Eq.~\eqref{chidef}. (See Eq.~\ref{eq:gammhatrelation} for relations to alternate definitions).
Hydrodynamics is a valid approximation for $\chi \gtrsim 1$. 
The solid line is for $\eta/s=2/4\pi$, and the band indicates 
the range $\eta/s=(1\leftrightarrow 3)/4\pi$, with $\eta/s=1/4\pi$ corresponding to the top of the band.   The multiplicity is shown for \PbPb{} at 
the LHC for a centrality range of 0-80\%~\cite{ALICE:2015juo}, and \AuAu{} at RHIC for a centrality range of 
0-60\%~\cite{PHOBOS:2010eyu}. Projections for the \OO{} run at the LHC are taken from~\cite{Brewer:2021kiv}.
\pPb{} data from the LHC are estimated from a recent analysis~\cite{Collaboration:2022xdv}.  The multiplicity from the geometry scan at RHIC is taken from~\cite{PHENIX:2018hho}.
}
\label{opacity}
\end{figure}

The first phenomenological importance of the hydrodynamic attractor is that it determines the energy density with 20\% accuracy during the initial stages from the measured charged particle multiplicity, almost irrespective of the underlying theory~\cite{Giacalone:2019ldn}.   
This level of precision can then significantly constrain models for the initial state, such as those built upon the color-glass condensate \cite{Gelis:2010nm}. The second phenomenological importance of the attractor solution is that it provides a criterion for the onset of hydrodynamics, a criterion which has been validated in detailed studies. Specifically, the hydrodynamic gradient expansion becomes appropriate at a time $\tau_{\rm hydro}$ when $w\simeq 1$ or larger.  
To translate this theoretical criterion into an experimental one,  we
first note that total entropy per rapidity in hydrodynamics is directly related to the hydrodynamic multiplicity~\cite{Kurkela:2018vqr} 
\begin{align}
   \frac{dS}{dy} =&\ b_{\rm hrg}  \frac{dN}{d\eta}    
\end{align}
where $b_{\rm hrg}\simeq 8.3$ is determined by the particle content of
the hadron resonance gas model and the lattice equation of state,
and has a 15\% uncertainty \cite{Kurkela:2018vqr}.
The  temperature at late times is determined by the constant entropy  or multiplicity
$dN_{\rm ch}/d\eta \propto \tau T^3$ leading to
\begin{align}
   w  \equiv \frac{\tau T}{4\pi \eta/s} =& 
   \ \frac{1}{4\pi \eta/s} \left(\frac{1}{N_0}\frac{dN_{\rm ch}}{d\eta} \right)^{1/3}   \left(\frac{\tau}{R} \right)^{2/3}
   \\
   \equiv &  \ \chi \left(\frac{\tau}{R} \right)^{2/3},
\end{align}
where we have defined the \emph{opacity}  of the system\footnote{
   There are 
   several definitions of this type of opacity in the literature. 
   In a  conformal relaxation time approximation,
   an opacity $\hat\gamma$ was defined from from
   the transverse energy at time zero~\cite{Kurkela:2019kip,Ambrus:2022qya,Ambrus:2022koq}.  Using the attractor
   solution $\hat \gamma$ can be related to the opacity defined here:
   \begin{equation} 
       \label{eq:gammhatrelation}
        \chi  =  C_{\infty}^{1/4} \left(\frac{5\hat \gamma}{4\pi}\right)^{8/9} \simeq 0.432\, {\hat{\gamma}}^{8/9}  .
   \end{equation}
   so that $\chi=1$ is $\hat\gamma\simeq 2.6$.
}
\begin{equation}
    \chi = \frac{1}{4\pi \eta/s} \left(\frac{1}{N_0}\frac{dN_{\rm ch}}{d\eta} \right)^{1/3}\,,   \qquad  N_0 \equiv \frac{\pi b_{\rm hrg} }{b_{\rm qgp}} \simeq 6.68 .
\end{equation}
We note that the multiplicity factor $N_0$ is primarily determined by the properties of the equation of state, which is known from  lattice QCD. Consequently the uncertainties 
in $N_0$ are only of order 20\%.
The system will have a strong hydrodynamic response if  $\tau_{\rm hydro}/R \ll 1$~\cite{Kurkela:2018wud}
\begin{equation}
    \frac{\tau_{\rm hydro}}{R}  = \frac{1}{\chi^{2/3} }  \propto \frac{1}{\sqrt{dN_{\rm ch}/d\eta} } \, ,
\end{equation}
which  amounts to a requirement that the opacity is large $\chi \gg 1$.
The final expression for  the opacity is
\begin{equation}
\label{chidef}
      \chi =  \left(\frac{2}{4\pi \eta/s}\right) \left(\frac{dN_{\rm ch}/d\eta}{54}  \right)^{1/3} .
\end{equation}
Figure~\ref{opacity} shows this opacity as a function of multiplicity and the range of 
systems explored in heavy-ion collisions. 

\noindent {\bf Transverse Expansion}. 
The attractor solution $\mathcal E(w)$ for the energy density is valid before the transverse expansion begins.  Viscous corrections in this regime are organized in inverse powers of $w$: \pagebreak
\begin{align}
   \mathcal E(w) =& 1 + \frac{C_1}{\chi}   \left(\frac{\tau}{R}\right)^{-2/3} + \frac{C_2 }{\chi^2}  \left(\frac{\tau}{R}\right)^{-4/3}  +  \ldots  \\ \nopagebreak[4]
   \Rightarrow & 1 + \frac{C_1}{\chi} +  \frac{C_2}{\chi^2} + \ldots\,.
 \end{align}
At the boundary of applicability,  when $\tau \simeq R$, the attractor $\mathcal E(w)$ is only a function of $\chi$.  Since the subsequent three-dimensional hydrodynamics must match with the attractor solution at early times, the transverse hydrodynamic response is only a function of the opacity $\chi \propto (dN_{\rm ch}/d\eta)^{1/3}$, and viscous corrections to the transverse flow are organized in powers of $1/\chi$~\cite{Basar:2013hea}.

This has a number of consequences. First, since
the opacity for a conformal system is only a function of multiplicity and not the radius,  
systems with different sizes but the same multiplicity should exhibit similar transverse flow.  
Experimentally, it is observed that at the same multiplicity the fluctuation driven $v_2$ and $v_3$ are the same in the \pPb{} and \PbPb{} to 5\% accuracy~\cite{CMS:2012qk,Basar:2013hea}. 
This seems to corroborate the conformal assumptions~\cite{Basar:2013hea}, provided a minimal random cluster model (with the number of clusters proportional to the multiplicity) is adopted for the geometric fluctuations.
We should emphasize here that there are corrections to the conformal characterization of the system, which remain to be fully clarified.

Additional corrections to the transverse response  stem from the pre-equilibrium evolution and are not captured by hydrodynamics even for $\chi \gg 1$. They are of order  $\tau_{\rm hydro}/R \propto 1/\chi^{3/2}$ and are suppressed compared to the first viscous correction, 
which is of order $1/\chi$~\cite{Kurkela:2018qeb,Kurkela:2019kip,Ambrus:2022qya}.
The non-equilibrium response modifies the initial conditions for hydro by an amount of order $\tau_{\rm hydro}/R$. 
Conversely, for $\chi>1$ the non-equilibrium dynamics plays a prominent role and hydrodynamics is not clearly applicable.

\subsubsection{Medium response of partonic excitations}
\label{sec:progress:mesoscopic:medium_response}

Jets, the collimated spray of particles resulting from a hard scattering, are typically used to probe the medium properties at the microscopic level. However, the response of the QGP to the hard scattered parton provides insight at the mesoscopic level and into the interplay between soft and hard physics. 
    
    Medium response refers to the excitation of the medium caused by the jet-medium interaction and plays a non-negligible role in the transport of jet energy and momentum.    In the jet-medium interaction, some jet momentum is transferred to medium constituents or recoil particles, which further propagate in the medium and interact with other medium constituents. 
    These recoil medium constituents carrying some of the jet energy and momentum undergo the hadronization process mostly through coalescence and contribute to low-momentum hadrons spreading around the jet axis. Since these soft hadrons are correlated with the jet, they, in principle, should be considered as part of the jet and not be subtracted as a background. The uncorrelated background can be computed theoretically through simulations of the same events without the triggered jet and subtracted from the events with the jet. Experimentally, such background subtraction without affecting hadrons from the medium response is still a challenge. 

    The medium response not only contributes to the jet reconstruction and jet-correlated observables but is also closely related to some interesting phenomena related to the properties of the QGP medium.
    Since recoil medium constituents go through further interaction, this part of the jet energy and momentum they carry will be spread to more recoil and radiated partons leading to thermalization of the energy and momentum lost by the leading jet partons .
    Thus, the medium response can address a fundamental question in investigating the QGP properties: how do non-equilibrium partons equilibrate in a QGP heat bath?  
    When a fluid gains energy and momentum from a source moving faster than the speed of sound, 
    a conical-shaped shock wave, the so-called Mach cone, forms.  The angle of the Mach cone is determined by the medium's sound velocity, while the shock wave's width is influenced by the shear viscosity~\cite{Neufeld:2008dx}. The jet-induced medium response, therefore, also reflects the  hydrodynamic transport properties of QGP. 
    Furthermore, through the interplay between the jet-induced medium response and background flow, jets have the potential to distinguish among different possible global flow patterns of the background QGP fluid~\cite{Tachibana:2015qxa,Yang:2021qtl,Tachibana:2020mtb}. 

\begin{figure}[htb]
\centering
\includegraphics[width=0.4\columnwidth]{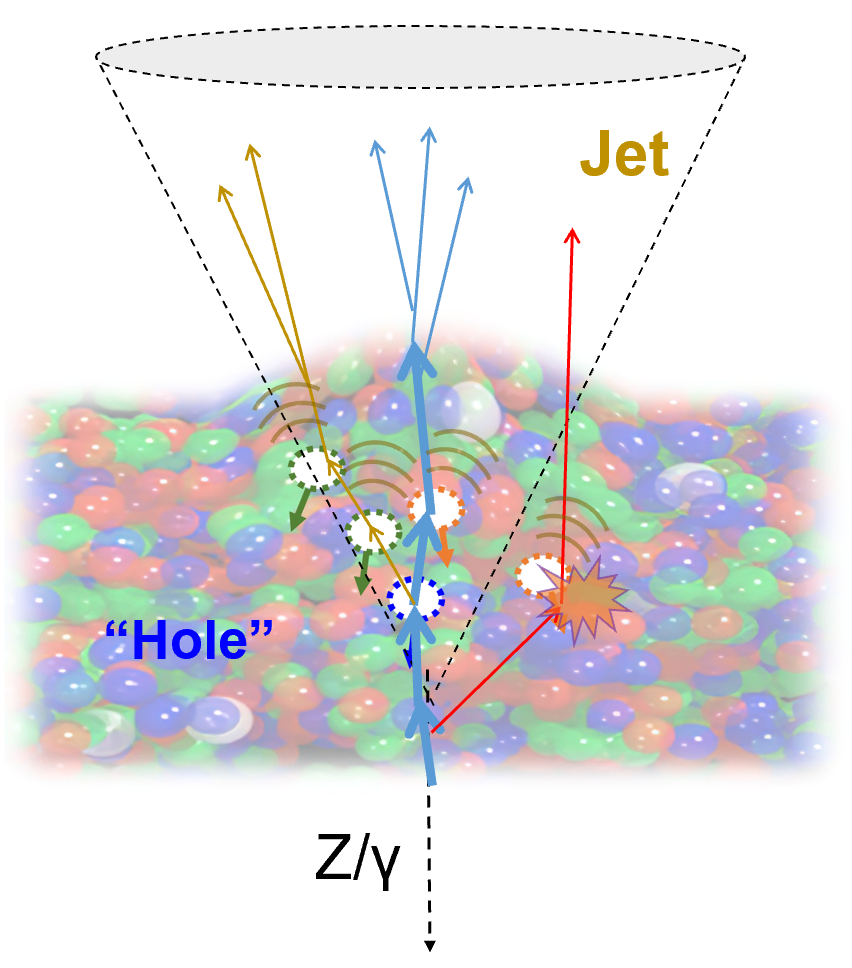}
\caption{Illustration of a parton traversing the medium and forming a jet opposite a direct photon or Z boson.}
\label{fig:medium-response-holes}
\end{figure}

    In some Monte-Carlo models of the in-medium jet shower,     the medium response is implemented as the transport of     recoil partons~\cite{Li:2010ts,Zapp:2012ak,Zapp:2013vla,Cao:2017hhk,Luo:2018pto,Park:2018acg} in a linear Boltzmann transport approximation. In this recoil prescription, the energy-momentum exchange between the jet and the medium is mediated by 2-to-2 parton scatterings. 
    The initial medium parton in these scattering processes is sampled from the thermal distributions. These medium partons, after the scattering, will propagate and interact with the medium just like other jet partons.
    The successive interactions between the recoil and the medium partons form a structural wake of recoils, which spreads like a Mach cone. 
    In addition to jet-induced medium response in the form of the Mach-cone-like excitation, jet-medium interaction kicks the medium parton into a recoil particle and leaves behind the ``holes'' (Figure~\ref{fig:medium-response-holes}), leading to a diffusion wake in the medium.

\begin{figure}[htb]
\centering
\includegraphics[width=0.42\columnwidth]{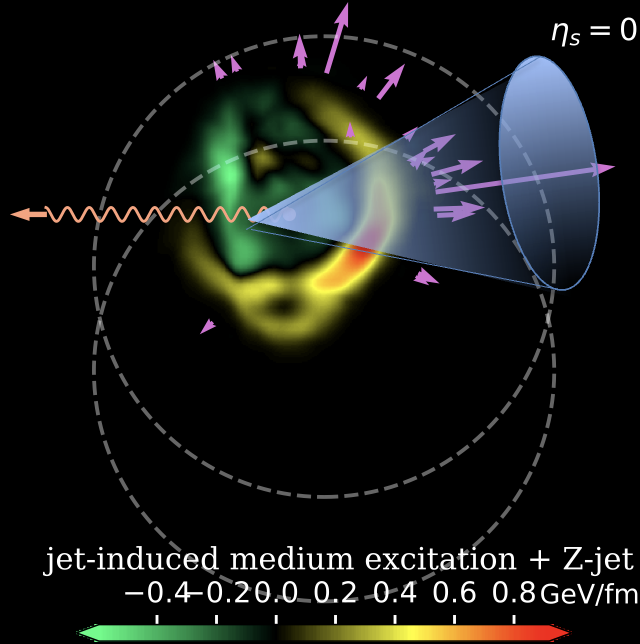}
\caption{Mach cone induced by a $Z$-triggered jet.}
\label{fig:machcone}
\end{figure}

\begin{figure}[htb]
\centering
\includegraphics[width=0.69\columnwidth]{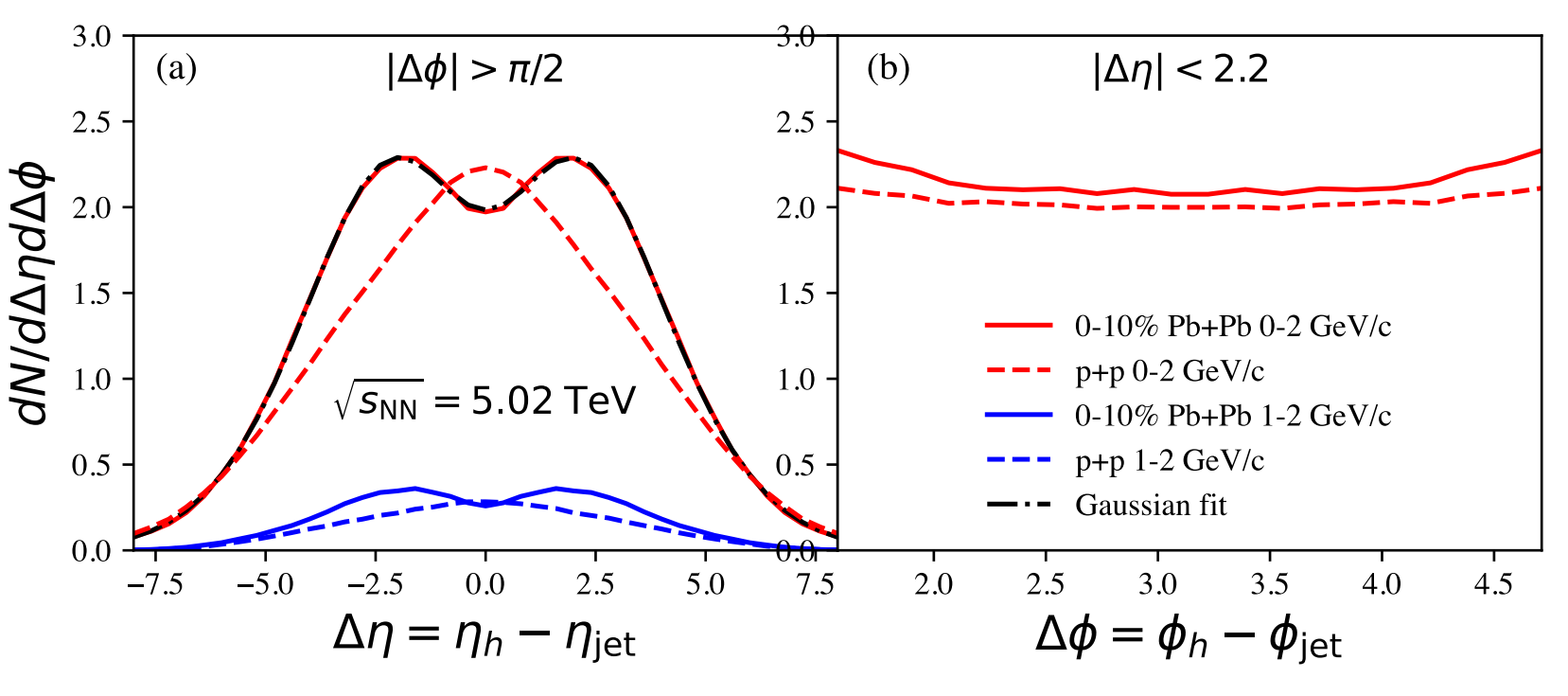}
\includegraphics[width=0.3\columnwidth]{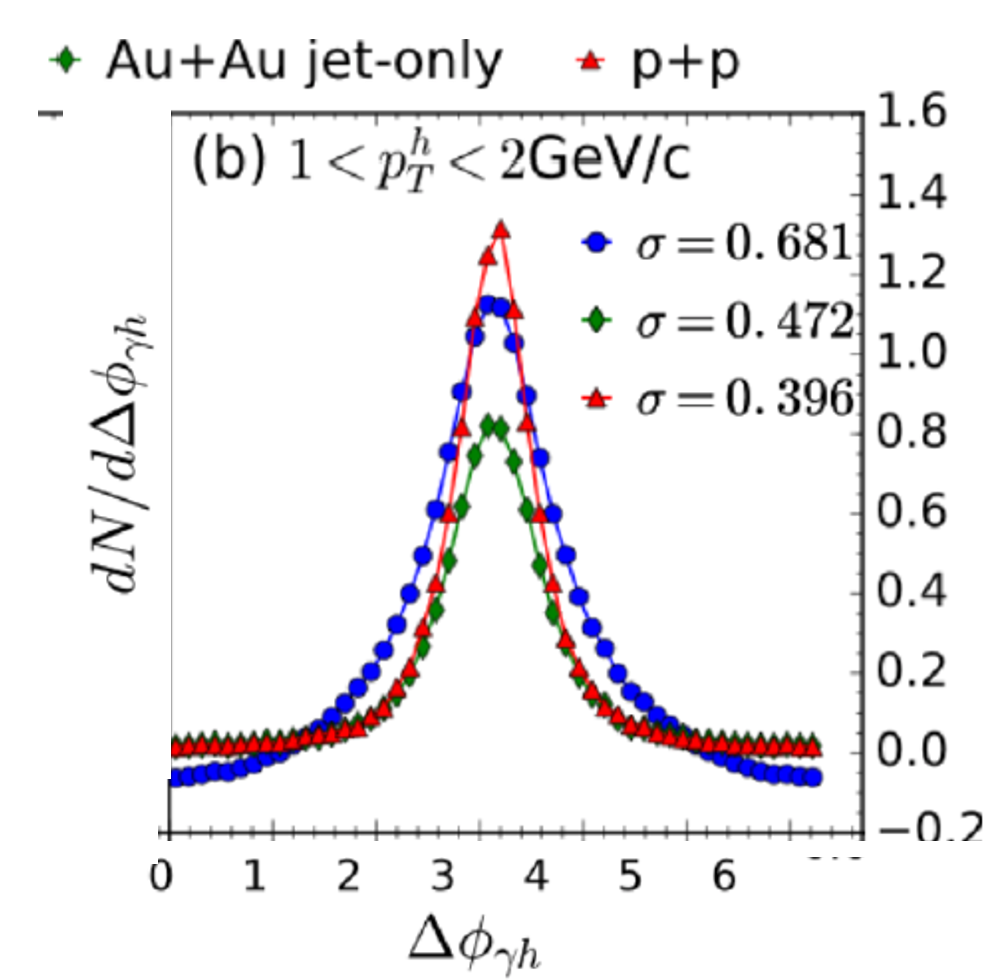}
\caption{CoLBT model results on the $\gamma$-jet-hadron correlation in both azimuthal angle and rapidity.}
\label{fig:3.1.2:jet-h-corr}
\end{figure}

    The approximation of the recoil prescription breaks down when the energy of the jet shower parton drops to the typical energy scale for the thermal medium constituents. The number of recoil partons becomes large beyond the linear Boltzmann transport approximation. 
    To extend the model’s applicability to this regime, one can solve the full Boltzmann equations directly. In a static and uniform medium, one can complete some of the integrals over the relative angles among jet, medium and recoil partons analytically in the collision kernel and solve the reduced Boltzmann equations numerically \cite{Schlichting:2020lef,Mehtar-Tani:2022zwf}. The solutions can illustrate the thermalization of recoiled partons and the formation of medium response, including the diffusion wake.
    One can also apply the hydrodynamic description to the soft modes of jet energy-momentum transport together as well as the bulk dynamics of the medium~\cite{Chaudhuri:2005vc,Renk:2005si,Satarov:2005mv,Neufeld:2008fi,Noronha:2008un,Qin:2009uh,Betz:2010qh,Neufeld:2011yh,Schulc:2014jma,Tachibana:2014lja,JETSCAPE:2020uew}. 
    This can be done by solving the hydrodynamic equation with the source term~\cite{Tachibana:2017syd,Okai:2017ofp,Chen:2017zte,Chang:2019sae,Tachibana:2020mtb,Casalderrey-Solana:2020rsj,Yang:2021qtl,Yang:2022nei,Pablos:2022piv,Yang:2022yfr}. 
    \begin{align}
    \nabla_{\mu}T_{\rm fluid}^{\mu \nu}(x)&=J^{\nu}(x).
    \label{eq:HydroEqWithSource}
    \end{align}
    Here $T^{\mu\nu}_{\rm fluid}$ is the energy-momentum tensor of the medium fluid including the thermalized part of jet energy and momentum, and $J^{\nu}$ is the source term representing the space-time profile of the energy-momentum deposition from the non-thermalized part of jets from the jet transport. One can describe the jet propagation and hydrodynamic evolution of the medium response in a coupled jet transport and hydrodynamic model \cite{Chen:2017zte,Chen:2020tbl,JETSCAPE:2020uew,Tachibana:2020mtb} where the source term in the hydrodynamic equations comes from jet transport in a medium whose evolution is governed by the hydrodynamic equations. The state of art approach is a concurrent simulation of the jet transport and hydrodynamic evolution coupled through a source term in the hydrodynamic equations constructed with partons from jet transport that are considered thermalized in the local medium \cite{Chen:2017zte,Chen:2020tbl,Yang:2022nei,Yang:2021qtl,Zhao:2021vmu}.
    The deposited energy and momentum propagate as a hydrodynamic flow of a Mach cone in the medium, which allows the final state bulk medium-derived hadrons to correlate with the jet. In the models, those jet-correlated medium hadrons are obtained at the same time as the hadrons in the whole bulk medium, without distinction, using the Copper-Frye formula \cite{Cooper:1974mv} as in the conventional hydrodynamic models. One can calculate hadron yield from the medium response by subtracting the same hydro events without jets (Figure~\ref{fig:machcone}). Energy-momentum from the medium response including the energy-momentum depletion due to the diffusion wake should be included in the jet reconstruction and calculation of jet profiles and fragmentation functions. Since one cannot unambiguously separate soft hadrons from medium response and radiation in and around the jet direction, the depletion of soft hadrons due to the diffusion wake is a unique phenomenon related to jet-induced medium response.  The jet-hadron correlations for soft hadrons in both azimuthal angle and rapidity are found to have some unique features due to transverse momentum broadening and diffusion wake in the medium response (Figure~\ref{fig:3.1.2:jet-h-corr}).

The medium response and medium recoil effects have been first studied by dijet missing transverse momentum measurements. 
Since then, the analyses of the jet-induced excitation included the broadening of the jet profile at a large radius~\cite{CMS:2018zze}, cone-size dependence of the jet suppression~\cite{CMS:2021vui}, and enhancement of low \pt particles in the jet fragmentation function~\cite{ATLAS:2018bvp} and jet-hadron angular correlation~\cite{ATLAS:2019pid}. 
Moreover, $Z$/$\gamma$ tagged jets are unique and clean probes because electroweak bosons do not participate in the strong interaction and serve as a tagger of the initial hard scattering. 
Measurements of $\gamma$-hadron correlations~\cite{ATLAS:2019dsv,CMS:2018jco,PHENIX:2020alr} have provided new insights into the medium response and medium recoil effects near the outgoing hard-scattered partons.
Recently, $Z$-hadron correlations have also been carried out, with the motivation to search for the reduction of the associated yield near the $Z$ bosons \cite{ATLAS:2020wmg,CMS:2021otx}. 

\subsubsection{Jet Modification in Small Systems}

With the overwhelming evidence of collectivity in small systems, it is natural to explore other signatures of QGP formation such as the modification of hard jets. However, there are several caveats to such an exploration. In the simplest possible terms, the radiative energy loss of a hard parton in a dense (almost static) medium may be expressed using the BDMPS formula~\cite{Baier:1996sk}, 
\begin{equation}
    \Delta E \propto \hat{q} L^2. \label{Baier_pocket_formula}
\end{equation}
Where, $\hat{q}$ is the jet transport coefficient described in the subsequent subsection, and $L$ is the length traversed by the parton. While more realistic expressions for energy loss in dynamical media will possess more dependencies, the basic fact remains that radiative energy loss of a hard parton is approximately proportional to the square of the length traversed by the hard parton.

\begin{figure}[!htb]
\centering
\includegraphics[width=0.325\textwidth]{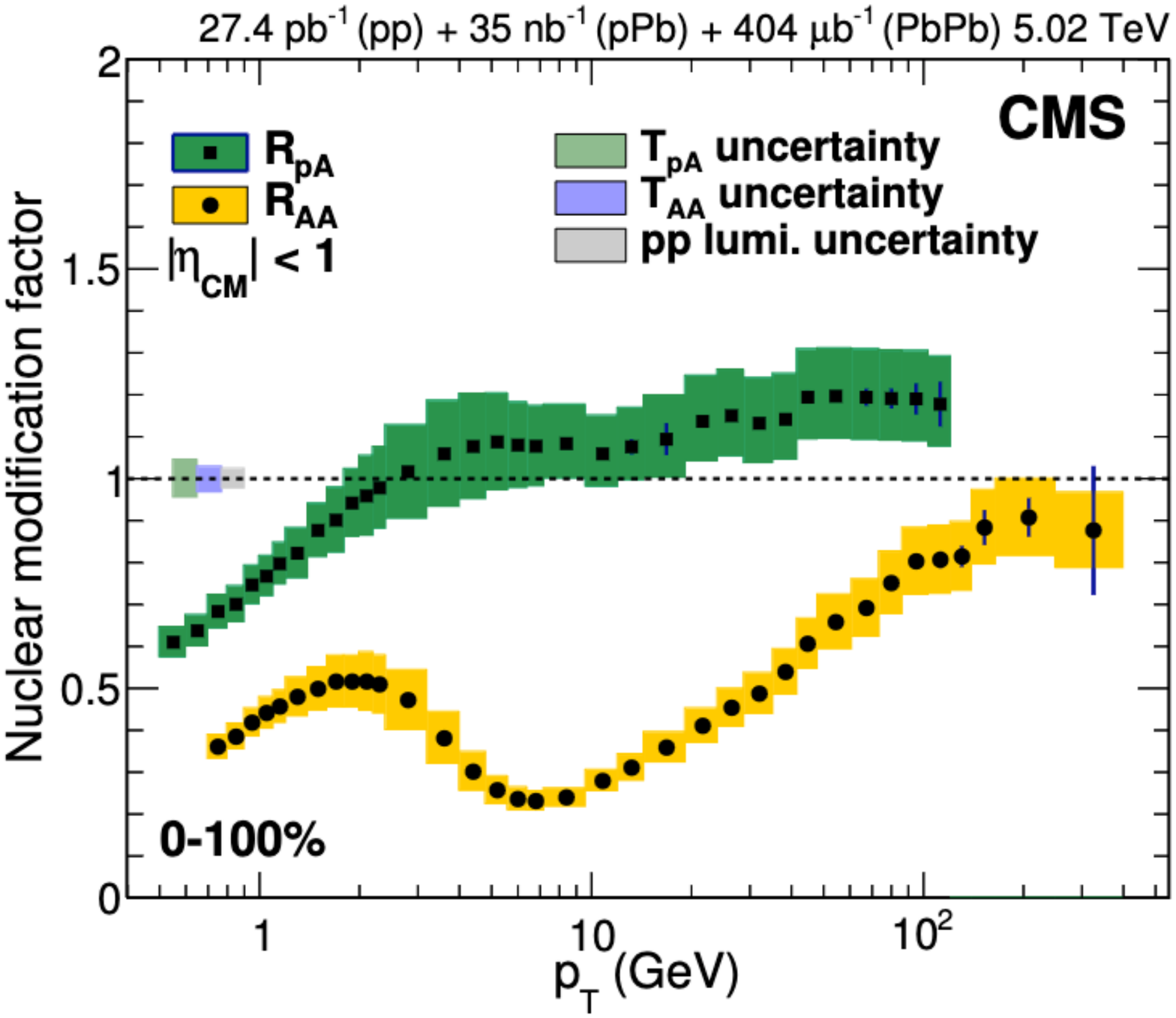}
\includegraphics[width=0.325\textwidth]{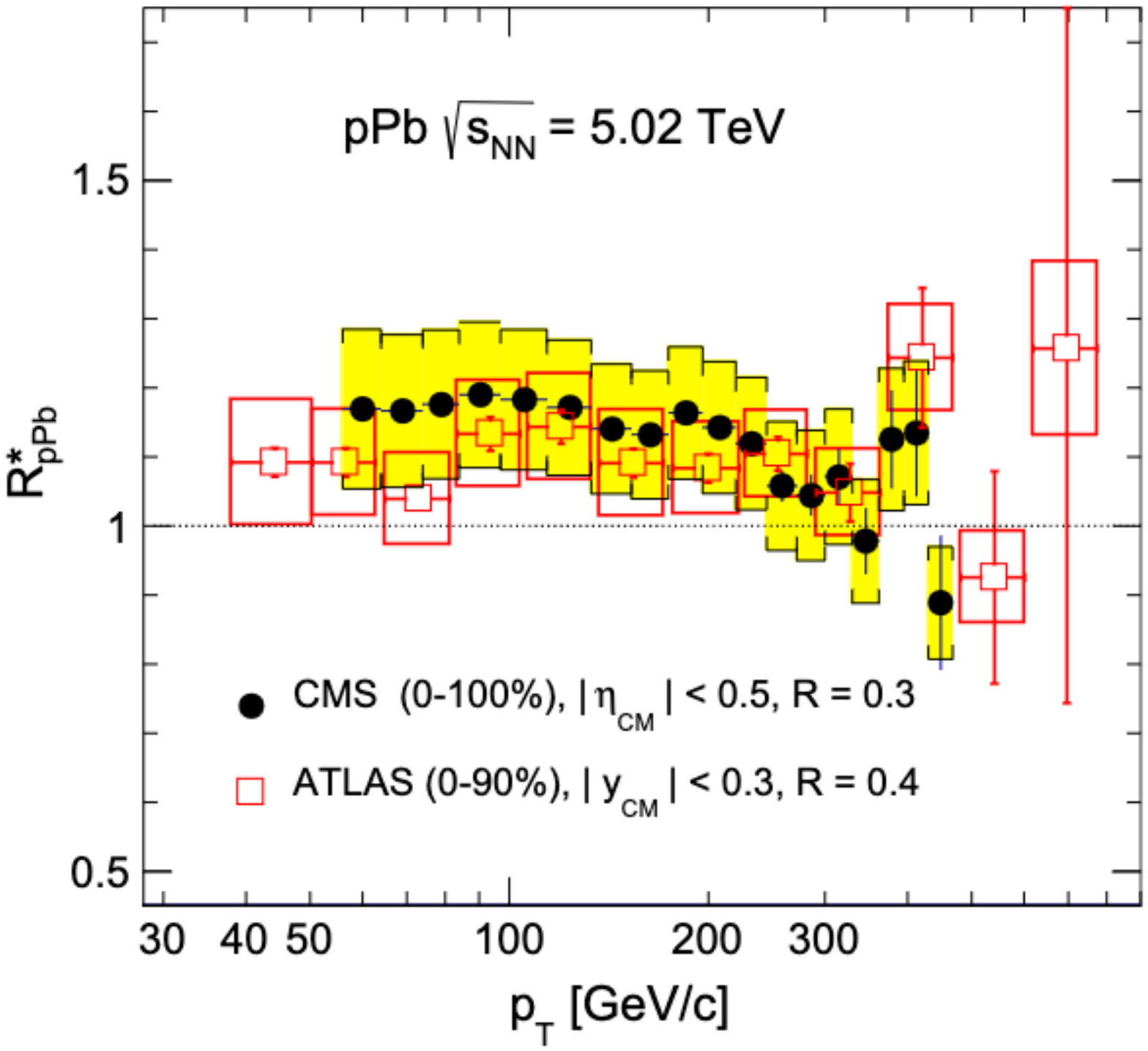}
\includegraphics[width=0.325\textwidth]{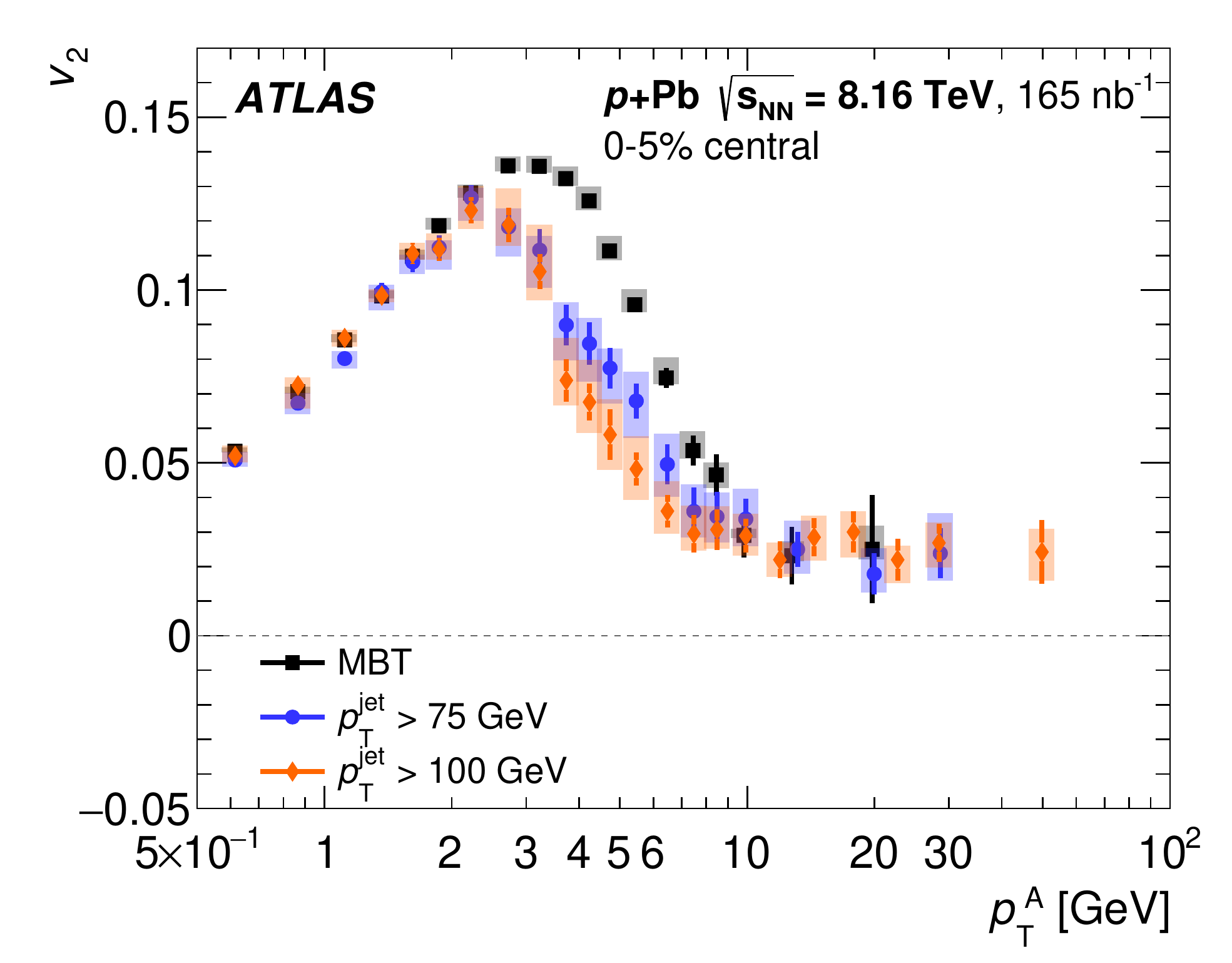}
\caption{(Color online) Left panel shows the minimum bias nuclear modification for charged hadrons in $p$-$Pb$ collisions ($R_{pA}$) at 5.02~TeV, compared to the minimum bias $R_{AA}$ in $Pb$-$Pb$ collisions at 5.02~TeV. Center panel shows the $R_{pPb}$ for jets in minimum bias collisions. The $R_{pPb}$ in both the left and center panels are consistent with unity indicating no modification of jets or high-$p_T$ hadrons in the medium created in $p$-$Pb$ collsions. The right panel shows the the second Fourier coefficient ($v_2$) of the azimuthal anisotropy of high-$p_T$ hadrons in 0-5\% most central collisions at $8.16$~TeV. This figure suggests a remnant anisotropy, even at $p_T \gtrsim 50$~GeV.}
\label{high_pt_pA}
\end{figure}

As a result, energy loss of hard partons is expected to drop rapidly with system size. In fact, the simplest scan with system size as observed in the centrality dependence of jet modification observables does reveal a pattern consistent with Eq.~\eqref{Baier_pocket_formula} (See Sect.~\ref{sec:progress:microscopic:JETSCAPE} for details and plots). Following this line of inquiry, several studies of the nuclear modification factors of jets and high-$p_T$ hadrons, in minimum bias $p$-$A$ collisions, have yielded results consistent with no modification~\cite{CMS:2016xef} (See the left and center panel in Fig.~\ref{high_pt_pA}). However, the hadron $v_n$ at high $p_T$ was measured to be non-zero~\cite{ATLAS:2014qaj, ATLAS:2019vcm} (right panel in Fig.~\ref{high_pt_pA}). In A+A collisions non-zero $v_n$ is interpreted as resulting from path-length dependent energy loss (see the right panel in Fig.~\ref{high_pt_pA}). This has caused great interest in the exploration of other observables~\cite{ATLAS:2022iyq, PHENIX:2018foa, ALICE:2015ppz} that could be more sensitive to small amounts of energy loss. This has also raised the question: In how small of a system can jet suppression be observed? We will revisit this question in \ref{sec:future:microscopic:jets}.

\clearpage 

\subsection{Microscopic I: Jets and leading hadrons}
\label{sec:progress:microscopic}

Collisions of protons and nuclei often engender hard scatterings with transverse momentum transfers $\delta p_T \gg \Lambda_{QCD}$ which lead to the formation of hard partons, with energies $E\gg Q$ the virtuality of the parton. At high enough energy, the virtuality $Q$ is itself much larger than $\Lambda_{QCD}$ ($Q\gg \Lambda_{QCD}$). The decay of these hard partons occurs via the repeated emission of progressively softer partons. Due to the condition $E\gg Q$, the ensuing spray of partons is collinearly  collated into jets of partons. It is important to note that a jet is defined by the jet-finding algorithm used to cluster the resulting fragmentation products into a jet. 
The large scales involved allow for the use of pQCD to calculate the cross section and various other properties of these jets~\cite{Sterman:1977wj,Ellis:1980wv,Bassetto:1983mvz,CTEQ:1993hwr}.

As these jets traverse the dense QGP formed in heavy-ion collisions, they are modified~\cite{Bjorken:1982tu}. 
The scattering in the medium leads to the production of more radiation from the jet, leading to more energy being radiated at larger angles, referred to as jet quenching~\cite{Wang:1992qdg,Baier:1996kr,Zakharov:1997uu}. 
Over the last two decades, there has been a tremendous amount of development in this field: At the start of the RHIC program, various formalisms, all based on pQCD were focused on the study of leading hadrons that emanated from the fragmentation of leading partons which suffered energy loss in the medium~\cite{Gyulassy:2000fs,Wang:2001ifa,Arnold:2002ja,Salgado:2002cd}. Over the first decade at RHIC, there arose a series of efforts to understand energy loss via the AdS/CFT conjecture~\cite{Maldacena:1997re}, both for light~\cite{Chesler:2008uy, Liu:2006ug} and heavy flavors~\cite{CasalderreySolana:2007qw}. With the onset of the LHC and the availability of high statistics jet and high-$p_T$ hadron data at very high energies (in the $100$'s of GeV), the focus has moved to a multi-stage process~\cite{Cao:2017zih,Caucal:2018dla} with an initial vacuum-like shower with a diminished coupling with the medium~\cite{Kumar:2019uvu, Mehtar-Tani:2011hma}, followed by the multiple scattering induced energy loss phase~\cite{Baier:1994bd,Baier:1996kr,Zakharov:1997uu,Arnold:2002ja}.

In the period after the last \LRP, the community has now moved far beyond the energy lost by the hard partons in the jet to a deep study of how the energy of various fragments of the jet is redistributed by the medium (jet substructure), to the reappearance of the energy in the medium from the recoil of the hard partons~\cite{Li:2010ts,He:2015pra, KunnawalkamElayavalli:2017hxo, JETSCAPE:2022jer,JETSCAPE:2023hqn}, to the appearance of a wake of the jet in the soft medium~\cite{Tachibana:2017syd,Chen:2017zte,Chang:2019sae,Tachibana:2020mtb}. Likewise experimental results now include a wide range of observables beyond jets and leading hadrons, such as jet azimuthal anisotropy~\cite{ATLAS:2021ktw}, groomed~\cite{ALargeIonColliderExperiment:2021mqf} and un-groomed~\cite{CMS:2018zze} jet substructure, coincidence observables~\cite{ALICE:2019sqi} etc.  

The orders of magnitude increase in the number and statistical precision of jet measurements have opened up extensive theoretical possibilities in the field. It is now possible to compare different approaches to various experimental observables. To make this process rigorous each energy loss module should be constrained to employ the same initial state to generate both the hard and soft sectors, lose energy in an identical fluid simulation, and be subjected to the same hadronization routine. This can only be achieved using an event generator framework, where modules that simulate different aspects of a collision can be replaced in an extensive end-to-end simulation. Since the last \LRP, such a framework called JETSCAPE is now available~\cite{Putschke:2019yrg}.

As an illustration of the abilities of such frameworks, we present recent results from JETSCAPE covering a wide range of jet, leading hadron, jet substructure and jet coincidence observables in the first subsection. Simulations carried out within these end-to-end generator frameworks typically do not contain the newest theories or exploratory approaches and still have not addressed the entirety of all jet based observables. As such, in the subsequent sections, each of these observables will be discussed in detail, highlighting new theoretical improvements, along with a discussion of experimental observables that have yet to be satisfactorily addressed by theory or simulation. The final subsection will address the somewhat new development of Bayesian analysis for jet based observables, and the ability to study the cross correlation between various parameters in data driven approaches. 

\subsubsection{Simultaneous results from end-to-end simulator frameworks}
\label{sec:progress:microscopic:JETSCAPE}

For years after the start of the RHIC and even the LHC program, (up to and including the last \LRP) the \emph{modus operandi} of most theorists, working in jet quenching, was to carry out calculations of individual observables. Since the onset of end-to-end simulation frameworks such as JETSCAPE, one can now focus on the calculation of \emph{events}. Observables are built by combining final state particles from these events. One can also look at observables from intermediate states in the evolution, to directly see how different stages of the evolution initiate and change each observable. 

In the following, we present recent results from JETSCAPE, where almost the entire range of jet based observables (with the exception of jet $v_2$) was addressed by a single series of simulations~\cite{JETSCAPE:2022jer,JETSCAPE:2023hqn} with only two jet based parameters tuned to data at one energy and centrality. 
These parameters are the normalization of the jet transport coefficient $\hat{q}$ and the transition scale $Q_{SW}$, between the high and low virtuality stages of a jet quenching simulation. The jet propagation is carried out on top of a pre-calibrated bulk simulation. The presented results  
used the T$_{\rm{R}}$ENTo model as the initial state for both bulk and jet evolution, followed by energy loss via a 2 stage (MATTER + LBT) simulation in a 2+1D (MUSIC) fluid dynamical simulation, followed by hadronization of the soft sector via Cooper-Fry and a Lund model based hadronization (Pythia) for the hard sector.

\begin{figure}[!htb]
\centering
\includegraphics[width=0.45\textwidth]{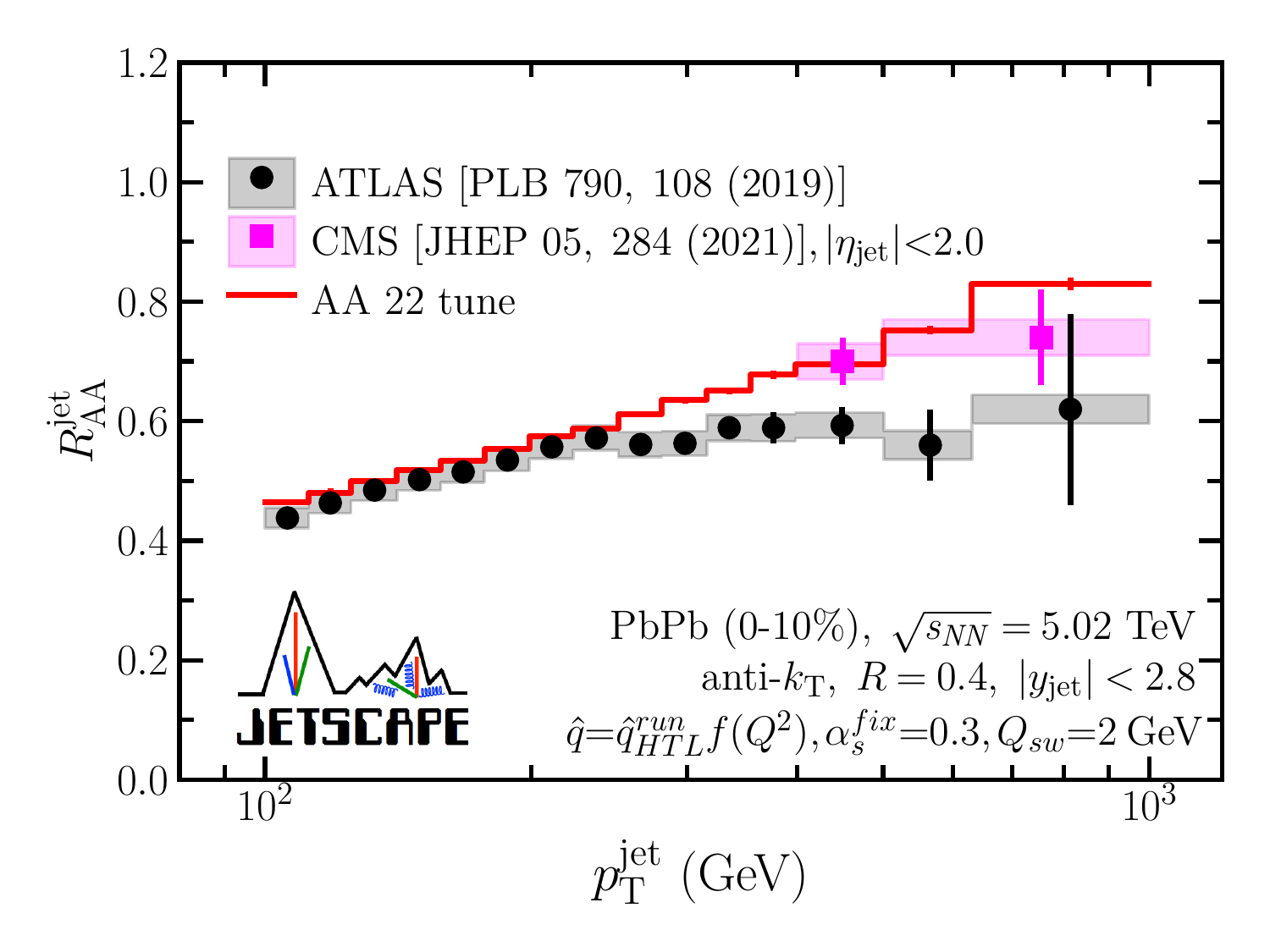}
\includegraphics[width=0.45\textwidth]{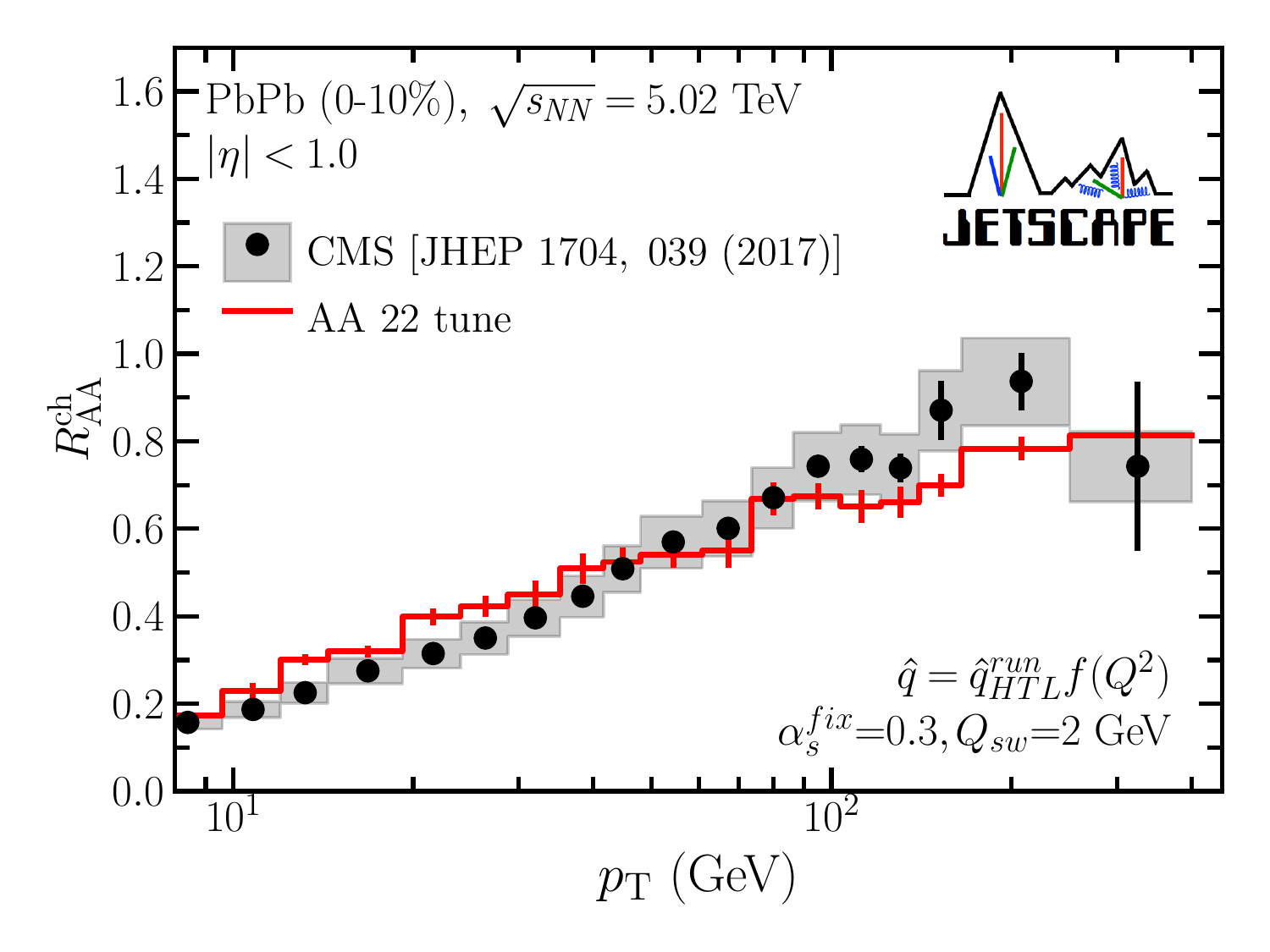}
\caption{(Color online) Nuclear modification factor for inclusive jets (left) and charged-particles (right) in most central (0-10$\%$) Pb+Pb collisions at $\sqrt{s_{\mathrm{NN}}}=5.02$~TeV 
from MATTER+LBT simulations within JETSCAPE. The fit to these data from ATLAS and CMS set the two parameters $\alpha_{\rm S}^{\rm fix} = 0.3$ and $Q_{\rm SW}=2$~GeV. All JETSCAPE results below will use these parameters. (Left) Results for inclusive jet $R_{\mathrm{AA}}$ 
with $R=0.4$ and $y_{\mathrm{jet}} < 2.8$, compared to ATLAS data~\cite{ATLAS:2018gwx} (black circles)
and 
CMS data for 
$\eta_{\mathrm{jet}} < 2.0$~\cite{CMS:2021vui}
(magenta squares). (Right) Results for charged-particle $R_{\mathrm{AA}}$ with $\eta<1.0$, compared to CMS data~\cite{CMS:2016xef}.} 
\label{jetscape-jet-hadron-5TeV-central}
\end{figure}

In these calculations, MATTER~\cite{Majumder:2013re, Cao:2017qpx} simulates the high-virtuality stage with coherence effects weakening the coupling with rising off-shellness of the parton~\cite{Kumar:2019uvu}. This is indicated with the factor $f(Q^2)$, which reduces with increasing $Q^2$ [$f(Q^2)$ is calculated from Ref.~\cite{Kumar:2019uvu}, \emph{not} determined from data]. This factor is combined with a jet transport coefficient calculated using running coupling in leading order HTL effective theory~\cite{Arnold:2008vd}:
\begin{align}
\hat{q}^\mathrm{run}_\mathrm{HTL} &= C_{a}\frac{50.484}{\pi} \alpha^\mathrm{run}_\mathrm{s}(Q_{\rm max}^2) \alpha^\mathrm{fix}_\mathrm{s} T^{3} \ln\left[\! \frac{2ET}{m^2_\mathrm{D}}\! \right],
\label{eq:type2-q-hatform}
\end{align}

The lower virtuality phase is simulated with LBT~\cite{He:2015pra}. The transition between MATTER and LBT is set at a $Q=2$~GeV by comparison with data. Both MATTER and LBT employ a similar recoil prescription where a parton is sampled from the local medium and then scattered off a jet parton. The recoil parton becomes a part of the evolving jet, while the holes left by the incoming partons are subtracted from the final observables, if they lie within jet cones.

\begin{figure}
    \centering
    \includegraphics[width=0.9\columnwidth]{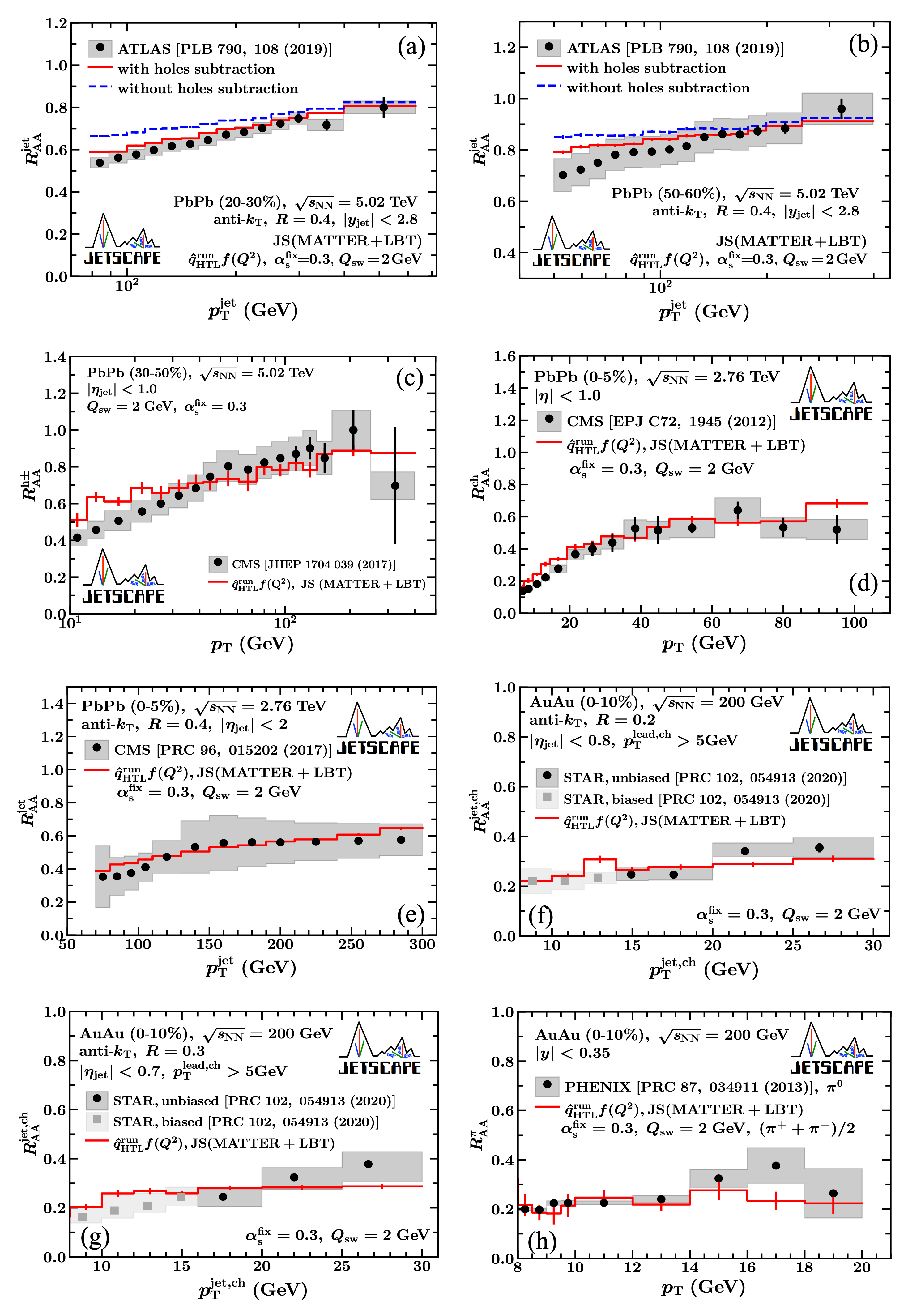}
    \caption{Red calculations have no further tuning after fitting inclusive jet $R_{\mathrm{AA}}$ and charged-particle $R_{\mathrm{AA}}$ at most central (0-10$\%$, $\sqrt{s_\mathrm{NN}}$=5.02~TeV) Pb+Pb  collisions and are compared to various measurements (black circles) across a variety of centrality classes, collision energies and species, and kinematic ranges. %
}
    \label{fig:jetscape-multipanel}
\end{figure}

In Figs.~\ref{jetscape-jet-hadron-5TeV-central}, and \ref{fig:jetscape-multipanel} %
we demonstrate that the jet and leading hadron $R_{AA}$ at all centralities  from top LHC to top RHIC energies can be simultaneously described by the same simulation with no refitting of parameters between energy or centrality. Unlike the results from the prior JET collaboration~\cite{JET:2013cls}, which required setting the normalization of $\hat{q}$ at RHIC and LHC separately, current simulations from JETSCAPE only require the normalisation and the transition scale to be determined at one energy and centrality.

In Fig.~\ref{jetscape-jet-hadron-5TeV-central}, the normalization of $\hat{q}$ determined by the parameter $\alpha_{\rm S}^{\rm fix} =0.3$, and the transition scale $Q_{\rm SW} = 2$~GeV are set by comparing with the data on inclusive jets (left) and leading hadrons (right). In Fig.~\ref{fig:jetscape-multipanel} (a) and (b) we show the comparison between simulation and ATLAS data~\cite{ATLAS:2018gwx} for $R_{\mathrm{AA}}$ of jets in semi-central events. These plots also demonstrate the effect of removing and not removing the holes created by the in-coming medium partons which are scattered into becoming a part of the jet.

In Fig.~\ref{fig:jetscape-multipanel} panels (c) and (d), the centrality and collision energy dependence of the leading hadron suppression is demonstrated in comparison with data from semi-central LHC collisions at $\sqrt{s_{\rm NN}} = 5$~TeV and central collisions at $2.76$~TeV. The collision energy dependence of jets %
is demonstrated with LHC and RHIC data in panels (e) and (f) of Fig.~\ref{fig:jetscape-multipanel} respectively. Finally jets and charged pions at RHIC energies are compared with JETSCAPE simulations in panels (g) and (h) of Fig.~\ref{fig:jetscape-multipanel}, all of whose parameters were determined in Fig.~\ref{jetscape-jet-hadron-5TeV-central}.

\begin{figure*}[htb!]
\centering
\includegraphics[width=0.84\textwidth]{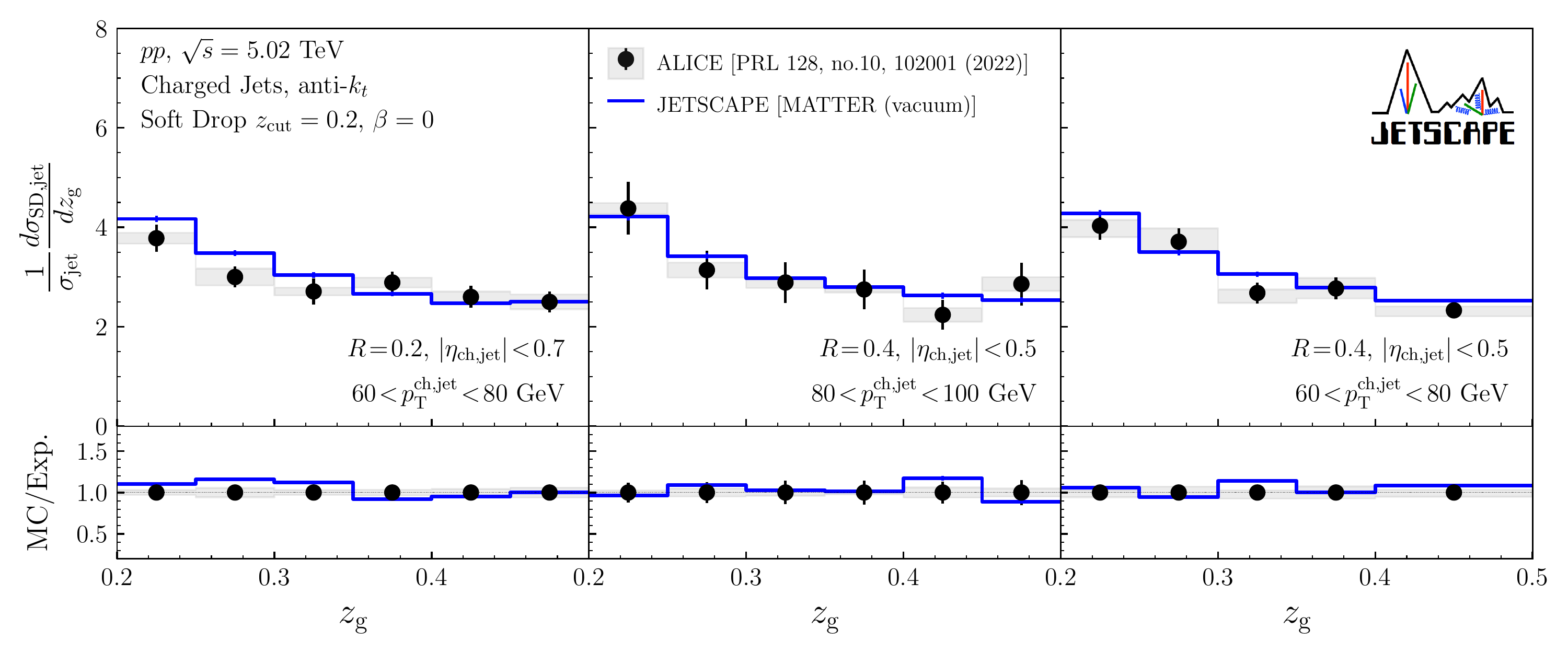}
\includegraphics[width=0.84\textwidth]{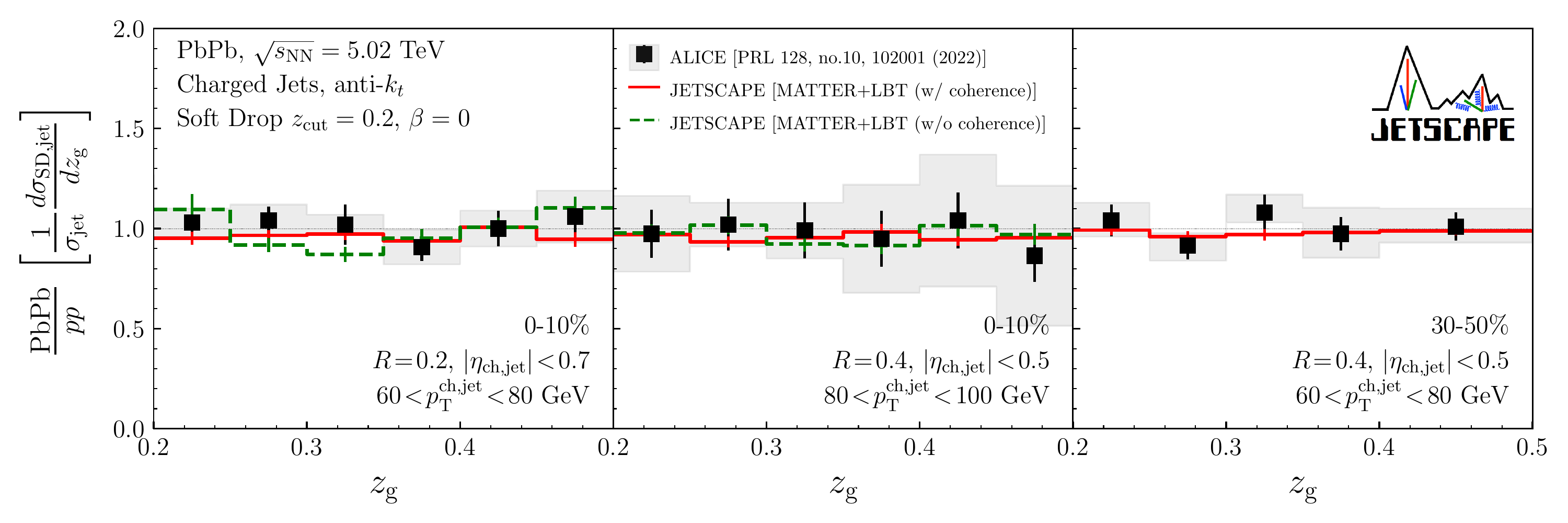}
\caption{(Color online) (Top) Distributions of soft dropped jet splitting fraction $z_{g} = \frac{\min[z_1,z_2]}{z_1+z_2}$ for charged jets 
in $p$+$p$ collisions at $\sqrt{s}=5.02$~TeV and the ratios 
for different jet cone size $R$, and $p^{\mathrm{ch,jet}}_{T}$ range. 
(Bottom) Ratios of $z_{g}$ distributions for charged jets 
between Pb+Pb and $p$+$p$ collisions at $\sqrt{s_{\mathrm{NN}}}=5.02$~TeV
for different centrality, jet cone size $R$, and $p^{\mathrm{ch,jet}}_{T}$ range. 
The lines show the results from JETSCAPE. The circles (top) and squares (bottom) with statistical error bars and systematic uncertainties bands are the experimental data from ALICE~\cite{ALargeIonColliderExperiment:2021mqf}.  
}
\label{fig:alice_zg}
\end{figure*}

The same set of simulations which generated all of the plots in Figs.~\ref{jetscape-jet-hadron-5TeV-central}-\ref{fig:jetscape-multipanel}, can now be reanalysed and compared to data on a variety of jet substructure observables. The utility of calculating events rather than observables allows for rigorous cross comparisons between observables as no new events are generated. In the following we present the simulation results for a handful of substructure observables,

\begin{figure*}[htb!]
\centering
\includegraphics[width=0.9\textwidth]{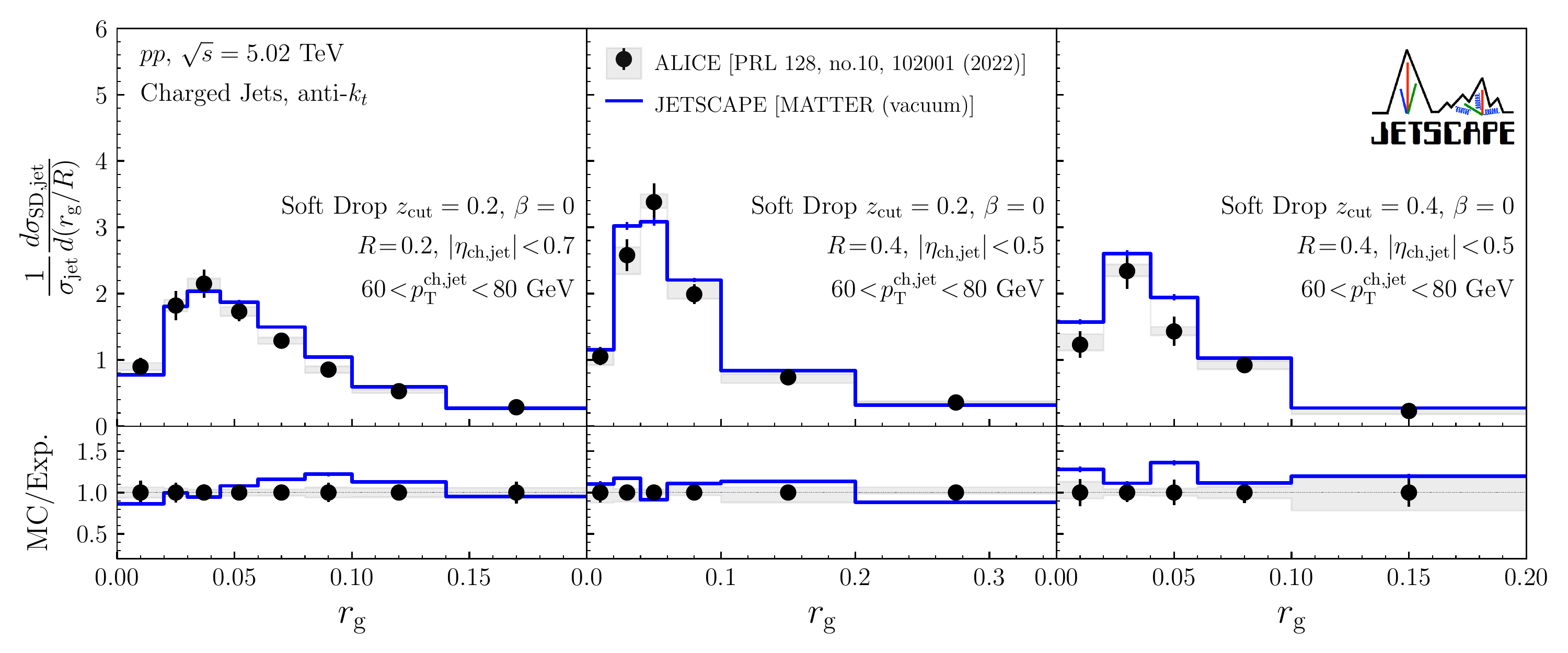}
\includegraphics[width=0.9\textwidth]{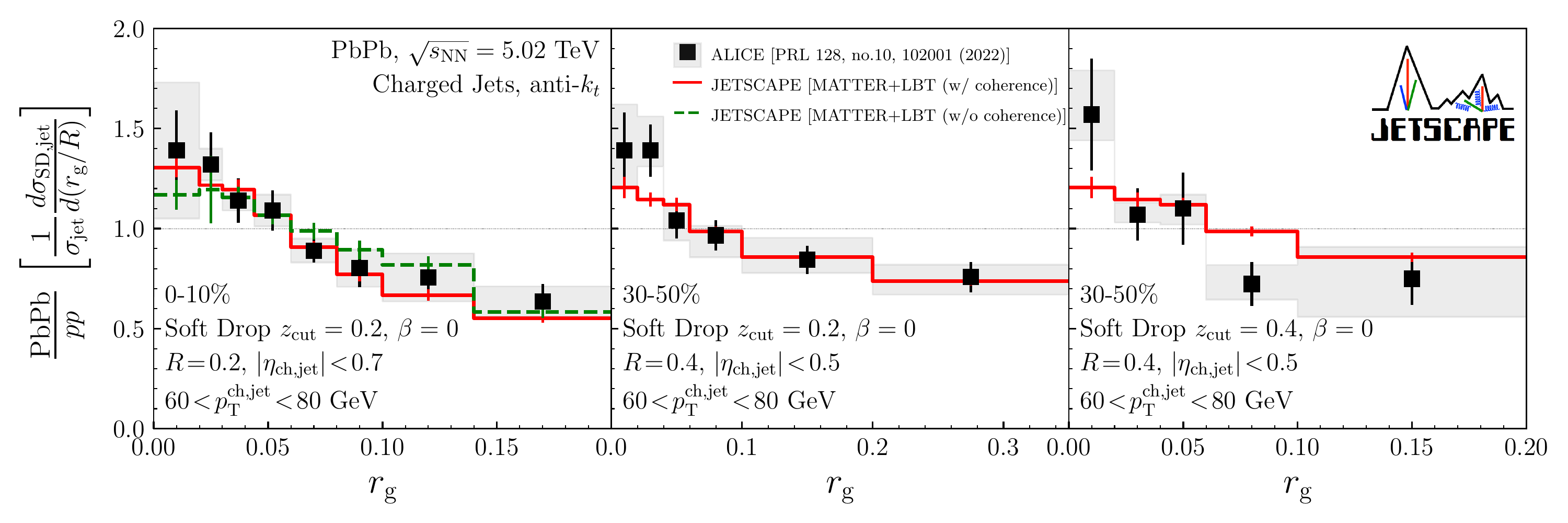}
\caption{(Color online) (Top) Distributions of jet splitting radius $r_{g}$ for soft dropped charged jets in $p$+$p$ collisions at $\sqrt{s}=5.02$~TeV and the ratios 
for different jet cone size $R$, and $p^{\mathrm{ch,jet}}_{T}$ range. 
(Bottom) Ratios of $r_{g}$ distributions for charged jets 
between Pb+Pb and $p$+$p$ collisions at $\sqrt{s_{\mathrm{NN}}}=5.02$~TeV
for different centrality, jet cone size $R$, soft drop parameter $z_{\mathrm{cut}}$, and $p^{\mathrm{ch,jet}}_{T}$ range. 
The lines show the results from JETSCAPE. The circles (top) and squares (bottom) with statistical error bars  and the experimental data from the ALICE Collaboration~\cite{ALargeIonColliderExperiment:2021mqf}, respectively. 
The bands indicate the systematic uncertainties of the experimental data. }
\label{fig:alice_rg_pp}
\end{figure*}

In the top panel of Fig.~\ref{fig:alice_zg}, a distribution of the soft drop momentum fraction~\cite{Larkoski:2014wba} variable $z_g$ calculated using jets from the same simulation in Fig.~\ref{jetscape-jet-hadron-5TeV-central} are compared with data from $p$-$p$ collisions measured by the ALICE experiment. The bottom panel presents the ratio of the distribution from quenched jets in $Pb$-$Pb$ collisions to that in $p$-$p$. The experimental data indicate no modification of this distribution, which is also demonstrated by the theoretical simulation. In the figures, the red solid line represents the default calculation, while the green dashed line represents simulations where the coherence effect is removed, i.e., there is no weakening of the interaction with the virtuality of the hard parton.

\begin{figure*}[htb!]
\centering
\includegraphics[width=\textwidth]{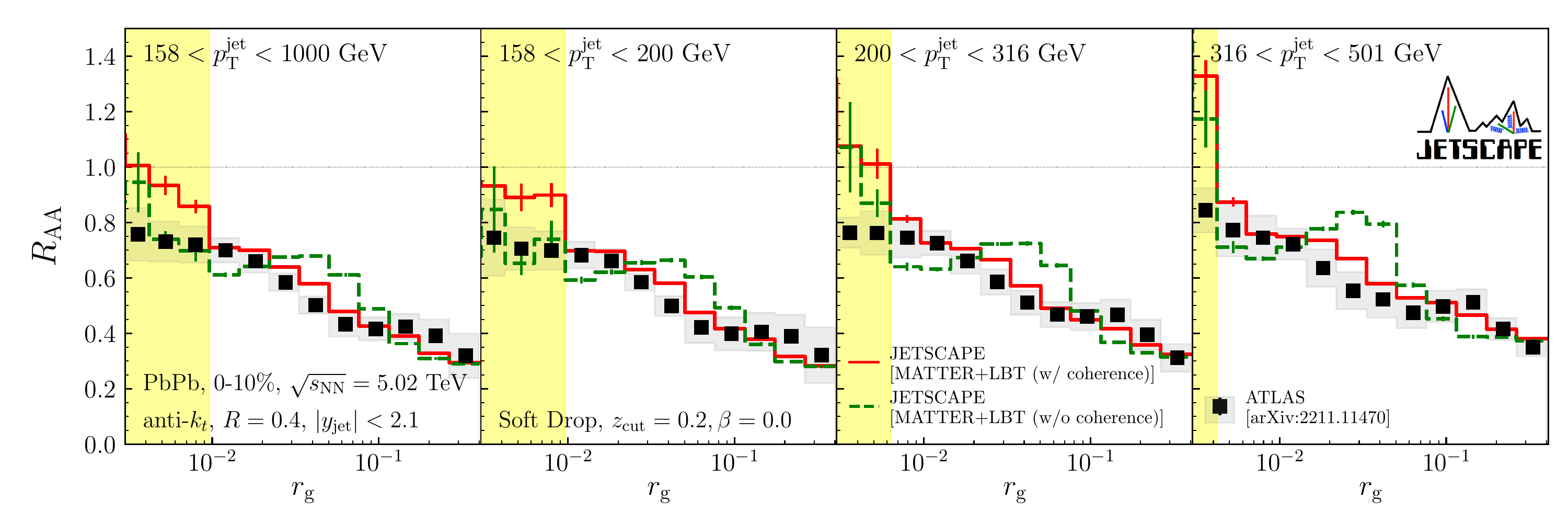}
\caption{(Color online) Nuclear modification factor $R_{\mathrm{AA}}$ in 0-10\% Pb+Pb collisions at $\sqrt{s_{\mathrm{NN}}}=5.02$~TeV, as a function of $r_{g}$ for jets with different $p_{T}^{\mathrm{jet}}$. 
The solid and dashed lines with statistical error bars show the results from the same events generated from Fig.~\ref{jetscape-jet-hadron-5TeV-central} with and without coherence effects, respectively. 
For comparison, the experimental data
from the ATLAS Collaboration~\cite{ATLAS:2022vii} are shown by squares with statistical errors (bars) and systematic uncertainties (bands). 
The yellow-shaded regions are the bins where the transverse momentum between the prongs is less than 1~GeV. }
\label{ATLAS_RAA_vs_rg}
\end{figure*}

In the top panel of Fig.~\ref{fig:alice_rg_pp}, the corresponding angular distribution of the two prongs ($r_g$) emanating from the soft drop procedure are presented. Measurements from ALICE are compared with predictions from JETSCAPE. The top panel represents the distribution in $p$-$p$, while the bottom panel presents the ratio between the distribution of quenched jets in $Pb$-$Pb$ and those in $p$-$p$. Unlike the case of the $z_g$ distribution, the $r_g$ distribution shows a clear narrowing of the angle between the two prongs formed as a result of the soft drop re-clustering procedure. Results including coherence effects, seem to be marginally better in comparisons with experimental data. 

A clear separation between calculations including and not including coherence effects is seen in the plot of the $R_{AA}$ as a function of $r_g$ as presented in Fig.~\ref{ATLAS_RAA_vs_rg}. This figure presents the nuclear modification factor for re-clustered jets that have passed the soft drop condition as a function of the angle between the eventual two prongs. The yellow band represents bins where the transverse momentum between the prongs is less than 1~GeV and therefore represent the non-perturbative region. In the region where pQCD based calculations should work, we see a clear preference for calculations which include coherence effects (red solid lines), as opposed to simulations without coherence effects (green dashed lines).

In the subsequent subsections, the current and exploratory aspects of the physics of inclusive jets, leading hadrons, jet substructure, jet coincidence and heavy flavor observables will be discussed in more detail. New theoretical directions will be identified, and experimental data without a satisfactory theoretical description explored. The current section will end with a discussion of aspects of Bayesian analysis for the hard sector of heavy-ion collisions. In the subsequent section, a discussion on non-jet based rare hard probes will be carried out.

\subsubsection{Jet quenching theory}
\label{sec:progress:microscopic:jet_quenching_theory}

In a seminal (unpublished) work \cite{Bjorken:1982tu}, Bjorken hypothesized that the formation of a quark-gluon plasma in high-energy hadronic collisions would be reflected in the suppression of jets due to collisional energy loss which was pointed out later to be negligible as compared to radiative parton energy \cite{Gyulassy:1990ye} to cause the jet quenching phenomenon. Such parton energy loss can also lead to significant suppression of large $p_T$ single inclusive hadron spectra as well as hadron spectra associated with a hard trigger (photon or hadron) as estimated within the HIJING Monte Carlo model \cite{Wang:1991hta,Wang:1992qdg} and the pQCD parton model \cite{Wang:1998bha,Wang:1996yh}.
This phenomenon was successfully observed in the early 2000s at RHIC in the quenching of high-energy single inclusive hadrons \cite{PHENIX:2001hpc,STAR:2002ggv}, dihadrons \cite{STAR:2002svs} and $\gamma$-hadron spectra \cite{STAR:2009ojv, PHENIX:ppg090}. A decade later, jet quenching was confirmed by the study of fully reconstructed jets at the LHC \cite{ATLAS:2012tjt}. 

This remarkable discovery has spurred a lot of theory and experimental research activity in the past two decades with the objective of using jets as a multi-dimensional tool to probe the properties of the quark-gluon plasma at various length scales. 

{\bf  Transverse momentum broadening and parton energy loss }

The theory of jet quenching is based on transverse momentum broadening and parton energy loss \cite{Mehtar-Tani:2013pia,Blaizot:2015lma,Qin:2015srf,Majumder:2010qh}: a high-energy parton produced early in a heavy-ion collision through a hard scattering process, before the plasma has formed, will subsequently propagate through the hot deconfined matter and undergo multiple scattering whose effect is to i) increase the transverse momentum of the energetic parton via a diffusive process in transverse momentum space:
\begin{align}
  \langle k_\perp^2\rangle \equiv \hat q L   
\end{align}
where $L$ is the medium length and $\hat q$ the diffusion coefficient aka the jet quenching parameter that encodes the properties of the hot QCD matter. The general assumption is that the scattering centers are independent from each other. This is justified so long as the in medium mean free path is much larger than the correlation length, and, %
the parton's mean lifetime is shorter than the mean free path in the medium. 

In addition to elastic processes, multiple scattering may trigger coherent gluon radiation \cite{Gyulassy:1993hr} for long-lived partons. Considering the dominant processes when the radiated gluon goes through coherent multiple scattering, the total radiated energy loss has quadratic path length dependence \cite{Baier:1996kr,Baier:1996sk,Zakharov:1996fv,Gyulassy:2000fs,Wiedemann:2000za,Wang:2001ifa,Arnold:2002ja}. During this quantum coherence time $t_f$ the radiated gluon accumulates a transverse momentum via diffusion $k_f^2\sim \hat q t_f$ which implies that $t_f\equiv 2\omega/k_\perp^2\sim \sqrt{\omega /\hat q }$ for massless partons. \footnote {The gluon radiation process off heavy quarks is more intricate given the new scale introduced into the system, i.e. the heavy quark mass. In that case, the coherence time becomes $t_f=\frac{2\omega}{k^2_\perp+y^2 M^2}$ where $y$ is the momentum fraction carried away by the radiated gluon and $M$ is the mass of the heavy quark~\cite{Abir:2015hta}.} Thus, the softer gluons in the soft radiation and multiple scattering limit have shorter formation times and therefore can be emitted quasi-instantly with constant rate 
\begin{align}
   \omega \frac{{\rm d}\Gamma}{{\rm d } \omega} = \frac{\alpha_s C_R}{\pi} \sqrt{\frac{\hat q }{\omega}}
\end{align} 
This rate is suppressed at large frequency due to the Landau-Pomeranchuk-Migdal (LPM) effect and its maximal suppression is reached at the gluon frequency $\omega_c\equiv \hat q L^2$ and minimum angle $\theta_c\equiv (\hat q L^3)^{-1/2}$.  While the light-flavor LPM effects have been calculated using Hard Thermal Loops (HTL) formalism \cite{Bellac:2011kqa, Arnold:2001ms,Arnold:2002ja,Arnold:2002zm}, thus providing HTL-resummed scattering and radiation rates, the same cannot be said for heavy flavors. However, the generalized HTL formalism to describe heavy flavor interaction with the QGP at lower virtualities has been devised by Caron-Huot \cite{Caron-Huot:2007rwy}. For more complete phenomenological studies to be done, that formalism needs to be applied in earnest, resulting in LPM-resummed rates for heavy flavor scattering and radiation, akin to Refs.~\cite{Arnold:2001ms,Arnold:2002ja,Arnold:2002zm}.

Early studies of high hadron suppression and jet quenching were based on single parton energy loss. However, a complete treatment of jet quenching requires going beyond this approximation. This will be discussed in what follows. 

\begin{figure}[htb]
\centering
\includegraphics[width=12cm]{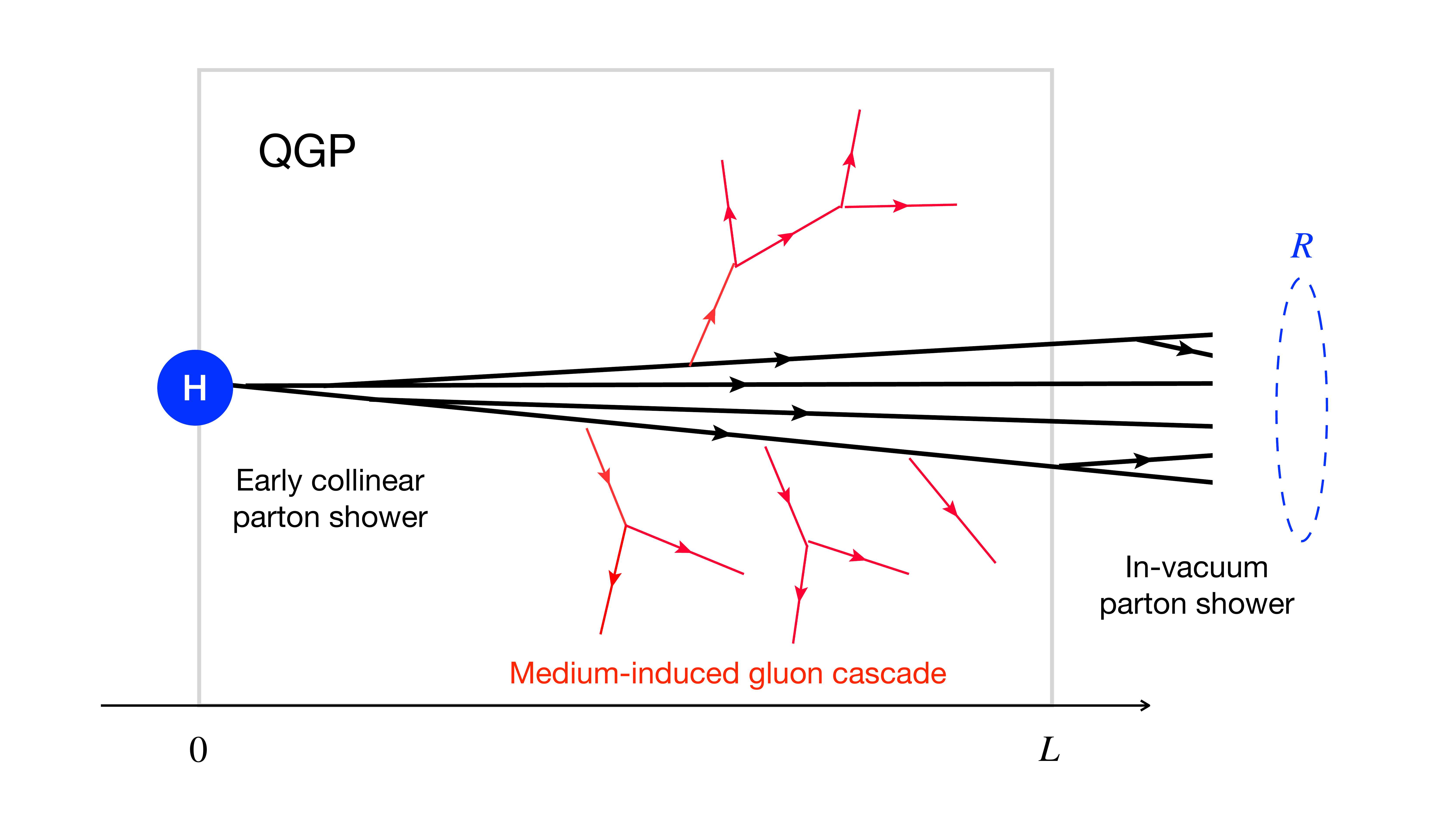}
\caption{(Color online) Sketch of the different stages of jet evolution in the quark-gluon plasma (see description in the text).}
\label{fig:parton-shower}
\end{figure}

{\bf  Jet quenching and color decoherence} 

Jets are coherent multi-partonic systems that extend in time and space, which requires to go beyond single parton energy loss. The question of how these complex objects interact and lose energy to the plasma and how their substructure is modified as a result of these final state interactions has been a central focus in the theory community. 
 
There are three distinct stages to jet evolution in the QGP \cite{Caucal:2020zcz} (cf. Figure~\ref{fig:parton-shower} for an illustration). The first stage corresponds to a hard and collinear parton cascade that forms due to the large virtuality associated with the hard scattering. This cascade develops early, mostly before energy loss processes take place \cite{Mehtar-Tani:2017web,Caucal:2018dla}. 
 The high-virtuality stage may also acquire medium-induced correction such as discussed in Ref.~\cite{Wang:2001ifa,Abir:2015hta,Sirimanna:2021sqx}.  
For a single gluon radiation of energy $\omega$ the time window associated with this stage is $1/{p_TR^2} \ll  t \ll  t_f \sim \sqrt{\omega/\hat{q} }$. Or equivalently, in terms of transverse momentum $Q^2\equiv (p_T R)^2 \ll   k_\perp^2\ll  k^2_f= \sqrt{\omega \hat q}$. \footnote{ To include gluon radiation off heavy quarks in these ordered inequalities, the following substitution is needed: $k^2_\perp\to k^2_\perp+y^2 M^2$, with $M$ the heavy quark mass and $y$ the gluon momentum fraction. The other channel involving heavy quarks is their pair production through $g\to Q+\bar{Q}$. In that case, the transverse momentum conditions is modified according to $k^2_\perp\to k^2_\perp+ M^2$, where $M$ is the heavy quark mass.}
In the second stage copious medium induced radiation off the jet color charges produced in the first stage take place with constant rate inside the medium of size $L$. Medium induced radiation is followed by a gluon and quark cascade \cite{Sirimanna:2022zje} that efficiently transports energy from fast color charges down to the plasma temperature scale where energy is dissipated \cite{Blaizot:2013vha,Blaizot:2013hx,Mehtar-Tani:2018zba,Mehtar-Tani:2022zwf}. In the last stage the fast charges that escape the plasma continue fragmenting until their vitality reaches non-perturbative scales where hadronization occurs. While focus herein is on describing the showering in the large-energy ($E$) eikonal limit, there is work improving eikonal results by supplying sub-eikonal $\mathcal{O}(1/E)$ corrections~\cite{Sadofyev:2021ohn,Andres:2022ndd}. Though some phenomenological implications of sub-eikonal effects have been studied in, e.g., \cite{Antiporda:2021hpk,Fu:2022idl},
 in the future such studies would benefit to be implemented within more realistic multi-stage dynamical model describing jet propagation in a nuclear medium. 
The assumption that all medium-induced splittings in the shower are independent may be questionable for splittings with relatively large formation times. In this case, there may be quantum interference between subsequent splits (beyond what can be refactored in an effective $\hat{q}$ parameters) that need to be carefully considered as explored in, e.g., Ref.~\cite{Arnold:2021pin,Arnold:2020uzm,Arnold:2023qwi}. Depending on the amount of interference between subsequent splits, it may be that Monte Carlo showering algorihtms, akin to those discussed herein, need to be revised to take into account such effects to achieve higher precision computations. 

An important feature of parton cascades is color coherence that plays a role beyond the leading order of single gluon radiation and thus needs to be accounted for in parton shower energy loss. It is a quantum interference effect that suppresses gluon radiation with wavelength larger than the size of the emitting system: for instance, gluon radiation  off of a quark antiquark in a color singlet state is suppressed at angles larger than the angle between the pair due to destructive interference leading to angular ordering of successive splittings. In the presence of a hot colored medium this property is expected to be altered as a result of rapid color precession of jet color charges \cite{Mehtar-Tani:2010ebp,Mehtar-Tani:2011hma}. Color coherence also affects medium-induced radiation similarly. Energy loss of a quark-antiquark pair depends on the extent to which the medium resolves the individual colors charges \cite{Casalderrey-Solana:2012evi}. If the antenna (quark-antiquark pair) opening angle is larger than the characteristic angle $\theta_c \equiv (\hat q L^3)^{1/2}$ angle the medium resolves the pair and thus the latter loses energy as independent color charges with intensity proportional to $2C_F$, in the opposite scenario energy loss is proportional to the total charge of the pair which is either vanishing in the case off a single state or proportional to $C_A$ in the case of gluon parent \cite{Mehtar-Tani:2017ypq}. A remarkable consequence of color coherence is that a jet with opening angle smaller than the medium angular scale $\theta_c \gg R$ loses energy as a single color charge given by the parent parton \cite{Mehtar-Tani:2017web}. 

\begin{figure}[htb]
\centering
\includegraphics[width=7cm]{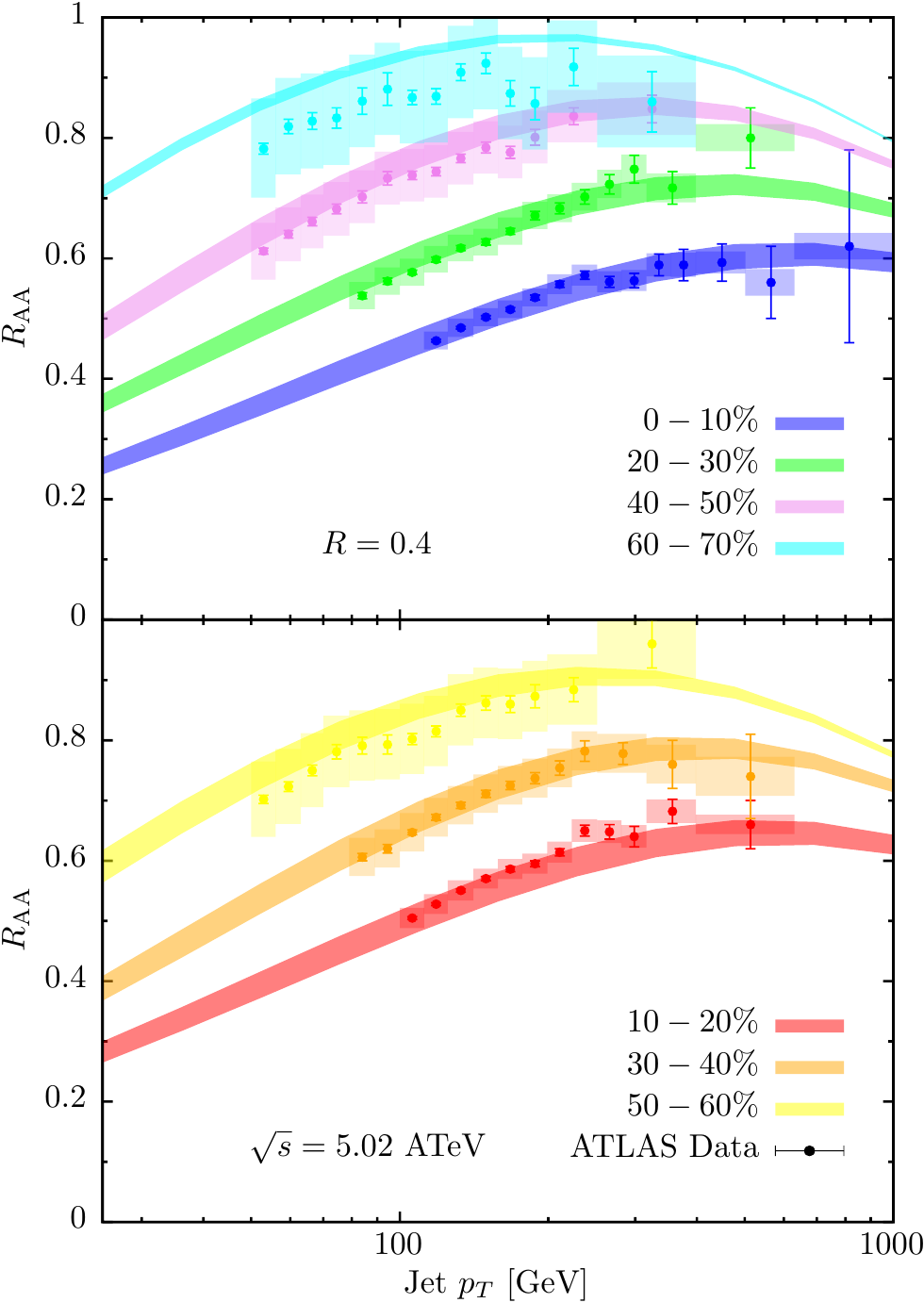}
\caption{
Analytic calculation of inclusive jet $R_{AA}$ in \PbPb{} collisions at $
\sqrt{s} = 5.02 \rm TeV$, compared to ATLAS data \cite{ATLAS:2018gwx}, for different
centralities. It includes resummation of energy loss effects from hard, vacuum-like emissions occurring in the medium and color coherence effects \cite{Mehtar-Tani:2021fud}.(Color online) }
\label{fig:allcent_R04}
\end{figure}

The study of the antenna system in the medium allowed to elucidate how the phase space of parton showers is affected by energy loss. As a phenomenological application, the physics of color (de)coherence was implemented at all order in perturbation theory in the leading logarithm approximation for inclusive jet spectrum and the corresponding nuclear modification factor \cite{Mehtar-Tani:2021fud} (cf. Figure~\ref{fig:allcent_R04}). It was also shown to yield an excess of soft particles inside the jet in a study of the jet fragmentation function \cite{Mehtar-Tani:2014yea,Caucal:2018dla}. Moreover, jet substructure observables were extensively studied to probe the resolution power of the medium \cite{Mehtar-Tani:2016aco,Caucal:2021cfb,Caucal:2019uvr}. The effect of the medium response is still to be investigated in these studies.

A lot of progress has been made recently to unify the various energy loss formalisms both analytically, with the so-called improved opacity expansion (IOE) \cite{Mehtar-Tani:2019tvy,Barata:2021wuf}, and numerically \cite{Feal:2018sml,Andres:2020vxs,Andres:2020kfg} paving the way for precision phenomenology in the future. A future prospect is to improve on the accuracy of such gluon cascades by systematically computing higher order corrections to medium-induced gluon splitting including full kinematics \cite{Sievert:2019cwq,Arnold:2021pin,Arnold:2020uzm}.

{\bf  Extractions of $\hat q$, higher order and quantum corrections}

It is now widely held that the transverse momentum jet transport coefficient $\hat{q}$, defined as the mean squared transverse momentum exchanged per unit length between a jet parton and the medium, encapsulates the dominant effect of the medium modification on a propagating jet. Thus, $\hat{q}$ reveals properties of the medium that can be probed via the quenching of hard jets. 

Calculations of jet observables, either via Monte-Carlo event generators or semi-analytical theoretical approaches, encode a range of assumptions within the form of $\hat{q}$ used and its dependence on the temperature of the medium, the scale and energy of the hard parton etc. For example, if one assumed that the QGP could be described using leading order HTL~\cite{Braaten:1989mz, Frenkel:1989br}, one would obtain the expression for $\hat{q}$ in Eq.~\eqref{eq:type2-q-hatform}, assuming running coupling and that the energy of the jet far exceeds the temperature scale of the plasma~\cite{Arnold:2008vd}. This form, both with and without additional scale dependence (coherence effects), has been widely used in phenomenological comparisons between experiment and simulation where the $\alpha_{\rm S}^{\rm fix}$, the coupling at the medium scale is used as a fit parameter (see the plots in Sec.~\ref{sec:progress:microscopic:JETSCAPE}). 

Along with the extractions of $\hat{q}$ from comparison with data, there have been several attempts to calculate this from first principles. Next-to-leading order calculations of $\hat{q}$, entirely within the assumptions of HTL theory~\cite{Caron-Huot:2008zna}, indicate large corrections to the leading order formula of Eq.~\eqref{eq:type2-q-hatform}. There have been attempts to estimate the soft sector of $\hat{q}$ using a effective 3-D theory on the lattice called Electro-static QCD (EQCD)~\cite{Panero:2013pla}. While these represent small corrections to the LO formula at temperatures $T \gtrsim 10 T_C$, they remain large corrections close to $T_C$. 

In an alternative approach, one could drop the assumption of an HTL plasma and define $\hat{q}$ using the operator product expansion~\cite{Majumder:2012sh} and extract $\hat{q}$ from the leading operator in the expansion, yielding~\cite{Kumar:2020wvb}:
\begin{align}
    \frac{\hat{q}}{T^3} \simeq \frac{4\pi \alpha_s }{N_C T^4}  \langle F^{+ j} F^{+}_j \rangle_{\mathrm{T-V}}. 
\end{align}
In the equation above, the $F^{+ j} F^{+}_j$ represents the local color summed gluon field strength correlator. The subscript $T-V$ indicates the thermal minus vacuum expectation. These operator products can now be calculated in Lattice-QCD. In Ref.~\cite{Kumar:2020wvb}, continuum extrapolated lattice calculations of $\hat{q}$, in both quenched and 3-flavor QCD, are presented in comparison with extractions from phenomenology, leading order HTL and EQCD calculations. 

\begin{figure*}[htb!]
\centering
\includegraphics[width=0.8\textwidth]{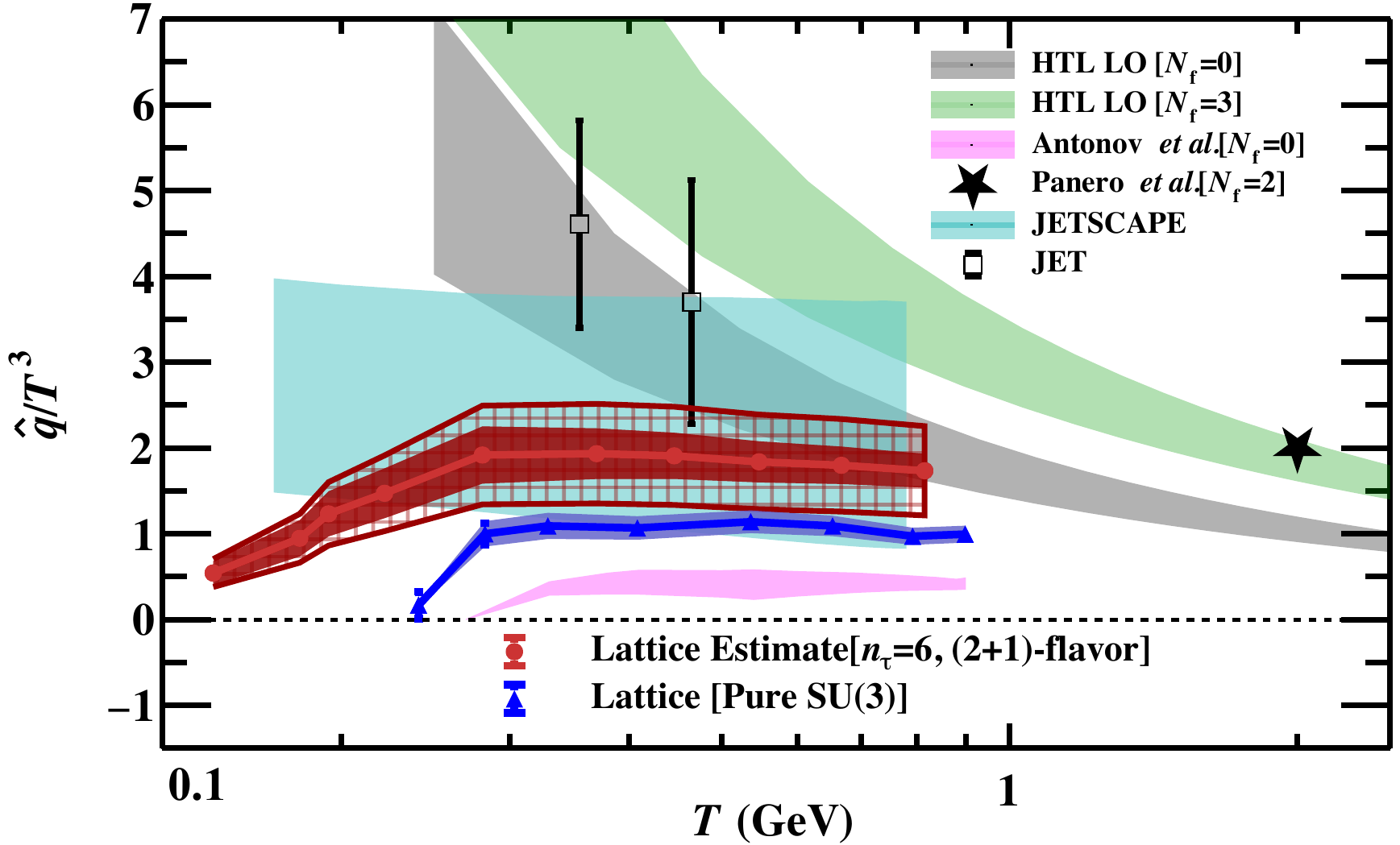}
\caption{(Color online) Cross comparisons of the jet transport parameter $\hat{q}$, obtained from many different approaches~\cite{Kumar:2020wvb}: Phenomenological comparisons with data from the JET and JETSCAPE collaborations, and first principles theoretical approaches from Lattice QCD, EQCD (Panero et al.), Hard thermal loop effective theory, and a stochastic vacuum model (Antonov et al.). }
\label{all_qhat}
\end{figure*}

The results of such cross comparisons are presented in Fig.~\ref{all_qhat}, where the results from the JET~\cite{JET:2013cls}, and JETSCAPE~\cite{JETSCAPE:2022ixz} collaborations are compared with first principles calculations in Lattice-QCD~\cite{Kumar:2020wvb}, Leading order HTL~\cite{Arnold:2008vd}, Electrostatic QCD~\cite{Panero:2013pla}, and a stochastic vacuum model at $N_f = 0$~\cite{Antonov:2007sh}. This plot represents the current state-of-the-art of our knowledge of the jet quenching parameter $\hat{q}$. The results from EQCD at temperatures below 0.5~GeV and next-to-leading order corrections to $\hat{q}$, within HTL perturbation theory are beyond the scale of the plot. All calculations tend to approach similar values at $T\gtrsim 10 T_C$, with the obvious separation between the quenched ($N_f=0$) and 3-flavor QCD results. 

In past few years, higher order quantum  corrections to transverse momentum broadening were investigated. It was in particular shown that they are enhanced by a large double logarithm of the medium size, i.e., $\alpha_s\ln^2 L$ \cite{Liou:2013qya,Blaizot:2013vha,Blaizot:2014bha,Iancu:2014kga}. When resumed to all orders they result in an anomalous scaling of transverse momentum broadening that reflects a super diffusive behavior 
\begin{align}
    \langle k_\perp^2\rangle \sim L^\gamma\, ,
\end{align}

where $\gamma \simeq 1+2\sqrt{\bar\alpha}$ and $\bar\alpha\equiv C_A\alpha_s/\pi$. In a recent series of articles a systematic approach was developed to compute higher order corrections and running coupling effects \cite{Iancu:2014sha,Caucal:2022fhc,Caucal:2021lgf,Ghiglieri:2022gyv}. Note that this anomalous scaling is a consequence of the renormalization of the jet quenching parameter and its scale dependence in the non-linear regime of multiple scattering \cite{Blaizot:2014bha}. At very high momentum transfer the jet quenching parameter obeys the DGLAP evolution equation \cite{Casalderrey-Solana:2007xns,Caucal:2022mpp}. 
One important outlook is to implement these quantum corrections to jet quenching observables that are currently based on the leading order formulation of $\hat q$.

 While historically $\hat{q}$ has been considered to have a temperature and momentum dependence, recent development suggests that its non-trivial virtuality dependence plays an important role in explaining experimental data \cite{Kumar:2019uvu}. The phenomenological implications of this virtuality dependence on heavy flavor quarks is investigated in the next section, while light flavors can be found in \cite{JETSCAPE:2022jer}. The virtuality dependence alone may not enough to fully explain heavy quark $R_{AA}$ \cite{JETSCAPE:2022hcb}, and phenomenological study in Ref.~\cite{JETSCAPE:2022hcb} points towards further theory development to obtain not only a temperature ($T$)- and virtuality($Q^2$)-dependent $\hat{q}$, but also a heavy quark mass $M$-dependent $\hat{q}$: i.e., $\hat{q}(T,Q^2,M)$. As one goes towards lower virtualities, a momentum $(p)$-dependent $\hat{q}$ should also be considered (e.g. Ref.~\cite{Xu:2018gux}, and reference therein), and is further affected by LPM-resummations.

\subsubsection{Phenomenology of Hard Heavy Flavors}

\begin{figure}[!h]
\begin{center}
\begin{tabular}{cc}
\includegraphics[width=0.55\textwidth]{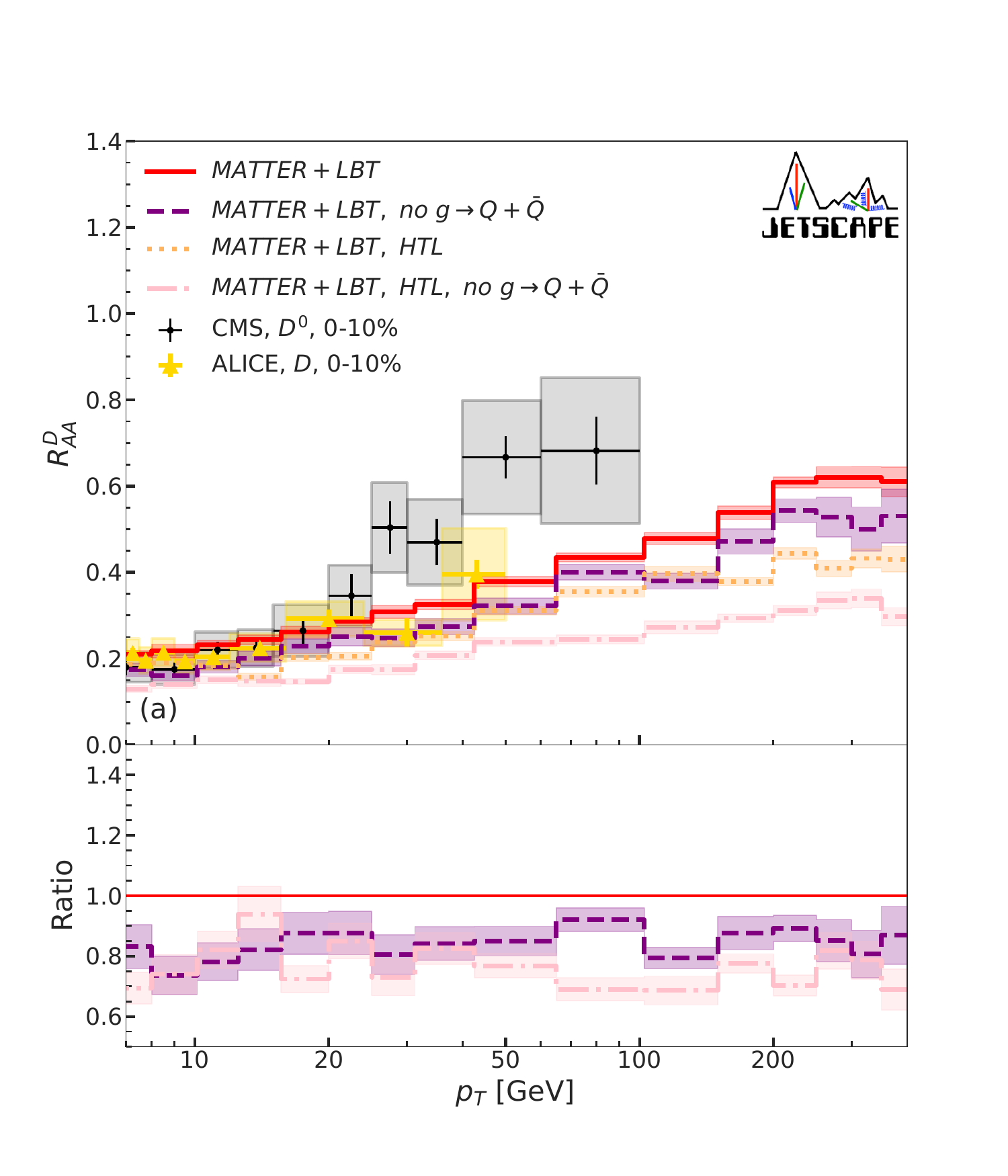}
\end{tabular}
\end{center}
\caption{(Color online) Nuclear modification factor for D-mesons in $\sqrt{s_{NN}}=5.02$ TeV \PbPb{} collisions at the LHC at 0-10\% centrality using a combination of higher-twist energy loss (MATTER) and linearized Boltzmann transport (LBT) \cite{JETSCAPE:2022hcb}. For comparison, simulations with a virtuality-independent $\hat{q}$ have the HTL designation in the legend. The dashed line in the ratio subplots divides MATTER+LBT no $g\to Q+\bar{Q}$ to MATTER+LBT, while the dotted-dashed line divides MATTER+LBT HTL no $g\to Q+\bar{Q}$ to MATTER+LBT HTL.}
\label{fig:MATTER_LBT_comp_gQQ}
\end{figure}
There has been wealth of phenomenological progress in simulating high-energy heavy flavor propagation within a nuclear medium incorporating many of the formal theoretical advances mentioned above. While the high-energy elastic $2\to 2$ scattering is common among many heavy-flavor energy loss formalisms \cite{Cao:2018ews,Ke:2019jbh}, the treatment of inelastic/radiative processes is where different phenomenological assumptions are used. The high virtuality portion of heavy quark energy loss in the QGP can be well described using the higher twist formalism\footnote{For light flavor, the higher twist result can be shown to agree with the approach by Gyulassy-Levai-Vitev (GLV) \cite{Gyulassy:1999zd,Gyulassy:2000fs} as discussed in, e.g., Ref.~\cite{Cao:2020wlm}. However, the Djordjevic and Gyulassy \cite{Djordjevic:2003zk} extension to the GLV formalism for heavy flavor has yet to be as thoroughly compared with the higher twist result.}. At lower virtuality, there are many formalisms for heavy quark radiation. For instance, the Linearized Boltzmann Transport (LBT) approach \cite{Cao:2016gvr,He:2015pra} uses higher-twist inspired radiation \cite{Abir:2015hta}; the approach developed at Duke \cite{Cao:2013ita,Cao:2015hia,Xu:2015iha} uses the improved Langevin dynamics and extends it within the Linear-Boltzmann-plus-Diffusion-Transport-Model LIDO \cite{Ke:2018jem,Ke:2019jbh}; the Nantes approach use $2\to 3$ inelastic processes \cite{Meggiolaro:1995cu,Aichelin:2013mra}. While all these approaches made important contributions towards understanding how hard heavy flavor quarks interact with the nuclear medium, the most recent advance combining different energy loss formalisms together. 

A recent comparison with experimental data has revealed that a multi-formalism/multi-scale approach is required to simultaneously describe light flavor observables from different $\sqrt{s_{NN}}$ collision energies \cite{Kumar:2019uvu}. In particular, the combination of high-virtuality/higher-twist formalism with the low-virtuality effective Boltzmann transport, both of which are being further improved (as discussed in the previous section), revealed to be a fruitful combination. On the heavy flavor side, the need for a multi-scale approach was investigated for the first time in Ref.~\cite{JETSCAPE:2022hcb}. In that study, an important additional ingredient was the phenomenological inclusion of in-medium production of heavy-flavor via $g\to Q+\bar{Q}$, incorporated within the high-virtuality energy loss simulation, which was combined together with low-virtuality Boltzmann transport approach before comparing with data.~\footnote{A recent calculation \cite{Attems:2022otp,Attems:2022ubu} has also explored the importance of heavy flavor pair production via $g\to Q+\bar{Q}$ in a static medium with constant temperature. It would be interesting to investigate how these results change using a multistage energy loss in a dynamical medium.} Heavy flavor pair production is most important for studies of charm energy loss, given that the gluon virtuality needed to produce charm quarks is the smallest among heavy quarks. Ref.~\cite{JETSCAPE:2022hcb} has found that dynamical charm pair production in the QGP plays an important role when describing D-meson, as depicted in Figure~\ref{fig:MATTER_LBT_comp_gQQ}. In that simulation, a phenomenological light flavor $\hat{q}(T,Q^2)$ was used to study the importance of having a virtuality dependent $\hat{q}$, given that the heavy flavor $\hat{q}(T,Q^2,M)$ is not yet known. These results show that a multi-scale energy loss calculation of heavy flavors is needed. The multi-scale (heavy quark mass and virtuality) dependence of $\hat{q}$ can be better assessed by investigating how D-meson $R_{AA}$ changes as a function of centrality. Indeed, the amount of QGP that heavy quarks have to traverse is reduced in more peripheral collisions, affecting the balance between high-virtuality energy-momentum exchange relative to low-virtuality Boltzmann transport, as seen in Ref.~\cite{JETSCAPE:2022hcb}. The high-precision measurements planned by the sPHENIX Collaboration, as well as the upcoming higher-statistic measurement from the LHC, will be crucial in pinning down the full functional dependence of $\hat{q}$, using Bayesian model-to-data comparisons.

\subsubsection{Jets, leading hadrons and coincidence measurements}
\label{sec:progress:microscopic:jets_and_leading_hadrons}

One of the signatures of the jet quenching phenomenon is the suppression of high-momentum charged hadrons.  There has been many measurements since the last \LRP with increasingly higher precision~\cite{CMS:2018yyx,CMS:2016xef,CMS:2012aa,ATLAS:2022kqu,ATLAS:2016xpn,ATLAS:2015qmb,ALICE:2021lsv,ALICE:2019hno}.%
A general trend is observed across all measurements.  These charged hadron measurements are complimented by those measuring the suppression of jets~\cite{CMS:2021vui,CMS:2018dqf,CMS:2016uxf,ATLAS:2018gwx,ATLAS:2014ipv,ATLAS:2012tjt,ALICE:2019qyj,ALICE:2015mjv,ALICE:2013dpt}.%
There is a good agreement between different experiments for anti-k$_\mathrm{T}$ jets with a distance parameter up to 0.4, with a slight tension at high \pt.  The modification of jet also manifests itself in \pt balance of back-to-back jets for both inclusive dijets~\cite{CMS:2015hkr,CMS:2012ulu,CMS:2011iwn,ATLAS:2022zbu,ATLAS:2017xfa,ATLAS:2010isq} and b dijets~\cite{CMS:2018dqf}.

\begin{figure}[ht!]
    \centering
    \includegraphics[width=0.45\textwidth]{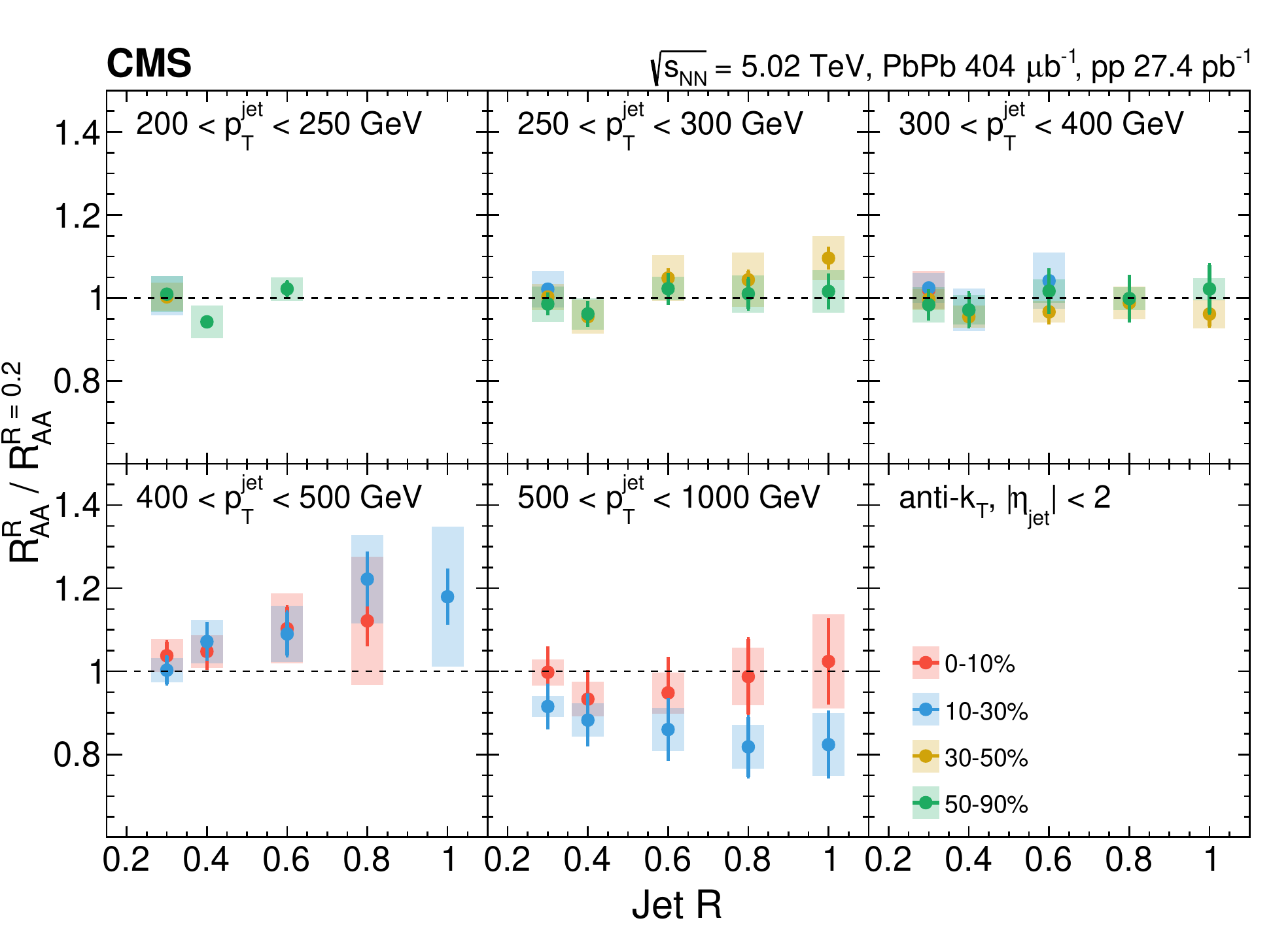}
    \includegraphics[width=0.35\textwidth]{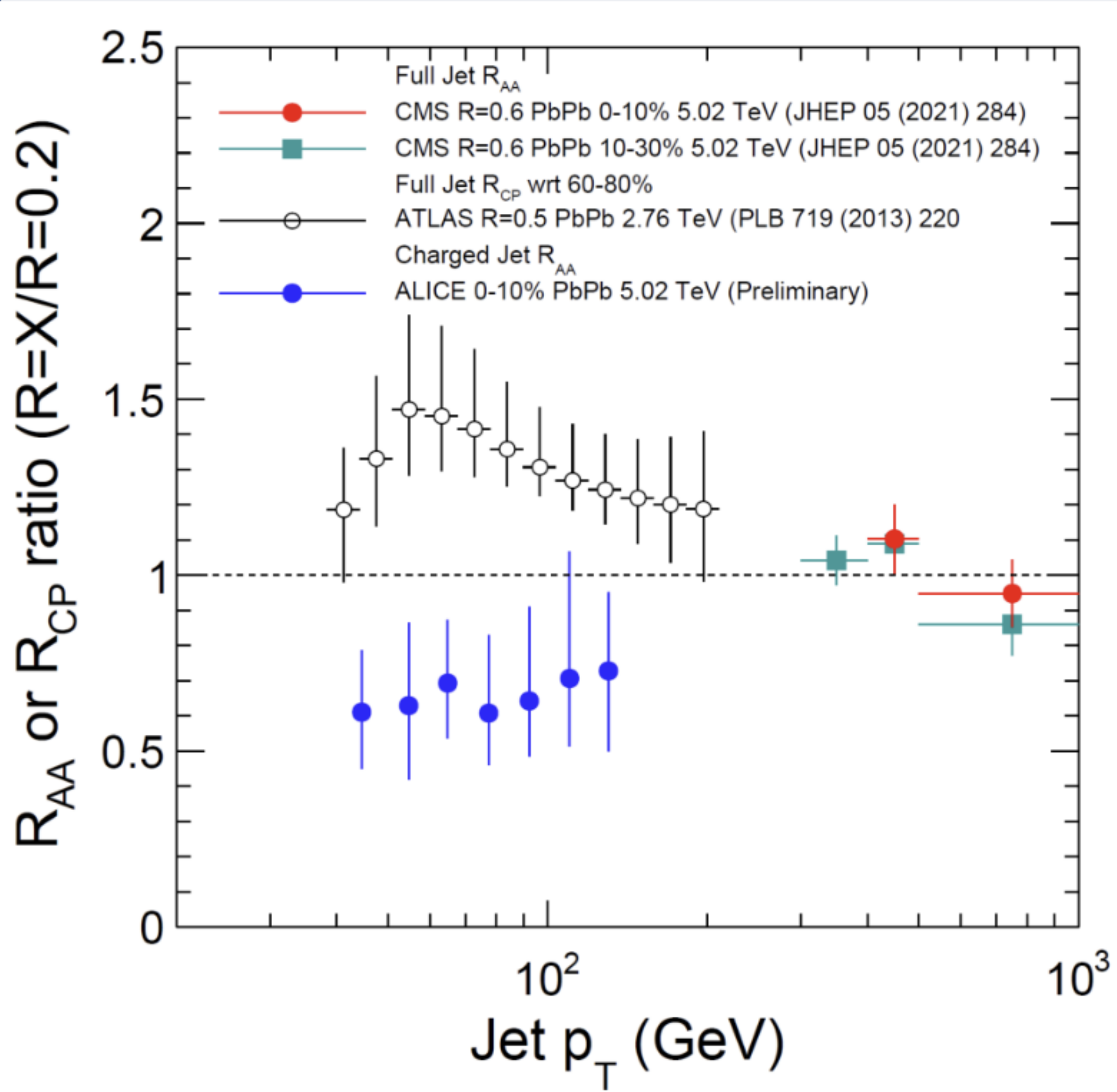}
    \caption{Jet \raa measurements with large distance parameter as measured by CMS~\cite{CMS:2021vui} (left) and ATLAS~\cite{ATLAS:2012tjt} compared to the machine learning extracted result from ALICE~\cite{ALICE:2023waz} (right).}
    \label{Figure:LargeRJetRAA}
\end{figure}

In recent years, there is an effort in different experiments to investigate the suppression of larger size jets up to a distance parameter of 1.0.  It poses a nontrivial experimental challenge.  Results for the larger jets may provide additional input to understanding the mechanism with which the high-energy parton interacts with the QGP.  For example, if the size is big enough that color decoherence can be probed, a larger suppression may be observed; on the other hand, energy recovery of smaller jets will reduce the amount of suppression.  The result reported by the CMS collaboration at high energy indicated that there is no strong dependence on the \raa from 0.2 to 1.0~\cite{CMS:2021vui}, as shown in the left panel of Figure~\ref{Figure:LargeRJetRAA}.  The situation at lower \pt is more interesting with ALICE reporting a new measurement showing a larger suppression of wide jets with a novel machine-learning based approach to jet reconstruction~\cite{ALICE:2023waz}, while the ATLAS \rcp result~\cite{ATLAS:2012tjt} indicate a different trend, as can be seen in the right panel of Figure~\ref{Figure:LargeRJetRAA}.  It will be exciting for future measurements to resolve this apparent tension.

\begin{figure}[ht!]
    \centering
    \includegraphics[width=0.3\textwidth]{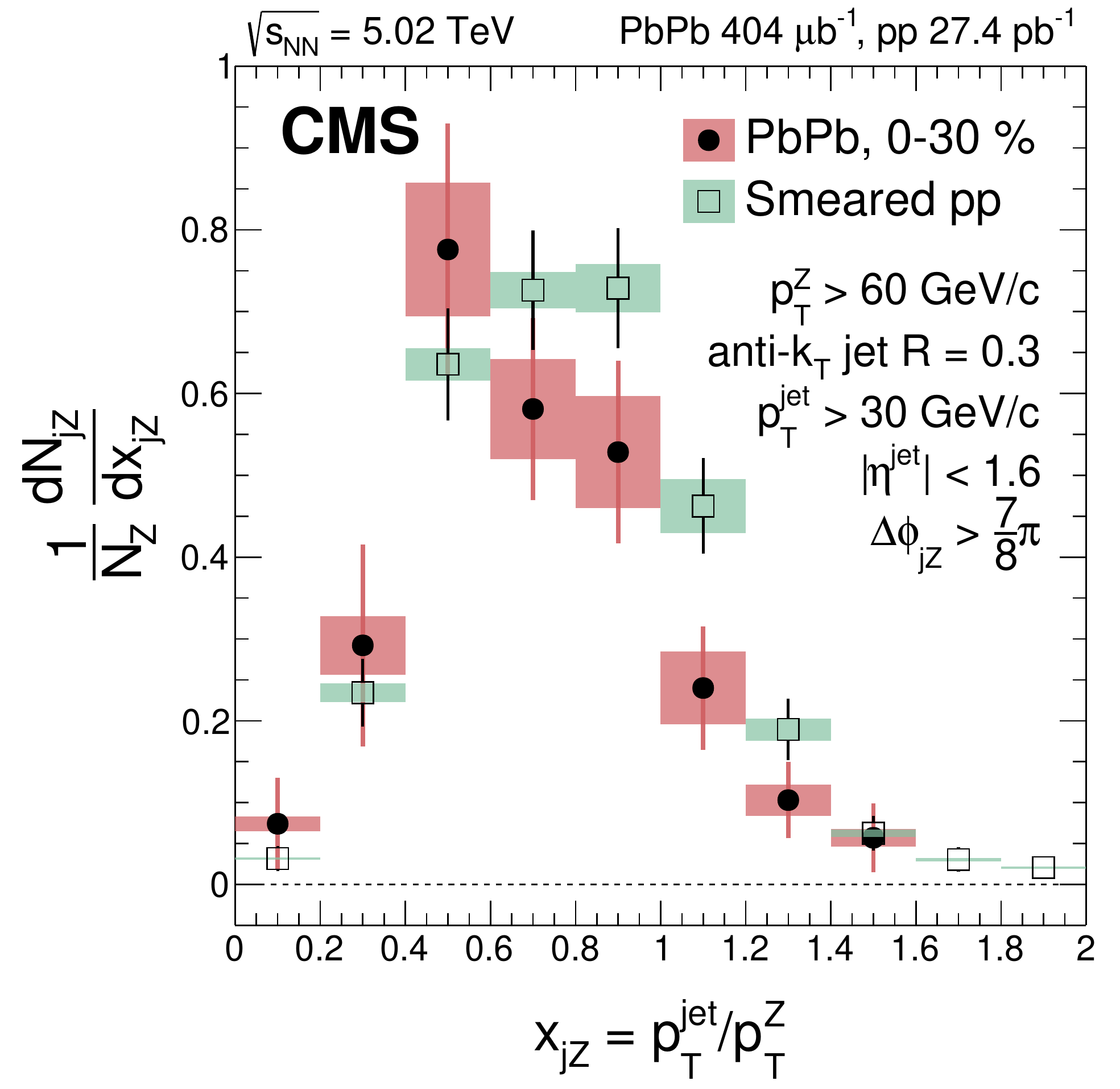}
    \includegraphics[width=0.4\textwidth]{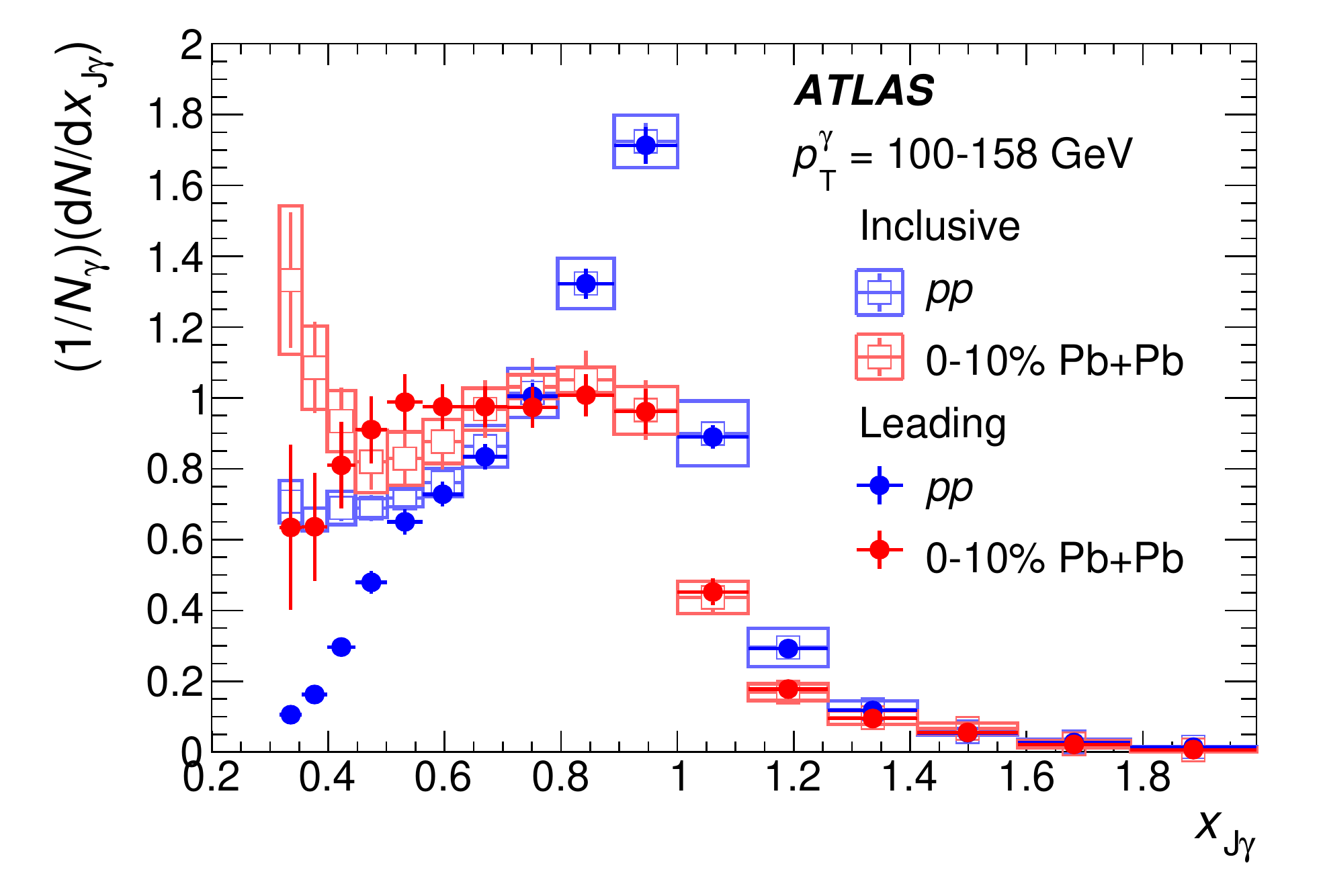}
    \caption{Example of measurements of jets balancing against a $Z$ boson or a photon~\cite{CMS:2017ehl,ATLAS:2018dgb}.}
    \label{Figure:ZGammaJetBalance}
\end{figure}

\begin{figure}[ht!]
    \centering
    \includegraphics[width=0.35\textwidth]{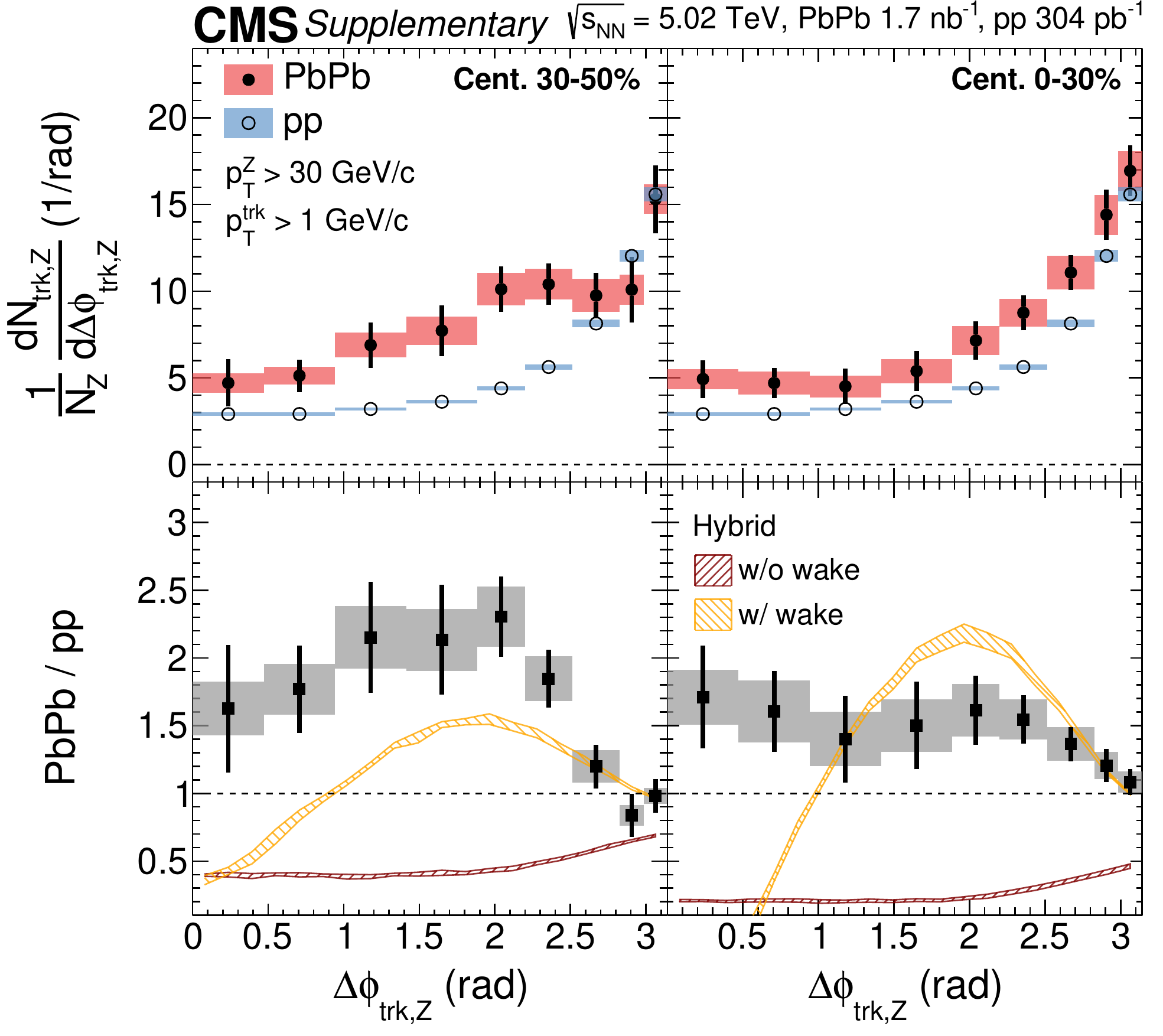}
    \includegraphics[width=0.28\textwidth]{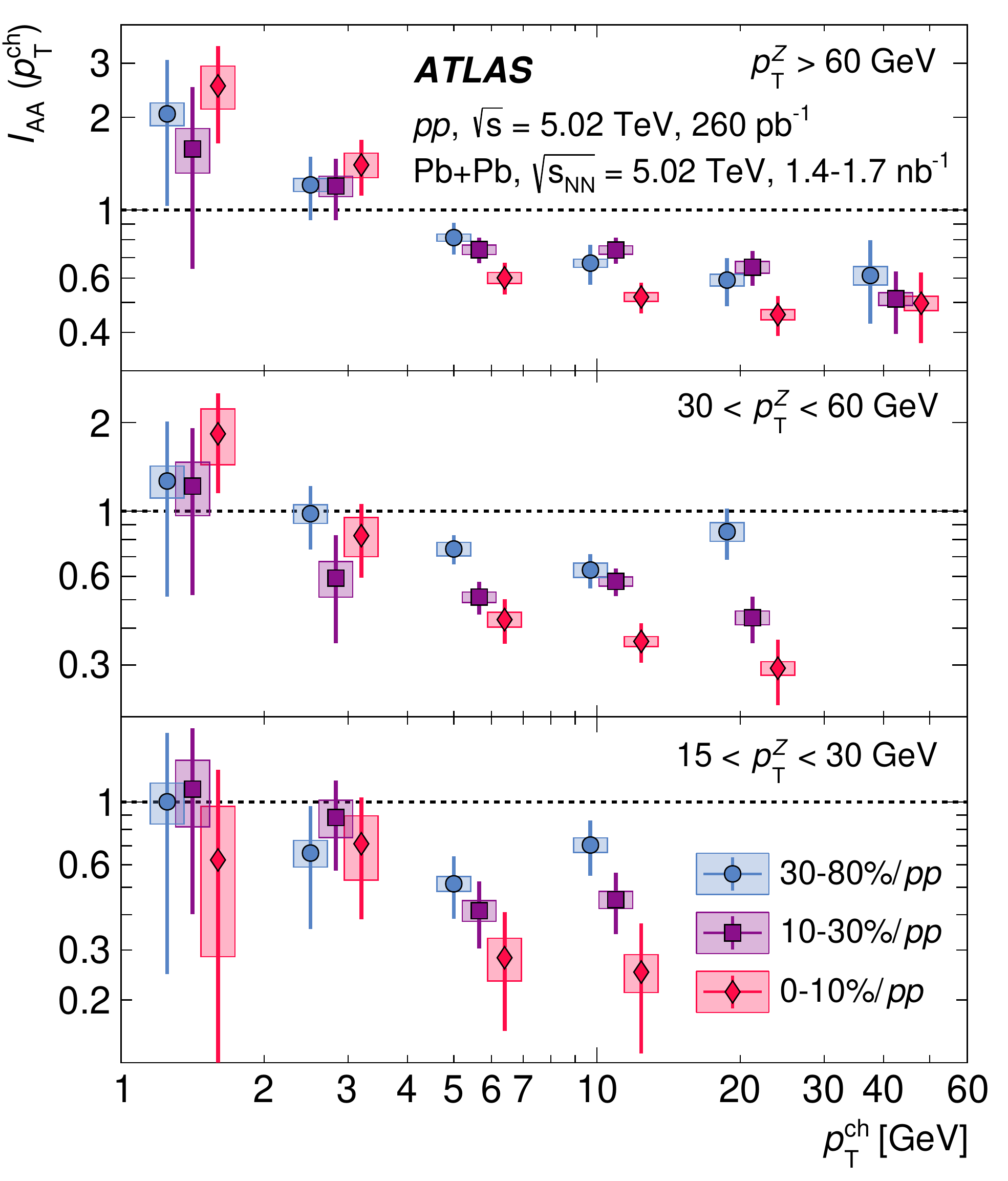}
    \includegraphics[width=0.32\textwidth]{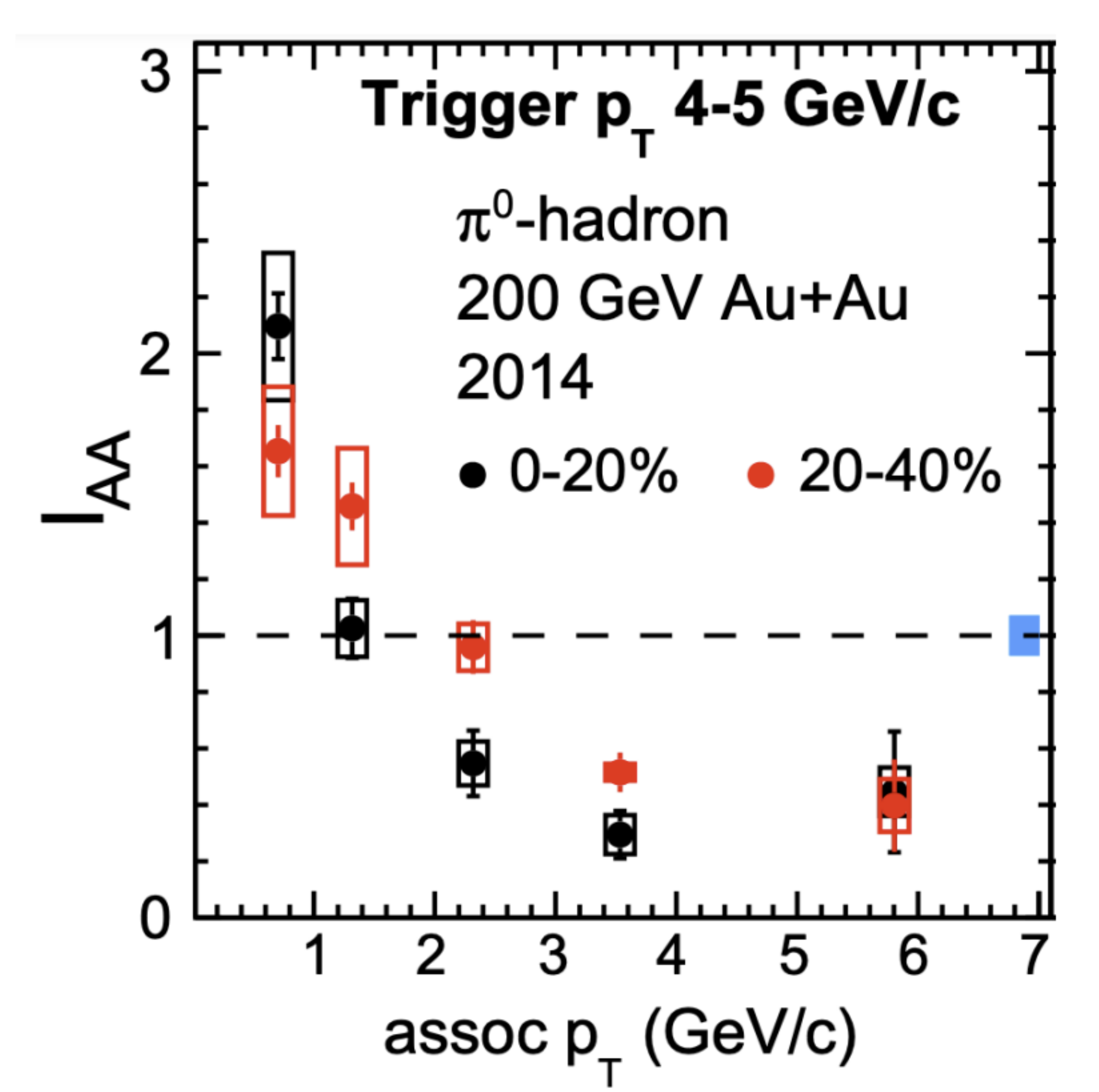}
    \caption{Example of measurements~\cite{CMS:2021otx,ATLAS:2020wmg,PHENIX:2020alr} of hadrons produced in association with a $Z$ boson (left, middle), or a $\pi^0$.}
    \label{Figure:ZGammaHadron}
\end{figure}

The energy loss of jets can also be established by measuring jets associated with a Z boson or a high energy photon~\cite{CMS:2017ehl,CMS:2017eqd,CMS:2012ytf,ATLAS:2018dgb}.  Relative to the transverse momentum of the $Z/\gamma$, a smaller \pt of the jet is seen in heavy-ion collisions compared to the reference $pp$ collisions, as shown in Figure~\ref{Figure:ZGammaJetBalance}. 
Events with these topologies are studied in more detail by looking into distribution of hadrons with respect to the Z boson~\cite{CMS:2021otx,ATLAS:2020wmg} or photon~\cite{PHENIX:2020alr}  As can be seen in Figure~\ref{Figure:ZGammaHadron}, it is found that there is also an excess of particles on the same side as the Z.  In addition to photons, jets and hadrons recoiling against a high-energy $\pi^0$ has been studied at RHIC, indicating a broadening of jet shower and a modification of the acoplanarity distribution when compared to the PYTHIA baseline.

Investigation of the distribution of hadrons inside jets provides insight on the jet quenching phenomenon.  There has been a number of measurements on the fragmentation function of inclusive jets~\cite{CMS:2014jjt,CMS:2012nro,ATLAS:2018bvp,ATLAS:2017pgl,ATLAS:2017nre,ATLAS:2014dtd,ALICE:2022vsz,ALICE:2020pga,ALICE:2018ype}, as well as Z/photon tagged jets~\cite{CMS:2018mqn,ATLAS:2019dsv}.  Together with the jet transverse momentum profile and jet-track correlation measurements~\cite{CMS:2022btc,CMS:2021nhn,CMS:2020geg,CMS:2018jco,CMS:2018zze,CMS:2016cvr,CMS:2016qnj,CMS:2015hkr,CMS:2013lhm,ATLAS:2019pid,ALICE:2019sqi,ALICE:2019whv} it is observed that the energy of the jet is transported away from the axis on average as soft particles, and it extends beyond $\Delta R \sim 1.0$.

\begin{figure}[ht!]
    \centering
    \includegraphics[width=0.3\textwidth]{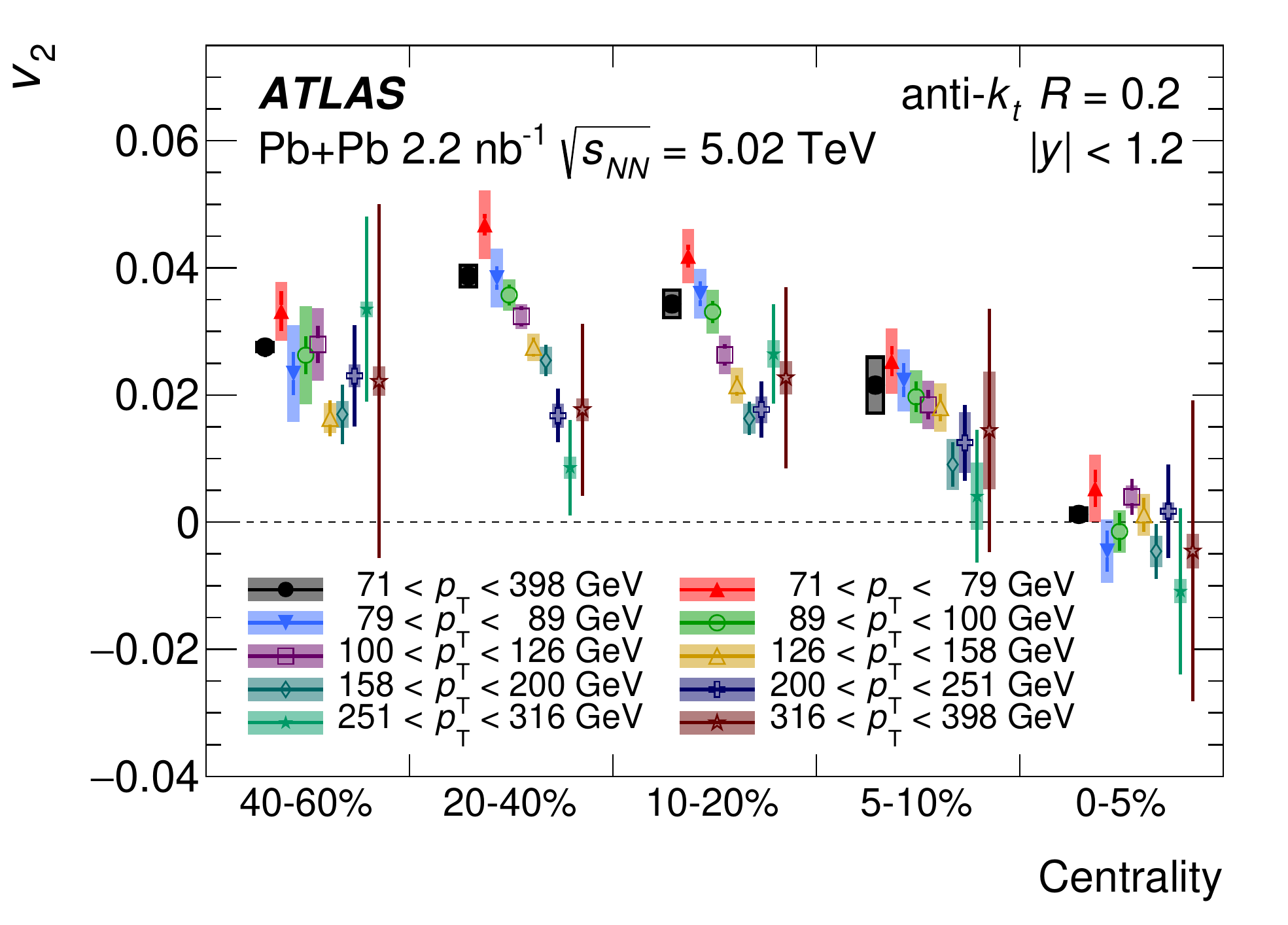}
    \includegraphics[width=0.3\textwidth]{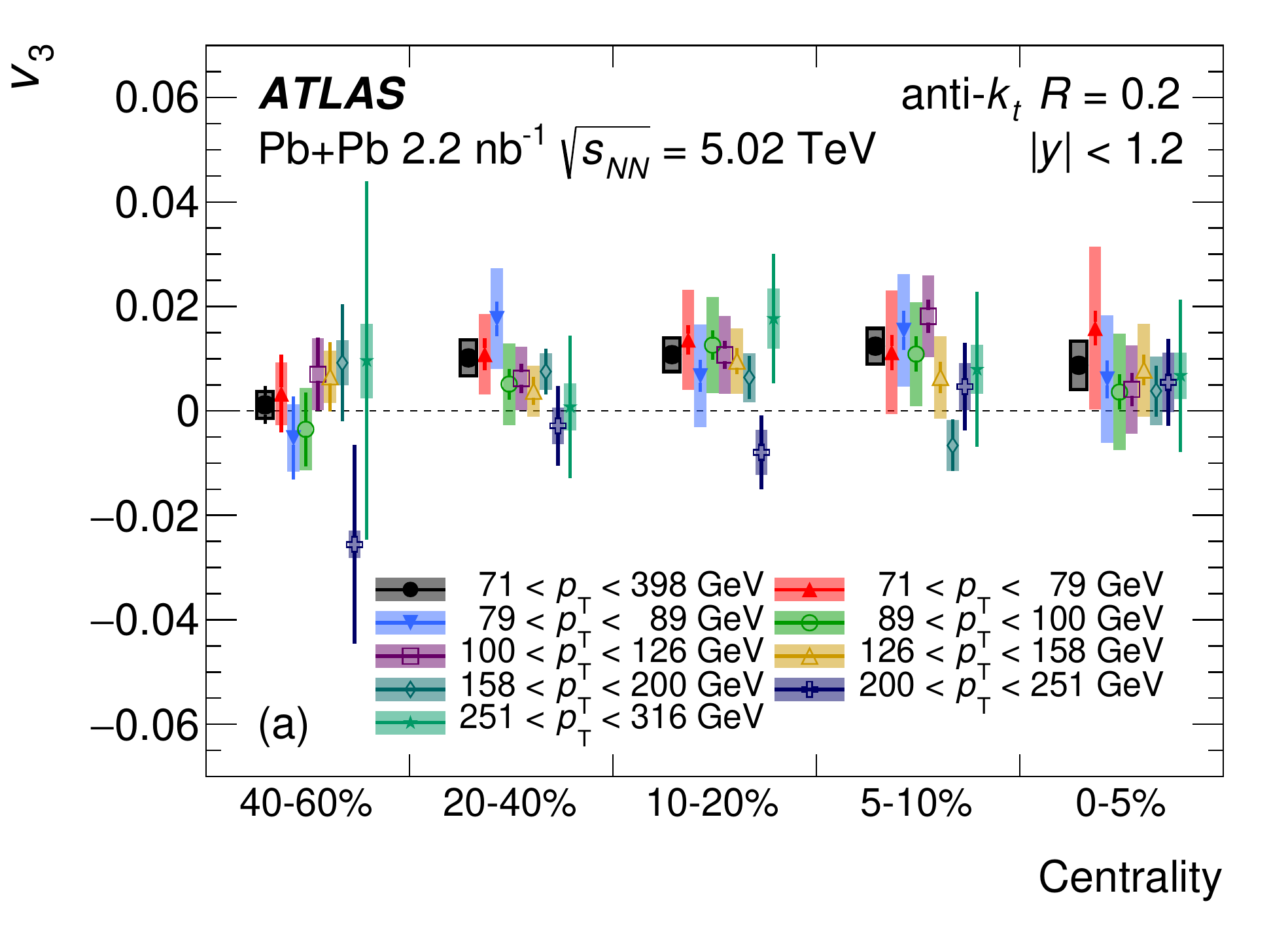}
    \includegraphics[width=0.3\textwidth]{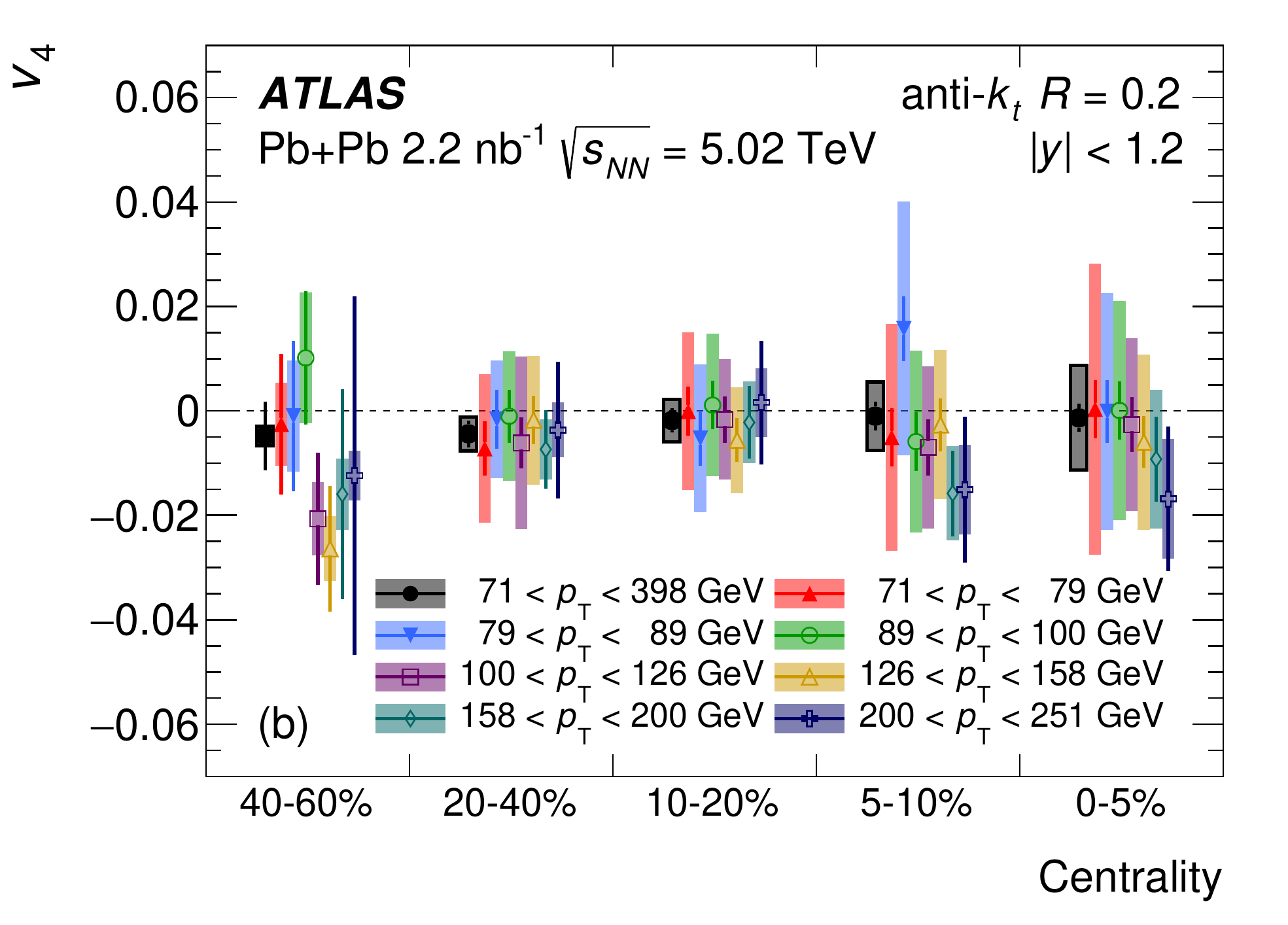}\\
    \includegraphics[width=0.6\textwidth]{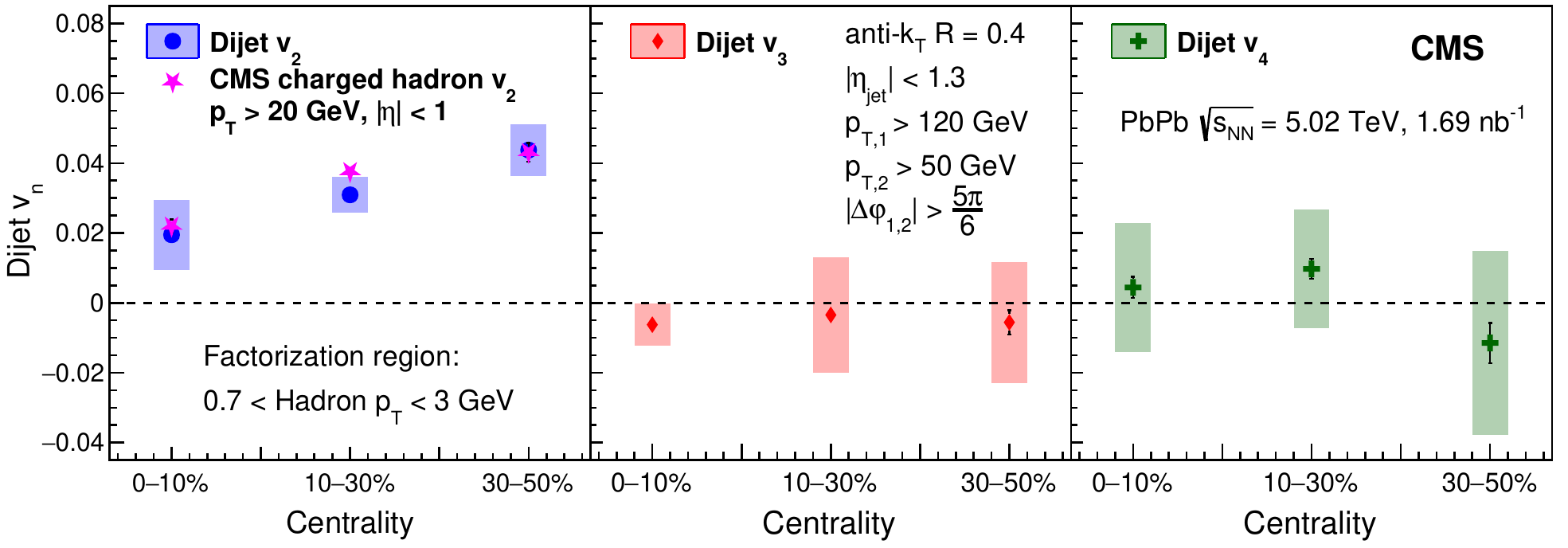}
    \caption{Jet and dijet $v_n$ measurements~\cite{CMS:2022nsv,ATLAS:2021ktw,ATLAS:2013ssy} at the LHC.}
    \label{Figure:JetVn}
\end{figure}

Correlations between hadrons and jets are also further studied by classifying the leading dijet in the events in different $x_j \equiv p_{T,2}/p_{T,1}$ bins~\cite{CMS:2021nhn}.  It is found that the transverse momentum profile of the subleading jet in a less balanced configuration is more modified compared to its counterpart in a more balanced dijet configuration.  Figure~\ref{Figure:JetVn} shows jet and dijet $v_n$ measurements~\cite{CMS:2022nsv,ATLAS:2021ktw,ATLAS:2013ssy} which correlate jet direction with all the hadrons in the event.  A $v_2$ signal is observed while $v_3, v_4$ are consistent with zero within uncertainties.

\subsubsection{Jet substructure}

\label{sec:progress:microscopic:jet_substructure}
Jets play a crucial role in the study of QGP tomography~\cite{Cunqueiro:2021wls, Apolinario:2022vzg}. It was already realized in the 2015 \LRP~\cite{Akiba:2015jwa} that jets can serve as multi-scale probes of the medium due to their rich internal substructure which encodes information about the jet evolution at various energy scales \cite{Andrews:2018jcm}. Average jet properties such as the energy profile (jet shape) as a function of the distance from the jet axis, and longitudinal momentum carried by individual particles within jet (fragmentation function) were studied at both RHIC~\cite{PHENIX:2020alr, PHENIX:2012aba, PHENIX:2010hgz, PHENIX:2010nlr, STAR:2021kjt, STAR:2020xiv, STAR:2017hhs, STAR:2016dfv, STAR:2009axa} and the LHC~\cite{CMS:2013lhm, CMS:2014jjt, CMS:2018mqn, CMS:2018jco, ATLAS:2019dsv, ATLAS:2018bvp, ATLAS:2014dtd, ALICE:2019whv, ALICE:2022vsz, ALargeIonColliderExperiment:2021mqf, ALargeIonColliderExperiment:2021mqf}. The results showed that a fraction of the original jet's energy was redistributed away from its axis due to jet-medium interactions and were carried predominantly by low energy particles. Jet substructure studies going beyond average quantities such as angularity~\cite{ALICE:2018dxf,ALICE:2021njq} and mass~\cite{ATLAS:2018jsv,ALICE:2017nij,CMS:2018fof} were also performed to utilize the more differential phase space mapped out by per-jet substructure fluctuations. With sPHENIX as the new detector at RHIC, upgrades to STAR and the detectors at the LHC (ATLAS, CMS, ALICE and LHCb) were planned and carried out since the 2015 \LRP to allow for precision jet substructure measurements in an environment where medium effects are expected to be significant. Therefore, we stand at the precipice of a quantitative jet substructure program aiming at elucidating the multi-scale, space-time evolution of jets and the QGP. This progress will depend on successful data-taking and the completion of the RHIC scientific mission in Runs 2023-2025, as well as utilizing the LHC Run 3 and Run 4 data for new high precision scans of the jet substructure phase-space.

The longitudinal and transverse momentum distributions of jet particles encode comprehensive information about the whole jet evolution history. This motivates an infinite class of jet substructure observables defined via inclusive sums of transverse momentum weighted angles among particles or jet axis to arbitrary powers~\cite{Larkoski:2013eya, Larkoski:2014pca}. On the other hand, motivated by the sequential gluon emissions predicted by theory, clustering algorithms define another class of jet substructure observables which are quantified by the information available in the clustering tree~\cite{Dreyer:2018nbf}. Both methods contain contributions from underlying events or even pileups in high-luminosity proton collisions which complicate the comparison with theoretical calculations. Therefore jet grooming to remove unwanted particles based on characteristic differences between signal jets and their backgrounds (typically soft and at large angles from the jet axis) needs to be performed. In addition, subtraction techniques based on clustering or per-particle identification were developed for high-multiplicity collisions such as in high-pileup proton-proton or central heavy-ion collisions~\cite{Bertolini:2014bba, Berta:2014eza}. %
By removing sensitivities to soft particles, originally motivated by background mitigation purposes, one can design jet observables which are dominated by perturbative processes~\cite{Larkoski:2014wba, Dreyer:2018tjj}. These observables then offer a robust approach to probe the wide range of emission phase space~\cite{Chien:2019osu} which are now being explored by the community. 

With the rise of modern machine learning development concurrently with the 2015 \LRP period~\cite{Guest:2018yhq, Albertsson:2018maf, Larkoski:2017jix, Haake:2018hqn}, jet substructure studies benefited from these powerful data analysis tools to extract the rich information contained in jets, deal with background corrections and detector effects. Frameworks to capture comprehensively jet information were proposed using different jet representations, including substructure observable bases~\cite{Thaler:2010tr,Datta:2017rhs, Komiske:2018cqr, Komiske:2017aww, Chien:2018dfn} and also de-clustering tree branching history~\cite{Dreyer:2018nbf} as a bottom-up approach. This naturally extends conventional, single-variable jet modification studies to correlations among multiple jet features. In particular, the correlation between jet quenching and substructure was explored to investigate how jets with different topology or substructure would lose energy differently. Machine learning tools were directly applied on jet representations to extract modification patterns first in simulation-based studies~\cite{Verma:2021ceh, Lai:2021ckt, Du:2021qwv, Du:2021brx, Du:2020pmp, Chien:2018dfn, Apolinario:2021olp, Liu:2022hzd}. Such efforts help not only define concrete observables to eventually test underlying jet quenching mechanisms in data, such as creating a direct connection to theory without the need for advanced methods, but also provide opportunities for searching modification patterns using data-driven methods. 

To map out jet evolution in more detail therefore requires measurements of multiple jet observables in a high-dimensional space. The jet clustering tree which provides a splitting history has been actively utilized in recent experimental and theoretical studies. 
Measurements of the individual, subsequent branching are also in progress. We see that later branching and narrow-angle ones start to deviate from the expectation of perturbative parton splitting and show a more democratic momentum sharing \cite{Chien:2021yol, Apolinario:2022guz}.  
On the other hand, machine learning techniques to extract the full multi-dimensional correlation among different substructure observables are applied in a cleaner environment \cite{Andreassen:2019cjw, H1:2021wkz}. It is promising to unfold multi-differential distributions in heavy-ion jet studies once we have experimental control over individual jet substructure observables. The necessity of precision in both experimental measurements and theoretical calculations will be an important aspect of the 2023 \LRP.

As described earlier, groomed jet observables can reduce the impact of soft, wide-angle radiation which is theoretically more challenging to be described. %
Fully corrected data can therefore be compared to theoretical calculations with higher accuracy. In general, a jet substructure observable is defined as a sum over contributions from both hard and soft particles in a jet. By restricting the jet constituents that contribute to the observable, as is what is done in grooming, one can design substructure observables with sensitivities to specific subsets of jet constituents. Here we refer to hard jet substructure as those observables which have reduced sensitivities to soft jet particles, while soft jet substructure either retains or has enhanced sensitivity to soft particles. 

\paragraph{Hard jet substructure}

One of the first measurements for hard jet substructure at the LHC was the soft-drop groomed momentum sharing $z_g$ \cite{CMS:2017qlm}. It is defined as the momentum fraction of the first soft branch in the declustering satisfying the condition,
\begin{equation}
    z = \frac{{\rm min}(p_{T,1},p_{T,2})}{p_{T,1} + p_{T,2}} > z_{\rm cut} \Big(\frac{R_{12}}{R}\Big)^{\beta}
\end{equation}
where $1,2$ label the two branches in the declustering tree and $z_{\rm cut}$ and $\beta$ are the soft-drop parameters. $R_{12}$ is the angle between the two branches and $R$ is the jet radius. The groomed jet radius $R_g$ is similarly defined as the angle between two such branches. There is less ambiguity in the origin of high-energy particles because they are primarily produced in hard scattering. These particles are more likely to survive soft-drop which removes wide-angle soft particles due to the angular-ordered declustering. Any observable based on the constituents of a groomed jet is potentially a hard jet substructure. However, it is also realized that the effect of grooming can differ among jets. For example, a jet with a large groomed jet radius would have little radiation removed, therefore retains the sensitivities to soft particles. 

\begin{figure}
    \centering
    \includegraphics[width=0.95\textwidth]{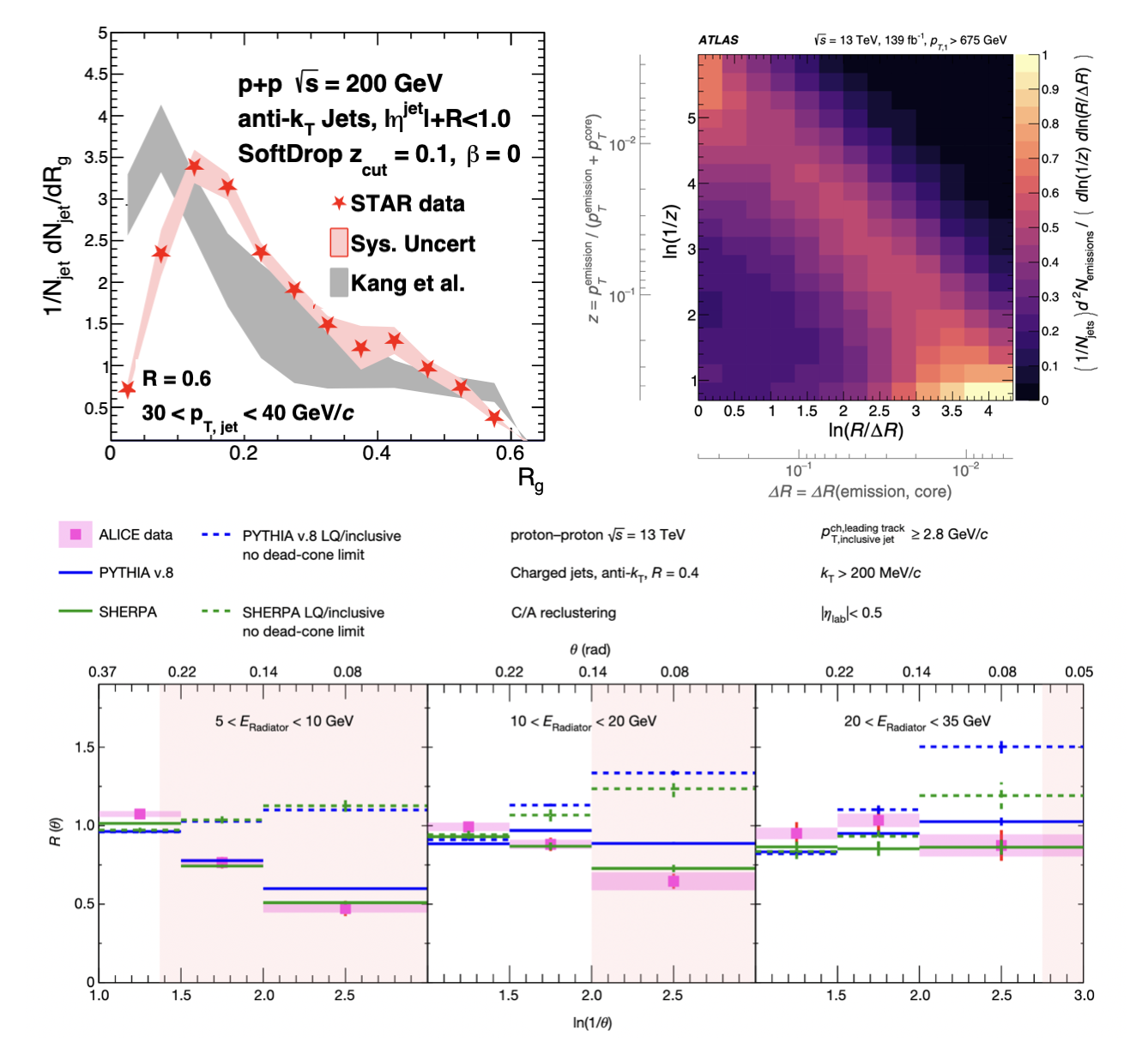}
    \caption{Top left: Groomed jet radius $R_g$ distribution for jets at RHIC energies in the red markers compared to a pQCD calculation in the gray shaded region~\cite{STAR:2020ejj}. Top right: Fully corrected primary Lund plane for high-momentum jets at the LHC with the angle and momentum fraction as the coordinates ~\cite{ATLAS:2020bbn}. Bottom: Observation of the dead-cone effect as a suppression of jets with small opening angle and tagged with a D0-meson, with varying emitter energies~\cite{ALICE:2021aqk}.}
    \label{fig:jss_pp}
\end{figure}

Figure~\ref{fig:jss_pp} highlights three such jet substructure measurements and their impact on our understanding of QCD. The top left panel shows the distribution of the opening angle of the first hard splitting of jets at RHIC energies from the STAR experiment, compared to a NLL pQCD calculation in the shaded gray band~\cite{STAR:2020ejj}. The comparison suggests the necessity of higher order calculation and non-perturbative contribution in order to describe the data. The top right panel shows the ATLAS measurement~\cite{ATLAS:2020bbn} of the Lund plane for high-momentum jets at the LHC, with specific regions of the phase-space sensitive to hadronization or perturbative emissions. The ALICE collaboration utilized the jet clustering tree and compared the opening angle distributions of inclusive jets and jets tagged with the $D^0$ meson. The dead-cone effect was observed in the suppression of narrow angle emissions as shown in the bottom panel of Figure~\ref{fig:jss_pp}.

\begin{figure}
    \centering
    \includegraphics[width=0.95\textwidth]{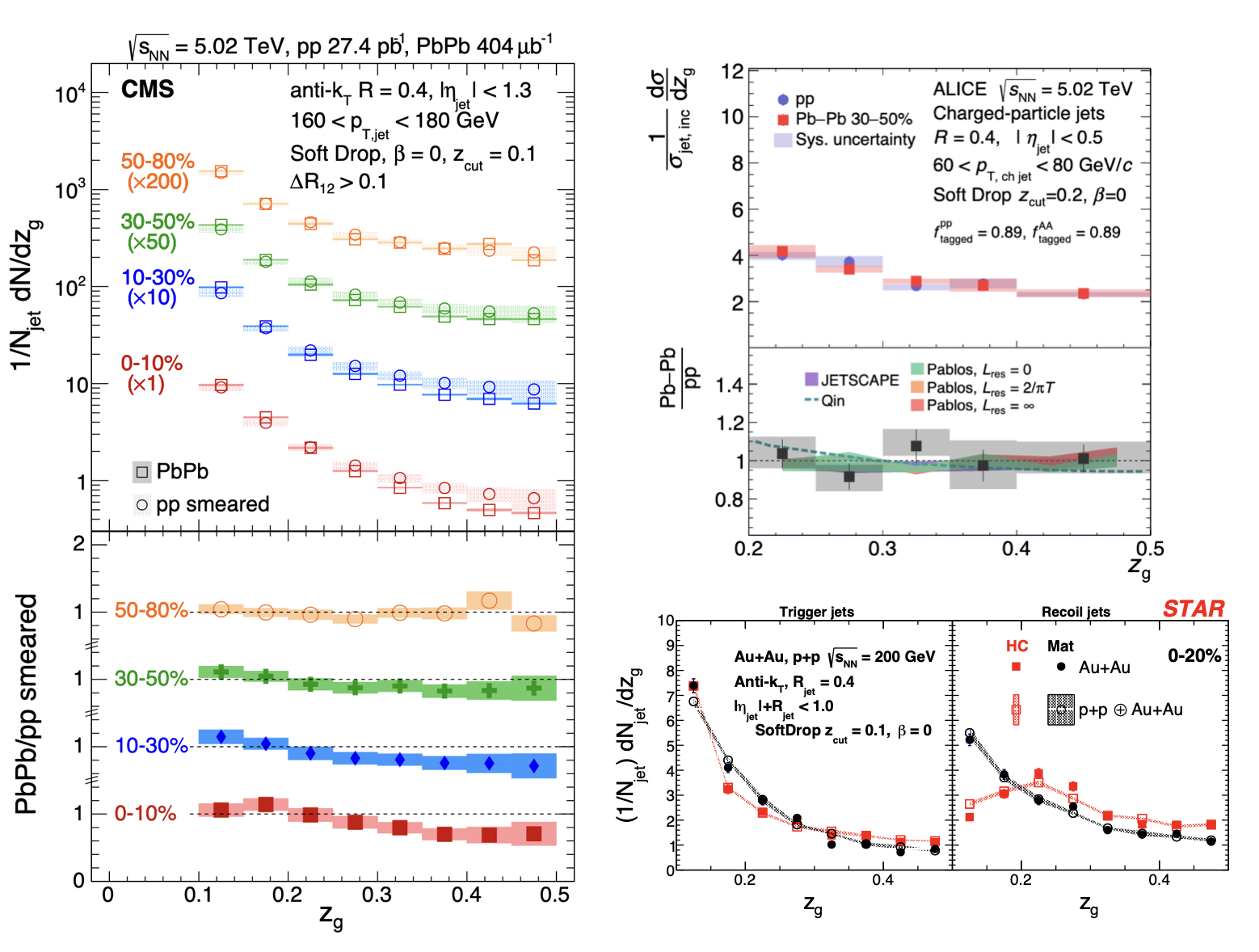}
    \caption{Left: CMS measurement~\cite{CMS:2017qlm} of the soft-drop $z_g$ for high-momentum jets at the LHC with different centrality selections in \PbPb{} collisions, compared to the vacuum baseline (open markers). The bottom panel shows the ratio between \PbPb{} and \pp{} results. Top right: Normalized $z_g$ distributions from the ALICE collaboration~\cite{ALargeIonColliderExperiment:2021mqf} with $z_{cut} = 0.2$ for low energy jets at the LHC. The bottom panel shows the ratio with the vacuum baseline. No significant modification is observed even for the most central collisions. Bottom right: Measurements of the normalized $z_g$ distributions for trigger (left panel) and recoil (right panel) jets from the STAR collaboration~\cite{STAR:2021kjt} for the most central \AuAu{} collisions compared to an embedded \pp{}\&\AuAu{} baseline. Red markers represent the HardCore di-jets and black markers represent the Matched di-jets. }
    \label{fig:aa_zg}
\end{figure}

Since the 2015 \LRP, there has been significant progress on quantifying the modification of jets in their substructure. The primary goal is to correlate jet suppression to substructure and quantitatively discriminate between proposed jet quenching mechanisms. With substructure observables, one can identify characteristic scales which are sensitive to jet-medium interactions and the QGP's microscopic properties. Measurements of jet shapes and masses~\cite{ALICE:2017nij, CMS:2018fof} suggest a narrowing of the jet core, while soft contributions such as medium responses to jets can still affect significantly the outer jet regions. Modifications of hard jet substructure will inform us of the onset of jet-medium interaction in the whole jet formation process. These complementary measurements highlight the multi-faceted nature of medium recoils and jet quenching effects; the jet mass seems to favor less recoils whereas the shapes and fragmentation function require a larger contribution in order to fit the data. In looking more into the convolution of the varying scales within jets, the first application of groomed jet substructure observables in heavy-ion collisions was the measurement of the momentum fractions at the first hard emission which survives soft-drop. The left panel of Figure~\ref{fig:aa_zg} presents the first such measurement of the $z_g$ distribution from the CMS collaboration~\cite{CMS:2017qlm} at the LHC with $z_{cut} = 0.1$ and $\beta = 0$. This measurement, along with the measurement of the groomed jet mass~\cite{CMS:2018fof}, showed small modifications which are consistent with suppression of hard emissions with $z_{g} \sim 0.5$. Little modification is seen when one imposes more stringent grooming criterion with a larger $z_{cut}$ as in the measurement from the ALICE collaboration~\cite{ALargeIonColliderExperiment:2021mqf} (top right panel of Figure~\ref{fig:aa_zg}), or selects jets with a hard constituent ($p_{T} > 2.0$ GeV) at RHIC by the STAR collaboration~\cite{STAR:2021kjt} (bottom right panel). 

There has also been much progress recently on revisiting energy correlators and connecting them to advanced theoretical tools of calculating correlation functions \cite{Chen:2020vvp,Andres:2022ovj}. Since these energy correlators are cross sections of angles weighed by particle energies, they are insensitive to soft particles and belong to hard jet substructure. The small-angle behavior of energy correlators may give a robust probe of low energy QCD dynamics and hadronization \cite{Chien:2021yol,Apolinario:2022guz}. 

At the current stage, the study of hard jet substructure helped initiate a more robust, hopefully, understanding of jet evolution in both vacuum and medium. For quenched jets, we observe little modification of some hard jet substructures in a variety of measurements. This leads us to the investigation of soft jet substructure which may be more sensitive to medium-induced effects which we will discuss next. 

\paragraph{Soft jet substructure}

Soft particles within a jet may be produced by a variety of sources such as vacuum-like or medium-induced soft emissions, correlated (typically referred to as medium recoil) and uncorrelated medium background particles. The quantitative description of these soft particles to disentangle their different origins is an active area of research. 

\begin{figure}
    \centering
    \includegraphics[width=0.95\textwidth]{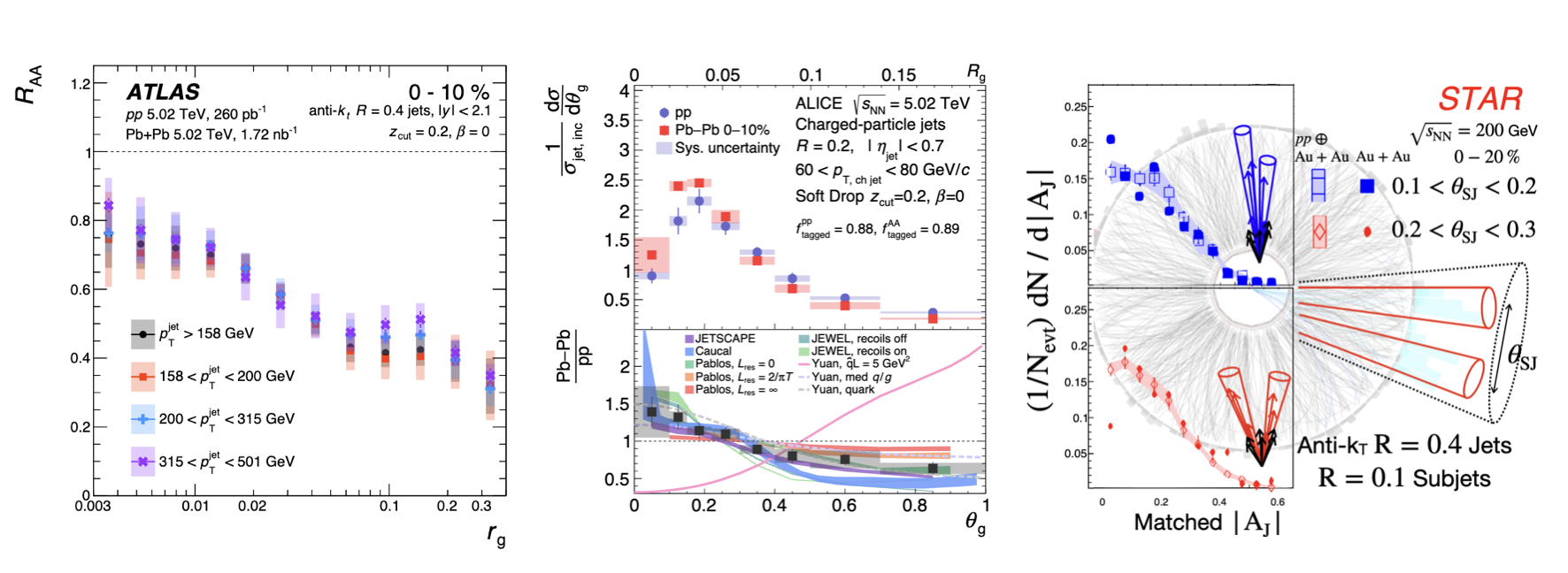}
    \caption{Left: Jet nuclear modification factor $R_{AA}$ for most central events as a function of the soft-drop groomed jet radius $R_{g}$ for varying jet momenta from the ATLAS collaboration~\cite{ATLAS:2022vii}. Middle: Normalized distributions of the groomed jet radius $R_{g}$ for jets from central heavy-ion collisions from the ALICE collaboration~\cite{ALICE:2022hyz} compared to the vacuum baseline and theoretical calculations. Right: Di-jet momentum asymmetry at RHIC from the STAR collaboration~\cite{STAR:2021kjt} for selected di-jets with varying sub-jet opening angles in central \AuAu{} collisions compared to an embedded vacuum reference.}
    \label{fig:aa_rg}
\end{figure}

Soft particles within jets generated due to medium interactions are evident from the excess of low-momentum particles, which are impacted by the heavy-ion background. There have recently been two approaches to study the correlation between the hard splitting angle and the degree of quenching at both RHIC and the LHC. Grooming techniques naturally introduce an angular observable at the selected hard splitting which was studied recently at the LHC as discussed in the previous section. By increasing the value of $z_{cut}$, one reduces the impact of the heavy-ion background in the selection of the splitting. It can also introduce a bias toward a later splitting which can affect the discussion on the space-time picture of jet quenching. The first such studies were done at the LHC as shown in the left and middle panel of Figure~\ref{fig:aa_rg} by measuring the nuclear modification factor as a function of the $R_{g}$ for different jet momenta~\cite{ATLAS:2022vii, ALICE:2022hyz}. Both measurements observe a variation of the quenching effects at an angular scale smaller than $0.1$. Such a study has the potential to point towards studying coherence vs decoherence energy loss by probing the coherence length of the plasma. With the bias of later splitting introduced by a harder grooming criterion, such measurements are complementary to the ones performed at RHIC as shown on the right panel of Figure~\ref{fig:aa_rg}. Since the energy scale of jets at RHIC is significantly lower, a jet selection bias called the HardCore selection is introduced which biases the surviving jets towards harder fragmentation in the recent STAR measurements~\cite{STAR:2021kjt}. %
The momentum asymmetry for these selected di-jet was measured differentially as a function of the opening angle of sub-jets with a momentum threshold as opposed to a momentum fraction requirement. The angular resolutions between $0.1 < \theta < 0.3$ show similar levels of momentum asymmetry caused by the loss of low momentum particles leading to the quantification of energy loss from a single color charge with an unmodified splitting structure. These studies suggest a consistent understanding of jet energy loss for a particular, biased jet population via a broad selection of jet and event geometry in experiment.

A critical challenge in the study of soft jet substructure is to appropriately separate the signal from the background. The medium-induced emissions within the jet cone as well as correlated and uncorrelated medium radiations can all affect soft jet substructure. We have seen experimental evidence of the recovery of momentum balance, i.e. the lost energy from the jet can be recovered in the soft sector. The extent to which one needs to go further away from the jet axis to recover the lost energy is highly dependent on the type of jets considered in the study. For inclusive jets at the LHC, we have seen that one needs to go to almost the entire hemisphere. On the contrary, for hard-fragmentation triggered jets at RHIC the situation is significantly different, with the quenched energy being recovered within a relatively narrow jet cone.

\subsubsection{Heavy-flavor tagged jets}

\label{sec:progress:microscopic:heavy_flavor_tagged_jets}

Studying jets originating from heavy flavor quarks (charm and bottom) provides additional insights to the jet quenching mechanism in the QGP. Due to their larger masses, heavy quarks are expected to lose less energy due to gluon radiation than light quarks, known as the dead-cone effect \cite{Dokshitzer:2001zm}. Such a QCD-inspired effect has been confirmed experimentally by the ALICE collaboration in \pp\ collisions, as illustrated in Figure~\ref{fig:jss_pp} \cite{ALICE:2021aqk}.

Identification of heavy flavor jets experimentally usually employs two approaches: i) exploitation of the distinct properties of heavy flavor hadrons and their decay topology, e.g., longer decay length and larger mass compared to light favor hadrons; ii) tagging with fully reconstructed heavy flavor hadrons or their leptonic decay daughters. 

As a reference for measurements of modifications to heavy flavor jets in the hot medium, nuclear modification factors for $b$-jets \cite{CMS:2015gcq,ALICE:2021wct} and $c$-jets \cite{CMS:2016wma} have been measured in \pPb\ collisions at $\sqrts$ = 5.02 TeV at the LHC, where a QGP of extended volume is not expected to be formed. The results are consistent with no significant modifications to the heavy flavor jet production in the cold nuclear matter. 

Modifications to $b$-jet yield in heavy-ion collisions compared to \pp\ collisions have been measured by CMS and ATLAS collaborations \cite{CMS:2013qak, ATLAS:2022fgb}, where a strong suppression of the $b$-jet production rate is observed. 
\begin{figure}[htbp]
\begin{minipage}{1.0\linewidth}
\centerline{\includegraphics[width=0.55\linewidth]{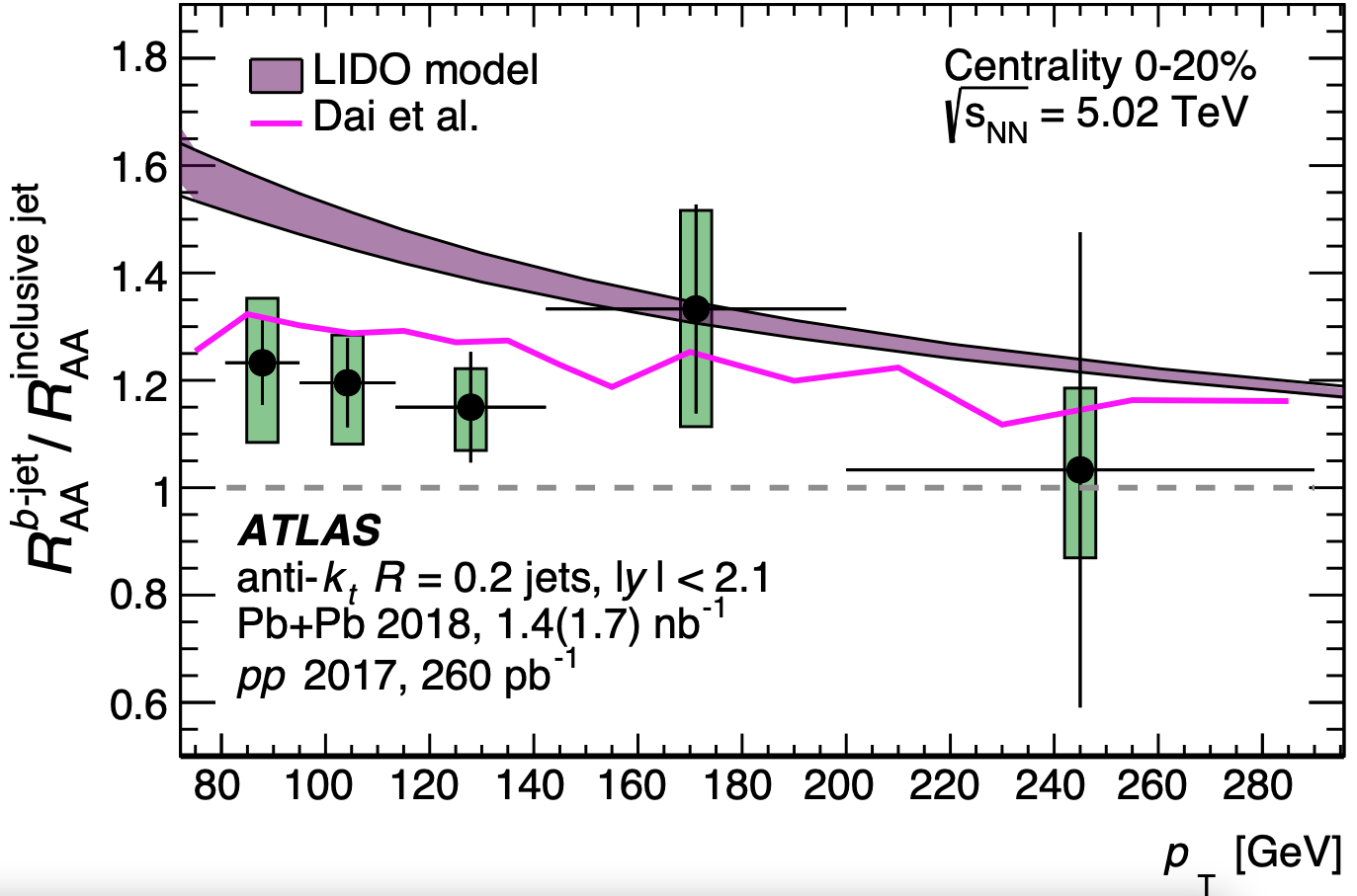}}
\end{minipage}
\caption[]{Ratio of $b$-jet \raa to that of inclusive jets as a function of \pt in 0-20\% \PbPb\ collisions at $\sqrts$ = 5.02 TeV \cite{ATLAS:2022fgb}.}
\label{fig:HFjets_LHC_bjet_Raa}
\end{figure}
To study the dependence of the medium-induced energy loss on parton's mass and color charge, Figure~\ref{fig:HFjets_LHC_bjet_Raa} shows the ratio of \raa for $b$-jets and inclusive jets. The former contains contributions from gluons splitting into $b\bar{b}$ pairs, and the latter is made up of jets mostly originating from light quarks and gluons. The ratio is consistently above 1, indicating that $b$-jets are less suppressed than inclusive jets. These results suggest that mass and color charge play an important role in determining the parton energy loss. 

\begin{figure}[htbp]
\begin{minipage}{0.475\linewidth}
\centerline{\includegraphics[width=0.85\linewidth]{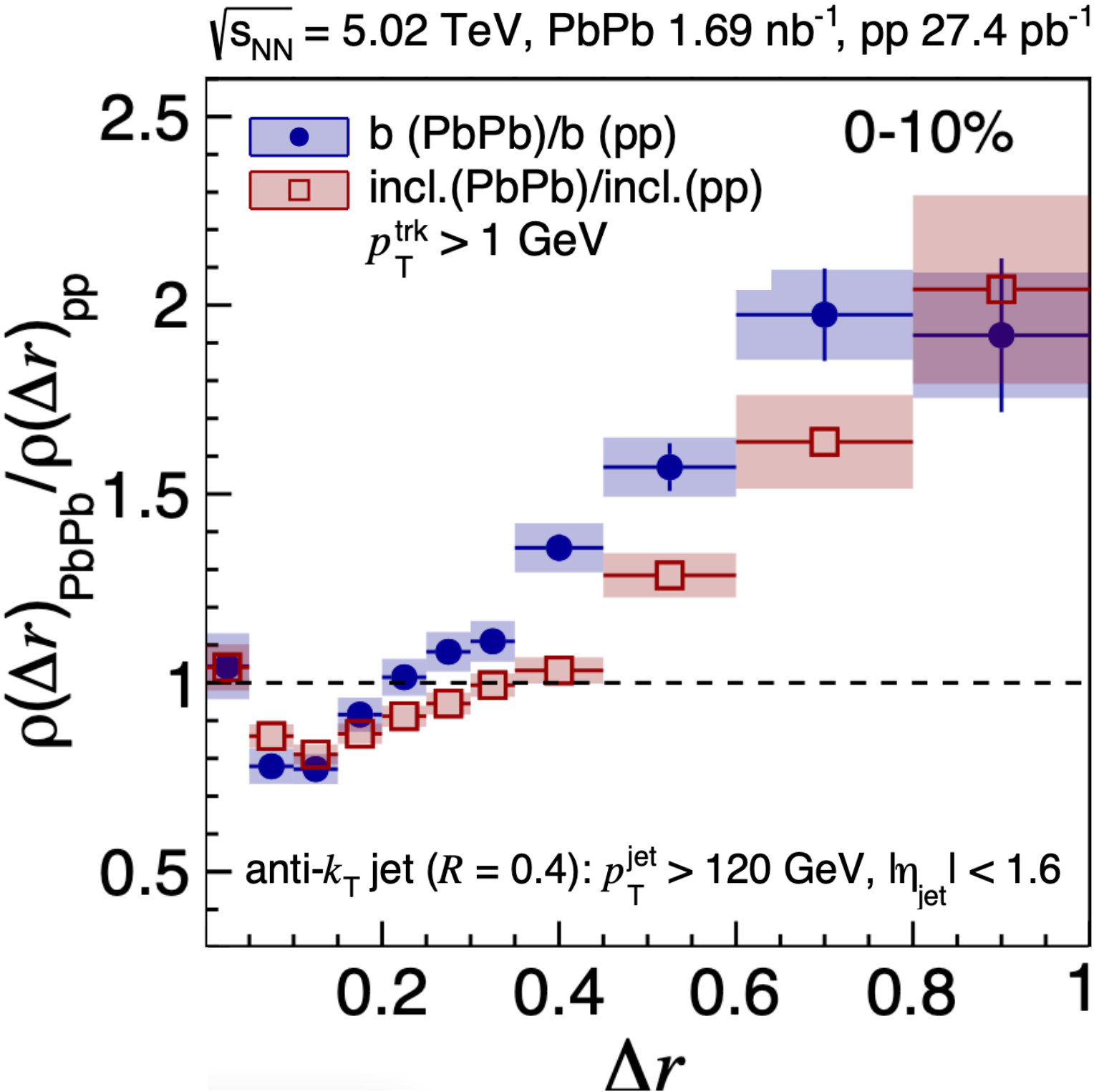}}
\end{minipage}
\begin{minipage}{0.525\linewidth}
\centerline{\includegraphics[width=0.9\linewidth]{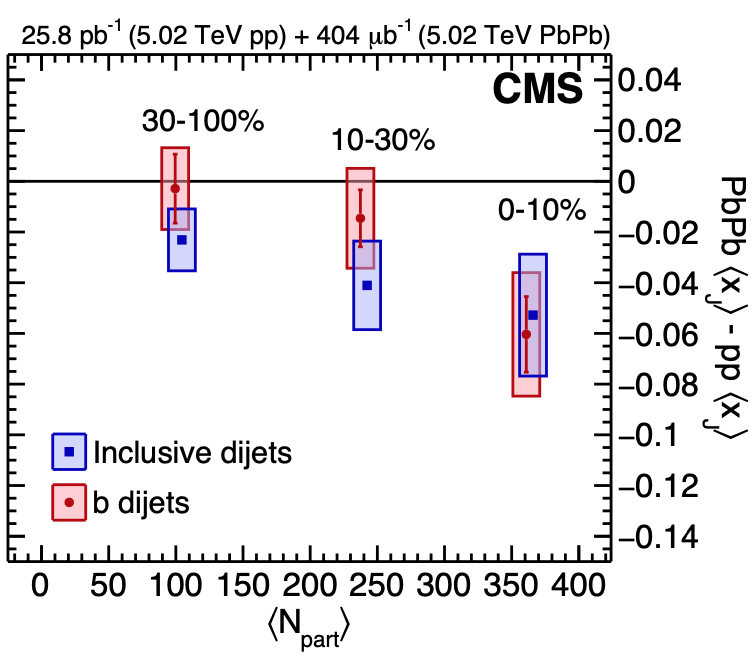}}
\end{minipage}
\caption[]{Left: ratio of jet shape distributions in \PbPb\ and \pp\ collisions at $\sqrts$ = 5.02 TeV for $b$-jets and inclusive jets \cite{CMS:2022btc}. Right: difference of dijet imbalance measured in \PbPb\ and \pp\ collisions at $\sqrts$ = 5.02 TeV for $b$-dijets and inclusive dijets \cite{CMS:2018dqf}.}
\label{fig:HFjets_LHC_bjet_substructure}
\end{figure}

Substructures of heavy flavor jets are used to probe the jet quenching phenomenon in more detail. The left panel of Figure~\ref{fig:HFjets_LHC_bjet_substructure} shows the ratios of jet shape distributions, {\it i.e.} normalized transverse momentum profile of charged particles with \pt $>$ 1 GeV/$c$ in jets, in central \PbPb\ and \pp\ collisions at $\sqrts$ = 5.02 TeV for $b$-jets (filled circles) and inclusive jets (open squares) \cite{CMS:2022btc}.
For both types of jets, a redistribution of energy to larger distances from the jet axis is seen, and the effect is stronger for $b$-jets than for inclusive ones. Figure~\ref{fig:HFjets_LHC_bjet_substructure}, right panel, shows the difference in $\langle x_{J}\rangle$ as a function of centrality between \PbPb\ and \pp\ collisions at $\sqrts$ = 5.02 TeV for $b$-dijets (circles) and inclusive dijets (squares) \cite{CMS:2018dqf}. Here, $\langle x_{J}\rangle=\langle p_{\rm{T,2}}/p_{\rm{T,1}}\rangle$, where $p_{\rm{T,1}}$ and $p_{\rm{T,2}}$ are the transverse momenta of leading and subleading jets in a dijet pair. In 0-10\% central collisions, $\langle x_{J}\rangle$ is shifted to a small value in \PbPb\ collisions compared to \pp\ collisions, indicating jet energy loss in the QGP. The amount of shift is compatible between $b$-dijets and inclusive dijets.

\begin{figure}[htbp]
\begin{minipage}{0.475\linewidth}
\centerline{\includegraphics[width=0.95\linewidth]{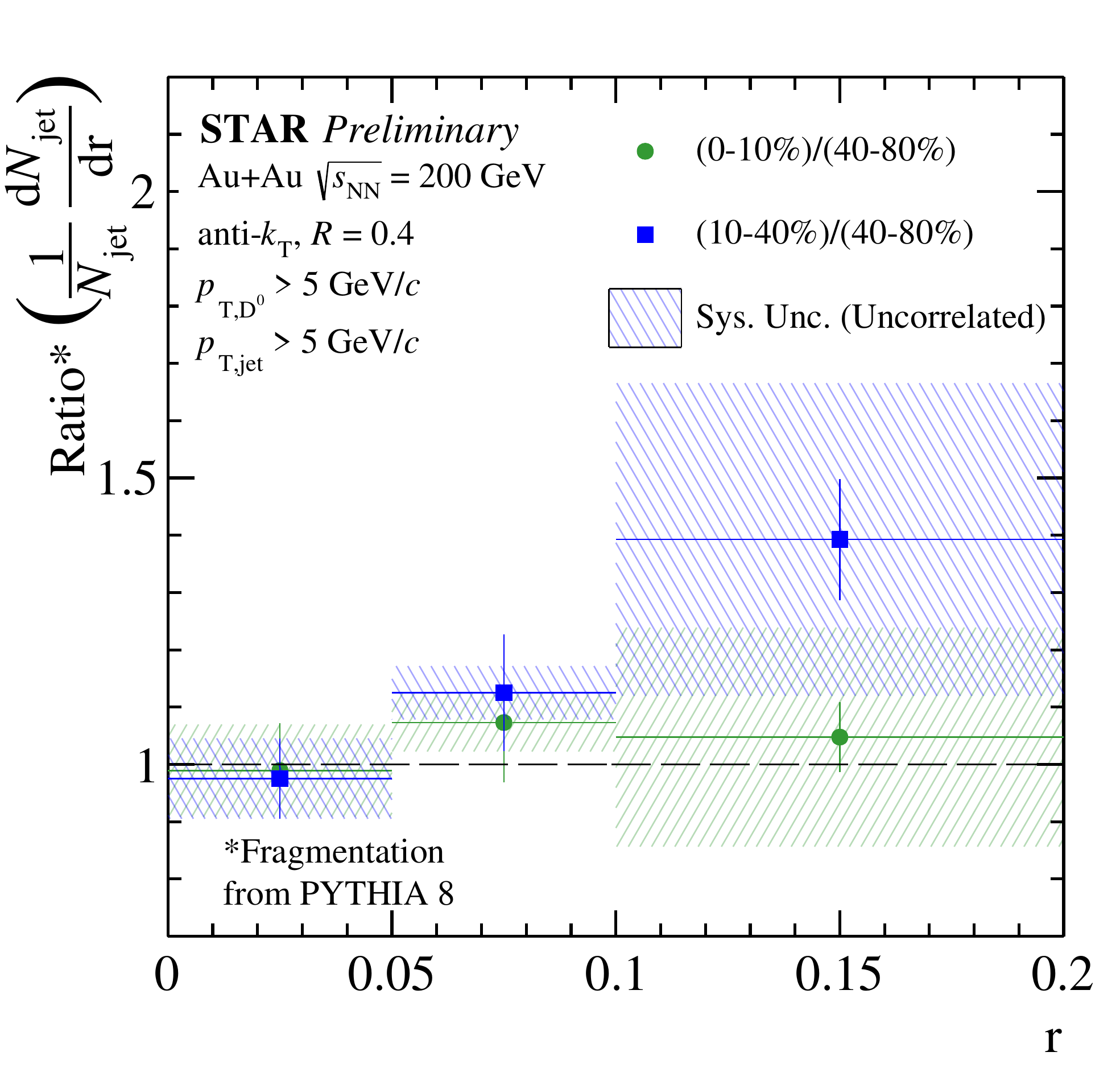}}
\end{minipage}
\begin{minipage}{0.525\linewidth}
\centerline{\includegraphics[width=0.65\linewidth]{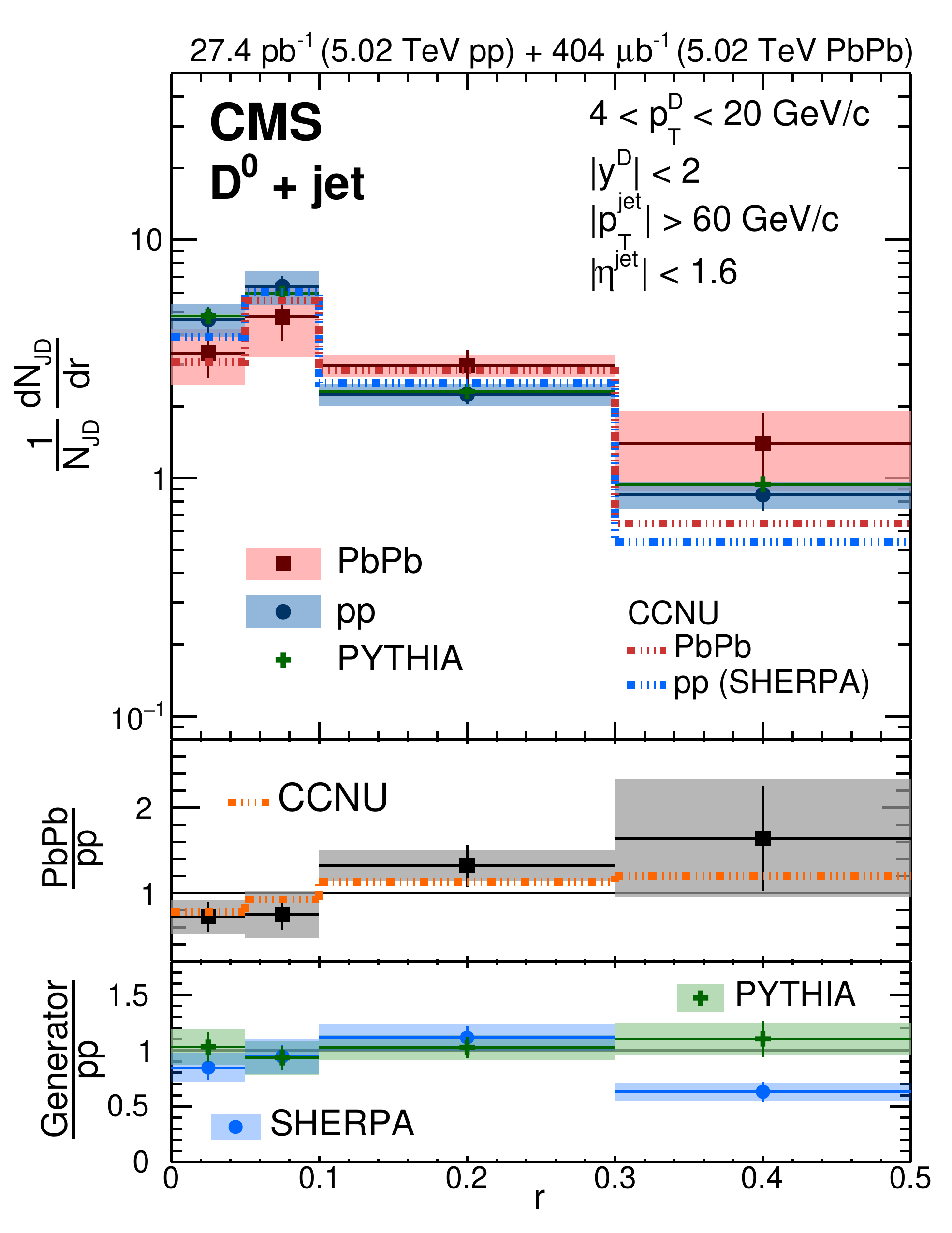}}
\end{minipage}
\caption[]{Measurements of the radial distribution for $D^{0}$ mesons with respect to the jet axis at RHIC (left) \cite{Roy:2022yrw} and the LHC (right) \cite{CMS:2019jis}.}
\label{fig:HFjets_cjet}
\end{figure}

The $c$-jet substructure is studied by measuring the radial profile of the $D^{0}$ mesons in jets, {\it i.e.} distribution of the distance between $D^{0}$ mesons and the jet axis. Figure~\ref{fig:HFjets_cjet} shows the ratios of the $D^{0}$ meson radial profiles in heavy-ion and \pp\ collisions at RHIC (left) and the LHC (right, middle panel) \cite{Roy:2022yrw, CMS:2019jis}.
At RHIC, the ratio in central collisions is consistent with unity within uncertainties for $p_{\mathrm{T},D^{0}} > 5$ GeV/$c$ and $p_{\rm{T,jet}} > 5$ GeV/$c$, while there is a hint of $D^{0}$ mesons diffusing away from the jet axis in heavy-ion collisions at the LHC for $4 < p_{\mathrm{T},D^{0}} < 20$ GeV/$c$ and $p_{\rm{T,jet}} > 60$ GeV/$c$. The observation at the LHC is quantitatively different from charged particle radial distributions in inclusive jets, where a suppression is seen for $p_{\mathrm{T,trk}} > 4$ GeV/$c$ and $r>0.1$ in \PbPb\ compared to \pp\ collisions \cite{CMS:2018zze}.

\subsubsection{Bayesian methods for hard probes}
With the proliferation of results in the hard sector in recent years described in this paper, analyses following a global parameter extraction approach have become increasingly important, in line with what was mentioned in the 2015 Long Range Plan.
Due to the nontrivial dependence of QGP transport properties on the experimental observables, many different effects are often entangled in the measurements.
A global analysis is well suited to extract the properties, for example (but not limited to) on the viscosities and $\hat{q}$, with precision and from increasingly larger set of measurements simultaneously.
It is complementary to many other studies utilizing a smaller set of data to understand the properties of QGP.
Many recent advances from the statistics community also allows rigorous and scalable analyses with well-controlled precision.
One of the popular approaches is the so-called Bayesian analysis approach~\cite{Bayesian1,Bayesian2,Bayesian3}.
In this approach, Bayesian statistics is used to describe the posterior probability density in the theory parameter space and thereby extract the ``best-fit'' parameters.

Bayes’ Theorem states that
\begin{align}
P(\vec{\theta} | \vec{x}) = \dfrac{P(\vec{x} | \vec{\theta}) P(\vec{\theta})}{P(\vec{x})},
\end{align}
where the term $P(\vec\theta|\vec{x})$ denotes the conditional probability density of theory parameter $\vec{\theta}$ condition on experimental data $\vec{x}$, and is usually referred to as the ``Bayesian posterior function.''
The term $P(\vec{x}|\vec\theta)$ is the ``Bayesian likelihood function'', and it is typically obtained through Monte-Carlo simulation methods where a set of events are simulated following a given input parameter set $\vec\theta$ and subsequently analyzed to compare with experimental input.
The computing cost to obtain the Bayesian likelihood function is nontrivial, and it is not computationally feasible to calculate for all possible input parameter $\vec\theta$.
The common approach to this problem is to calculate the Bayesian likelihood function for only a selected set of parameter points (typically referred to as ``design points''), and perform interpolation, for example with Gaussian process emulators~\cite{GaussianProcess}, to obtain the function for the full parameter space.
It has been observed that one can obtain reasonable performance with a much smaller number of design points compared to a naive grid-like approach where the number of points scales with the volume of the parameter space.
Finally, a popular approach to analyze the Bayesian posterior function in recent work is to utilize Markov-Chain Monte-Carlo (MCMC) methods~\cite{MCMC,Foreman-Mackey_2013} to obtain projected parameter density functions in one or two dimensions.

Many recent works~\cite{JETSCAPE:2021ehl,Fan:2022ton,Xie:2022ght,Xu:2017obm,Xie:2022fak} for the hard probes have been focusing on the extraction of the $\hat{q}$ parameter and charm diffusion coefficient $D_s$.
Even with the analysis methods to drastically reduce the computation cost needed, the computing requirement is still considerable.
Therefore many recent results adopted the approach to separate the soft and the hard sector: the parameters corresponding to the bulk properties are taken as input to the calculation and are not treated as part of the target parameter space for the analyses.
Many recent advances in the Monte-Carlo generators (for example the JETSCAPE/XSCAPE framework~\cite{Putschke:2019yrg}) allow a more accurate description of a wider class of jet observables that are sensitive to treatment of the correlated medium response.  For example the multi-staged approach employed by ref.~\cite{JETSCAPE:2021ehl,Fan:2022ton} attempts to describe the parton energy loss with the MATTER~\cite{Kordell:2017hmi} and the LBT~\cite{Cao:2016gvr,He:2015pra} models each specialize in different ranges of parton virtuality, allowing more flexibility in the modeling of the jet quenching process.  Xie et al~\cite{Xie:2022ght} employed an information field~\cite{Bialek:1996kd,Ensslin:2013ji,LemmField1999} based approach in the extraction of the $\hat{q}$ parameter to explore potential new ways to extract $\hat{q}$ without an explicit functional assumption.
Current analyses also highlighted a need for an improved experimental uncertainty reporting.

The scope of analyses is usually limited by the computational capacity.
There has been a number of ongoing efforts to improve the current Bayesian analysis methods in the context of parameter extraction in heavy-ion physics, focusing on ways to reduce computational requirements for a similar performance.
As an example, the active learning approach~\cite{settles2009active} iteratively includes new design points to maximally reduce uncertainties from the interpolation procedure.
Approaches such as transfer learning~\cite{Liyanage:2022byj} or multi-fidelity~\cite{MultiFidelity} utilize existing interpolated functions as the base either from related analysis or from lower fidelity models which do not cost as much to run.  The full simulations then only need to capture the difference between the full model and the base model, thereby reducing computing needs.
A sustained effort in research on methodology improvements is vital for the future physics program for global parameter extraction, where parameter spaces that are much larger than those probed in the current iteration of analyses need to be explored, and which would not be possible without significant improvement to the analysis method.

\subsection{Microscopic II: Quarkonia, open heavy flavor, electromagnetics and bound states}

\subsubsection{Theory: heavy flavor}

\label{sec:progress:microscopic:theory_heavy_flavor}
From a theoretical perspective, heavy flavor probes provide a unique opportunity to learn about the in-medium QCD force and its manifestation in the binding of heavy quarks and in inelastic or elastic reaction rates
involving heavy quarks. Below we discuss theory developments related open heavy flavor and quarkonia.

{\bf Open Heavy Flavor.} In the deconfined QGP, heavy quarks produced at the initial stage of a heavy-ion collision propagate through the QGP, during which their momenta change and their energies are lost. By studying the nuclear modification factors in the production of open heavy flavor hadrons, we can probe how the QGP modifies the transport of heavy quarks and thus studying the QGP transport properties. Cold nuclear matter effects (CNM) such as the nPDF also affect the heavy flavor hadron production. Yields of different heavy flavor mesons and baryons are also sensitive to the hadronization mechanism, which can be studied in other collisions such as $e^-e^+$, $e^-p$ and \pp{} collisions. Hadronization, however, may be modified
by the presence of hot medium, see e.g. Refs. \cite{Mannarelli:2005pz,Ravagli:2007xx}.
This is consistent with the lattice study of charm fluctuations and correlations
\cite{Mukherjee:2015mxc} as discussed below.

At low transverse momentum, the dominant energy loss mechanism of heavy quarks is elastic scattering, while at high-transverse momentum, radiative energy loss becomes dominant. Radiative energy loss can be studied perturbatively and is similar to the energy loss of a high-energy parton, except for the finite mass effect. An effective description of heavy quark dynamics inside the QGP is based on a Langevin equation with diffusion and drag, as well as a radiation term added. The heavy quark diffusion coefficient encodes the QGP transport property that can be probed by studying open heavy flavor hadrons. The heavy quark diffusion coefficient has been studied perturbatively, nonperturbatively via AdS/CFT~\cite{Herzog:2006gh,Gubser:2006qh,Casalderrey-Solana:2006fio,Andreev:2017bvr}, lattice QCD methods~\cite{Banerjee:2011ra,Ding:2012sp,Francis:2015daa,Brambilla:2020siz,Altenkort:2020fgs},  and the T-matrix approach which uses some input from 
lattice QCD \cite{Liu:2016ysz,Liu:2017qah}. It was also determined from experimental data via Bayesian analysis~\cite{Xu:2017obm,Cao:2018ews}.
In the large heavy quark mass limit, the heavy quark diffusion
coefficient is obtained from the spectral function of the chromoelectric field strength correlator~\cite{Casalderrey-Solana:2006fio,Caron-Huot:2009ncn}
\begin{align}
\kappa_{\rm fund} =  \frac{g^2}{3N_c} {\rm Re} \int d t\, \big\langle {\rm Tr}_{\rm c}[ U(-\infty,t) E_i(t) U(t,0) E_i(0) U(0, -\infty)] \big\rangle_T \,,
\end{align}
where $E_i$ is the $i$th component of the chromoelectric field and $U$ denotes a timelike Wilson line in fundamental representation. 
There was significant progress made in the lattice QCD calculations using this approach 
\cite{Francis:2015daa,Altenkort:2020fgs,Brambilla:2020siz,Brambilla:2022xbd,Banerjee:2022uge,Banerjee:2022gen}. These calculations have been so far limited to quenched QCD. The results obtained by different methods and different groups are consistent and the lattice results hint toward a temperature dependence of the heavy quark diffusion coefficient. The mass suppressed contribution to the heavy quark diffusion coefficient that can be expressed in terms of chomo-magnetic field strength correlator~\cite{Bouttefeux:2020ycy,Laine:2021uzs}, was estimated very recently independently by two groups~\cite{Brambilla:2022xbd,Banerjee:2022uge}, leading to the same result of $10-20\%$ correction for bottom quark and $20-30\%$ correction for charm quark compared to the case of infinitely heavy quark for temperature around $1.5T_c$. The first lattice calculation of the
heavy quark diffusion coefficient in 2+1 flavor QCD appeared very recently \cite{Altenkort:2023oms}.
Strictly speaking in QCD the heavy quark diffusion coefficient describes the diffusion of
the conserved net heavy quark number, and by definition this 
diffusion coefficient only depends on the temperature.
To understand the production of heavy flavor hadrons in heavy ion collisions one needs to
introduce an effective momentum dependent diffusion coefficient in the Langevin dynamics \cite{Moore:2004tg}.  
In the zero momentum limit this effective diffusion coefficient reduces to the usual heavy quark diffusion
coefficient that is defined in QCD. The effective heavy quark diffusion coefficient including its temperature and momentum dependence has been calculated non-perturbatively in the T-matrix approach \cite{Liu:2016ysz,Liu:2017qah}.

It was pointed out the charm fluctuations and charm baryon number correlations obtained on the lattice can provide information about the nature of charm degrees of freedom close to the QCD crossover~\cite{Mukherjee:2015mxc}. Current lattice data hint toward the existence of charm mesons and baryons above the crossover temperature~\cite{Mukherjee:2015mxc}. This was confirmed in calculation of 
the charm fluctuations and charm light quark flavor correlations within the T-matrix approach \cite{Liu:2021rjf},
which agree well with the lattice results.
Within the T-matrix approach the in-medium spectral functions of the $D$ meson in QGP have been obtained \cite{Liu:2021rjf}.
Furthermore,  in-medium masses of charm mesons have been directly from lattice QCD meson correlation functions and it was found that corresponding masses may shift downward by about 1\% close to the crossover~\cite{Aarts:2022krz}.

{\bf Quarkonium.} Quarkonia, bound states of $c\bar{c}$ (charmonium) and $b\bar{b}$ (bottomonium) pairs, have played a unique role in probing the properties of the QGP as they are sensitive to medium effects. Hot medium effects include plasma screening of the heavy quark-antiquark potential, dissociation and recombination. In different temperature regimes, these hot medium effects can be related to different transport properties of the QGP. Quarkonia of different sizes are affected differently by these hot medium effects and thus providing a powerful probe of the QGP properties at different scales. Quarkonium suppression in heavy ion collisions is also sensitive to the feed-down structure, e.g., $\psi$(2S)$\rightarrow$\jpsi, $\Upsilon$(3S)$\rightarrow$$\Upsilon$(2S)$\rightarrow$$\Upsilon$(1S) and $\chi_b$(1P)$\rightarrow$$\Upsilon$(1S), as well as cold nuclear matter effects.

For quarkonium production at low transverse momentum, the description of quarkonium in-medium dynamics varies in different temperature regimes. Our understanding of quarkonium dynamics inside the QGP was greatly advanced by the application of the open quantum system framework (recent reviews can be found in Refs.~\cite{Rothkopf:2019ipj,Akamatsu:2020ypb,Sharma:2021vvu,Yao:2021lus}), since the last \LRP. 

An open quantum system is a system of interest coupled with an environment that is integrated out. The total Hamiltonian can be written as the sum of the system Hamiltonian, the environment Hamiltonian and their interaction $H=H_S+H_E+H_I$. The combination of the system and the environment evolve together unitarily and time-reversibly
\begin{align}
\frac{d\rho(t)}{dt} = -i[H, \rho(0) ] \,,
\end{align}
where the initial density matrix is assumed to be factorized $\rho(0)=\rho_S(0)\otimes \rho_E(0)$. If we only focus on the time evolution of the system by tracing out the environment degrees of freedom, we will arrive at a non-unitary time evolution equation that is time-irreversible. For the application to quarkonium inside the QGP, the system contains heavy quark-antiquark pairs that can be either in the color singlet or the color octet, while the QGP serves as the environment. The QGP is usually assumed to be at thermal equilibrium with a temperature $T = 1/\beta$ and thus $\rho_E(t) = e^{-\beta H_E}/{\rm Tr}(e^{-\beta H_E})$ is time independent.

\begin{figure}[t]
\centering
\includegraphics[width=0.8\textwidth]{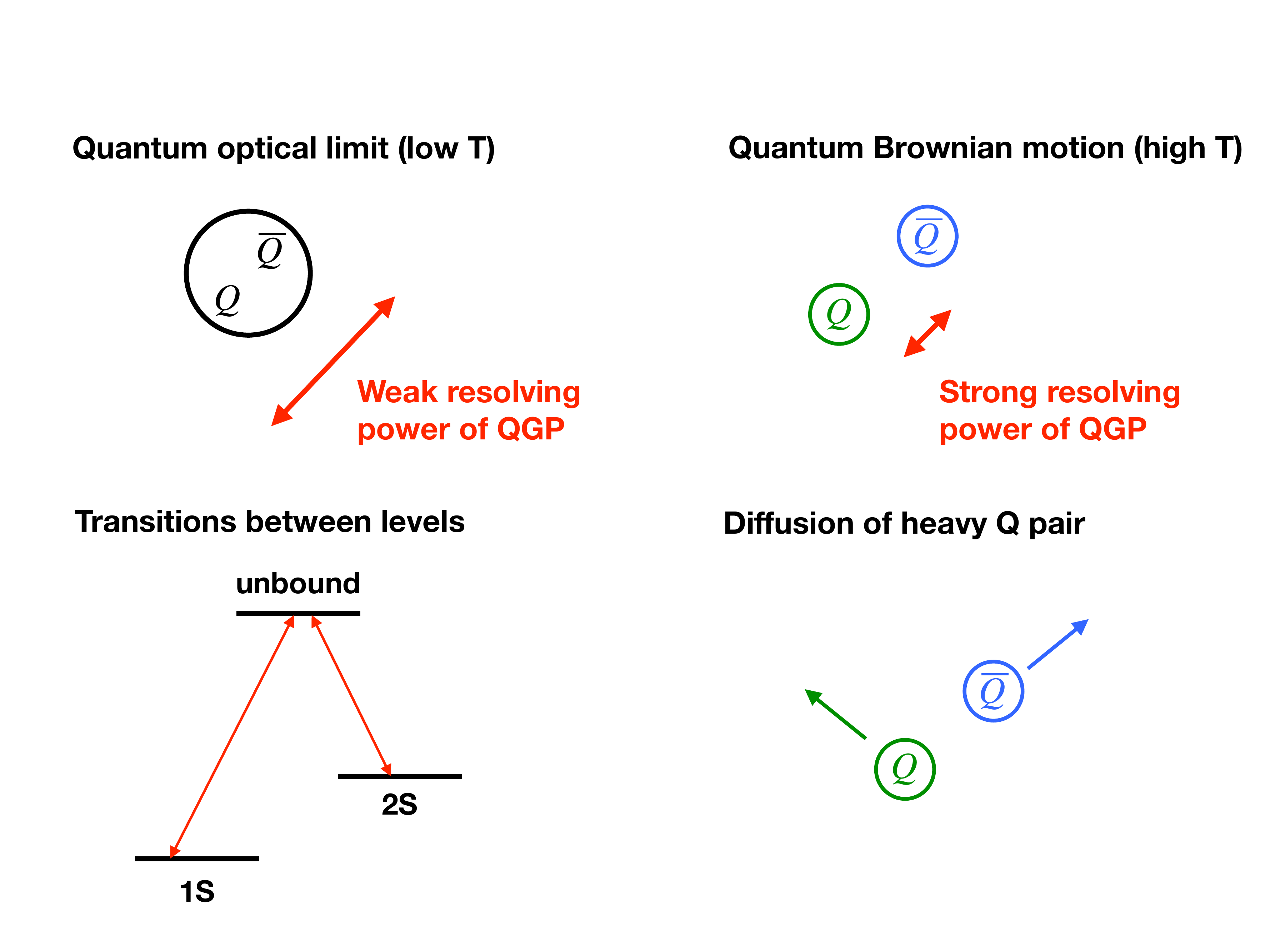}
\caption{Intuitive illustration of the quantum optical (left) and the quantum Brownian motion (right) limits. At low temperature, the resolving power of the QGP is too weak to see the internal structure of a quarkonium state and the eigenstates of the quarkonium relative motion form a good basis for computation, in which the quarkonium in-medium dynamics is described by transitions between eigenstates at different energy levels. At high temperature, the resolving power of the QGP is strong enough to see the internal structure of quarkonium and the QGP is interacting with the heavy quark and antiquark almost independently. As a result, both heavy quarks diffuse, which leads to decoherence of the quarkonium wavefunction. (Figure courtesy of Xiaojun Yao.)}
\label{fig:3.4.1_two_limits}
\end{figure}

The non-unitary and time-irreversible time evolution equation of the system can be greatly simplified in two limits. One limit is called the quantum Brownian motion limit, valid at high temperature and the other is the quantum optical limit, valid at low temperature. An intuitive illustration of these two limits is shown in Figure~\ref{fig:3.4.1_two_limits}. Both the quantum Brownian motion limit and the quantum optical limit of quarkonium transport have been investigated in recent studies. In the quantum Brownian motion limit, the time evolution of the open system is described by a Lindblad equation~\cite{Akamatsu:2012vt,Akamatsu:2014qsa,Katz:2015qja,Brambilla:2016wgg,Blaizot:2017ypk,Kajimoto:2017rel,Akamatsu:2018xim,Miura:2019ssi,Brambilla:2020qwo,Brambilla:2021wkt,Akamatsu:2021vsh,Brambilla:2022ynh,Miura:2022arv}, a non-unitary generalization of the Schr\"odinger equation. So far only the Lindblad equations for one heavy quark antiquark pair have been studied due to limited computing power, which can be improved by future developments of computing tools and machine learning techniques. In practice, many studies used the semiclassical limit of this Lindblad equation that corresponds to a Langevin equation. In the quantum optical limit, a Boltzmann equation for quarkonium dissociation and regeneration was derived by applying a Wigner transform to the density matrix and performing a gradient expansion~\cite{Yao:2018nmy,Yao:2020eqy}. The Boltzmann equation that has been widely used in quarkonium phenomenology (see e.g.~\cite{Rapp:2017chc,Du:2019tjf}) is one approximation of the more general quantum approach. Motivated by the quantum approach, recent studies started to look into the interplay between the open heavy quark dynamics and the quarkonium dissociation and recombination processes, since both of them are part of the same density matrix. Some of these studies treat the open heavy quark dynamics as diffusion in a Langevin equation~\cite{Blaizot:2017ypk,Du:2022uvj}. Other studies constructed a coupled set of Boltzmann equations for open heavy flavors and quarkonia to study quarkonium thermalization and production~\cite{Yao:2017fuc,Yao:2018zrg,Yao:2018sgn,Yao:2020xzw}. In all these studies, both spatial and momentum correlations between the unbound heavy quark antiquark pair is accounted for in recombination. Boltzmann equations with dissociation and recombination have also been used to study $X(3872)$~\cite{Wu:2022blx}, $B_c$~\cite{Wu:2023djn} and $\Xi_{cc}$~\cite{Yao:2018zze} production in heavy ion collisions.

Important new theoretical developments that are obtained or motivated from using the open quantum system framework include: (1) the understanding of quarkonium dissociation as a result of quantum decoherence of the quarkonium wavefunction~\cite{Akamatsu:2011se,Rothkopf:2013kya}; (2) the microscopic construction of the recombination term that is on the same theoretical foundation as, and thus consistent with dissociation in both quantum and classical transport approaches~\cite{Yao:2018nmy,Blaizot:2018oev,Akamatsu:2018xim,Miura:2019ssi,Yao:2020kqy,Yao:2020eqy}, which were absent in previous transport studies~\cite{Yan:2006ve,Liu:2009nb,Zhao:2010nk,Liu:2010ej,Zhao:2011cv,Song:2011xi,Du:2015wha,Zhou:2016wbo,Chen:2017duy,Zhao:2017yan,Du:2017qkv,Zhao:2021voa,Li:2022qxg}; and (3) the construction of a novel transport coefficient for a heavy quark-antiquark pair entangled in color~\cite{Brambilla:2017zei,Yao:2020eqy}, whose generalization to finite frequency also contains important information on quarkonium in-medium dynamics~\cite{Yao:2020eqy}.

With these advances, our current understanding of low-$p_T$ quarkonium in-medium dynamics is as follows: at high temperature, the $Q\bar{Q}$ pair is unbound, each of the heavy quarks diffuses in the QGP, while interacting with each other via some real potential. The real potential only lasts for a short time period, which is related to the imaginary part of the potential. The diffusion dynamics is determined by the same transport coefficient $\kappa_{\rm fund}$ as in the open heavy flavor transport. As the QGP expands and cools down, color and spatial correlations between the $Q\bar{Q}$ pair cannot be neglected any more, which affect the diffusive dynamics and the transport coefficient that is relevant becomes $\kappa_{\rm adj}$ and $\gamma_{\rm adj}$, which are defined by
\begin{align}
\kappa_{\rm adj} + i\gamma_{\rm adj} = \frac{ g^2 T_F}{3 N_c} \int d t\, \big\langle \ml{T} E^a_i(t) W^{ab}(t,0) E^b_i(0) \big\rangle_T \,,
\end{align}
where $W$ denotes a time-like Wilson line in adjoint representation.
Finally, in the low temperature regime, quarkonium dynamics can no longer be described in the diffusion picture, rather, it is described in terms of transitions between different bound and unbound states. The QGP property that determines the rates of these transitions (dissociation and recombination) is given by the finite frequency generalization of $\kappa_{\rm adj}$:
\begin{align}
[g_E^{++}]^>(\omega) = \frac{ g^2 T_F}{3 N_c} \int d t\, e^{-i\omega t} \big\langle E^a_i(t) W^{ab}(t,0) E^b_i(0) \big\rangle_T \,.
\end{align}
The difference between the fundamental and adjoint versions of the chromoelectric correlators is not in their representation, but in the operator ordering~\cite{Eller:2019spw,Scheihing-Hitschfeld:2022xqx}. At next-to-leading (NLO) order, the spectral functions of these two chromoelectric field correlators differ by a temperature independent piece~\cite{Burnier:2010rp,Binder:2021otw,Scheihing-Hitschfeld:2022xqx}
\begin{align}
\label{eqn:3.4.1.delta_rho}
\Delta \rho(\omega) = \frac{T_F g^4(N_c^2-1) \pi^2}{3 (2\pi)^3}  |\omega|^3 \,,
\end{align}
where $\Delta \rho(\omega)$ denotes the difference between these two spectral functions. The most striking feature of Eq.~\eqref{eqn:3.4.1.delta_rho} is the non-oddness in frequency. The sPHENIX program at RHIC will provide a unique opportunity to probe this object at finite frequency.

In practical calculations, the transition between different temperature regimes can be determined by lattice studies of correlation functions in  Euclidean time, which are Laplace transforms of the spectral functions and can tell in-medium quarkonium masses and width as well as the corresponding melting temperatures, see e.g. Refs. \cite{Bazavov:2009us,Mocsy:2013syh} for reviews. The main challenge of reconstructing the spectral functions from the lattice results on the correlation function is the limited extent of the Euclidean time direction, $\tau<1/T$, with $T$ being the temperature, and the limited number of temporal grid points $N_{\tau}$. The lattice calculations of the heavy flavor probes significantly matured since the last \LRP. It was realized that to deal with the limited Euclidean time extent at $T>0$ in the study of bottomonium properties at high temperatures it is advantageous to use optimized meson operators that are mostly sensitive to the meson state of interest~\cite{Larsen:2019zqv,Larsen:2019bwy}. Therefore, unlike in the previous lattice QCD calculations that 
used point-like meson operators and could not provide reliable information on the in-medium mass 
and width of different quarkonium states, now it is possible to directly obtain this information from
lattice correlators. Using the above approach and 2+1 flavor $N_{\tau}=12$ lattices the in-medium masses and widths for $\Upsilon(1S),~\Upsilon(2S),~\Upsilon(3S),
\chi_b(1P)$ and $\chi_b(2P)$ states have been estimated~\cite{Larsen:2019zqv}. The mass shift with respect to the vacuum values turned out to be insignificant within the estimated errors, but the thermal width of different bottomonia states was found to increase with increasing temperature. Furthermore, the hierarchy of the thermal width of different bottomonium states appears to follow the hierarchy of their sizes~\cite{Larsen:2019zqv}. 

The study of the complex $Q\bar Q$ potential at non-zero temperature also requires the reconstruction of the spectral function of Wilson line correlation function. The position of the dominant peak in this spectral function defines the real part of the potential, while the width gives the imaginary part of the potential~\cite{Rothkopf:2011db}. Lattice calculations with $N_{\tau}=12$ were performed very recently~\cite{Bala:2021fkm}. This calculation indicates that the real part of the potential is not screened and is about the same as in the vacuum~\cite{Bala:2021fkm}. If confirmed by future lattice calculations
wit larger $N_{\tau}$ this finding will make the revision of many models for quarkonium production in heavy ion collisions necessary. The imaginary part of the potential is quite sizable and increases with the temperature and with the separation between the quark and anti-quark~\cite{Bala:2021fkm}. 
It should be noted, however, that in lattice QCD one calculates the 
energy of a static quark anti-quark pair. This energy 
contains the  energy of the static quark and anti-quark, as well
as the interaction energy between the static quark and anti-quark. It is possible
both of these components are temperature dependent but their sum turns to be temperature independent. This possibility was discussed in Refs. \cite{Zhao:2010nk}.

Finally, by studying spatial bottomonium correlation functions it is possible to provide constraints on the melting temperature of different bottomonium states. In general the relation
of the spatial meson correlation function to the spectral function is quite involved (see e.g. Ref. \cite{Mocsy:2013syh}). However, at high temperatures it is easy to estimate the spatial correlation function
corresponding to unbound heavy quark and anti-quark and compare it to the lattice result. Such comparison
for the ground state bottomonium has been performed and it was found that for $T>500$ MeV
the corresponding lattice meson correlator agrees with the one of unbound quarks. Thus 
the 1S bottomonium state melts for $T>500$ MeV \cite{Petreczky:2021zmz}.
A similar comparison for the $\chi_b$ states indicates that these states will melt at $T \simeq 350$ MeV \cite{Petreczky:2021zmz}.

\subsubsection{Experiment: quarkonia}
\label{sec:progress:microscopic:experiment_quarkonia}
Along the progresses made from the theoretical side as detailed in the previous section, progresses on quarkonium measurements have also been made from the experimental side, which will be highlighted in the following.

\begin{figure}[htbp]
\begin{minipage}{0.50\linewidth}
\centerline{\includegraphics[width=0.95\linewidth]{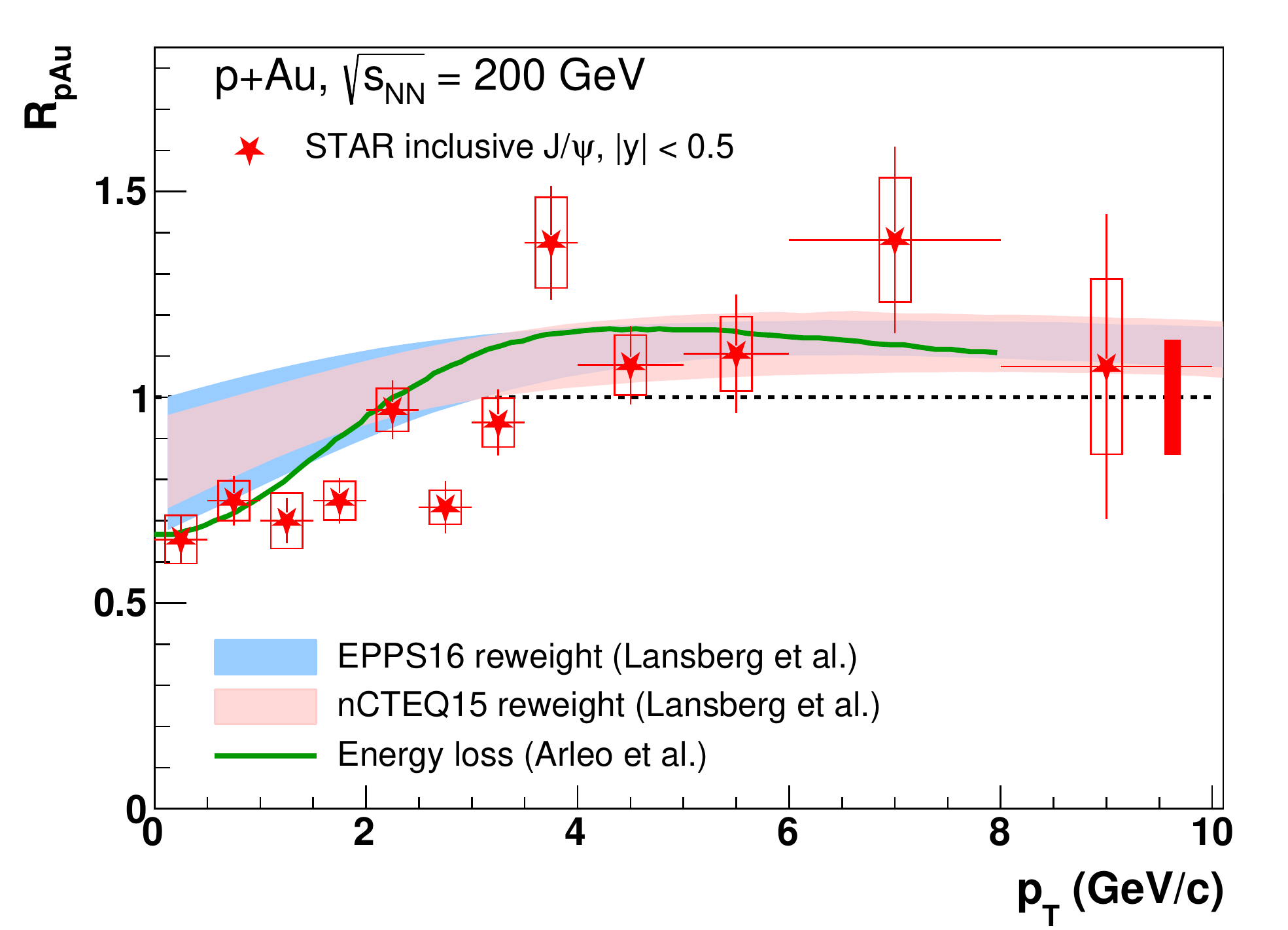}}
\end{minipage}
\begin{minipage}{0.50\linewidth}
\centerline{\includegraphics[width=0.95\linewidth]{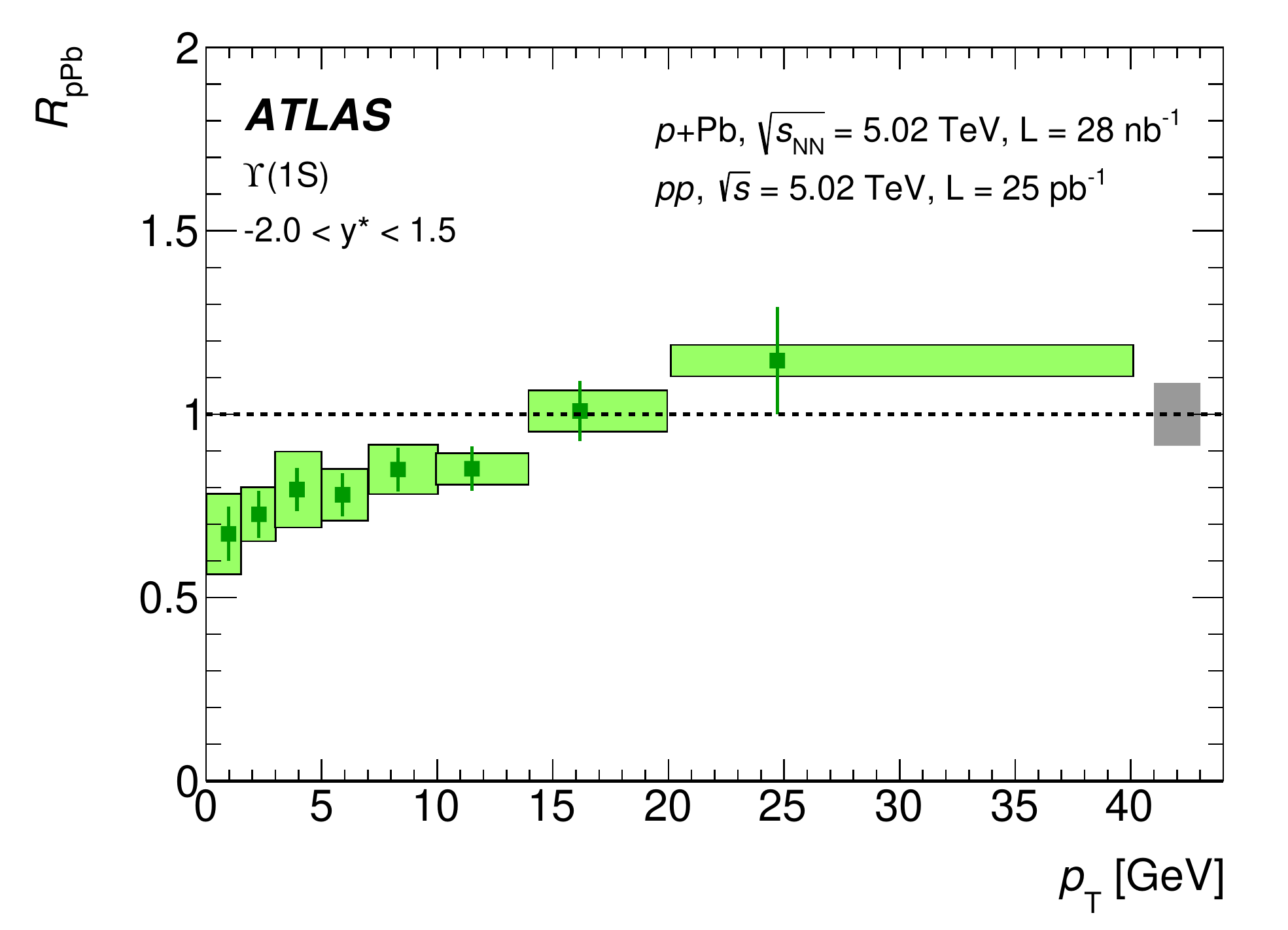}}
\end{minipage}
\caption[]{Measurements of \rpa\ as a function of \pt\ for \jpsi\ in \pAu{} collisions at $\sqrt{s_{\rm NN}}$ = 200 GeV (left) \cite{STAR:2021zvb} and for $\Upsilon$(1S) in \pPb\ collisions at $\sqrt{s_{\rm NN}}$ = 5.02 TeV (right) \cite{ATLAS:2017prf}.}
\label{fig:Quarkonia_RpA_vs_pt}
\end{figure}

One can quantify the CNM effects, including both initial and final state effects \cite{Du:2018wsj}, experimentally by measuring quarkonium production in \pA\ collisions, in which a QGP of extended volume as that observed in heavy-ion collisions is not expected to be formed \cite{PHENIX:2013pmn,PHENIX:2019brm, PHENIX:2022nrm, STAR:2021zvb, LHCb:2013gmv, LHCb:2016vqr, LHCb:2018psc, ATLAS:2017prf, CMS:2017exb, CMS:2018gbb, CMS:2022wfi, ALICE:2013snh, ALICE:2014cgk, ALICE:2016sdt, ALICE:2019qie, ALICE:2021lmn}. Figure~\ref{fig:Quarkonia_RpA_vs_pt} shows measurements of the nuclear modification factor (\rpa) for \jpsi\ in \pAu{} collisions at $\sqrt{s_{\rm NN}}$ = 200 GeV by the STAR experiment (left) \cite{STAR:2021zvb} and for $\Upsilon$(1S) in \pPb\ collisions at $\sqrt{s_{\rm NN}}$ = 5.02 TeV by ATLAS (right) \cite{ATLAS:2017prf}.
A clear suppression of about 30\% is seen at low \pt, which could mimic the dissociation effect and therefore needs to be taken into account when interpreting similar measurements in heavy-ion collisions. As \pt\ increases, \rpa\ also increases and becomes consistent with unity within uncertainties at high \pt. Theoretical calculations including latest nPDF sets \cite{Lansberg:2016deg,Kusina:2017gkz} or coherent energy loss \cite{Arleo:2013zua}, as shown in the left panel of Figure~\ref{fig:Quarkonia_RpA_vs_pt}, can qualitatively describe data. To further study the
\begin{figure}[htbp]
\begin{minipage}{0.49\linewidth}
\centerline{\includegraphics[width=0.95\linewidth]{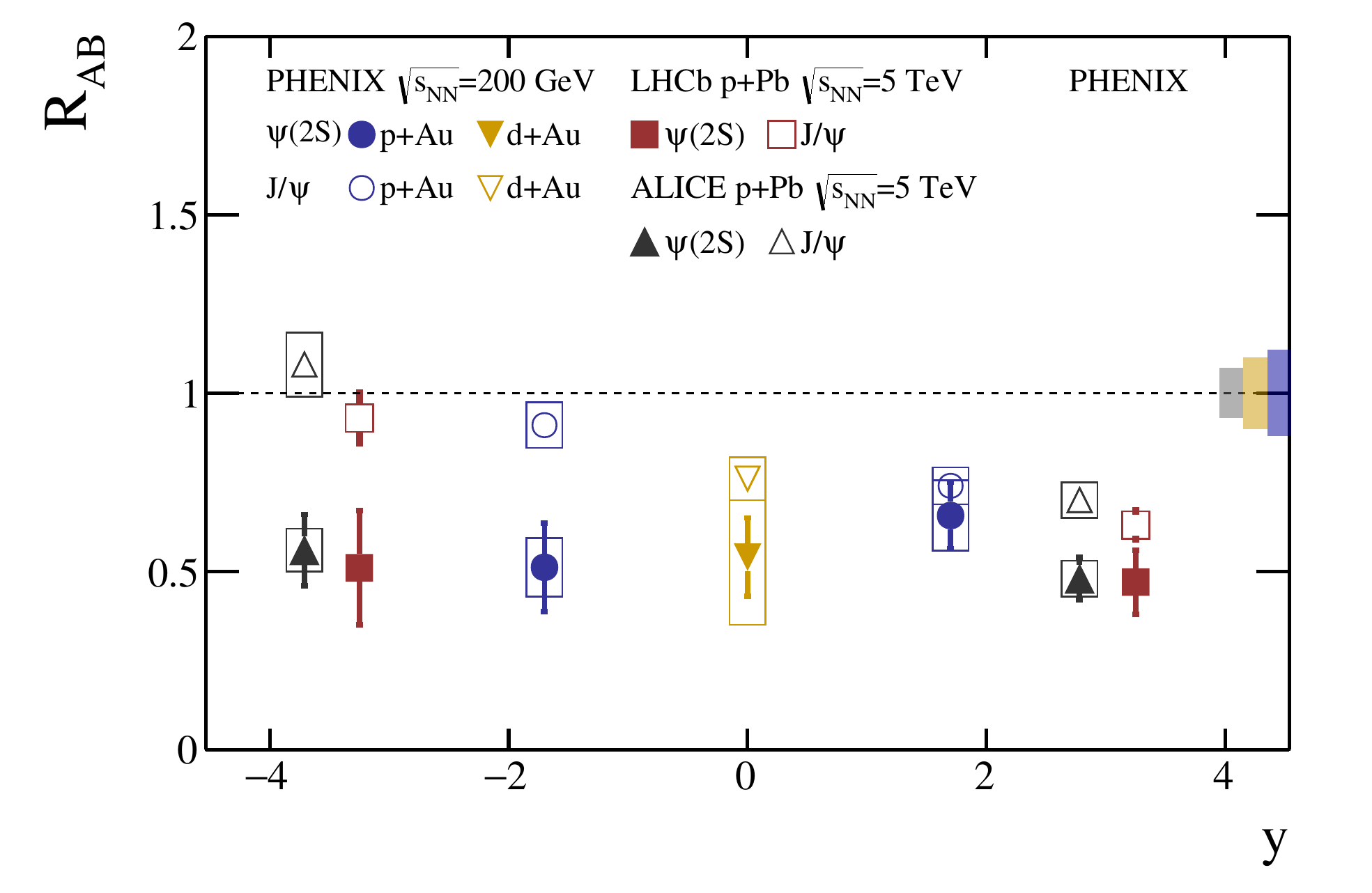}}
\end{minipage}
\begin{minipage}{0.49\linewidth}
\centerline{\includegraphics[width=0.95\linewidth]{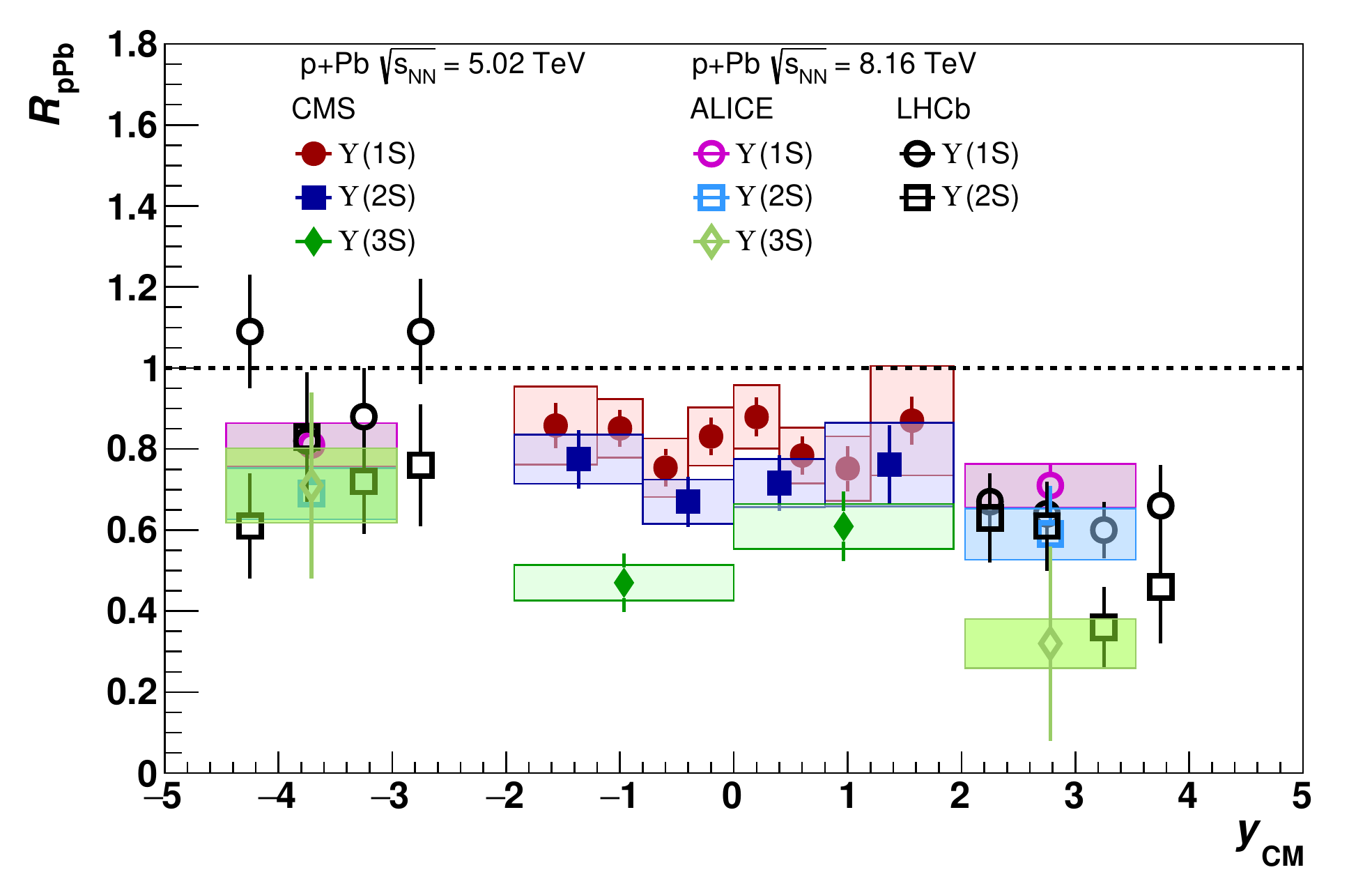}}
\end{minipage}
\caption[]{Charmonia (left) and bottomonia (right) \rpa\ as a function of rapidity \cite{PHENIX:2013pmn,PHENIX:2022nrm,ALICE:2014cgk,LHCb:2016vqr,ALICE:2013snh,LHCb:2013gmv,CMS:2022wfi,LHCb:2018psc,ALICE:2019qie}.}
\label{fig:Quarkonia_RpA_vs_y}
\end{figure}
CNM effects, measurements of charmonia (\jpsi, $\psi(2S)$) and bottomonia ($\Upsilon$(1S), $\Upsilon$(2S), $\Upsilon$(3S)) \rpa\ as a function of rapidity are shown in the left and right panels of Figure~\ref{fig:Quarkonia_RpA_vs_y}, respectively. For charmonia \cite{PHENIX:2013pmn,PHENIX:2022nrm,ALICE:2014cgk,LHCb:2016vqr,ALICE:2013snh,LHCb:2013gmv}, the level of suppression for $\psi(2S)$ is similar to that of \jpsi\ at forward p-going direction indicating the dominance of the initial-state effects, while $\psi(2S)$ is more suppressed than \jpsi\ at backward A-going direction, which is likely resulted from final-state effects, such as comover breakup \cite{Ferreiro:2014bia}, that affect \jpsi\ and $\psi(2S)$ differently. For bottomonia \cite{CMS:2022wfi,LHCb:2018psc,ALICE:2019qie}, all three $\Upsilon$ states are suppressed across the probed rapidity range with the excited states more suppressed than the ground state, which again points at final-state effects. 

\begin{figure}[htbp]
\begin{minipage}{0.475\linewidth}
\centerline{\includegraphics[width=0.95\linewidth]{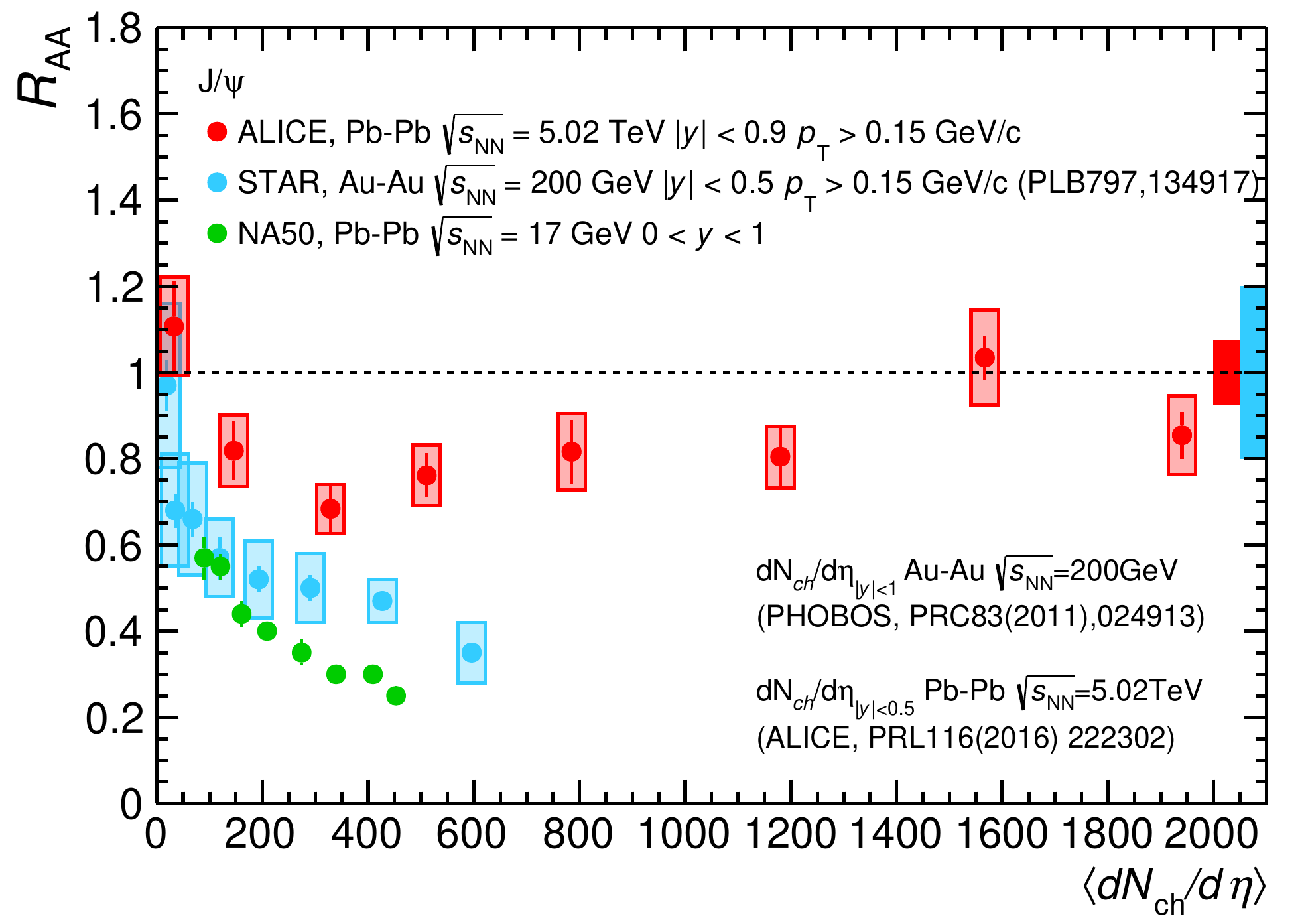}}
\end{minipage}
\begin{minipage}{0.525\linewidth}
\centerline{\includegraphics[width=0.95\linewidth]{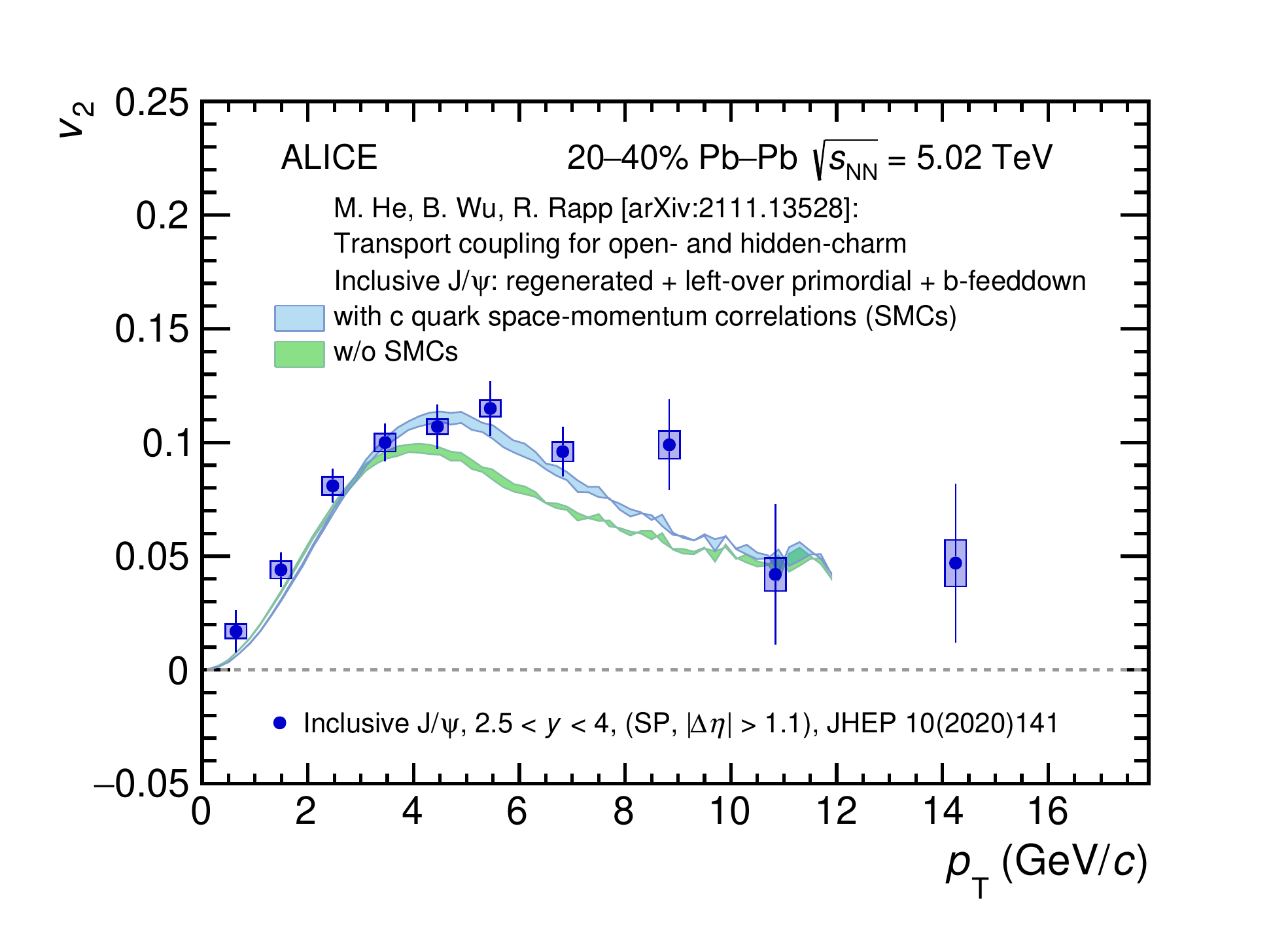}}
\end{minipage}
\caption[]{Left: \jpsi\ \raa\ at low \pt\ as a function of charged particle multiplicity in heavy-ion collisions by NA50, STAR and ALICE \cite{NA50:2004sgj,STAR:2019fge,ALICE:2022wpn}; Right: \jpsi\ $v_{2}$ as a function of \pt\ in 20-40\% \PbPb\ collisions $\sqrt{s_{\rm NN}}$ = 5.02 TeV \cite{ALICE:2020pvw}, compared with a transport model calculation \cite{He:2021zej}. }
\label{fig:Quarkonia_LHC_Jpsi}
\end{figure}

Measurements of the \jpsi\ \raa\ as a function of charged particle multiplicity in heavy-ion collisions by NA50, STAR and ALICE are shown in the left panel of Figure~\ref{fig:Quarkonia_LHC_Jpsi} \cite{NA50:2004sgj,STAR:2019fge,ALICE:2022wpn}. 
If the dissociation is the only effect in play, one would expect a larger suppression in the \jpsi\ yield, and thus a smaller \raa, in collisions of higher energy where a hotter QGP is expected to be formed {\it i.e.}, the \jpsi\ \raa\ should decrease with increasing collision energy \cite{Rapp:2017chc}. However, the opposite is observed experimentally, which points to the increasing contribution of recombination at higher energy. At the LHC, the \jpsi\ \raa\ is also seen to increase from peripheral to central collisions, which is also likely caused by the enhanced recombination in central collisions because of the larger number $c\bar{c}$ pairs produced in these collisions. The right panel of Figure~\ref{fig:Quarkonia_LHC_Jpsi} shows \jpsi\ $v_{2}$ as a function of \pt\ measured in \PbPb\ collisions at $\sqrt{s_{\rm NN}}$ = 5.02 TeV by ALICE. The \jpsi\ $v_{2}$ increases with \pt\ and reaches about 0.1 around \pt\ = 5 GeV/$c$. The \jpsi\ mesons produced during initial hard scatterings are not expected to acquire significant flow through thermalization. Such a large $v_{2}$ signal at low to intermediate \pt\ can only be explained by the dominance of regenerated \jpsi\ mesons inheriting the elliptic flow of the constituent charm quarks which likely have reached local thermalization in the QGP. Such a scenario is confirmed by the good agreement between data and a transport model calculation \cite{He:2021zej}, shown in the right panel of Figure~\ref{fig:Quarkonia_LHC_Jpsi}, in which \jpsi\ at low to intermediate \pt\ is mostly produced through recombination of diffusing charm and anti-charm quarks in the hydrodynamically expanding fireball. It becomes clear that both the dissociation and recombination contributions, besides other effects such as CNM effects and feeddown structure, are needed to provide a satisfactory explanation of \jpsi\ production in heavy-ion collisions from low to high energies.  

\begin{figure}[htbp]
\begin{minipage}{0.50\linewidth}
\centerline{\includegraphics[width=0.95\linewidth]{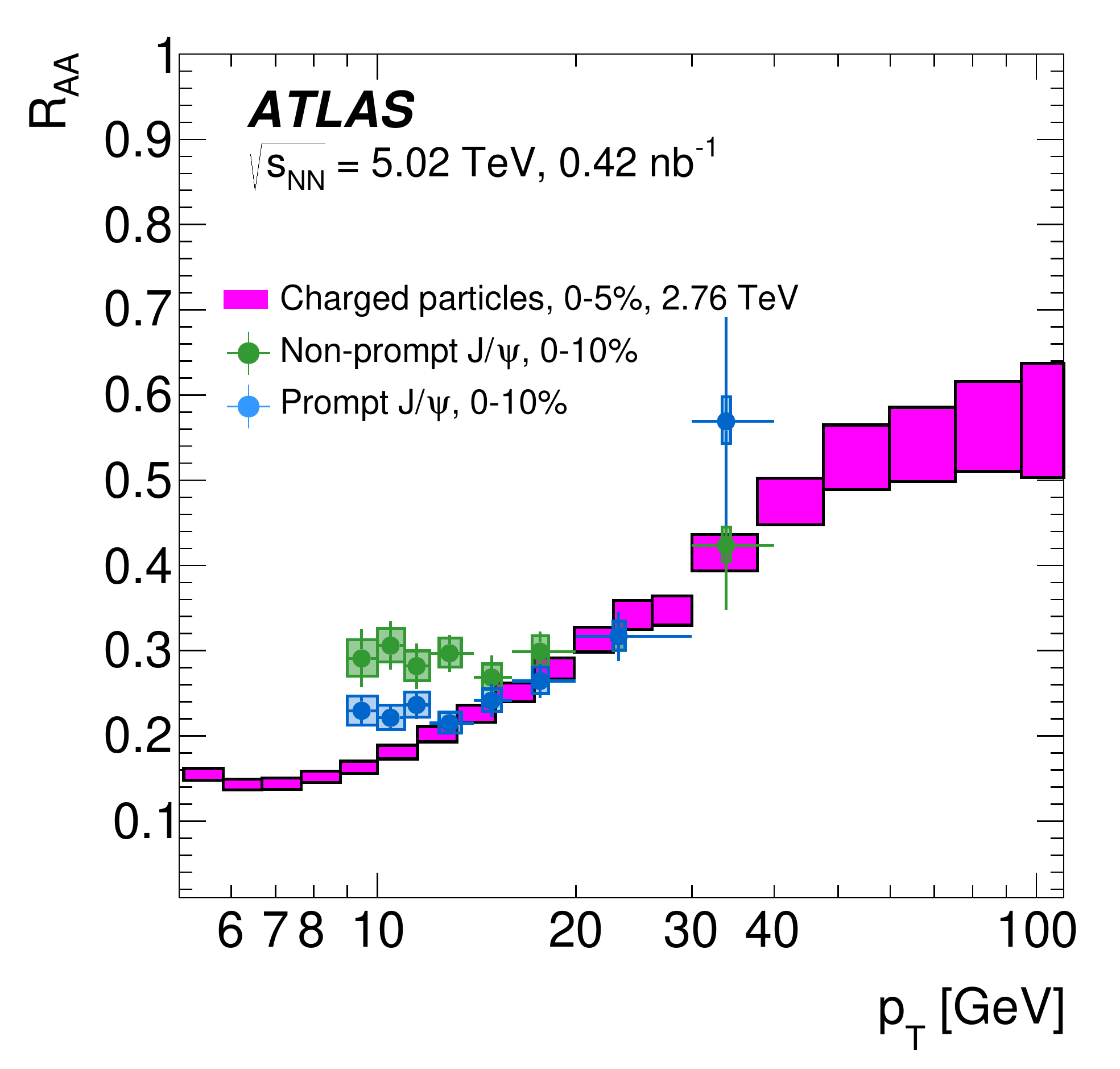}}
\end{minipage}
\begin{minipage}{0.50\linewidth}
\centerline{\includegraphics[width=0.95\linewidth]{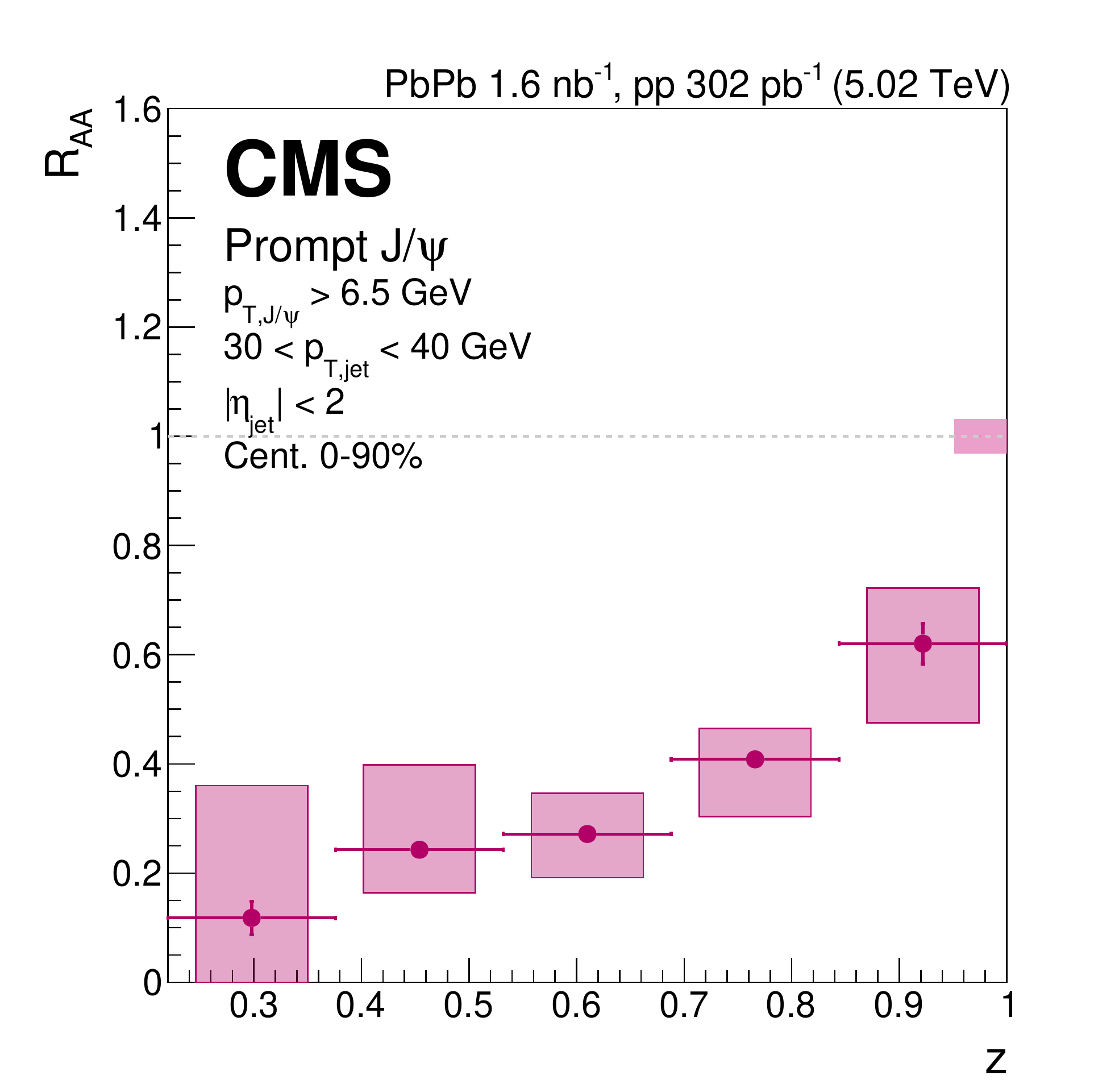}}
\end{minipage}
\caption[]{Left: comparison of \raa\ as a function of \pt\ for high-\pt\ prompt \jpsi\, non-prompt \jpsi\ and charged hadron \cite{ATLAS:2018hqe}. Right: \raa\ of \jpsi\ mesons contained in jets as a function of the jet momentum fraction carried by the \jpsi\ \cite{CMS:2021puf}.}
\label{fig:Quarkonia_LHC_Jpsi_2}
\end{figure}

The high-energy heavy-ion collisions at the LHC also open the door for measuring \jpsi\ at high \pt. The left panel of Figure~\ref{fig:Quarkonia_LHC_Jpsi_2} shows the ATLAS measurement of the prompt \jpsi\ \raa\ from 9 to 40 GeV/$c$ compared to that of charged hadrons, which are seen to coincide each other above 12 GeV/$c$. One possible explanation is that high-\pt\ \jpsi\ is dominantly produced through parton fragmentation, and the suppression seen in the \jpsi\ yield is a reflection of the energy loss of the parent parton before fragmentation, the same mechanism responsible for high-\pt\ charged hadron suppression. 
The parton energy loss mechanism could also explain the finite $v_{2}$ observed for \jpsi\ above 10 GeV/$c$, as shown in the right panel of Figure~\ref{fig:Quarkonia_LHC_Jpsi}, by invoking its path-length dependence. This idea is further explored by the CMS experiment by studying high-\pt\ \jpsi\ contained in jets \cite{CMS:2019ebt, CMS:2021puf}. It was found that about 85\% of \jpsi\ with energies above 15 GeV within the rapidity range of $|\eta|<1$ are contained in jets, {\it i.e.}, there are hadronic activities nearby the \jpsi, with energies above 19 GeV in \pp\ collisions at $\sqrt{s}$ = 8 TeV. This observation corroborates that high-\pt\ \jpsi\ is dominantly produced during parton fragmentation. The right panel of Figure~\ref{fig:Quarkonia_LHC_Jpsi_2} shows that the level of suppression for high-\pt\ \jpsi\ yields in jets increases with decreasing $z$, the fraction of jet \pt\ carried by \jpsi. Theoretical calculations based on leading power non-relativistic QCD show that the fraction of directly produced \jpsi\ from gluon fragmentation varies between 70-90\% from 10 to 100 GeV/$c$ \cite{Zhang:2022rby}.  This opens the door for obtaining an enriched sample of gluon-initiated jets by tagging with high-\pt\ \jpsi, which can be used to study gluon energy loss mechanism and compared to quark-initiated jets. 

\begin{figure}[htbp]
\begin{minipage}{0.49\linewidth}
\centerline{\includegraphics[width=0.95\linewidth]{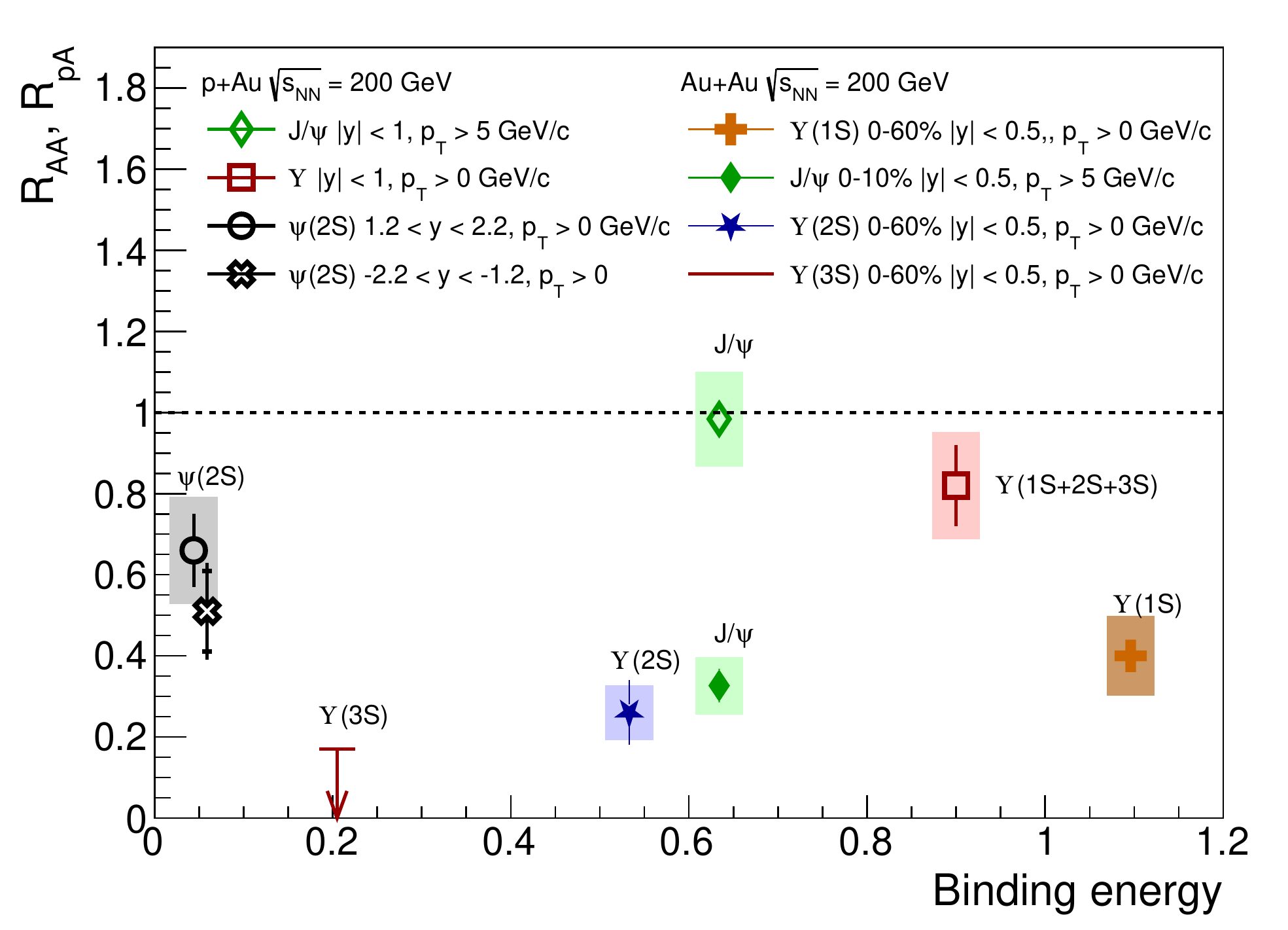}}
\end{minipage}
\begin{minipage}{0.49\linewidth}
\centerline{\includegraphics[width=0.95\linewidth]{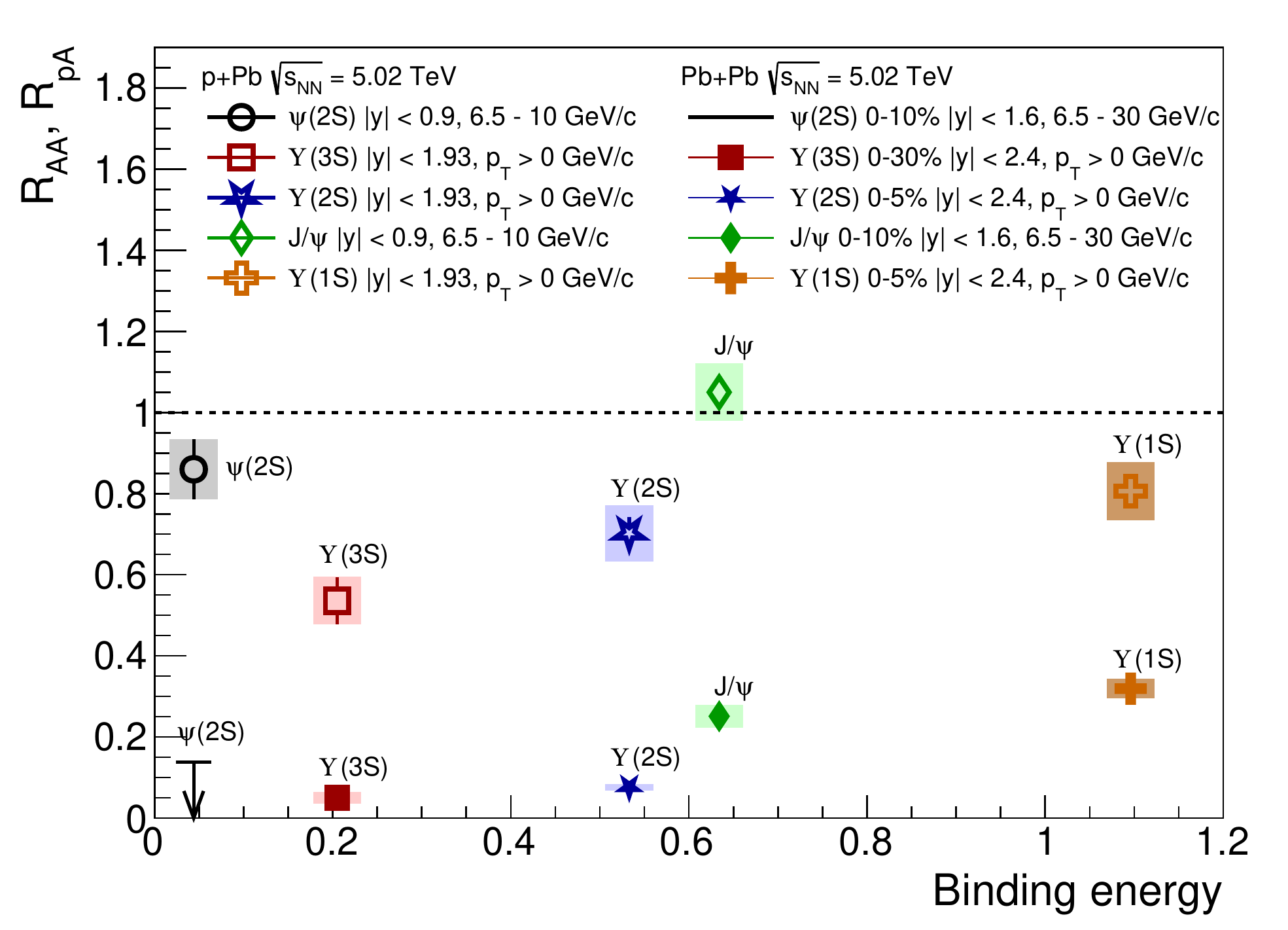}}
\end{minipage}
\caption[]{Quarkonium \raa\ (filled symbols) and \rpa\ (open symbols) as a function of binding energy measured at RHIC (left) and the LHC (right) \cite{PHENIX:2022nrm,STAR:2019fge,STAR:2022rpk,Ye:2017fwv,Zhang:2021snx,CMS:2018zza,CMS:2022rna,CMS:2017uuv,CMS:2022wfi,CMS:2017exb,CMS:2018gbb}.}
\label{fig:Quarkonia_Summary}
\end{figure}

The full power of utilizing quarkonium suppression to infer the transport and thermodynamic properties of the QGP arises from the ability to measure quarkonia of different sizes. Figure~\ref{fig:Quarkonia_Summary} shows the \raa\ (filled symbols) of charmonia (\jpsi, $\psi$(2S)) and bottomonia ($\Upsilon$(1S), $\Upsilon$(2S), $\Upsilon$(3S)) in heavy-ion collisions at RHIC (left) and the LHC (right) as a function of the binding energy, compared to corresponding measurements in \pA\ collisions (open symbols) \cite{PHENIX:2022nrm,STAR:2019fge,STAR:2022rpk,Ye:2017fwv,Zhang:2021snx,CMS:2018zza,CMS:2022rna,CMS:2017uuv,CMS:2022wfi,CMS:2017exb,CMS:2018gbb}.
Here, the binding energy is calculated as the difference between twice the lightest heavy quark meson mass and the quarkonium mass. For charmonia, measurements above 5 (6.5) GeV/$c$ are shown at RHIC (LHC) to minimize complications from CNM effects and recombination, except that the $\psi(2S)$ \rpa\ for 200 GeV \pAu{} collisions is shown for $\pT>0$ GeV/$c$ which is the only available measurement at RHIC. A suppression of the bottomonium states is seen in \pA\ collisions, while high-\pT\ \jpsi\ is not. In heavy-ion collisions, a sequential suppression pattern depending on the binding energy is clearly seen, in line with the expectation that more loosely bound states are more likely to be dissociated in the QGP. Such a sequential suppression pattern also points to the importance of accounting for the feeddown contributions from excited states to ground states, which are included in the experimentally measured yields, when interpreting these data. A comprehensive comparison to model calculations, which include all the effects that affect the measured quarkonium yields, is needed to extract the QGP properties.

\subsubsection{Experiment: open heavy flavor}
\label{sec:progress:microscopic:experiment_open_heavy_flavor}

Some of the requirements needed for the quark-gluon plasma characterization are: i) the medium needs to have a relatively large volume; ii) the probe needs to be formed and not shower in the vacuum before the QGP formation; and have different mass and momentum scales. Relativistic heavy-ion collisions produce QGP with volumes on the order of the size of the colliding nucleus. Heavy quarks (charm and bottom) have a formation time $<$0.07 fm/c, way before the formation of a thermalized QGP which is supposed to take between 0.3 and 1.5 fm/c \cite{Liu:2012ax}. The energy that a quark loses when crossing the QGP medium depends on its mass $m_q$. The gluon radiation formed when an incident quark crosses the QGP medium has a cone with minimum angle $\theta=m_q/E_q$, meaning that the dead radiation cone is larger for heavy quarks \cite{Dokshitzer:2001zm}. 

Heavy quarks are predominantly formed from gluons in high-energy collisions. At RHIC energies flavor excitation and pair creation are the dominant processes as indicated in measurements of muon pairs from heavy flavor decays measured by PHENIX \cite{PHENIX:2018dwt}. Gluon splitting is dominant at LHC energies.

\begin{figure}[htb]
    \centering
    \scalebox{0.6}{
    \input{figures/open_heavy_flavor/lambdac_D0.tex}}
    \caption{$\sigma\left(\Lambda_c\right)/\sigma\left(D^0\right)$ ratio measured in $e^+e^-$  \cite{CLEO:1990unu,ARGUS:1988hly,Gladilin:2014tba}, $ep$ \cite{ZEUS:2010cic,ZEUS:2013fws}, 
    \pp{}, \pPb{}, \AuAu{} and \PbPb{} collisions \cite{LHCb:2013xam,LHCb:2018weo,ALICE:2020wfu,STAR:2019ank,ALICE:2021bib,LHCb:2022ddg,CMS:2019uws}. }
\label{fig:lambda_c}
\end{figure}

One of the most interesting recent observations is the modification of the charm fragmentation fraction observed in $pp$ collisions when compared to $e^+e^-$ and $ep$ collisions as indicated in the study performed by the ALICE collaboration \cite{ALICE:2021dhb}. The modification infers the enhancement of charmed baryons which is compensated with a suppression of charmed mesons. This effect is also observed between $pp$ and heavy-ion collisions where the total $D^0$ cross-section per binary collision measured by STAR is not conserved in \AuAu{} collisions \cite{STAR:2018zdy}, despite the expected conservation of the number of charm quarks observed in semi-leptonic decays \cite{PHENIX:2004ggw}. 
The $\Lambda_c$ baryon enhancement relative to $D^0$ mesons was observed in several collision systems by several experiments at LEP, DESY, RHIC and LHC as shown in Figure~\ref{fig:lambda_c}. The \pt integrated results obtained from $e^+e^-$ \cite{CLEO:1990unu,ARGUS:1988hly,Gladilin:2014tba} and $ep$ \cite{ZEUS:2010cic,ZEUS:2013fws} are consistent and shown in the figure as an average with the uncertainty band covering the combined statistical and correlated uncertainties. The $\Lambda_c/D^0$ yield ratio is systematically larger in hadronic collisions. One of the most successful descriptions of the baryon enhancement is color reconnection \cite{Christiansen:2015yqa} where the underline partons in the proton contribute to the hadron formation. In large systems, such as central \AuAu{} and \PbPb{} collisions, hadron coalescence inside QGP further enhances charmed baryon fractions \cite{Minissale:2015zwa}. The same effects may also happen with bottom quark fragmentation. Other baryon formations are expected to be further studied. First studies on $\Lambda_b/B^0$ fractions were reported by LHCb in pPb collisions \cite{LHCb:2019avm}. The observation of more exotic baryons, such as $\Omega_{cc}$ and charmed tetraquarks (Section~\ref{sec:progress:microscopic:exotic_hadrons}), are on target for the Run3 LHC.

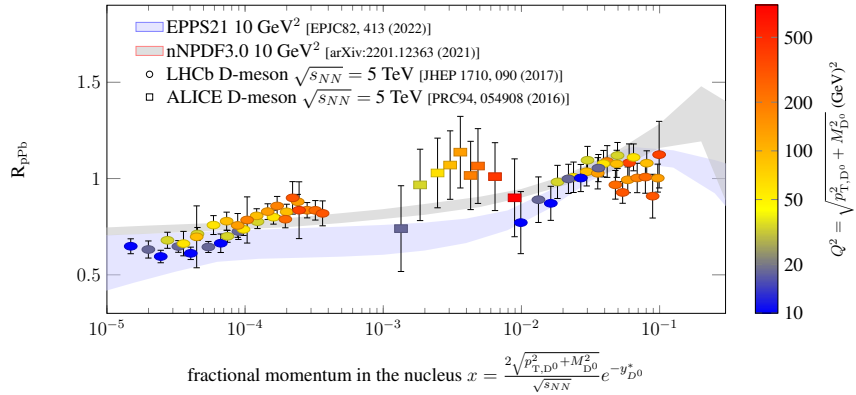
\begin{figure}[htb]
    \centering
    \scalebox{0.6}{
    \begin{tikzpicture}
\begin{semilogxaxis}
			[xscale=2.0,
  			yscale=1.2,
  			xshift=3cm,
		    xmin=1e-5, xmax=0.3,
    		ymin=0.3, ymax=1.9,
            font=\large,
    		xlabel={fractional momentum in the nucleus $x=\frac{2\sqrt{p_{\rm T,D^0}^2+M_{\rm D^0}^2}}{\sqrt{s_{NN}}}e^{-y^*_{D^0}}$},
    		ylabel=R$_{\rm pPb}$,
            colorbar,
            colorbar style={font=\large,
                            ylabel={\small $Q^2=\sqrt{p_{\rm T,D^0}^2+M_{\rm D^0}^2}$ (GeV)$^2$},
                            colorbar shift/.style={xshift=-1cm, yshift=1.2cm, yscale=1.2},
                            ymode=linear,ytick={1,1.3,1.7,2,2.3,2.7}, 
                            yticklabels={10, 20, 50, 100, 200, 500}},
            legend style={at={(0.4,0.98)},anchor=north,draw=none},
            legend cell align={left}]
    \addplot [name path=upper,draw=none,forget plot] table[x=xB, y=max_rpa] {figures/open_heavy_flavor/epps21.dat};
    \addplot [name path=lower,draw=none,forget plot] table[x=xB, y=min_rpa] {figures/open_heavy_flavor/epps21.dat};
    \addplot +[fill=blue!20!white, opacity=0.5] fill between[of=lower and upper];
    \addplot [name path=upper,draw=none,forget plot] table[x=xB, y=maxRpPb] {figures/open_heavy_flavor/nNPDF30.dat};
    \addplot [name path=lower,draw=none,forget plot] table[x=xB, y=minRpPb] {figures/open_heavy_flavor/nNPDF30.dat};
    \addplot +[fill=gray!50!white, opacity=0.5] fill between[of=lower and upper];
    \addplot+[black,only marks, scatter, point meta=explicit, mark=*,
  			mark options={scale=1,fill=white,draw=black},
  			error bars/.cd,
            y dir=both,y explicit] 
    		table [x=xB, meta expr=log10(\thisrow{Q2}), y=RAA, y error=RAA_err] {figures/open_heavy_flavor/raa_xb-lhcb.dat};
  \addplot+[black,only marks, scatter, point meta=explicit, mark=square*,
  		 	mark options={scale=1,fill=white,draw=black},
  			error bars/.cd,
            y dir=both,y explicit] 
    		table [x=xB, meta expr=log10(\thisrow{Q2}), y=RAA, y error=RAA_err] {figures/open_heavy_flavor/raa_xb-alice.dat};
   \legend{{EPPS21 10 GeV$^2$ {\footnotesize [EPJC82, 413 (2022)]}}, {nNPDF3.0 10 GeV$^2$ {\footnotesize [arXiv:2201.12363 (2021)]}} , {LHCb D-meson $\sqrt{s_{NN}}=$ 5 TeV {\footnotesize [JHEP 1710, 090 (2017)]}}, {ALICE D-meson $\sqrt{s_{NN}}=$ 5 TeV {\footnotesize [PRC94, 054908 (2016)]}}}
\end{semilogxaxis}
\end{tikzpicture}}
    \caption{Partonic momentum fraction dependency of $D^0$ nuclear modification factor in $pPb$ collisions obtained by LHCb \cite{LHCb:2017yua} and ALICE \cite{ALICE:2016yta} along with the most recent nuclear parton distribution functions.}
    \label{fig:HF_bjorken_x}
\end{figure}

At the same time, heavy flavor yields can be modified by initial effects in the nucleus such as nuclear shadowing, initial state energy loss and gluon saturation effects. A recent overview of initial-state effects can be found in \cite{Albacete:2017qng}.
Heavy flavor nuclear modification factor in p(d)+A collisions has been measured at RHIC and LHC. Figure~\ref{fig:HF_bjorken_x} shows the extension of the fractional momentum $x$ of $D^0$ meson nuclear modifications measured at LHC. The fraction of the proton longitudinal momentum $x$ is an approximation assuming a 2$\rightarrow$1 process. The new set of data coming from the forward coverage of LHCb are responsible to finally provide a constraint to the nuclear PDFs at small-$x$ and small momentum transfer $Q^2$ region as also shown in Figure~\ref{fig:HF_bjorken_x}. 

\begin{figure}[htb]
    \begin{subfigure}{0.26\linewidth}
    (a)\\
        \includegraphics[width=1.0\linewidth]{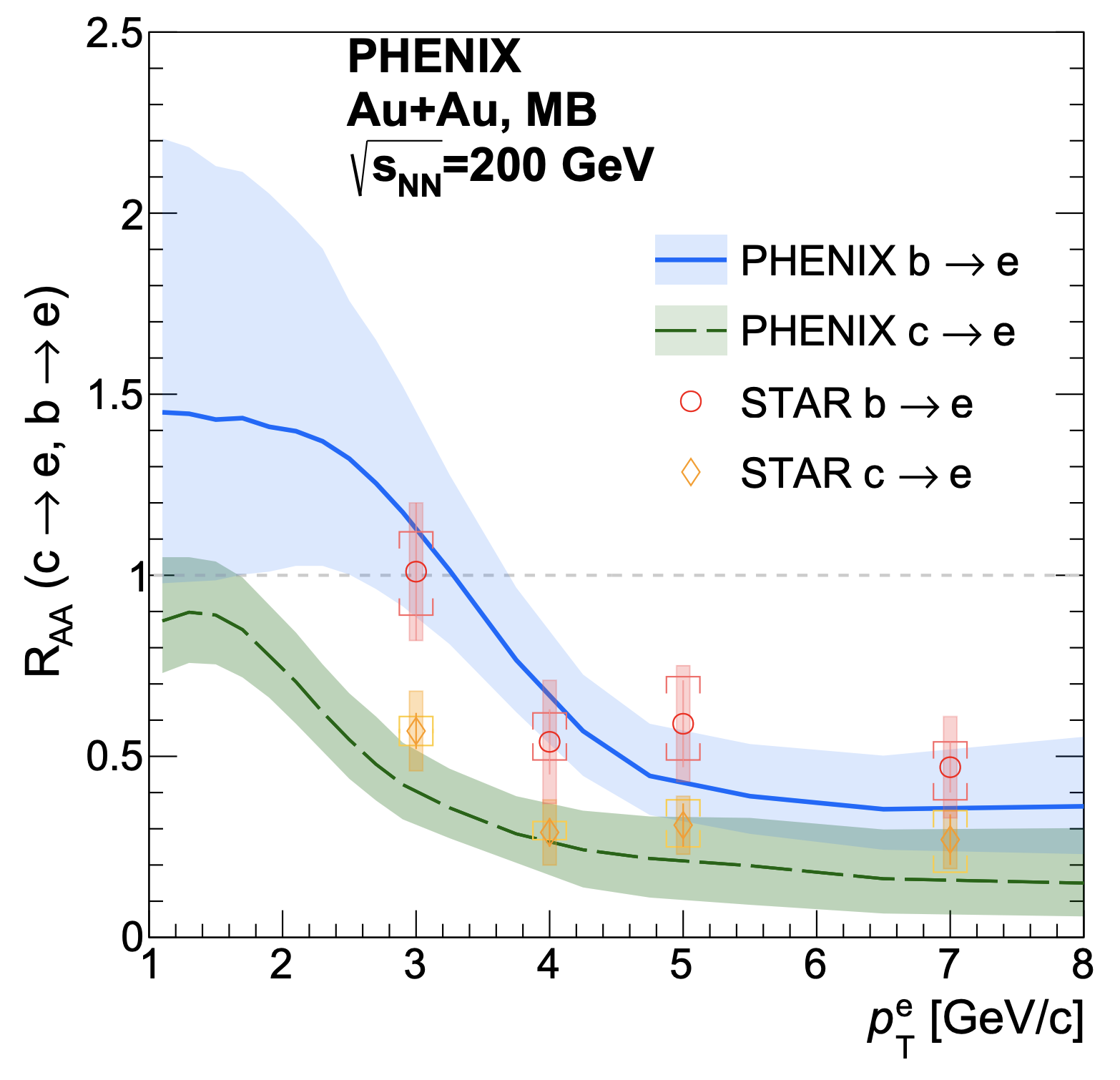}
    \end{subfigure}
    \begin{subfigure}{0.39\linewidth}
    (b)\\
        \includegraphics[width=1.0\linewidth]{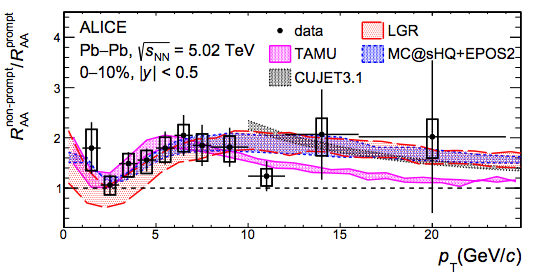}
    \end{subfigure}
    \begin{subfigure}{0.33\linewidth}
    (c)\\
        \includegraphics[width=1.0\linewidth]{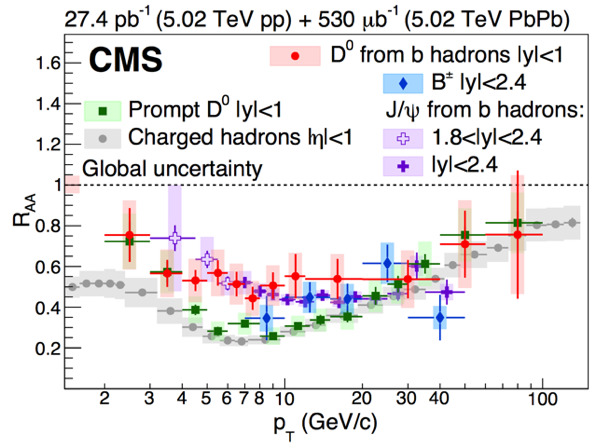}
    \end{subfigure}
    \caption{Nuclear modification factor of charm and bottom hadrons: (a) Electrons from semi-leptonic decays of charm and bottom hadrons measured by PHENIX \cite{PHENIX:2022wim} and STAR \cite{STAR:2021uzu}. (b) Non-prompt (from B-meson decays) over prompt $D^0$ nuclear modification factor ratio in \PbPb{} collisions from ALICE along with models \cite{ALICE:2022tji}. (c) Light hadron (charged hadron), charm hadron (prompt $D^0$) and bottom hadrons (non-prompt $D^0$, non-prompt \jpsi~ and B-mesons) nuclear modification factor in \PbPb{} collisions from CMS\cite{CMS:2018bwt}.}
    \label{fig:RAA_HF}
\end{figure}

Systematic evidence for a mass hierarchy on the quark energy loss in QGP $\frac{dE_{ud}}{dx}>\frac{dE_c}{dx}>\frac{dE_b}{dx}$ have been presented by experiments at RHIC and LHC in a variety of channels \cite{STAR:2018zdy, STAR:2021uzu, PHENIX:2022wim, ALICE:2022tji, CMS:2018bwt,ATLAS:2018ofq} as summarized in Figure~\ref{fig:RAA_HF}. Most nuclear modifications differences between $c$ and $b$ quarks happen in the intermediate \pt and reduce or vanish at high \pt. This is one of the major experimental accomplishments in the field of Hot QCD thanks to the investment in silicon vertex detectors which enable these measurements. New horizons were opened with the first observation of top quarks in \PbPb{} collisions by CMS \cite{CMS:2020aem}. The measured cross-section is consistent with binary $pp$ scaling. 

Relative abundances of $D^0$, $D^{\pm}$ and $D^{*+}$ are unmodified as observed by ALICE \cite{ALICE:2021rxa}, except strange D-meson yield $D_s^+$ which is enhanced in \AuAu{} and \PbPb{} collisions at RHIC and LHC when compared to other D-mesons as reported by STAR \cite{STAR:2021tte} and ALICE \cite{Acharya:2018hre}. Similar enhancement is also observed for $B_s^0$ in \PbPb{} collisions as measured by CMS \cite{Sirunyan:2018zys}. However, no apparent strangeness enhancement is observed in small systems as reported by ALICE \cite{ALICE:2019fhe}. The results suggest a hadron recombination in the QGP medium created in these collisions. 
Similar effects would also cause the enhancement of $B_c$ mesons formed by a charm and a bottom quark, an unlikely formation in $pp$ hard scattering processes but favored in QGP. 

\begin{figure}
    \begin{subfigure}{0.31\linewidth}
        (a)\\
        \includegraphics[width=0.9\linewidth]{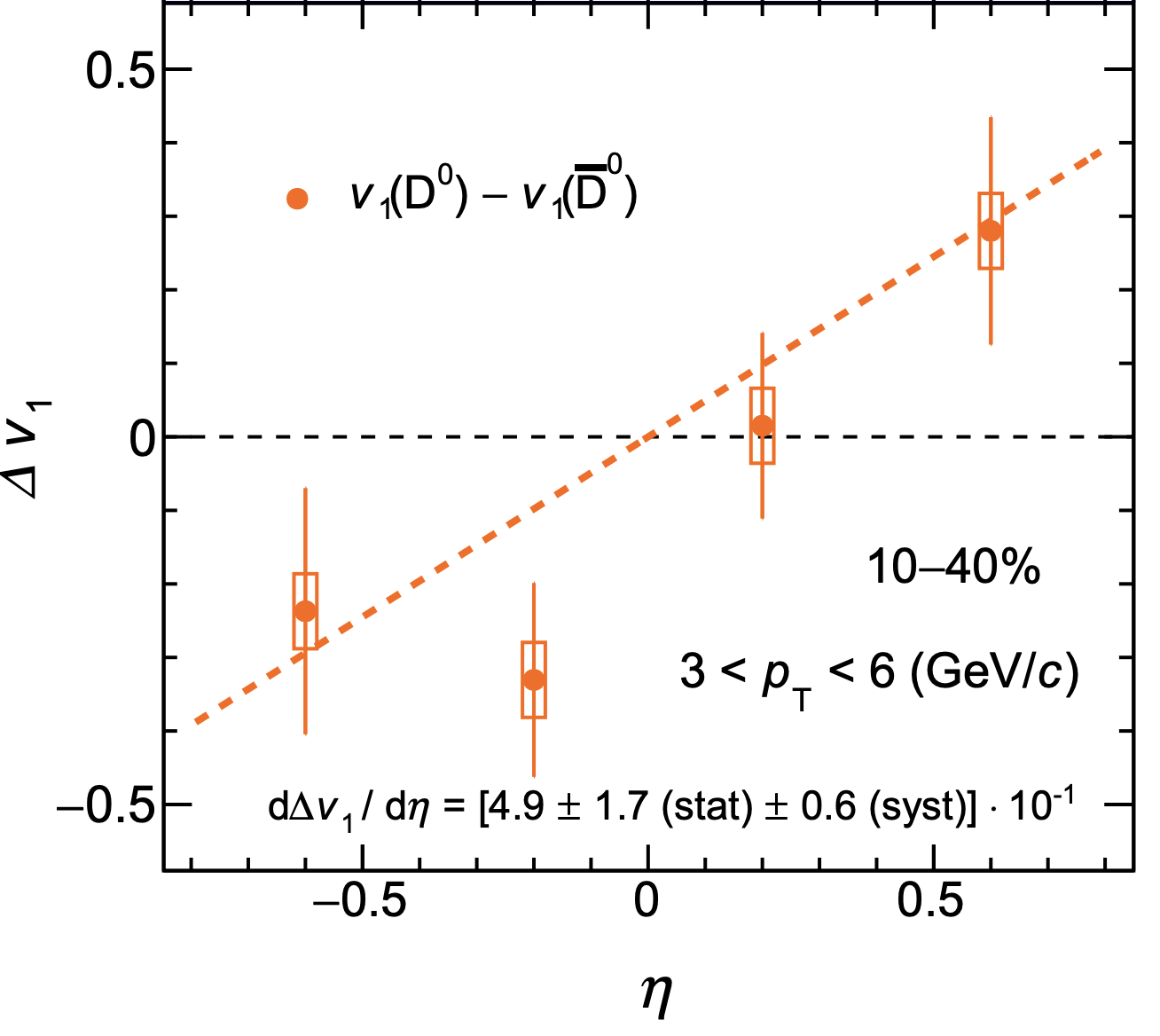}
    \end{subfigure}
    \begin{subfigure}{0.4\linewidth}
        (b)\\
        \includegraphics[width=1.0\linewidth]{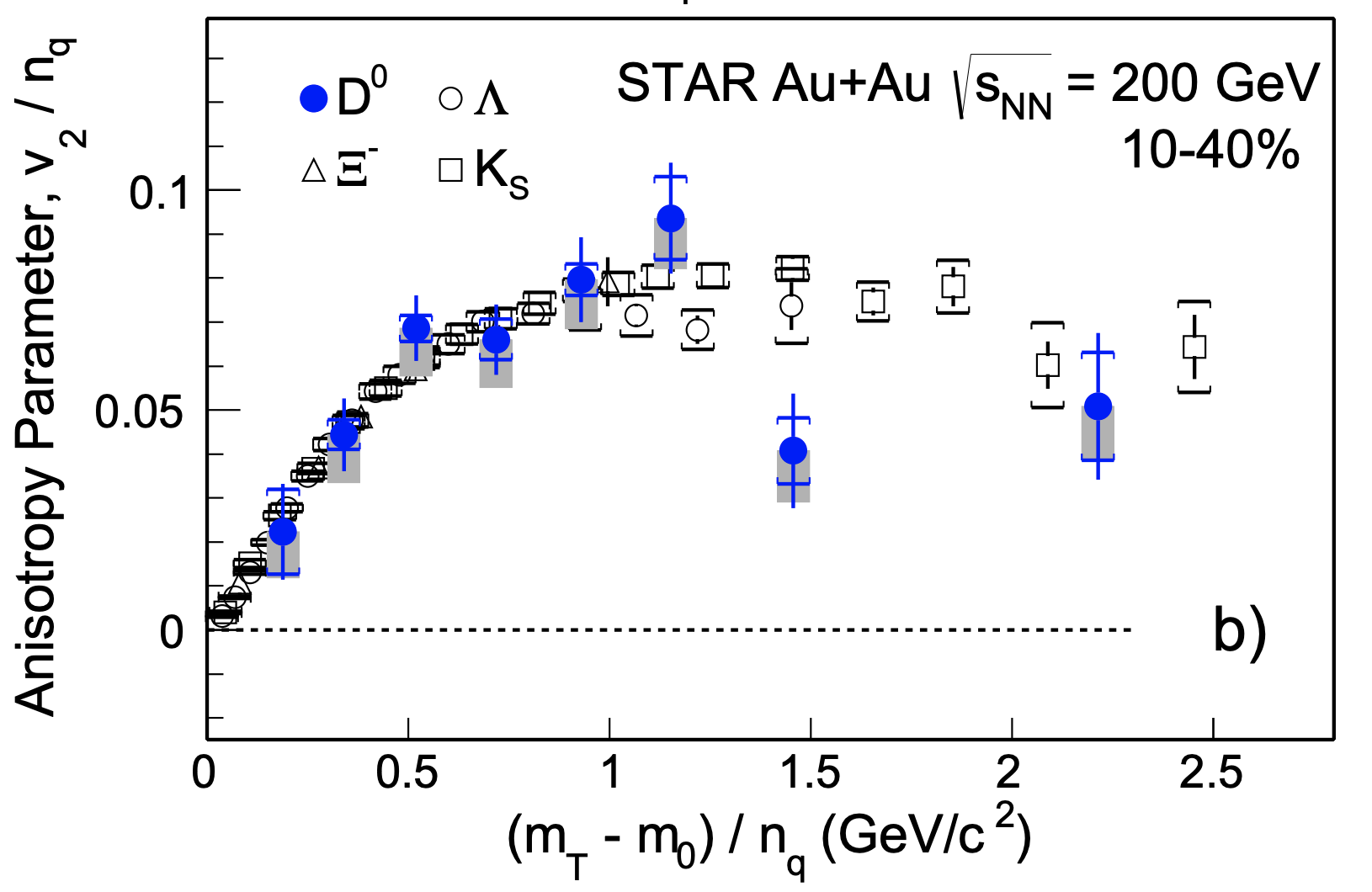}
    \end{subfigure}
    \begin{subfigure}{0.27\linewidth}
        (c)\\
        \includegraphics[width=1.0\linewidth]{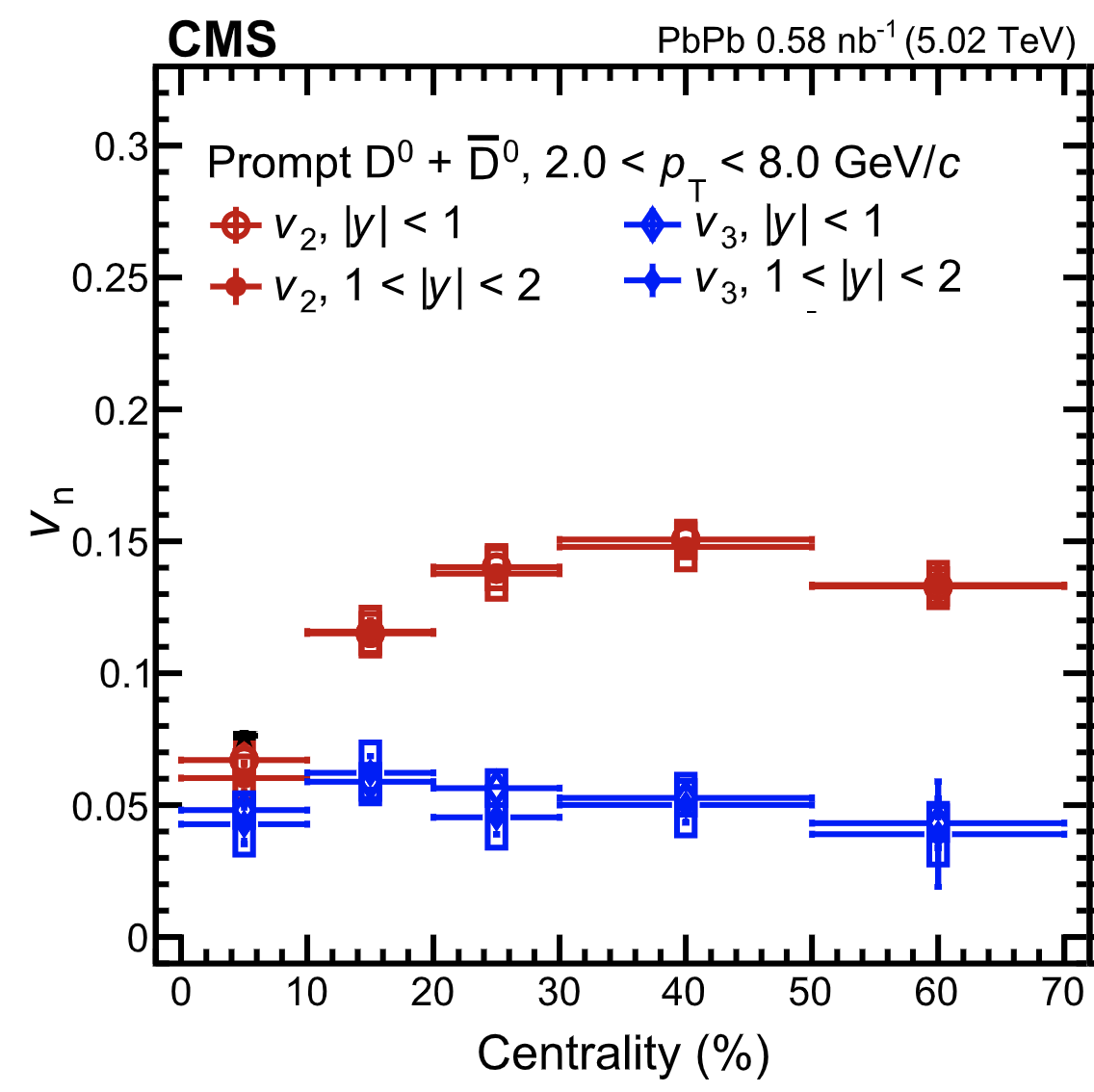}
    \end{subfigure}    
    \caption{(a) $D$-meson direct flow charge asymmetry measured by ALICE \cite{ALICE:2019sgg}. (b) $D^0$ elliptic flow over number of valence quarks compare lighter hadrons measured by STAR \cite{STAR:2017kkh}. (c)Centrality and rapidity dependence of $D^0$ elliptic and triangular flow from CMS \cite{CMS:2017vhp}.}
    \label{fig:D0_vn}
\end{figure}

\begin{figure}[htb]
    \centering
    \scalebox{0.65}{
    \begin{tikzpicture}
   \begin{axis}[
        name=lattice,
        width=15cm,
        height=8cm,
        legend columns=3,
        font=\large,
	    xmin=0.8, xmax=3.2,
        ymin=0, ymax=40,
        xlabel=$T/T_c$,
        xtick={1,1.5,2.0,2.5,3.0},
        xticklabels={1,1.5,2.0,2.5,3.0},
        ylabel=$2\pi D_s T$,
        legend style={at={(1.0,1.0)},anchor=north east, draw=none, font=\footnotesize,
        legend cell align={left}}]
        \addplot[only marks, blue, mark options={mark=*, scale=2},
                    error bars/.cd, x dir=both, x explicit, y dir=both, y explicit,] table [x=T, y=DsT, y error=error] {
                    T  DsT  error
                    1.04  5.6  2.47
                    1.09  7.11  3.06
                    1.24  5.63  2.71
                    1.51  5.065 2.05
                    1.96  9.825 5.635
                    };
        \addlegendentry {Banerjee \textit{et al.}(2012)\cite{Banerjee:2011ra}}
        \addplot[only marks, green!50!black, mark options={mark=pentagon*, scale=2},
                    error bars/.cd, x dir=both, x explicit, y dir=both, y explicit,] table [x=T, y=DsT, y error=error] {
                    T  DsT  error
                    1.46  1.79  1.27
                    2.2   2.01  1.13
                    2.9   2.27  1.136
                    };
        \addlegendentry {Ding \textit{et al.}(2012)\cite{Ding:2012sp}}
        \addplot[only marks, black, mark options={mark=square*, scale=2},
                    error bars/.cd, x dir=both, x explicit, y dir=both, y explicit,] table [x=T, y=DsT, y error=error] {
                    T  DsT  error
                    1.1   4.498  1.965
                    1.5   6.59  2.93
                    3.0   12.84  7.1  
                    };
        \addlegendentry{Brambilla \textit{et al.}(2020)\cite{Brambilla:2020siz}}
        \addplot[only marks, orange, mark options={mark=diamond*, scale=2},
                    error bars/.cd, x dir=both, x explicit, y dir=both, y explicit,] table [x=T, y=DsT, y error=error] {
                    T  DsT  error
                    1.5  5.5  1.3
                    };
        \addlegendentry{Francis \textit{et al.} (2015) \cite{Francis:2015daa}}
        \addplot[only marks, red, mark options={mark=triangle*, scale=2},
                    error bars/.cd, x dir=both, x explicit, y dir=both, y explicit,] table [x=T, y=DsT, y error=error] {
                    T  DsT  error
                    1.9  5.6  5.4
                    };
        \addlegendentry{Brambilla \textit{et al.}(2019)\cite{Brambilla:2019tpt}}
    \addplot[red, mark=none, line width=2.0] table[x=T, y=DsT] {
        T  DsT
        1.03 4.89
        1.13 5.12
        1.24 5.56
        1.32 6.04
        1.41 6.65
        1.49 7.26
        1.57 7.85
        1.65 8.45
        1.72 9.05
        1.81 9.81
        1.88 10.43
        1.94 10.94
        2.02 11.61
        2.12 12.37
        2.22 13.14
        2.31 13.75
        2.40 14.36
        2.46 14.79
        };
    \addlegendentry{QPM-Catania - LV\cite{Das:2015ana}}

    \addplot[blue, mark=none, line width=2.0]table[x=T, y=DsT]{
        T DsT
        1.05 3.56
        1.13 3.71
        1.21 3.99
        1.31 4.38
        1.40 4.80
        1.49 5.29
        1.58 5.83
        1.67 6.38
        1.75 6.91
        1.84 7.41
        1.92 7.90
        2.01 8.41
        2.10 8.94
        2.19 9.43
        2.27 9.88
        2.36 10.31
        2.44 10.74
        };
    \addlegendentry{QPM-Catania - BM\cite{Das:2015ana}}

    \addplot[violet, style={dashed}, line width=2.0]table[x=T, y=DsT]{
        T  DsT
        1.05 3.35
        1.33 4.74
        1.42 5.38
        1.50 5.98
        1.59 6.64
        1.67 7.21
        1.76 7.72
        1.85 8.28
        1.94 8.83
        2.03 9.30
        2.12 9.74
        2.21 10.20
        2.30 10.68
        2.38 11.09
        };
    \addlegendentry{PHSD \cite{Song:2015sfa}}

    \addplot [name path=upper,draw=none,forget plot] table[x=T, y=DsTmin] {
    T DsTmin
    1.00 1.94
    1.50 3.31
    2.00 4.81
    2.48 6.38
    };
    \addplot [name path=lower,draw=none,forget plot] table[x=T, y=DsTmax] {
    T DsTmax
    1.00 2.75
    1.50 4.63
    2.00 7.06
    2.48 10.06
    };
    \addplot +[fill=yellow!50!black, opacity=0.5] fill between[of=lower and upper];
    \addlegendentry{Duke (Bayesian)\cite{Xu:2017obm}}

    \addplot [name path=upper,draw=none,forget plot] table[x=T, y=DsTmin] {
    T DsTmin
    1.09 1.69
    1.30 1.94
    1.50 2.06
    1.80 2.38
    2.00 2.63
    2.10 2.63
    };
    \addplot [name path=lower,draw=none,forget plot] table[x=T, y=DsTmax] {
    T DsTmax
    1.09 3.06
    1.30 3.50
    1.50 3.69
    1.80 4.25
    2.00 4.63
    2.10 4.88
    };
    \addplot +[fill=black, opacity=0.2] fill between[of=lower and upper];
    \addlegendentry{MC@sHQ \cite{Andronic:2015wma}}

    \addplot [name path=upper,draw=none,forget plot] table[x=T, y=DsTmin] {
    T DsTmin
    1.00 2.56
    1.50 2.50
    2.00 2.56
    2.48 2.63
    };
    \addplot [name path=lower,draw=none,forget plot] table[x=T, y=DsTmax] {
    T DsTmax
    1.00 4.13
    1.50 4.19
    2.00 4.06
    2.48 4.19
    };
    \addplot +[fill=gray, opacity=0.5] fill between[of=lower and upper];
    \addlegendentry{Ads/CFT \cite{Horowitz:2015dta}}

    \addplot [name path=upper,draw=none,forget plot] table[x=T, y=DsTmin] {
    T DsTmin
    1.00 11.25
    1.10 12.63
    1.20 14.00
    1.30 15.69
    1.40 17.13
    1.50 18.19
    1.60 18.88
    1.70 19.50
    1.80 19.94
    1.90 20.19
    2.00 20.44
    };
    \addplot [name path=lower,draw=none,forget plot] table[x=T, y=DsTmax] {
    T DsTmax
    1.00 13.88
    1.10 15.31
    1.20 17.31
    1.30 19.13
    1.40 20.81
    1.50 22.31
    1.60 23.25
    1.70 23.88
    1.80 24.25
    1.90 24.56
    2.00 24.75
    };
    \addplot +[fill=green!50!black, opacity=0.5] fill between[of=lower and upper];
    \addlegendentry{T-Matrix V=F\cite{Riek:2010fk}}

    \addplot [name path=upper,draw=none,forget plot] table[x=T, y=DsTmin] {
    T DsTmin
    1.00 2.94
    1.10 3.50
    1.20 4.06
    1.30 5.00
    1.40 5.75
    1.50 6.44
    1.60 7.19
    1.70 7.88
    1.80 8.44
    1.90 8.81
    2.00 8.94
    };
    \addplot [name path=lower,draw=none,forget plot] table[x=T, y=DsTmax] {
    T DsTmax
    1.00 4.13
    1.10 4.69
    1.20 5.50
    1.30 6.31
    1.40 7.31
    1.50 8.19
    1.60 9.00
    1.70 9.81
    1.80 10.50
    1.90 10.88
    2.00 11.06
    };
    \addplot +[fill=green, opacity=0.5] fill between[of=lower and upper];
    \addlegendentry{T-Matrix V=U\cite{Riek:2010fk}}

    \addplot[black, style={dashed}, line width=2.0]table[x=T, y=DsT]{
    T DsT
    1.03 19.72
    1.09 21.03
    1.14 21.83
    1.18 22.64
    1.20 23.19
    1.28 24.48
    1.34 25.41
    1.40 26.56
    1.44 27.30
    1.51 28.59
    1.57 29.57
    1.63 30.53
    1.68 31.31
    1.74 32.07
    1.83 33.25
    1.89 34.19
    1.98 35.35
    2.05 36.38
    2.13 37.42
    };
    \addlegendentry{pQCD LO $\alpha(T)$\cite{Moore:2004tg}}
  \end{axis}
  \begin{axis}[
    at={(lattice.south east)},
    width=5cm,
    ytick=\empty,
    height=8cm,
    font=\large,
    ymin=0, ymax=32,
    xtick={0.2,5.0},
    xlabel = $\sqrt{s_{NN}}$ (TeV/$c$),
    legend style={at={(0.0,1.0)},anchor=north west, draw=none, font=\footnotesize,
    legend cell align={left}}]
        \addplot[only marks, blue!30!white, mark options={mark=square*, scale=2.5},
                    error bars/.cd, x dir=both, x explicit, y dir=both, y explicit,
                    error bar style={line width=10pt}]
                    table [x=sqrts, y=DsT, y error=error] {
                    sqrts  DsT  error
                    0.2  7  5
                    };
        \addlegendentry{STAR \cite{STAR:2017kkh}}
        \addplot[only marks, red!30!white, mark options={mark=square*, scale=2.5},
                    error bars/.cd, x dir=both, x explicit, y dir=both, y explicit,
                    error bar style={line width=10pt}]
                    table [x=sqrts, y=DsT, y error=error] {
                    sqrts  DsT  error
                    5  3  1.5
                    };
        \addlegendentry{ALICE \cite{ALICE:2021rxa}}
  \end{axis}
\end{tikzpicture}
    }
    \caption{Spatial diffusion coefficient $2\pi D_s T$ obtained from lattice computations (points), transport models (lines and bands) and constrained by experimental data: STAR uses models SUBATECH \cite{Nahrgang:2013xaa}, TAMU \cite{He:2019vgs}, Duke \cite{Cao:2015hia}, LBT \cite{Cao:2016gvr}, PHSD \cite{Song:2015sfa}, and ALICE uses TAMU, MC@sHQ+EPOS2 \cite{Nahrgang:2013xaa}, LIDO \cite{Ke:2018jem} ,LGR \cite{Li:2019lex}, Catania \cite{Plumari:2019hzp} at T=155 MeV.}
    \label{fig:diffusion_coefficient}
\end{figure}
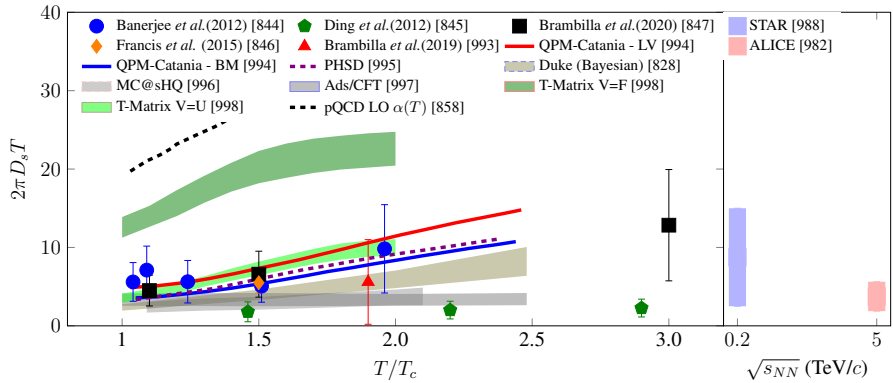

Preliminary results from PHENIX \cite{Hachiya:2019stg} and STAR \cite{Kelsey:2020bms} with electron decays indicate both charm and bottom quarks flow  with the bulk medium. The elliptic flow parameter $v_2$ reduces with the mass of the quark constituent. Elliptic $v_2$ and triangular flow $v_3$ of $D^0$ have been measured by STAR \cite{STAR:2017kkh} \cite{Lomnitz:2017ivg}(prelim.), ALICE \cite{ALICE:2021rxa} and CMS \cite{CMS:2017vhp} (Figure~\ref{fig:D0_vn}-c). The elliptic flow of $D^0$ scales with light hadrons when the number of valence quarks and the quark mass dependence is accounted for as seen in Figure~\ref{fig:D0_vn}-b. The observed combination of $R_{AA}$ and flow parameters is combined with current phenomenology models to determine the heavy flavor transport in QGP. The spatial diffusion coefficient $2\pi D_s T_c$ has been constrained using RHIC and LHC data as summarized in Figure~\ref{fig:diffusion_coefficient} along with lattice and transport model calculation efforts.

Relativistic heavy-ion collisions are expected to produce magnetic fields up to $eB \sim \left(m_{\pi}\right)^2 \approx 10^{14}$~T at its first instants, probably the largest magnet fields in nature \cite{Zhong:2014cda}. This field can alter the direct flow measured by the direct flow $v_1$, of positive and negative charged particles around the reaction plane. The charm quarks are formed at the earliest stages
of the collision, and  therefore will have to overcome much larger magnetic fields than charged particles. The asymmetry measured by STAR shows no charge asymmetry within the uncertainties \cite{STAR:2019clv} and ALICE shows two standard deviations linear dependence with $\eta$ shown in Figure~\ref{fig:D0_vn}-a \cite{ALICE:2019sgg}. The linear slope is around three orders of magnitude larger than for charged particles. The extension to this measurement to large rapidities in forward detectors can provide a larger lever arm to measure a larger charge asymmetry $v_1$ in the coming years.

\subsubsection{Exotic hadronic bound states}
\label{sec:progress:microscopic:exotic_hadrons}

\begin{figure}[ht]
\centering
\includegraphics[width=0.9\textwidth]{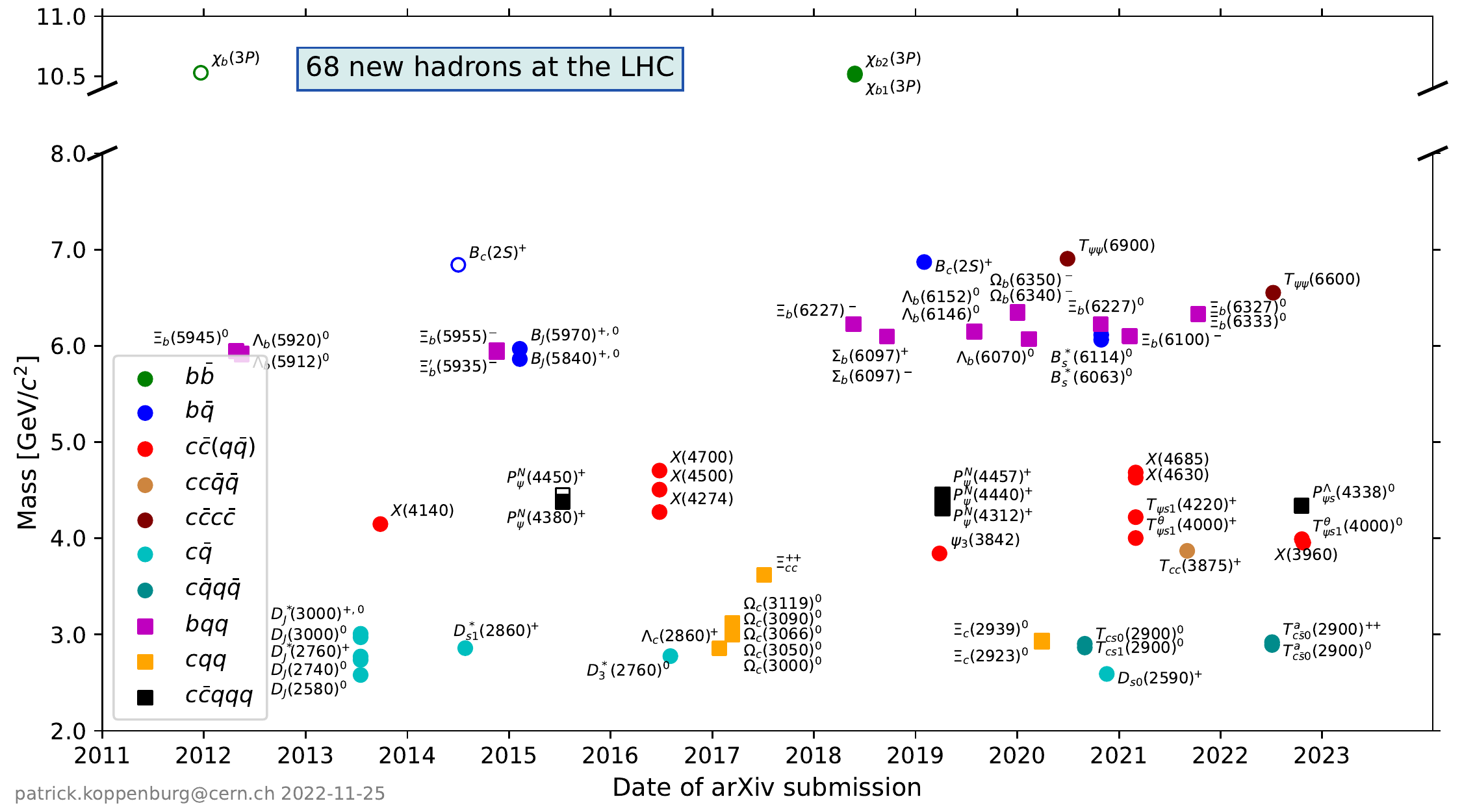}
\caption{New hadrons discovered at the Large Hadron Collider \cite{LHCb-FIGURE-2021-001-report,koppenburg_website}.}
\label{fig:Koppenburg_masses}
\end{figure}

The last decade of measurements at the Large Hadron Collider led to the discovery of more than 60 new hadrons (see Figure~ \ref{fig:Koppenburg_masses}).  Some of these newly observed particles are mesons, baryons, or higher quarkonia states that are expected from quark model calculations.  However, many of these new hadrons display properties that cannot arise from conventional combinations of two or three valence quarks.  These exotic hadrons are widely thought to consist of four or more quarks and maybe the tetra- and pentaquark states were first predicted in the earliest formulations of the quark model \cite{Gell-Mann:1964ewy, Zweig:1964ruk}.

The QGP provides a source of deconfined quarks which can coalesce into these exotic multiquark states at freeze-out, thereby providing a new testing ground for models of quark recombination at an extended scale of a number of constituent quarks \cite{ExHIC:2010gcb}.  Exotic hadrons are susceptible to color screening in the deconfined plasma, and measurements of their production rates can provide information on their radii and binding energy, which remain poorly understood.  In smaller collision systems, such as \pA{} or high multiplicity \pp{} collisions, interactions with the nucleus and/or other particles produced in the event can disrupt the formation of these bound states, which can also give new insight into their internal structure.  These phenomena cannot be accessed through studies of $b$ hadrons, where most exotic hadrons have been discovered in the decay products.

\begin{figure}[ht]
\centering
\includegraphics[width=0.6\textwidth]{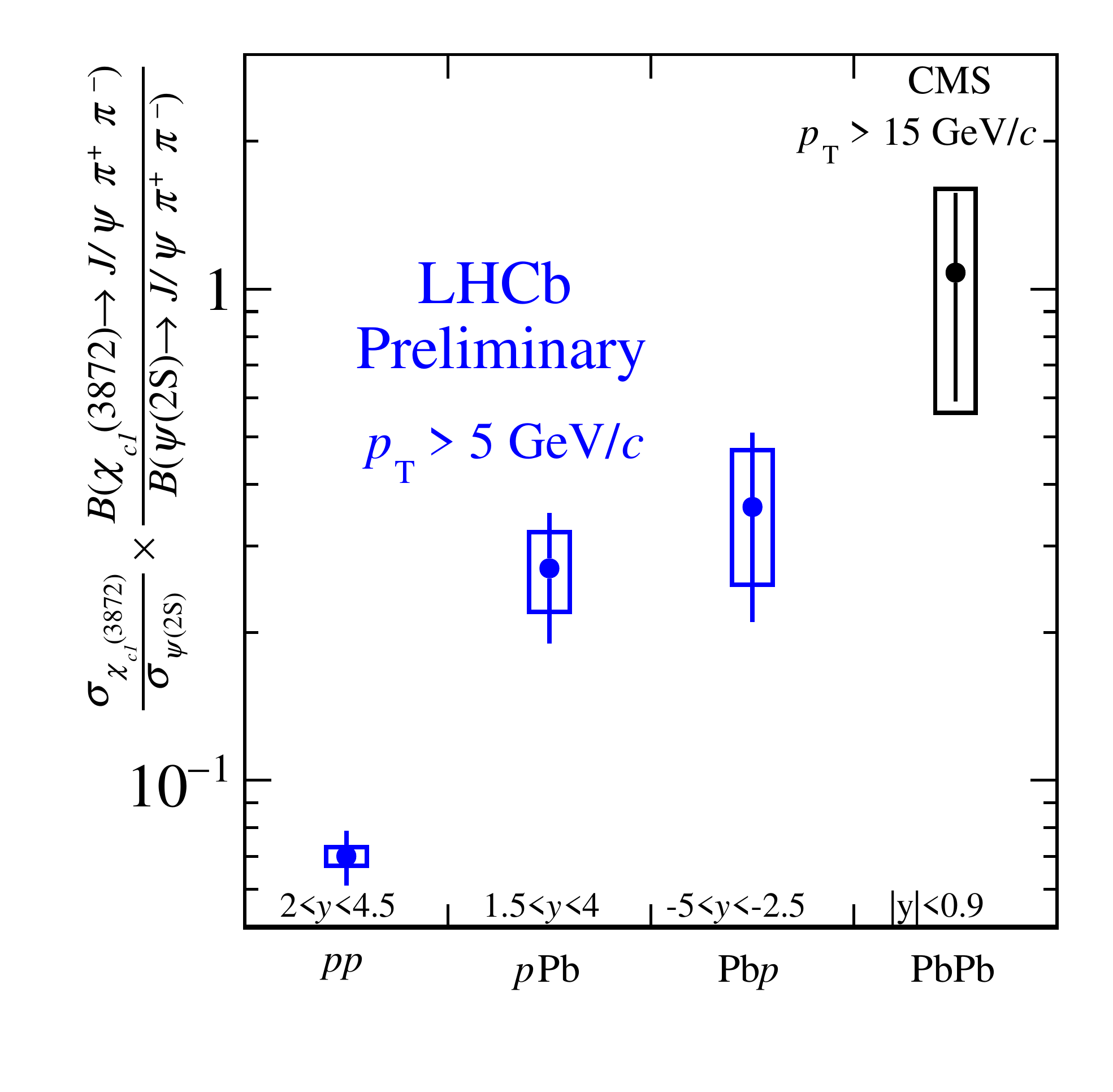}
\caption{The ratio of X(3872) to $\psi(2S)$ cross sections in various collision systems, as measured by LHCb \cite{LHCb_X_pPb} and CMS \cite{CMS:2021znk}.}
\label{fig:X3872overpsi2s}
\end{figure}

Recently, the first measurements of the exotic hadron X(3872) in heavy-ion collisions have become available.  The CMS collaboration measured a comparison between the exotic X(3872) and the conventional charmonium state $\psi(2S)$ in \PbPb{} collisions at  5.02 TeV and found that the X(3872)/$\psi(2S)$ ratio is enhanced a factor of $\sim$10 as compared to \pp{} collisions, although significant uncertainties on the data preclude drawing firm conclusions \cite{CMS:2021znk} (see Figure~\ref{fig:X3872overpsi2s}).  In response, several recent theoretical calculations have considered X(3872) production in \PbPb{} collisions using models that are widely successful at describing the production of conventional hadrons.  Calculations using the AMPT model followed by quark coalescence at the freeze-out show that an X(3872) modeled as a bound state of two $D$ mesons should be enhanced in central \PbPb{} collisions much more than a compact tetraquark \cite{Zhang:2020dwn}.  In contrast, a recent transport calculation comes to the opposite conclusion \cite{Wu:2020zbx}.  These models are widely successful at explaining conventional charmonium, but their application to an exotic state has provided new challenges.    

In smaller collision systems, the LHCb experiment has observed that the X(3872)/$\psi(2S)$ ratio drops as a function of multiplicity in \pp{} collisions at 8 TeV, which was interpreted in terms of final-state interactions among co-moving particles disrupting the X(3872) hadrons.  This data was interpreted to show support for both compact \cite{Esposito:2020ywk} and molecular models \cite{Braaten:2020iqw} of X(3872) structure.  Preliminary LHCb results from $p$Pb collisions at 8.16 TeV show an indication of enhancement of the  X(3872)/$\psi(2S)$ ratio, which falls between the values measured in \pp{} and \PbPb{} collisions \cite{LHCb_X_pPb}, which may indicate different effects on breakup and the hadronization process coming into play.  The variation in the X(3872)/$\psi(2S)$ ratio shows that the effect governing the production of the exotic X(3872) and the conventional $\psi(2S)$ differ as system size changes.

\subsubsection{Electroweak processes}
\label{sec:progress:microscopic:hard_electroweak}

Electroweak (EW) probes such as high-$p_\mathrm{T}$ direct photons, $W$ and $Z$ bosons, top quarks, and high-mass di-leptons from Drell-Yan production are valuable probes of the initial state of the nuclear collision system before it forms a Quark-Gluon Plasma. Significant experimental and theoretical progress has been made since the previous Long-Range Plan, with first measurements of the top quark and high-mass Drell-Yan production in nuclear collision systems, substantially more detailed studies of high-$p_\mathrm{T}$ direct photon, $W$, and $Z$ production, and the incorporation of these data into global analyses.

Traditionally, these measurements have been valuable for understanding so-called ``cold nuclear matter’’ effects on the rate of perturbative parton-parton scatterings, as commonly encoded in the modification of parton densities in the nuclei (nPDFs). Such measurements provide critical context needed to quantitatively interpret final state effects in nuclear collisions, such as parton energy loss or quarkonia dissociation in QGP. Because the precision of any extraction of QGP properties is automatically limited by the knowledge of initial state effects, data from electroweak probes are an integral part of the entire experimental ``Hot QCD'' physics program. Since the last \LRP, electroweak boson data has gone from demonstrating the presence of nuclear PDF modifications to providing precision information for global nPDF analyses, with the experimental uncertainties currently smaller than the theoretical ones in some kinematic regions. Some recently released nPDF which rely on this data include nCTEQ15~\cite{Kovarik:2015cma}, EPPS21~\cite{Eskola:2021nhw}, and nNNPDF3.0~\cite{AbdulKhalek:2022fyi} (with, e.g., specific discussions of the impact of LHC EW data in Refs.~\cite{Kusina:2016fxy,Kusina:2020lyz}).

Additionally, since nPDF effects are not expected to be strongly centrality dependent, the centrality-dependent yields of these objects are seeing renewed use as a proxy to test the centrality selection framework, and in particular the estimation of geometric parameters such as $T_\mathrm{AA}$, in \pA{} and \AA{} events. We summarize the recent progress and some future expectations here. 

Note that measurements of boson-tagged jet energy loss are discussed above in Section~\ref{sec:progress:microscopic:jets_and_leading_hadrons}, and measurements of thermal photons and lower-mass di-lepton probes are instead discussed earlier in Section~\ref{sec:progress:macroscopic:thermal_em_probes}.

{\em Direct photons.} Direct photons can probe the broadest kinematic. Because they are massless, they can be used to probe nPDFs at low $Q^2 \sim \left(p_\mathrm{T}^\gamma\right)^2$, whereas the high-mass bosons are restricted to $Q^2 \gtrsim m_{W/Z}^2$. However, they come with significant experimental challenges (e.g. in the removal of backgrounds from neutral meson decays) and theoretical challenges (in the proper treatment of isolation criteria and fragmentation photon processes in calculations~\cite{Chen:2022gpk}). Significant progress has been made as a result of the large luminosity of LHC Run 2 \PbPb{} and $p$+Pb data, and from continued improvements in the techniques to measure photons in a heavy-ion environment, including their identification and isolation at low \pt. 

\begin{figure}[t]
    \centering
        \includegraphics[width=0.40\textwidth]{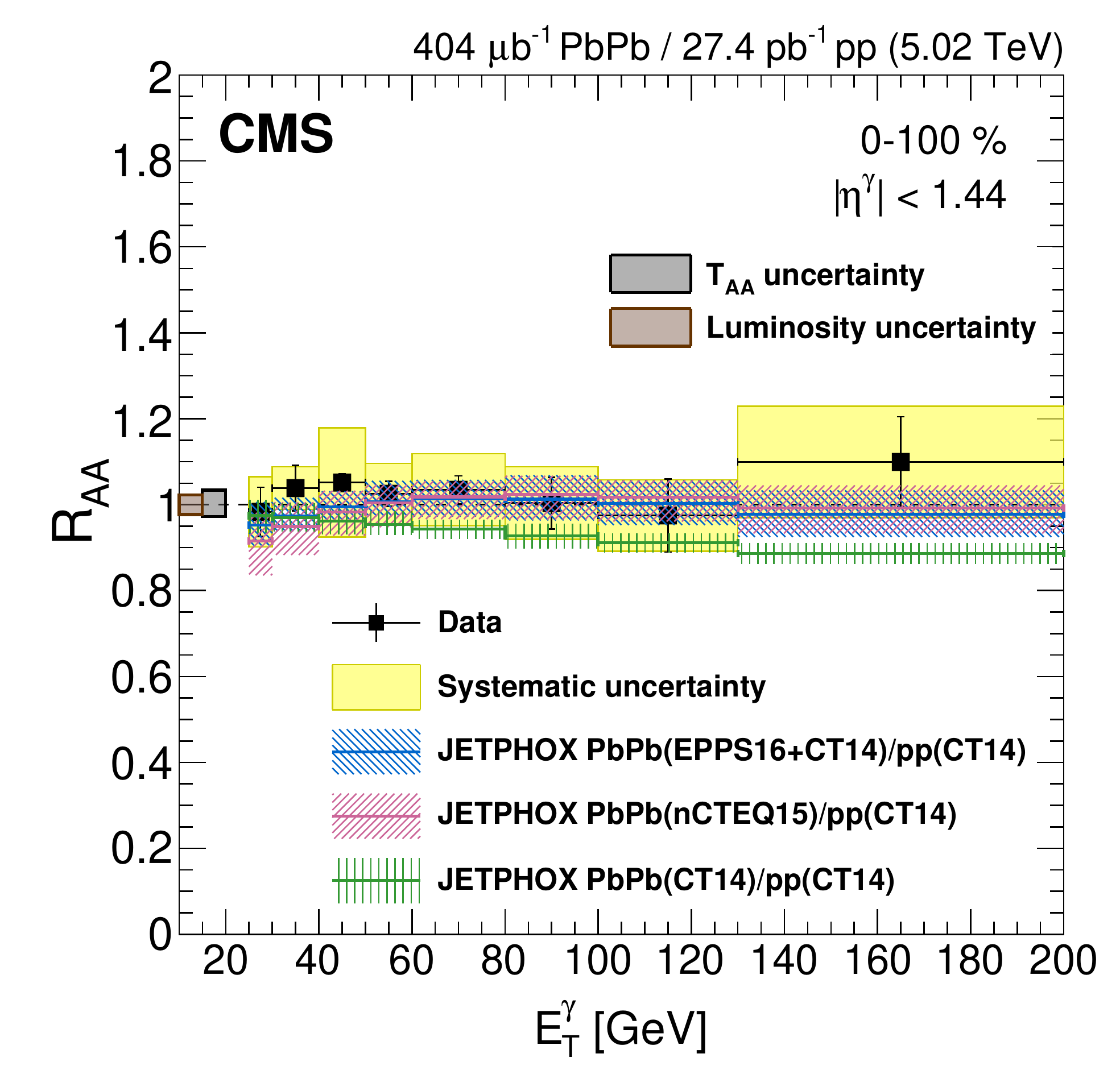}
        \includegraphics[width=0.53\textwidth]{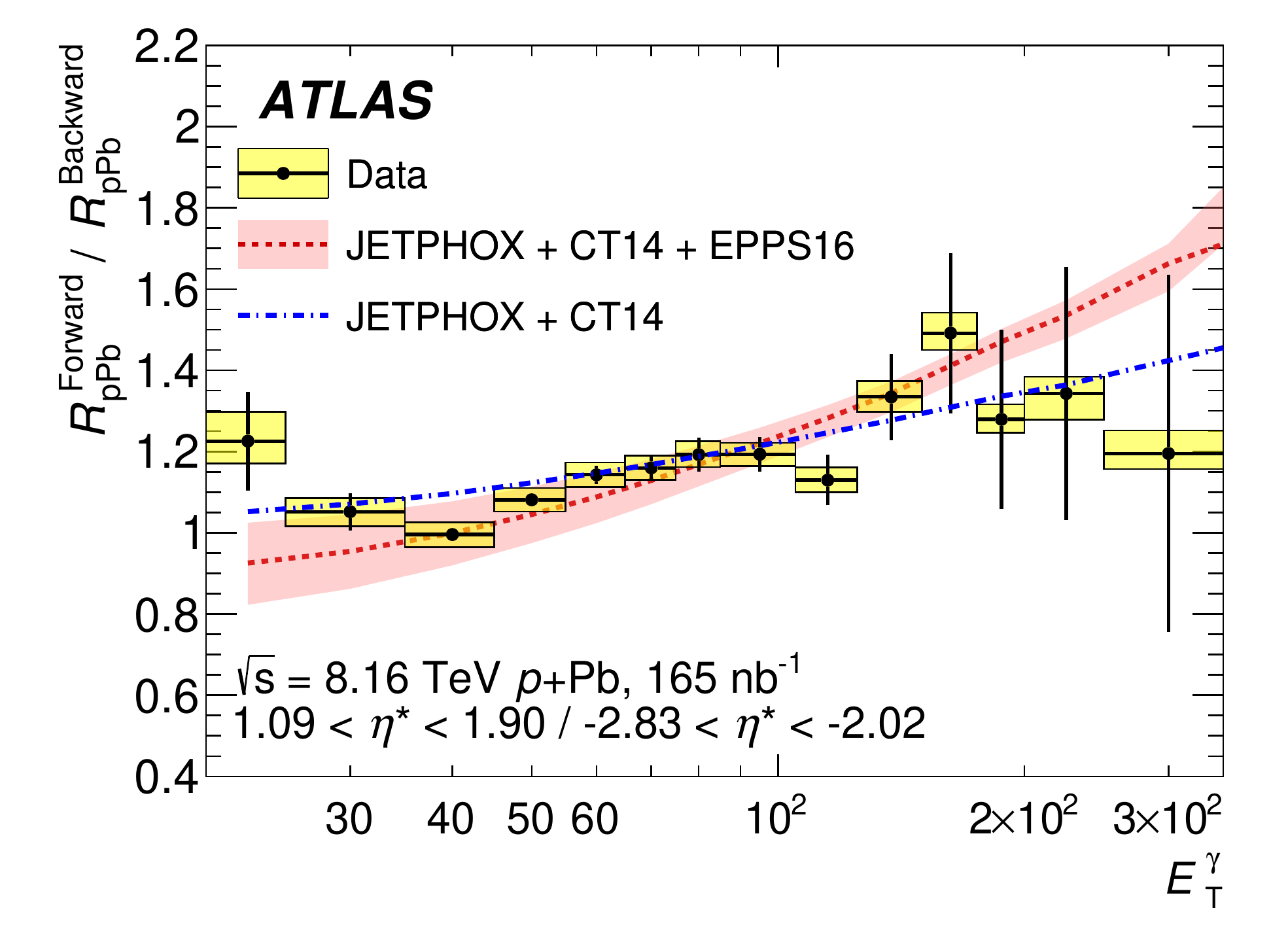}
 \caption{Measurements of isolated photon production in nuclear collisions: nuclear modification factor in \PbPb{} collisions by CMS~\cite{CMS:2020oen} (left), forward/backward ratio of nuclear modification factor in $p$+Pb collisions by ATLAS~\cite{ATLAS:2019ery} (right).}
    \label{EW:photons}
\end{figure}

The large luminosity has enabled improved measurements in \PbPb{} collisions, demonstrated by the latest CMS measurement of the direct photon $R_\mathrm{AA}$ at mid-rapidity~\cite{CMS:2020oen} (left panel Figure~\ref{EW:photons}). At the same time, technical improvements in older data have allowed us to extend the kinematic range where photon measurements can be made, bringing a sensitivity to a new kinematic region in nPDF modifications. The measurement by ATLAS in Run 1 data extends the coverage of direct photons to the forward rapidity region ($\left|\eta\right| = 1.52$--$2.37$)~\cite{ATLAS:2015rlt}. Finally, while the ATLAS and CMS measurements are typically restricted to the range $p_\mathrm{T} \gtrsim 20$~GeV, the final Run 1 measurement by ALICE~\cite{ALICE:2015xmh} can extend significantly lower, testing the comparison to pQCD-based expectations in the region $p_\mathrm{T} \gtrsim 5$~GeV. It should be noted, however, that the measurements in ATLAS and CMS are of isolated photons, whereas in ALICE the measurement is inclusive of all direct photons, potentially probing somewhat different physics.

In addition to the measurements above in \PbPb{}, ATLAS has produced the first isolated photon measurement in $p$+Pb collisions at the LHC~\cite{ATLAS:2019ery}. Here, the large acceptance is utilized for separate measurements in the forward-, mid-, and backward-rapidity regions, and to construct the ratios between them to cancel common uncertainties (right panel Figure~\ref{EW:photons}). Coupled with the wide $p_\mathrm{T}^\gamma$ range, the double-differential measurement results in a broad probe of the shadowing, anti-shadowing, and EMC regions. The ATLAS $p$+Pb measurement is the first direct photon measurement to be included in global nPDF extractions~\cite{AbdulKhalek:2022fyi}.

{\em $Z$ bosons.} $Z$ bosons are measured in their di-lepton decay modes to electrons or muons, which are unmodified by their resulting passage through the QGP. Although $Z$ bosons are the rarest of the electroweak boson probes, these channels have the significant advantage that the $Z$ yield can be measured with high purity (measurements have achieved QCD backgrounds levels of $\approx1$\% in the $Z$ mass window) and high precision (since excellent muon or $e/\gamma$ capability is a key design principle for the EW HEP programs at the LHC). 

Measurements of $Z$ production have been made in the high-luminosity 8.16~TeV $p$+Pb data by ATLAS~\cite{ATLAS:2015mwq}, CMS~\cite{CMS:2021ynu}, and LHCb~\cite{LHCb:2022kph}, and these are key contributors to global nPDF fits. In fact, this channel has even been argued to be sensitive to the modification of the strange and charm content in bound nucleons ~\cite{Kusina:2016fxy}. As a new development, the CMS measurement further reports on di-lepton production through the Drell-Yan process ($q\bar{q}\to{Z}/\gamma^*\to{l}{l}$) over the invariant mass range $m_{ll} > 15$~GeV. This first measurement demonstrates that the heavy-flavor background for this process at the LHC can be overcome, and thus opens a new potential channel for future study.

\begin{figure}[t]
    \centering
        \includegraphics[width=0.50\textwidth]{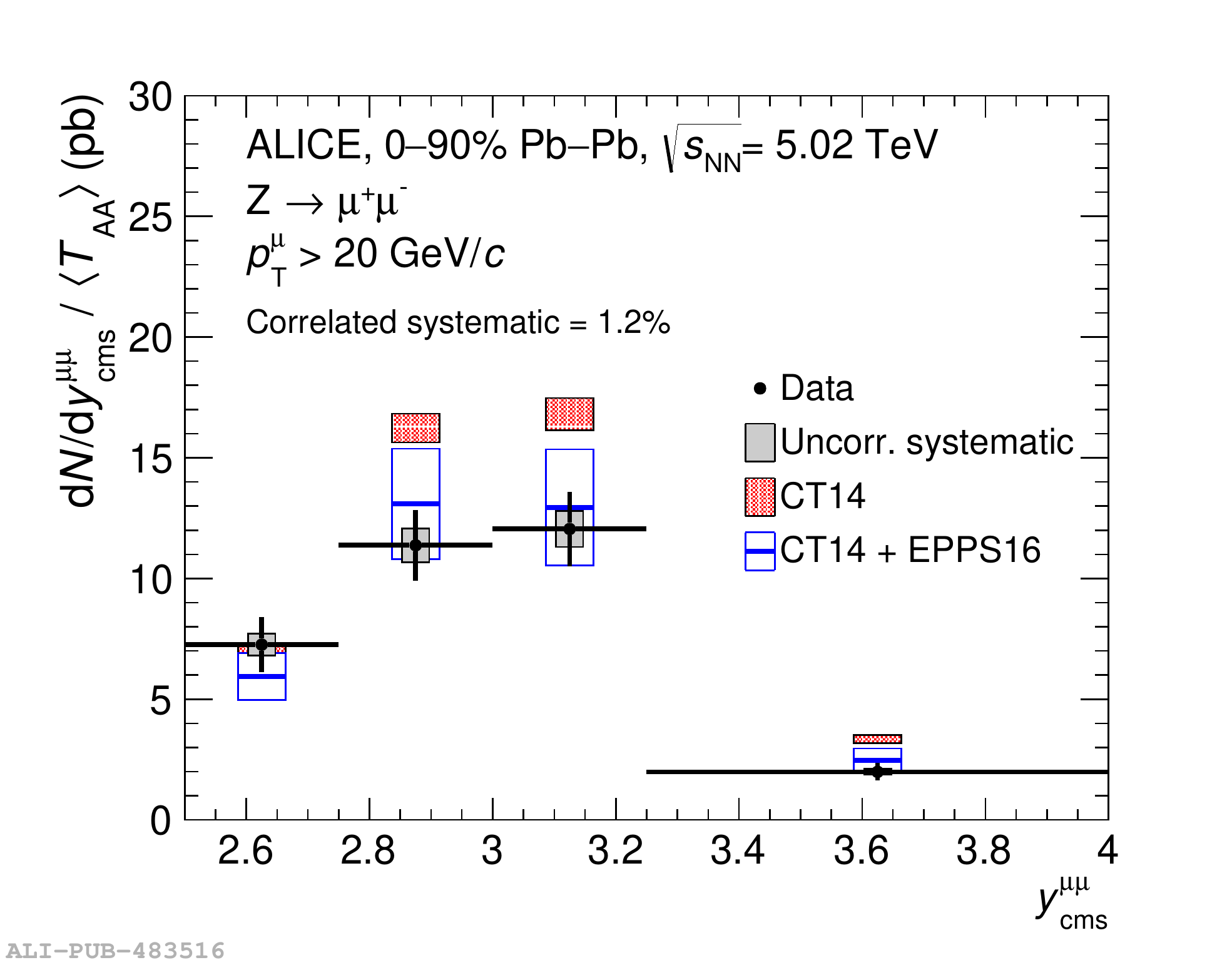}
        \includegraphics[width=0.42\textwidth]{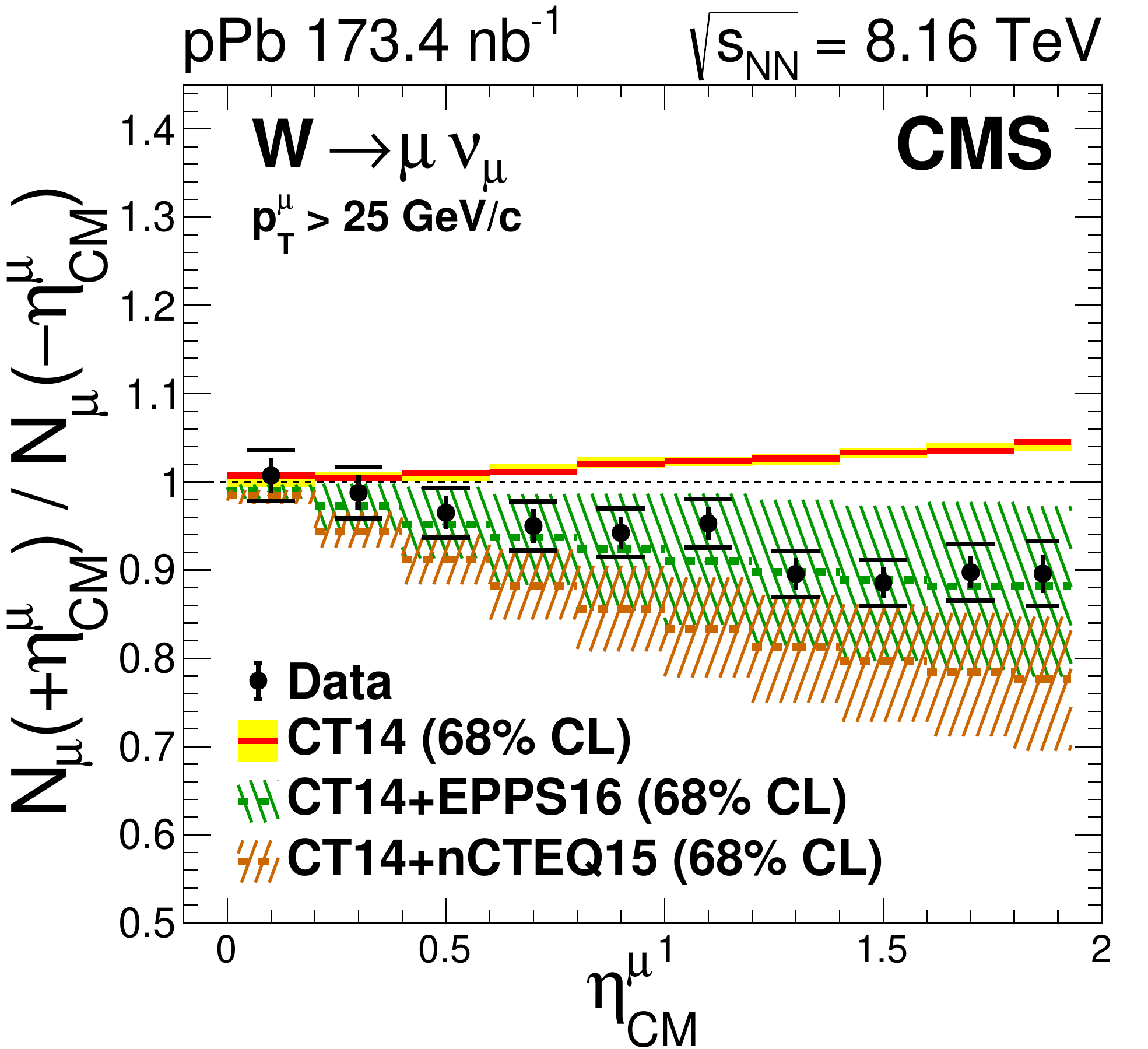}
    \caption{Measurements of heavy electroweak boson production in nuclear collisions: nuclear modification factor for $Z$ production in \PbPb{} collisions at forward rapidity by ALICE~\cite{ALICE:2020jff} (left), forward/backward ratio of $W^\pm$ production as a function of muon pseudorapidity in $p$+Pb collisions by CMS~\cite{CMS:2019leu} (right).
    }
    \label{EW:ZW}
\end{figure}

In \PbPb{} data, measurements by CMS have demonstrated that the azimuthal anisotropy for $Z$ bosons is compatible with zero~\cite{CMS:2021kvd} and those by ATLAS are approaching the precision needed to give input to nPDF fits~\cite{ATLAS:2019maq}. Both experiments have also used the high-luminosity \PbPb{} data to explore the centrality dependence of $Z$ production, discussed below. Finally, measurements in \pPb{} and \PbPb{} are reported in the far forward region ($\left|\eta^\mu\right| > 2.5$) by ALICE~\cite{ALICE:2020jff} (left panel of Figure~\ref{EW:ZW}). Although they feature significant uncertainties, they are complementary to the ones by ATLAS and CMS above and thus set the only constraints in an extreme kinematic region where nPDF modifications may be large. 

{\em $W$ bosons.} $W^\pm$ bosons are significantly more abundant than $Z$ bosons. They are produced dominantly via $q\bar{q}$ annihilation and are thus particularly useful for setting constraints on the nPDF modification for light quarks. So far, they have been measured in their leptonic decay channel ($W\to\nu{l}$), where one can have confidence that the decay products are not affected by the collision system (such as through jet quenching). Given the large resolution on the missing-$E_\mathrm{T}$ from the unmeasured neutrino, the results are typically reported as a function of electron or muon kinematics only, resulting in a potentially weaker correlation to the hard-scattering kinematics. 

CMS has recently reported a high-statistic measurement in 8.16 TeV $p$+Pb collisions~\cite{CMS:2019leu}, where the data most sensitive to nPDF modification are the forward-backward ratios as a function of muon pseudorapidity (right panel of Figure~\ref{EW:ZW}). The data rejects the null hypothesis (i.e. of no nPDF modification) with large significance, and shows a preference for some global fits (such as EPPS16) over others (such as nCTEQ15). ATLAS has measured $W^\pm$ production in 5.02~TeV~\cite{ATLAS:2019ibd} \PbPb{} collisions, with fine selections on event centrality. These results are discussed in the sub-section below. Finally, as with the measurements of $Z$ bosons above, ALICE has measured $W^\pm$ boson production in both \pPb{} and \PbPb{} in the forward region ($\left|\eta^\mu\right| > 2.5$)~\cite{ALICE:2022cxs}, which is kinematically complementary to those by ATLAS and CMS.

{\em Centrality dependence of EW boson production.} The $1/T_\mathrm{AA}$-scaled yields of EW bosons, or their nuclear modification factor $R_\mathrm{AA}$, in centrality-selected events serve as a good test of the overall centrality determination and geometric modeling of heavy-ion collisions, for which the experiments use the Monte Carlo Glauber model plus an accompanying model of particle production~\cite{dEnterria:2020dwq}. Since EW boson yields are not expected to experience significant energy loss effects, deviations from $T_\mathrm{AA}$ scaling can therefore diagnose mis-modeling of the collision geometry or the presence of trivial or non-trivial correlations which affect the centrality classification of events with a hard sub-process.

These studies have been particularly motivated by the observation in ALICE that the charged particle $R_\mathrm{AA}$ strongly decreases in very peripheral ($>80$\%) \PbPb{} events~\cite{ALICE:2018ekf}. One potential explanation is a jet veto effect, wherein a low-multiplicity selection suppresses hard process rates, modeled by HG-Pythia~\cite{Loizides:2017sqq}. However, it remains to be seen whether the effect depends on the particular method of estimating $T_\mathrm{AA}$ or whether other effects, such as the presence of ultra-peripheral collisions in the peripheral event sample, may contribute.

\begin{figure}[t]
    \centering
    \includegraphics[width=0.42\textwidth]{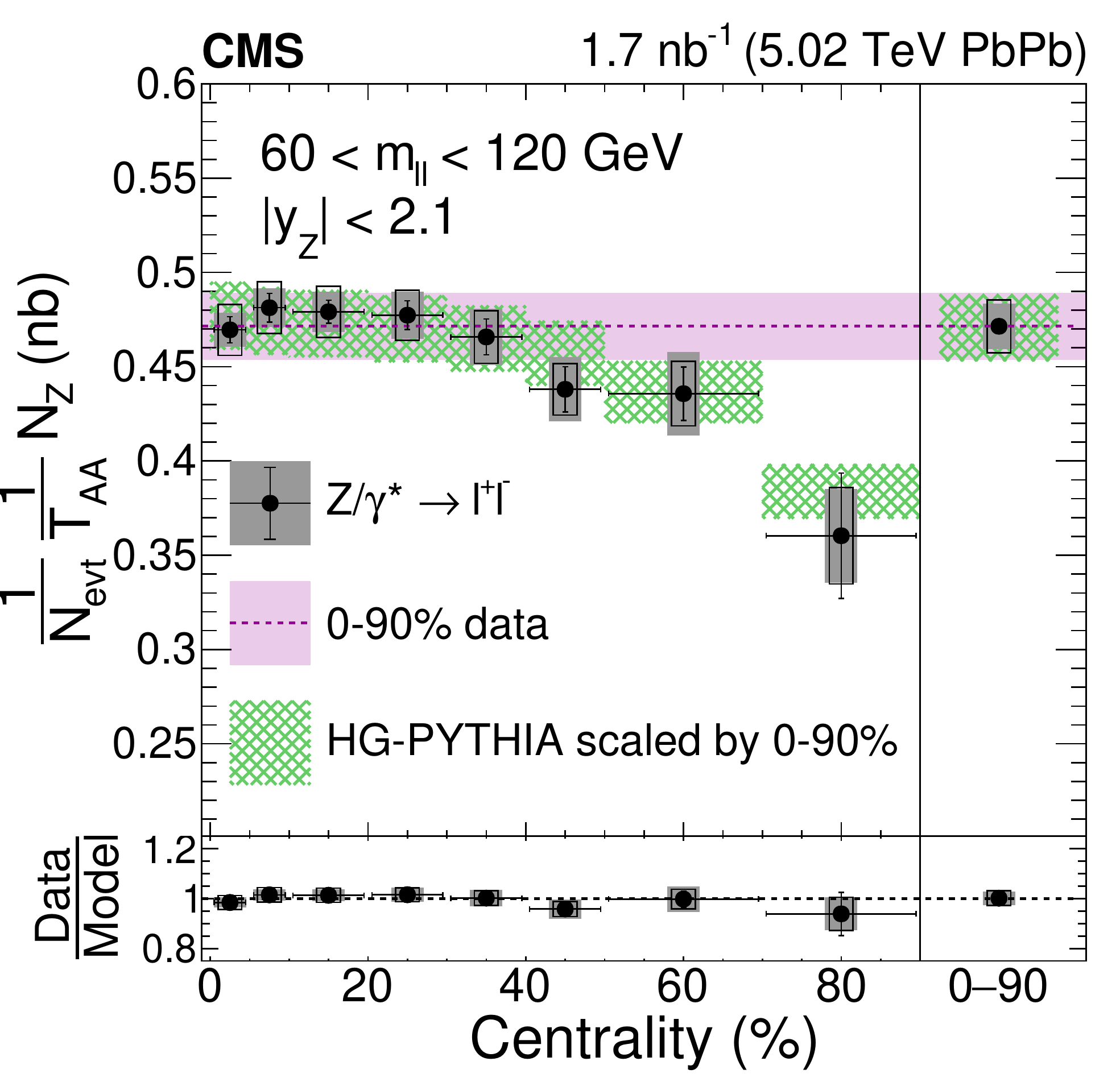}
    \includegraphics[width=0.50\textwidth]{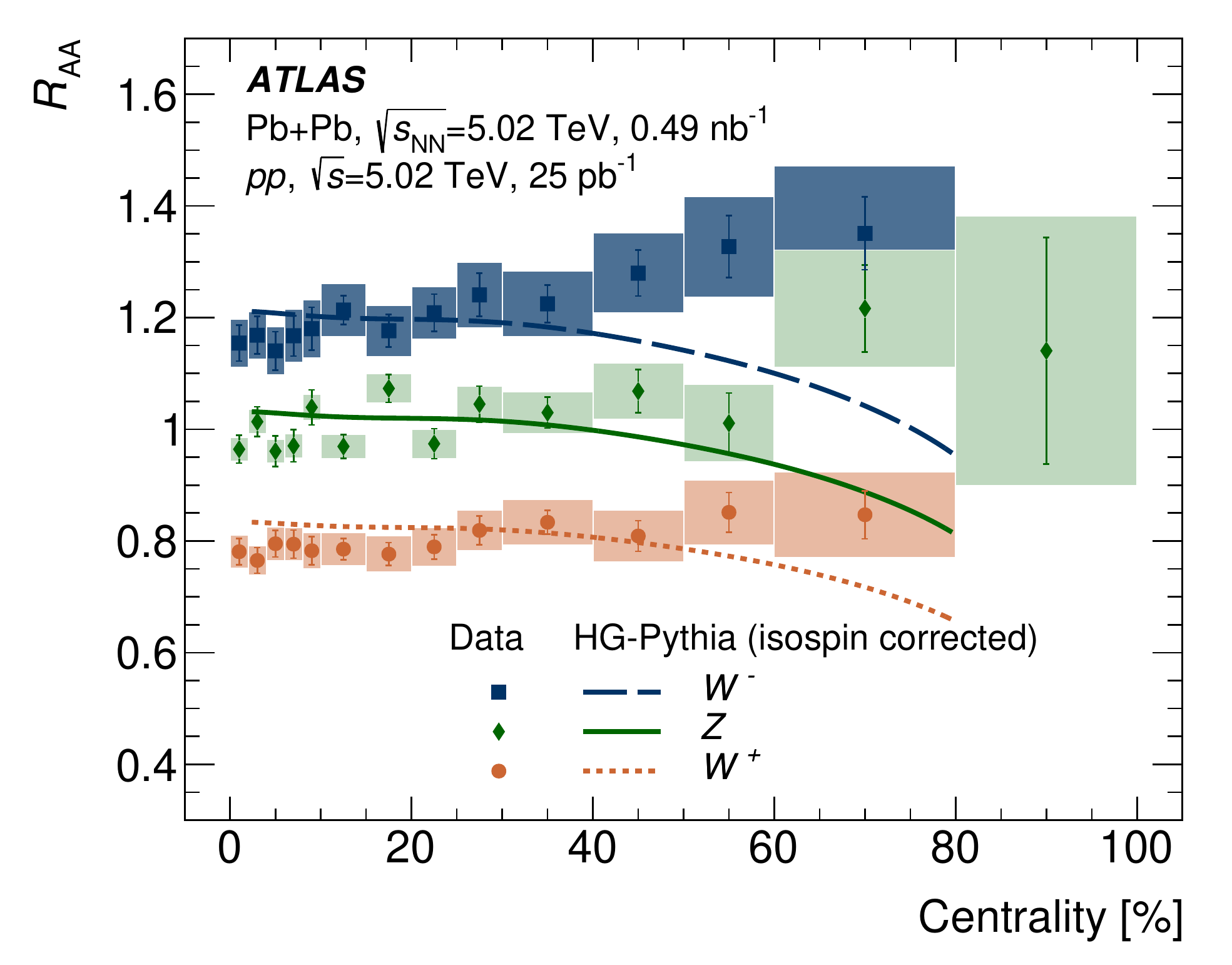}
    \caption{Centrality dependence of $Z$ production from CMS (left) and $Z$, $W^+$, $W^-$ production from ATLAS (right) in \PbPb{} collisions, both compared to the expectation from the HG-Pythia model~\cite{Loizides:2017sqq}.
    }
    \label{EW:centrality}
\end{figure}

The $1/T_\mathrm{AA}$-scaled yields for $Z$ bosons measured by CMS (left panel of Figure~\ref{EW:centrality}) systematically drop below unity in more peripheral events, compatible with the HG-Pythia prediction and thus supporting this interpretation. However, the $R_\mathrm{AA}$ values for $Z$ and $W^\pm$ bosons measured by ATLAS (right panel of Figure~\ref{EW:centrality}) systematically rise in more peripheral events. Some suggested explanations of the effect observed by ATLAS include a shadowing of the nucleon--nucleon cross section in heavy-ion collisions~\cite{Eskola:2020lee} or a larger than expected neutron skin effect~\cite{Jonas:2021xju}. Future work will be needed to resolve the picture. Nevertheless, if the uncertainties on $Z$ or $W$ yields become comparable to those on the traditional extraction of $T_\mathrm{AA}$, one option for future measurements may be to use these yields as the {\em de facto} control on the event geometry.

Additionally, such centrality-dependent studies should be repeated in asymmetric systems such as $p$+Pb collisions. In these small systems, the association between experimentally-selected centrality and underlying geometry is more difficult~\cite{ALICE:2014xsp,Perepelitsa:2014yta}, and color fluctuation effects~\cite{Alvioli:2014eda,Alvioli:2017wou} or additional QCD effects~\cite{Bzdak:2014rca,Kordell:2016njg} may contribute in novel ways to the correlation between hard process kinematics and the centrality signal. For example, a measurement of the centrality dependence of $Z$ production in 5.02~TeV $p$+Pb collisions by ATLAS~\cite{ATLAS:2015mwq} revealed a significant sensitivity to choices in the modeling of the nucleon--nucleon interaction~\cite{Loizides:2016djv,ATLAS:2015hkr}, such as whether to allow ``Glauber-Gribov'' cross section fluctuations.

{\em Top quarks.} An intriguing development since the last \LRP{} is the first measurements of $t\bar{t}$ production in heavy-ion collisions at the LHC. In $p$+Pb collisions~\cite{CMS:2017hnw}, $t\bar{t}$ production was observed by CMS by requiring an isolated lepton plus at least four jets, with a varying requirement on the number of observed $b$-jets to delineate background- and signal-dominated regions. While statistically limited, $t\bar{t}$ production rates potentially offer an alternative way to probe the nuclear modifications on large-$x$ gluons, since they are dominantly formed through gluon--gluon fusion. 

CMS has also found evidence of $t\bar{t}$ production in \PbPb{} collisions~\cite{CMS:2020aem}. In the \PbPb{} system, $t\bar{t}$ detection is made more complicated by the quenching of the high-energy $b$ and lighter quarks at the end of the top quark decay chain. This raises the possibility, therefore, of using $t\bar{t}$ events to probe the parton-QGP interaction, rather than the initial state. Indeed, the delayed decays of top quarks have been proposed as a novel way to understand the time dependence of jet quenching~\cite{Apolinario:2017sob} with future data.

\subsubsection{Ultra-peripheral Collisions}
\label{sec:progress:microscopic:ultraperipheral_collisions}

Disentangling initial-state and final-state effects in heavy-ion collisions is important for properly quantifying the properties of the QGP. 
One promising tool to reveal the intrinsic property of heavy nuclei at their initial state is through photon-induced interactions, commonly known as the \textit{ultra-peripheral collision} (UPC). Typically, the UPC takes place when the impact parameter between the two colliding nuclei is greater than the sum of their radii. The interaction is initiated by one or multiple photons emitted from the fast-moving charged ions, and only photons interact with the other nucleus. For a review of UPC, see Refs.~\cite{Bertulani:2005ru,Contreras:2015dqa,Klein:2017nqo,Klein:2019qfb,Klein:2020fmr} and references therein. The physics prospects of UPC studies in Run 3\&4 with the CERN LHC experiments have been recently reported in~\cite{Citron:2018lsq,Klein:2020nvu,Hentschinski:2022xnd}. 

There are generally three types of UPC physics process: i) inclusive production; ii) semi-inclusive and/or jet production; iii) exclusive production. In the past decade, most of the UPC measurements focused on exclusive production, dominated by diffractive vector meson production. However, since the last \LRP, there are increasing number of studies in both theory and experiment on jet photoproduction and inclusive particle photoproduction.
As of now, it is widely realized that UPCs can be extremely illuminating to understand the initial-state condition of heavy-ion collisions. %

\paragraph{Vector meson photoproduction} 
\label{subsubsec:vm}

\subparagraph{Vector meson in heavy nuclei} A general picture of vector meson (VM) photoproduction is the following. The quasi-real photon emitted by a nucleus fluctuates into a quark-antiquark pair, known as the color dipole, which forms a vector meson and scatters off the target nucleus. The interaction between the dipole and target is via a two-gluon exchange, leaving the event topology with a large rapidity gap between the VM and the target. In a leading order QCD calculation, the cross section of this interaction is expected to scale as the square of the gluon density, $\left[x G(x,Q^{2})\right]^2$, which makes it an ideal probe to the nuclear Parton Distribution Function (nPDF). Note that in a recent study done at Next-to-Leading Order~\cite{Eskola:2022vaf}, this expectation was found to be sensitive to quark distribution, which is one of the major theoretical achievements since the last long-range plan. 

\begin{figure}[thb]
\includegraphics[width=2.3in]{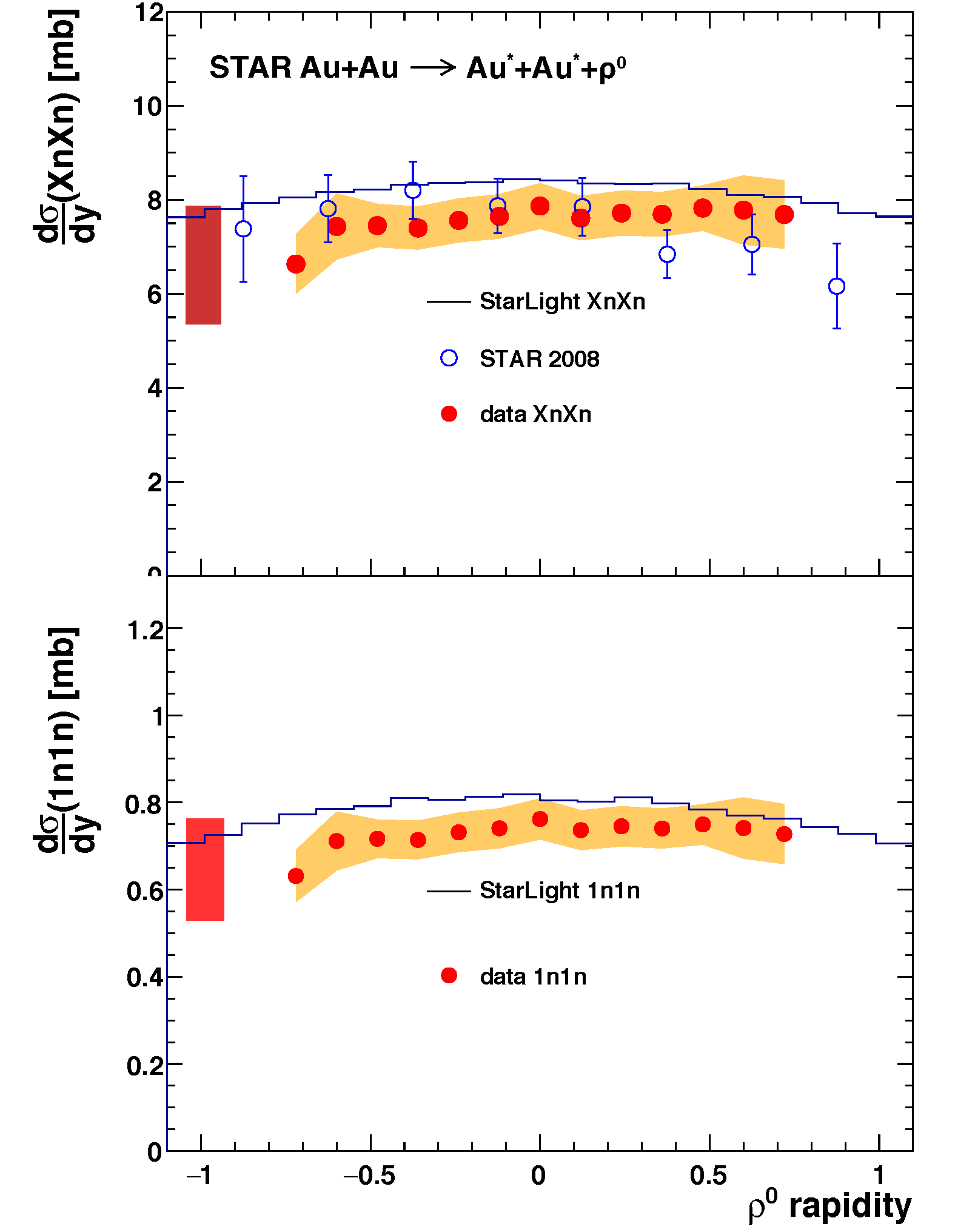}
\includegraphics[width=2.3in]{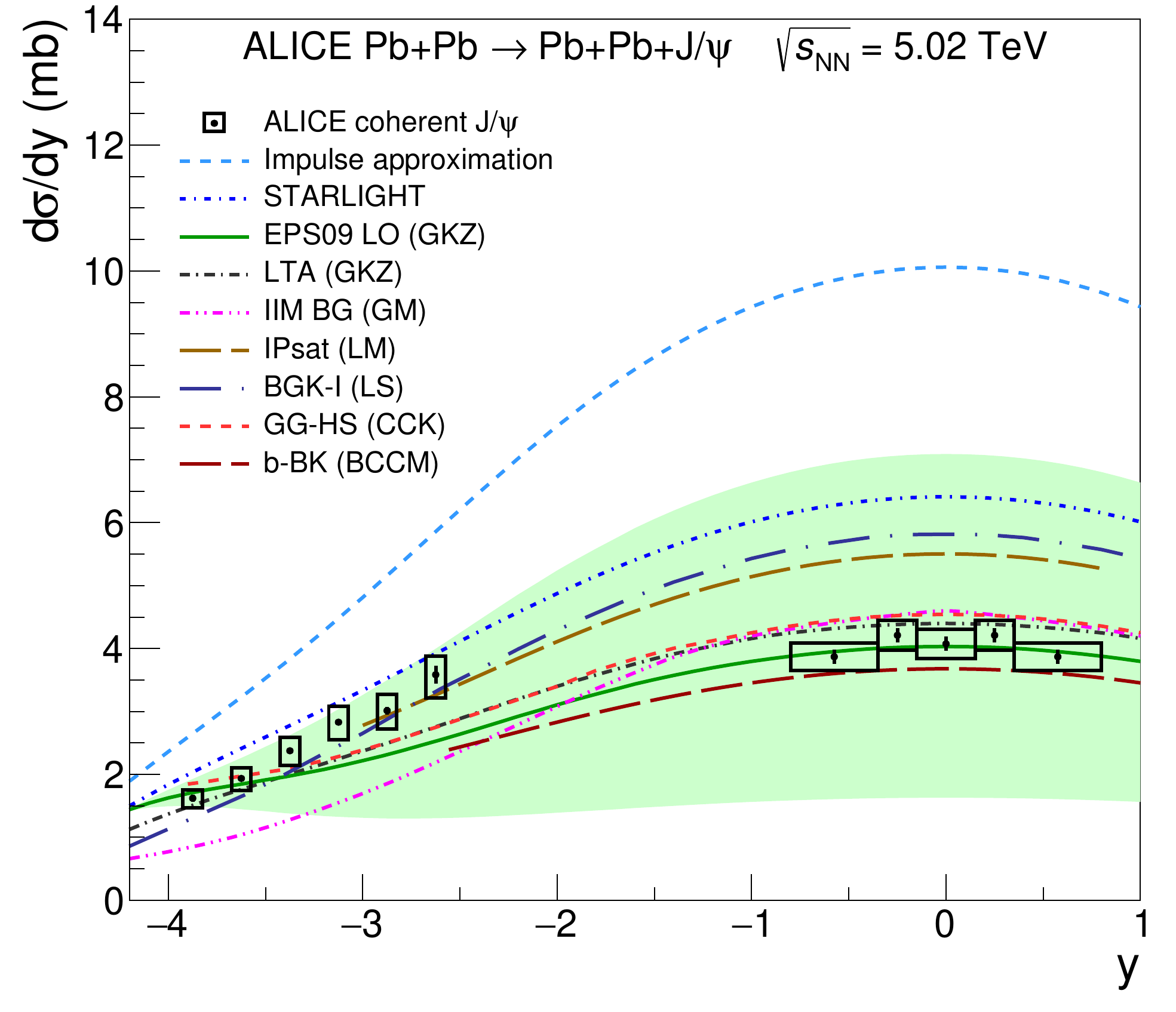}
  \caption{ \label{fig:3.5:figure_3} Left: coherent $\rho^{0}$ photoproduction, $d\sigma/dy$ as a function of $y$, in Au$+$Au UPC at $\sqrt{s_{_{\rm NN}}}=200$~GeV at RHIC using the STAR detector~\cite{STAR:2017enh}. The notation ``1n'' and ``Xn'' are neutron emission class measured by the ZDC, where ``1n'' denotes one neutron and ``Xn'' represents more than one neutron. Right: the same measurement based on $J/\psi$ particle done by the LHC ALICE experiment at both central and forward rapidity in Pb$+$Pb UPCs at $\sqrt{s_{_{\rm NN}}}=5.02$~TeV\cite{ALICE:2021gpt}. }
\end{figure}

In the past decade, especially since the last NP long-range plan, one of the most important studies for initial-state condition of heavy-ion collisions is the coherent $\rho^{0}$ and $J/\psi$ photoproduction at RHIC and the LHC, respectively. 
In Figure~\ref{fig:3.5:figure_3}, the differential cross section $d\sigma/dy$ as a function of $y$ for $\rho^{0}$ (left) and $J/\psi$ (right) photoproduction measured by STAR (RHIC) and ALICE (LHC) are shown, respectively. Similar measurements of coherent $J/\psi$ photoproduction at the LHC have been made by LHCb~\cite{LHCb:2021bfl,LHCb:2022ahs} and CMS~\cite{CMS:2016itn} as well. Theoretical and Monte Carlo models are compared with both data.
It is unambiguously found by all LHC experiments that the coherent $J/\psi$ photoproduction cross section in heavy nuclei is significantly suppressed compared to that of a free nucleon, shown by the difference with respect to the Impulse Approximation. Figure~\ref{fig:3.5:figure_3} right shows a large number of theoretical models compared with the data; these models can be categorized into two major physics models - i) nuclear shadowing model and ii) gluon saturation model. 

For i), the shadowing model is based on nuclear PDFs which, by construction, capture some of the nuclear modification present in lepton scattering data, as well as a dynamical modeling of shadowing through Leading Twist Approximation, which is based on Gribov-Glauber theory, QCD factorization, and HERA diffractive PDFs. One key feature of this model is that it does not have any suppression in a free nucleon. On the other hand for ii), gluon saturation models handle this problem differently, such that by a first principle argument of unitarity, gluon density cannot be infinity~\cite{Accardi:2012qut}. As the energy of the initial state nucleon and nucleus increases in heavy-ion collisions, the growth of gluon density would slow down and naturally, it is expected to be suppressed with respect to a free nucleon. This should be present in both nucleon and nuclei, except that it is easier to observe this non-linear dynamics in heavy nuclei at the same given $x$, quantitatively determined by the saturation scale $Q_s$~\cite{Accardi:2012qut}. 
Understanding the commonalities and differences has become one of the most important and urgent questions in order to narrow down what is responsible for such a large nuclear suppression seen in the data. The answer to this question has a direct impact to our understanding of the initial-state conditions of heavy-ion collisions.

Furthermore, differential cross section measurements of $p^{2}_{\rm T}$ in UPC VM photoproduction have been another major milestone since the last long-range plan, due to the higher statistic sample became available at both RHIC and the LHC. For coherent VM photoproduction, since the virtuality of the photon is small, the momentum transfer $-t$ can be approximated by the $p^{2}_{\rm T}$ of the VM, which is a Fourier transform of the source in the coordinate space, e.g., gluons. Therefore, with the momentum transfer $-t$ measured in UPCs, we have gained more access to the initial-state condition of heavy-ion collisions in an impact parameter dependent way. The momentum transfer $-t$ distribution of $\rho^{0}$ and $J/\psi$ particle in coherent photoproduction have been measured at RHIC~\cite{STAR:2017enh} and at the LHC~\cite{ALICE:2021tyx}, respectively. At RHIC, despite issues related to $\rho^{0}$, e.g., its soft mass scale and the overwhelming incoherent background, the measurement has been a major step forward by showing a proof-of-principle study. The data has been Fourier transformed from momentum space to the impact parameter space of partons. The first indication of diffractive structures, e.g., those in Ref.~\cite{Accardi:2012qut} predicted at the EIC, has been observed. Similar measurement has been done at the LHC~\cite{ALICE:2021tyx} but with much limited $-t$ range, due to similar challenges as seen at RHIC. %

For other VM photoproduction in heavy nuclei, there are measurements of $\psi(2S)$, where the ratio to $J/\psi$ has been studied. No significant deviation from that of a free proton has been observed~\cite{ALICE:2021gpt}. There is no measurement yet using $\phi$ or $\Upsilon$ in nucleus-nucleus UPC. The experimental challenge for $\phi$ is the soft kaon daughters from $\phi$ decay, which is below $\sim100$~MeV. For $\Upsilon$, the cross section is much smaller and more data is needed. However, UPC $\Upsilon$ in \pPb{} collisions has been measured for the first time by the CMS experiment~\cite{CMS:2018bbk}.

\begin{figure}[thb]
\centering
\includegraphics[width=4in]{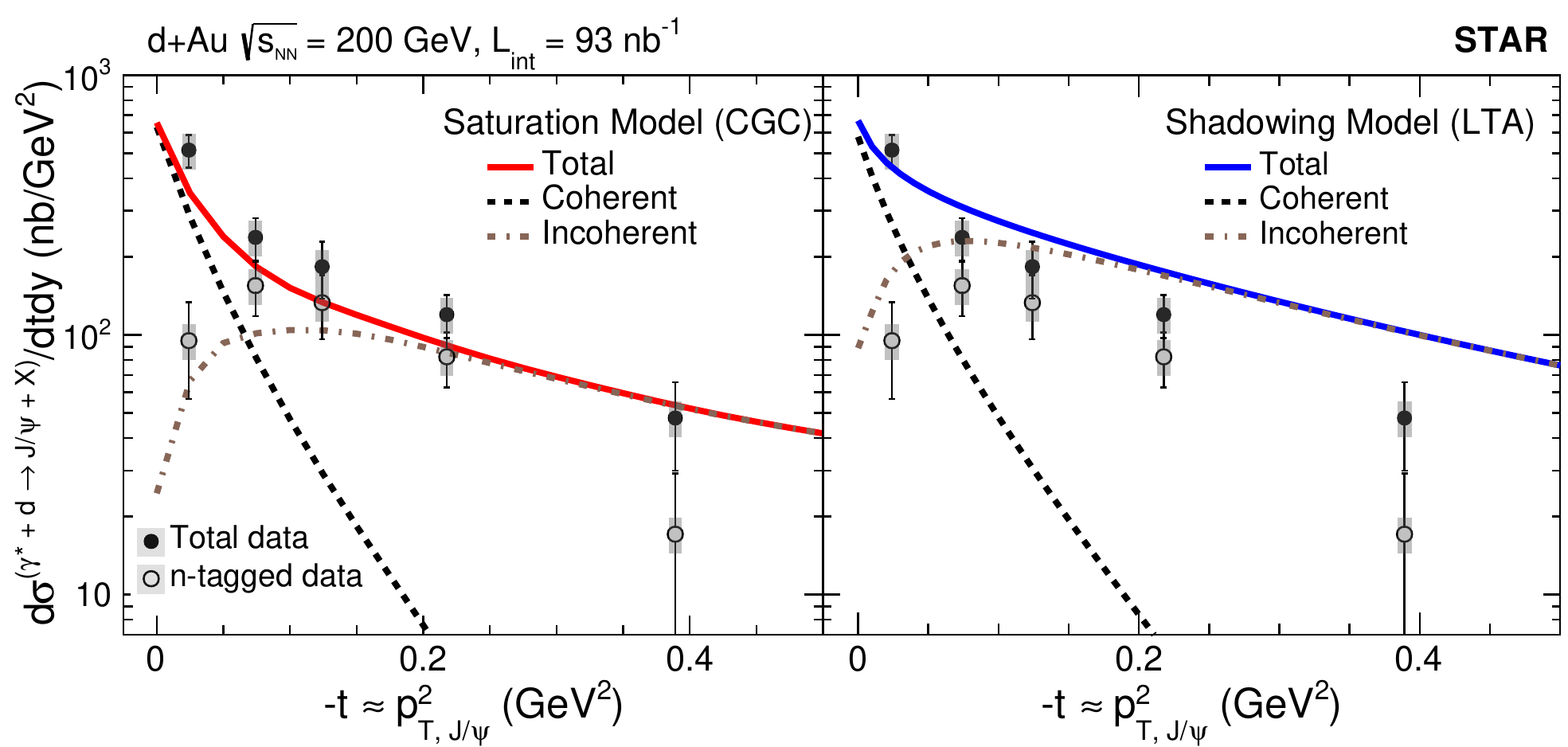}
  \caption{ \label{fig:3.5:figure_5} Differential cross section, $d\sigma/dtdy$, as a function of $-t$ of $J/\psi$ photoproduction in $\gamma^\ast+$d collisions. Color Glass Condensate and nuclear shadowing model Leading Twist Approximation are compared with data~\cite{STAR:2021wwq}. }
\end{figure}

\subparagraph{Vector meson in proton and light ions}
UPC $J/\psi$ in \pPb{} collisions have been measured by the ALICE experiment at 5.02 TeV, where the lead nucleus is the photon emitter and the proton is the target~\cite{ALICE:2014eof,ALICE:2018oyo}. The measurements in UPC have been a complementary study to those at H1 and ZEUS experiments at \ep{} collider HERA, which covers a wider photon-proton center-of-mass energy $W$ range. The quantitative energy dependence of the $J/\psi$ photoproduction in proton, measured by these studies, have been an important input to study the gluon saturation phenomena. 

Besides the energy dependence, another important lesson learned from $J/\psi$ photoproduction in proton is the initial-state fluctuation of hot-spots~\cite{Mantysaari:2016ykx}, which has significantly impacted the understanding of small system collective dynamics in heavy-ion collisions. Although the initial study~\cite{Mantysaari:2016ykx} was performed based on HERA data, recent measurement on d$+$Au UPC $J/\psi$ photoproduction has further confirmed this phenomenon at RHIC energies. In Figure~\ref{fig:3.5:figure_5}, differential cross sections $d\sigma/dtdy$ of $J/\psi$ photoproduction in d$+$Au UPC at center-of-mass energy $\sqrt{s_{_{\rm NN}}}=200$~GeV are shown. Two theoretical models, CGC and LTA discussed previously,
are compared with data. The CGC calculation is found to describe the STAR data of incoherent production better, and sub-nucleonic fluctuation plays an important role in their model. In addition, authors of the nuclear shadowing model have found that the mismatch of this \dAu{} UPC data could be caused by the overestimation of proton dissociation at RHIC energy~\cite{Guzey:2022qvc}. Nevertheless, photoproduction of $J/\psi$ in the simplest nuclear system, deuteron, has provided important insights to disentangle models and the first snapshot of gluon momentum (Fourier transform of the spatial coordinate) distributions inside the deuteron. This study has a close connection to those proposed at the EIC with spectator tagging technique~\cite{Tu:2020ymk,Jentsch:2021qdp}. 

\begin{figure}[thb]
\includegraphics[width=2.3in]{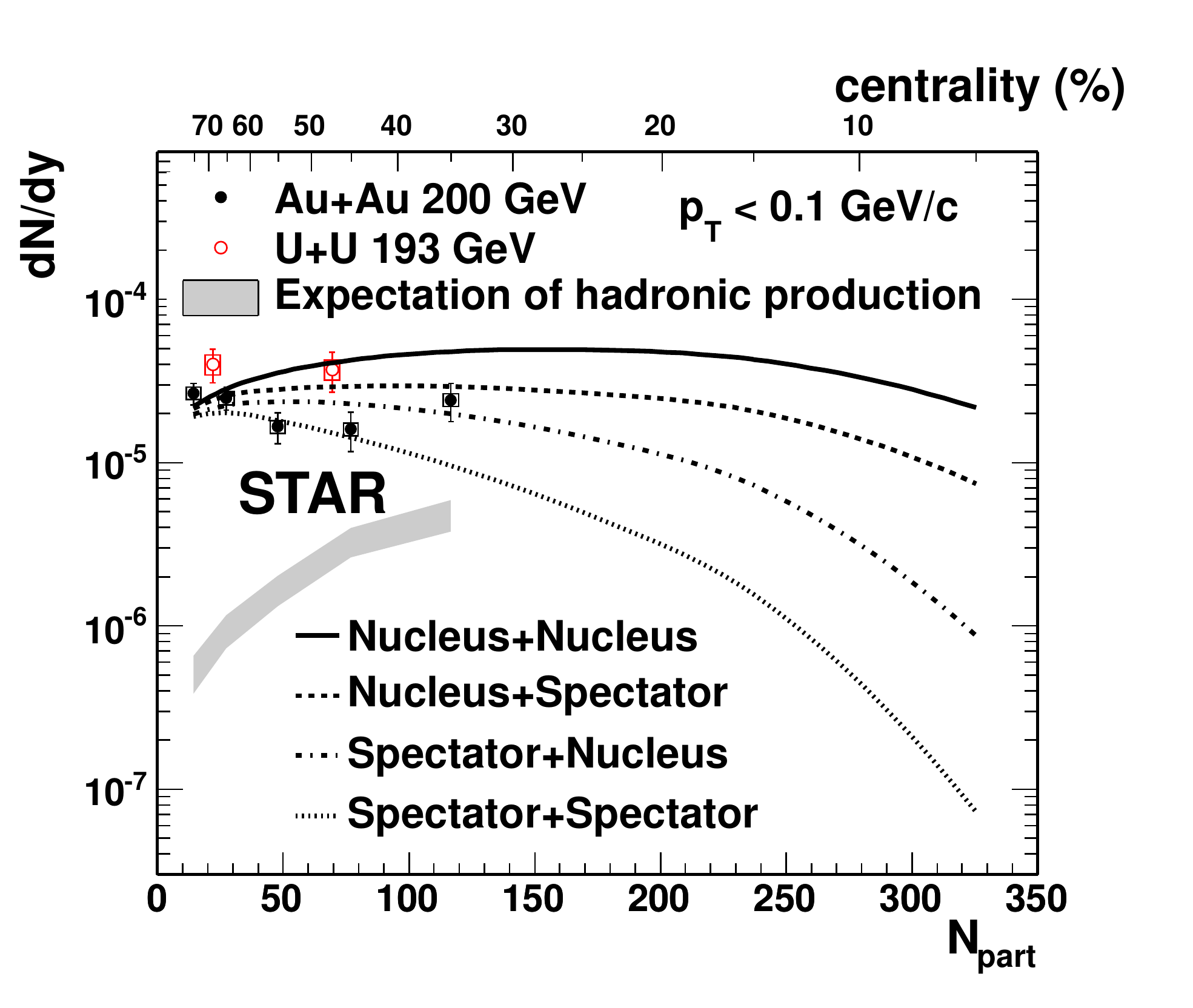}
\includegraphics[width=2.5in]{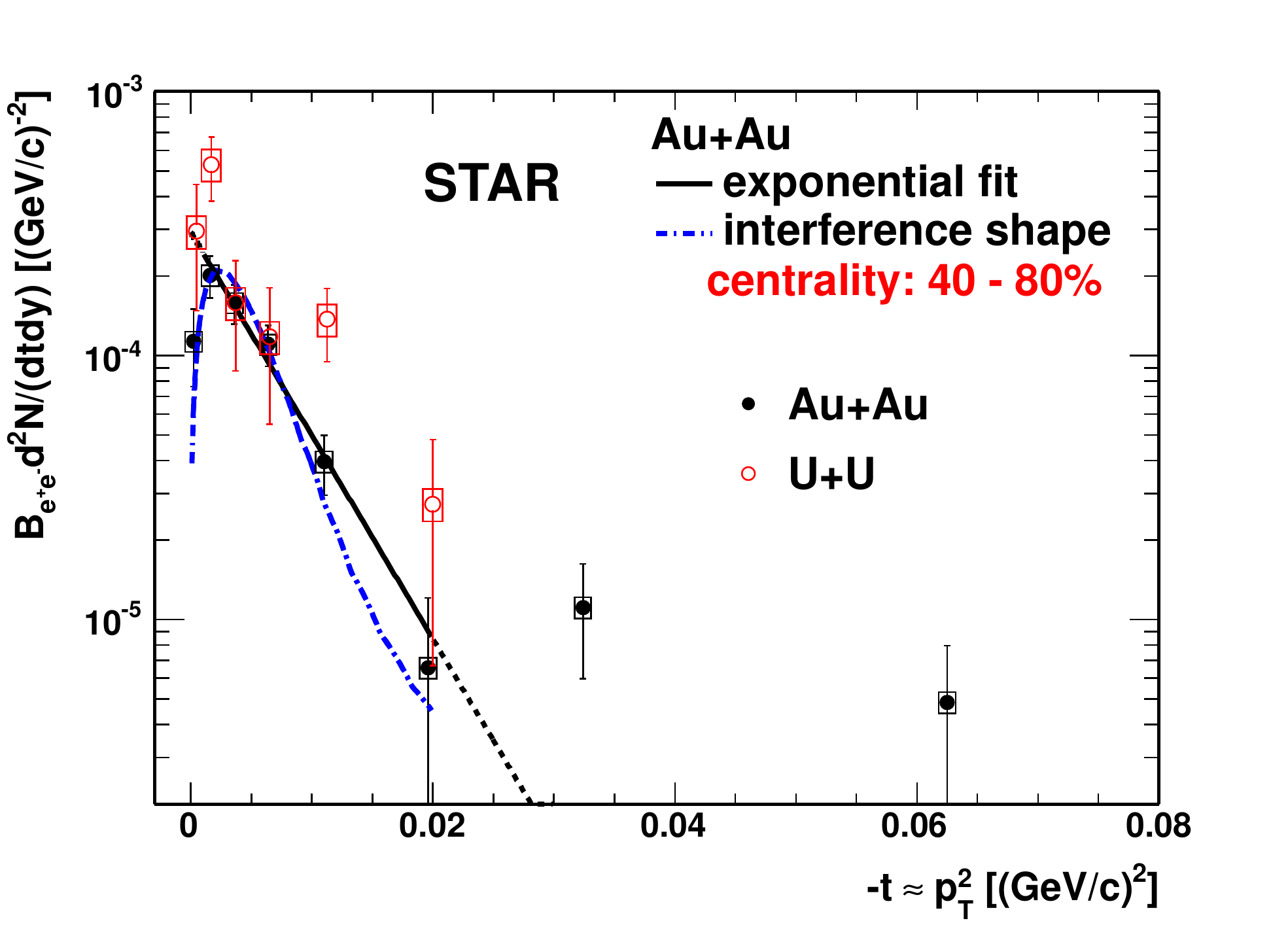}
  \caption{ \label{fig:3.5:figure_6} (left) STAR measurements of excess yields of $J/\psi$ as a function of number of participants ($N_{\rm{part}}$) for Au$+$Au collisions at $\sqrt{s_{_{\rm NN}}}$ = 200 GeV and U$+$U collisions at $\sqrt{s_{_{\rm NN}}}$ = 193 GeV~\cite{STAR:2019yox}. (right) STAR measurements of excess yields of $J/\psi$ as a function of the negative momentum transfer squared ($-t$) for Au$+$Au collisions at $\sqrt{s_{_{\rm NN}}}$ = 200 GeV and U$+$U collisions at $\sqrt{s_{_{\rm NN}}}$ = 193 GeV~\cite{STAR:2019yox}. The data are shown together with theoretical predictions for coherent $J/\psi$ photoproduction in a hadronic medium~\cite{Zha:2017jch}.    }
\end{figure}

\subparagraph{Vector meson in hadronic medium}
Conventionally, vector meson production from coherent photon induced processes is only visible and studied in UPCs, in which the two colliding nuclei stay intact to meet the coherent requirement. However, recent measurements of $J/\psi$ photoproduction were observed in hadronic heavy-ion collisions at both RHIC~\cite{STAR:2019yox} and the LHC~\cite{ALICE:2015mzu}.
To investigate the potential production mechanism behind the anomaly enhancements, STAR made differential measurements of the observed excesses compared to the \pp{} reference for $J/\psi$~\cite{STAR:2019yox}. The left panel of Figure~\ref{fig:3.5:figure_6} shows the excesses of $J/\psi$ as a function of the number of participants ($N_{\rm{part}}$) for \AuAu{} collisions at $\sqrt{s_{_{\rm NN}}}$ = 200 GeV and \UU{} collisions at $\sqrt{s_{_{\rm NN}}}$ = 193 GeV. The observed excesses reveal no significant centrality dependence, which is beyond the expectation of hadronic production.  The differential negative momentum transfer squared ($-t$) distribution is shown in the right panel of Figure~\ref{fig:3.5:figure_6}, which reveals the distribution of interaction positions for the production process. The slope of the $-t$ distribution is similar to that observed in UPC $J/\psi$ photoproduction~\cite{STAR:2017enh}. 
Furthermore, the first data point ($-t < 0.001$ (GeV/c)$^2$) is significantly lower than the extrapolation of the exponential fit, which is consistent with the destructive interference for coherent photoproduction in heavy-ion UPCs. These observations suggest that the anomaly excesses in peripheral heavy-ion collisions possess characteristics of diffractive production, and indicate the existence of the coherent photon-nucleus production mechanism in hadronic nuclear collisions.

Klusek-Gawenda and Szczurk first considered this question theoretically, and built up a phenomenological model of coherent photoproduction~\cite{Klusek-Gawenda:2015hja} in heavy-ion collisions with nuclear overlap, which can reasonably describe the excesses observed at ALICE. In their model, the production processes are  assumed to be exactly the same as those in UPC except for a modification to the photon flux to ignore the overlapping region. However, in this approach, the modification of photon flux is not unambiguous and it avoids the question of how the coherent requirement is satisfied in collisions with nuclear overlap. In Refs.~\cite{Zha:2017jch,Zha:2018jin,Zha:2018ytv}, they argue that although the whole nucleus cannot keep the coherent requirement, the spectator nucleons, free from hadronic interactions, could still act coherently for photoproduction, which should be attributed to the much longer lifetime of the spectator fragment compared to that of photoproduction processes. Therefore, the authors considered that the coherent processes occur either coupling to the whole nucleus or only the spectators, which  results in four configurations. As revealed in Figure~\ref{fig:3.5:figure_6}, the predictions from ~\cite{Zha:2017jch} can describe the STAR measurements reasonably well, and the data seem to favor the spectator coupling scenario.    
The coherent produced $J/\psi$s are formed in the initial stage of the collisions, and therefore interact with the latter formed QGP from the nuclear overlapping region, which makes it a potentially powerful tool to probe QGP. The $J/\psi$s from coherent photoproduction are concentrated at very low transverse momentum, in which the regeneration contribution should be negligible. With sufficient statistics, this makes the coherent produced $J/\psi$s a clear probe for the study of color screening effect in QGP. 

Currently, there are also limitations that prevent us from utilizing the additional produced $J/\psi$s to infer QGP properties. The first is that, due to the rare production rate, the experimental measurements lack precision at semi-central and central collisions. Recently, there have been new preliminary measurements from ALICE~\cite{ALICE:2022zso} and STAR~\cite{Li:2022ncj} collaborations with improved precision; however, they are not sufficient. The second is that the baseline to determine the modification factor can only be obtained from theoretical predictions, which are model dependent and cannot be directly extracted from the $\gamma$+p measurements. 

\paragraph{New directions in UPCs}
\label{subsubsec:new}

\subparagraph{Electron and muon pair production}

The production of lepton pairs in ultraperipheral collisions ($\gamma\gamma \to l^+ l^-$, sometimes called the Breit-Wheeler process)  serves to calibrate theoretical calculations of the nuclear photon fluxes, and also provide a clean environment to study their impact parameter dependence, which comes into play in a variety of processes.  
Measurements of exclusive lepton pair cross sections at the LHC in both muon~\cite{ATLAS:2022srr} and electron~\cite{ATLAS:2022srr} have shown good agreements with theory calculations and contribute to ongoing discussions of the role of higher-order Coulomb processes.
In 2018, measurements of electromagnetic production of lepton pairs in hadronic heavy-ion collisions by the STAR~\cite{STAR:2018ldd} and ATLAS~\cite{ATLAS:2018pfw} collaborations showed transverse momentum broadening inconsistent with traditional ``equivalent photon approximation'' (EPA) predictions, although consistent in terms of total cross section. Both initial-state and final-state~\cite{Klein:2018fmp,Klein:2020jom} effects were investigated. Measurements by the STAR~\cite{STAR:2019wlg} and CMS~\cite{CMS:2020skx} collaborations indicated significant broadening of the lepton pair momentum results from the initial photon flux, supporting the initial-state effect hypothesis, although not ruling out potential medium interactions proposed previously by the STAR and ATLAS collaborations.

Theoretical progress from lowest order QED calculations indicates that the kinematic distribution of the initial photon flux contains strong impact parameter dependence~\cite{Zha:2018tlq,Vidovic:1992ik}. The transverse momentum distribution of these calculations matches well with measurements ultra-peripheral and peripheral collisions, while the traditional EPA calculations (STARLight) show tension in the form of a narrower distribution. Recent ATLAS results~\cite{ATLAS:2022vbe} of transverse momentum and acoplanarity $\alpha$ ($\pi\alpha \simeq P_\perp / M_{ll}$) in central Pb+Pb collisions show that the full QED calculation can describe the depletion at $\alpha\simeq0$ better than the photon Wigner function ~\cite{Klein:2020jom}. This difference may be caused by the breakdown of the real-photon approximation in the  photon Wigner function at the extreme phase space, when both photon $k_{\perp}$($^{<}_{\sim}4$ MeV) approach zero and the conserved transition current could not be approximated as two real-photon vertex function~\cite{Wang:2022ihj,Vidovic:1992ik}.

In addition, ultra-relativistic nuclei produce a highly Lorentz contracted radial electric field emanating from the nucleus, with a magnetic field circling the nucleus. Both the electric and magnetic field are almost entirely Lorentz-contracted into the plane perpendicular to the direction of motion. Therefore, at any given point, the fields appear as a nearly transverse linearly polarized electromagnetic wave. The significant azimuthal dependence of Breit-Wheeler process in ultra-peripheral and peripheral collisions~\cite{STAR:2019wlg} are in good agreement with both SuperChic3 predictions, and with the lowest-order QED~\cite{Li:2019sin}, which predict a $\cos{4\phi}$ modulation of approximately $-17\%$ (for the kinematic ranges applicable for $e^+e^-$ measurements in ultra-peripheral collisions in STAR).

\subparagraph{Tau $g-2$}

The Breit-Wheeler process is able to produce pairs of tau leptons $(\gamma\gamma \rightarrow  \tau^+\tau^-)$, providing a unique opportunity to measure the anomalous magnetic moment of the tau lepton: $a_\tau = (g_\tau - 2)/2$~\cite{delAguila:1991rm}. Due to the larger mass of the tau, measurement of the anomalous magnetic moment provides stringent tests of fundamental predictions from QED and potentially probes physics beyond the standard model~\cite{Beresford:2019gww,Dyndal:2020yen}.  In several channels, radiative corrections to $a_\tau$ scale as $m^2_\tau/m^2_\mu$ which is $\sim280$ times larger than for the muon, perhaps making these modifications observable in changes to the measured cross sections. This approach to $a_\tau$ provides multiple benefits: the $Z^2$ enhancement of the coherent photon field leads to large cross sections, the produced events are quite clean, and allow the distinct topology of tau decays to be identified, and the unique event characteristics allow efficient triggering and selection. In several recent papers\cite{ATLAS:2022ryk,CMS:2022arf}, ATLAS and CMS have measured $\tau$ decays (in one channel in CMS, and in three in ATLAS), and the results for $a_\tau$ have already reached the precision of previous constraints from $e^+e^-$ experiments~\cite{DELPHI:2003nah}.

\subparagraph{Light-by-Light (LbyL)}

The elastic scattering of photons off of themselves is well known to be forbidden in classical electrodynamics, but it has long been predicted to occur via quantum processes -- primarily box diagrams involving charged fermions and bosons, but also BSM particles that couple to photons. However, direct observation of this process was only made in LHC Run 2  (2016-2018) by the large LHC experiments~\cite{ATLAS:2017fur,ATLAS:2019azn,CMS:2018erd}.
Measurements benefit from the widest possible rapidity acceptance and the lowest $p_T$ selection for the final state photons.  The limited virtuality and transverse momentum of the initial state photons provides a strong handle against the primary backgrounds, misidentified electron-positron pairs, and central exclusive production of photons via gluon-initiated processes.  This clean signal has been used for systematic searches for new physics. The most notable results have been for axion-like particles (ALPs)~\cite{CMS:2018erd}; the latest ATLAS data provide the most stringent limits to date~\cite{ATLAS:2020hii}, far exceeding those set by electron-positron machines, and even the LHC \pp{} program, in the range $5<m_a<100$ GeV.

\subparagraph{Quantum interference and the neutron skin} 
Vector meson photoproduction is of interest in quantum information systems. Photoproduction acts like a two-source interferometer, with production possible on either of the nuclei. Near mid-rapidity and at low $p_T$, these two possibilities are indistinguishable, and so interfere with each other~\cite{Klein:1999gv}. The two processes are related by a parity transformation; vector mesons have negative parity, so the interference is destructive. Because the two production positions are well separated in space, with a typical median impact parameter of 20 to 40 fermi or more, depending on the beam energies and species and any requirements on the presence of neutrons from nuclear excitation \cite{Baltz:2009jk}, the system is very close to a two-source interferometer. Since the typical lifetime of the $\rho$ meson is less than the time required for particles to travel this distance, any interference must involve the products of the vector meson decay, so this is an example of the Einstein-Podolsky-Rosen paradox~\cite{Einstein:1935rr}. The STAR Collaboration confirmed this interference in 2009~\cite{STAR:2008llz}.  Double vector meson production is also possible \cite{Klein:1999qj,Klusek-Gawenda:2013dka}; this allows for even more complex quantum correlations \cite{Baur:2003ar}. These studies require significantly improved luminosity and trigger efficiency than in current experiments, but are attractive possibilities for LHC Runs 3 and beyond. 

\subparagraph{Dijets and open charm} 
Looking ahead, exploring new probes of gluons will be important to reduce the systematic uncertainties in transforming vector meson cross sections into parton distributions \cite{Adeluyi:2011rt,Citron:2018lsq}.  This is especially important since recent NLO calculations of vector mesons have a surprisingly large contribution from quark PDFs, and also an unexpectedly large scale uncertainty \cite{Eskola:2022vpi}.  Photoproduction of dijets or open heavy quarks proceed via single gluon exchange, so should be less subject to these theoretical uncertainties.  In addition, the $Q^2$ is set by the pair or dijet invariant mass, so it is possible to probe parton distributions over a wide range of $Q^2$ with a single process.   
Specifically, the system of jets reflects the initial scattering kinematics: the transverse momentum sum ($H_T$) correlates with $Q$, the multijet mass and rapidity can be combined to determine the fractional photon energy $z_\gamma$ and the nuclear Bjorken $x_A$, the fractions of the per-nucleon beam momentum carried by the photon and nuclear parton.  Using these variables, ATLAS has measured (in a preliminary form \cite{ATLAS-CONF-2022-021}) fully corrected cross sections which directly measure the nuclear parton structure (nPDFs) — providing access to some of the important early physics of the EIC. These events are triggered and selected using a one-sided topology of forward neutron production observed in the ATLAS ZDCs. They compared this with a Pythia + STARlight simulation. In the future, it should be possible to extend both the Bjorken$-x$ and $Q^2$ range significantly by looking at softer jets (for lower $x$ and $Q^2$) and with more data (to get to higher $Q^2$).  Going further, it may be possible to use dijet events to directly measure the Wigner distribution \cite{Hatta:2016dxp}.  This takes advantage of the fact that there are two momentum scales in the problem - the $p_T$ of the individual jets, and the $p_T$ of the dijet system. 

Dijet events have also been observed by ATLAS in events with no activity in either ZDC (0n0n), and the distributions have been found to resemble expectations from diffractive dijet (photon-pomeron) production~\cite{Guzey:2016tek}.  Diffractive dijet production is sensitive to the gluon distributions in the nuclei, as well as their polarization, which is expected to lead to distinctive angular decorrelation effects~\cite{Dumitru:2018kuw,Mantysaari:2019csc}. CMS has studied this effect~\cite{CMS:2022lbi} in events with two jets accompanied by rapidity gaps in both directions, defined by charged particles, as well as energy vetos in the forward calorimeters.  By a measurement of the angle separating the dijet vector momentum sum ($Q_T$) and different ($P_T$), they calculate the mean $\langle\cos(2\phi)\rangle$ which is sensitive to dijet decorrelation effects for jet $p_T > 30$ GeV, $Q_T < 25$ GeV, and dijet $P_T > Q_T$.  The data are found to be drastically lower than expectations from RAPGAP~\cite{Jung:1993gf}, which describes HERA data without gluon polarization.  Its magnitude at lower $Q_T$ is reproduced by models~\cite{Hatta:2020bgy} which assume the decorrelation induced by soft radiation, but the same model does not rise with $Q_T$ as rapidly as the CMS data.  This measurement is complementary to the STAR dilepton studies in probing polarization effects of gluons, as well as photons~\cite{STAR:2019wlg,STAR:2022wfe}.

During LHC Run 4 and beyond, photoproduction of dijets may also be studied in the ALICE FoCal \cite{ALICE:2020mso}, allowing for measurements of parton distributions of protons and nuclei down to very low $x$ values.

Using a similar approach as jets, it is possible to get to still lower $x$ and $Q^2 \gtrsim 2$ GeV$^2$ by studying open charm, and maybe heavier quark hadrons.  The cross sections for charm are large \cite{Klein:2002wm}, so statistics should be ample for reconstructing a number of different mesons and even meson pairs (the meson from both the $c$ and the $\overline c$).  Bottom hadrons are also accessible, and even top production may be within reach \cite{Klein:2000dk,Goncalves:2006xi,Goncalves:2017zdx}.  The latter do not hadronize before they decay, so would require different methods to study.  There is also information to be had in pairs with large rapidity gaps between them \cite{Huayra:2019iun}. 

\subparagraph{Collectivity in inclusive photonuclear events} 
High-energy photoproduction processes can be studied using UPCs in nuclear collisions and at the EIC. The specific interest is on the inclusive $\gamma+$A processes, but unlike at the EIC, the photons involved in UPCs are quasi-real. The first measurements in this direction was initiated by measuring harmonic anisotropy $v_n$(\pPb{}) and $v_n$($\gamma$+Pb) at the LHC by the ATLAS collaboration~\cite{ATLAS:2021jhn}. Once the non-flow (correlations due to conservation effects) are removed, a signature of long-range correlations is observed in $\gamma$+Pb processes. The trend of the ATLAS data has been explained by CGC calculations~\cite{ATLAS:2021jhn,Shi:2020djm}. 
For UPCs at top RHIC energies one expects the maximum energy of the quasi-real photon to be approximately $E_\gamma\approx 3$ GeV. The typical range of the center-of-mass energy of the photon-nucleon system will therefore be $W_{\gamma N}\approx 40$ GeV. Therefore, Au$+$Au collisions at $\sqrt{s_{_{\rm NN}}}=200$ GeV will provide access to the $\gamma+$Au process at 40 GeV center-of-mass energy. The specific interest is high-activity inclusive $\gamma+$Au process to search for collectively. If collectivity is observed in $\gamma$+Au processes, it can provide a way to explore the creation of a many-body system exhibiting fluid behavior in photon-induced processes~\cite{Zhao:2022ayk}. A recent calculation in Ref~\cite{Zhao:2022ayk} assume $\gamma+A$ processes are equivalent to collisions of vector mesons with ions ($\rho$+A collisions) and describe first measurements of harmonic coefficients $v_n$ in photonuclear processes measured by the ATLAS collaboration~\cite{ATLAS:2021jhn}. The hypothesis of the $\gamma+A$ process as a $\rho+A$ collision and the formation of a fluid-dynamic medium can be tested at RHIC in a data-driven way. This can be done by comparing measurements in $\gamma$+Au processes at $W_{\gamma N}\approx40$ GeV and in $d+$Au collisions at $\sqrt{s_{_{\rm NN}}}=39$ GeV. 

\subparagraph{Search for baryron junctions in photonuclear events} 
Photonuclear processes can also be used to study the origin of baryon stopping and baryon structure in general. One proposed mechanism to explain baryon stopping is the baryon junction: a nonperturbative Y-shaped configuration gluons which is attached to all three valence quarks. The baryon stopping at RHIC energy range is ideal for such a study. 
Preliminary data with \AuAu{} collisions at $\sqrt{s_{_{\rm NN}}}=54.4$ GeV shows strong baryon stopping in enriched $\gamma+$Au events with a strong dependence on rapidity. This important observation provides the necessary impetus for further exploration using various available data sets. In particular, it would best interesting to test if this strong rapidity dependence of the $\bar{p}/p$ yield is consistent with the picture of baryon junction that predicts an exponential dependence of stopping with rapidity of form $\exp(-\alpha (y-Y_{\rm beam}))$ with $\alpha\approx0.5$. Extending these measurements with high statistics $\gamma$+Au-rich event samples, using Run 2023 and 2025 data on \AuAu{} collisions at $\sqrt{s_{_{\rm NN}}}=200$ GeV, will enable differential measurements of di-hadron correlations with different combinations of triggers and associated transverse momenta. These measurements will help search for collectivity, in addition to test the baryon-junction conjecture. Similar measurements can be done at the LHC although the net-proton yield and $(1-\bar{p}/p)$ are significantly smaller than at lower energies.

\clearpage

\subsection{Interdisciplinary}
\label{sec:progress:interdisciplinary}

The complex nature of studying quark-gluon plasma in heavy-ion collisions naturally lends itself to connections to other fields, either through a common set of tools or a mutual interest in related physics.

Connections include applications of holography in a number of different contexts, the study of cosmic rays~\cite{Albrecht:2021cxw}, exploration of applications of quantum computing to hot QCD~\cite{Barata:2021yri,deJong:2021wsd,Barata:2022wim,DeJong:2020riy}, as well as the study of antimatter~\cite{STAR:2015kha,STAR:2019wjm,Chen:2018tnh}.

Other prominent interdisciplinary topics related to studies of hot QCD are highlighted below.

\subsubsection*{Nuclear structure}
\addcontentsline{toc}{subsubsection}{Nuclear structure}

Connections with low-energy nuclear structure have been highlighted in Section~\ref{sec:progress:macroscopic:initial_cond}, in particular the ability to probe nuclear deformation parameters~\cite{Bally:2022vgo}. Other connections with nuclear structure include attempts to constrain the neutron skin of heavy nuclei through diffractive photoproduction in ultraperipheral collisions~\cite{ALICE:2015nbw,STAR:2017enh}, a result that would be of high interest for the nuclear astrophysics community.

\subsubsection*{Quantum electrodynamics and physics beyond the standard model}
\addcontentsline{toc}{subsubsection}{Quantum electrodynamics and physics beyond the standard model}

Ultraperipheral collisions have been used to study quantum electrodynamic processes, including light-by-light scattering
and the anomalous magnetic moment ($a = (g-2)/2$) of the tau lepton,
both of which have potential to search for physics beyond the standard model (see Section~\ref{sec:progress:microscopic:ultraperipheral_collisions}).

\subsubsection*{Phase diagram and relativistic fluid dynamics in astrophysics}
\addcontentsline{toc}{subsubsection}{Phase diagram and relativistic fluid dynamics in astrophysics}

Significant overlap exists with the astrophysics community. The nuclear equation of state at large baryon chemical potential and the presence of a critical point and a first-order phase transition plays a large role in the study of neutron stars~\cite{Wei:2021veo}. A wide range of astrophysical observables could provide insights in the equation of state at finite $\mu_B$, including supernova explosions showing a second burst of neutrinos \cite{Sagert:2008ka,Ouyed:2009dr,Nakazato:2013iba,Fischer:2017lag,Kuroda:2021eiv,Jakobus:2022ucs}, neutron-star mergers showing different ejecta \cite{Prakash:2021wpz,Fujimoto:2022xhv,DiClemente:2021dmz}, waveforms \cite{Most:2018eaw,Weih:2019xvw} or peak frequencies \cite {Blacker:2020nlq} in postmerger gravitational wave signal, neutron-star cooling \cite{Wong:2004md,Negreiros:2010tf,Sedrakian:2015qxa,Wei:2020saq}, glitches in their spin evolution \cite{2011RAA....11..679X,Singha:2022ubj}, gravitational wave emissions from individual stars \cite{Pitkin:2011sc,Mallick:2020bdc,Constantinou:2021hba,Bai:2021wrh,Bratton:2022kba}, and twin stars, two stars which possess approximately the same masses but with one being much more compact than the other, signaling a strong phase transition in their interior \cite{Alford:2013aca,Dexheimer:2014pea,Benic:2014jia,Christian:2019qer,Nandi:2020luz,Tan:2021nat}.
The multimessenger study of neutron star mergers with electromagnetic and gravitational waves further provides a connection with both hot and dense nuclear matter and with relativistic hydrodynamics~\cite{Most:2022wgo,Sorensen:2023zkk}. Applications of relativistic hydrodynamics in both astrophysical and collider experiments have led to a surge of interest in the mathematical and physical foundations of the theory.

\subsubsection*{Machine learning}
\addcontentsline{toc}{subsubsection}{Machine learning}

The computationally demanding simulations of hot QCD physics, combined with the enormous experimental data sets involved, have provided a fertile ground for applications of machine learning techniques. An area of particular interest has been quantification of uncertainties~\cite{Phillips:2020dmw} to help constrain systematically the properties of strongly-interacting quark-gluon plasma. Machine learning applications have also been used widely to accelerate model-to-date comparisons and in analysing experimental signatures from large data sets~\cite{Pang:2021vwl}, for example.

\clearpage

\section{Hot QCD facilities}
\label{sec:facilities}

There are two facilities in the world capable of creating the quark-gluon plasma in ultra-relativistic heavy-ion collisions, the Relativistic Heavy Ion Collider (RHIC) at Brookhaven National Laboratory (BNL) in New York and the Large Hadron Collider (LHC) at CERN in Switzerland. The impressive versatility and advances in the performance of these accelerators have enabled a wealth of experimental measurements and insights to the QGP since the last \LRP{} which are discussed in Section~\ref{sec:progress}. This section will focus on the facilities and detector upgrades that have enabled these scientific accomplishments to be realized.  In particular, RHIC successfully completed the Beam Energy Scan II and built sPHENIX as recommended in last \LRP{}, while the LHC experiments collected data in Run 2 and were upgraded in preparation for Run 3. Improvements to data acquisition capabilities at the experiments are essential for utilizing the unprecedented luminosities achieved at RHIC and LHC to achieve the most precise measurements which are capable of constraining the theoretical models. Progress on experimental measurements continues into this next decade as the new sPHENIX experiment together with the upgraded STAR experiment usher in a new era of RHIC measurements and with the start of Run-3 at the LHC.

\subsection{RHIC}

Since the previous \LRP{}, RHIC has exceeded expectations delivering the luminosity needed for the crucial BES-II, as well as enabled a detailed study of the CME from collisions of isobars. Several upgrades to STAR will maximize the measurements achieved from these collisions. Meanwhile PHENIX completed data taking in 2016 in preparation for construction of sPHENIX, which is now the cusp of taking data.  This section highlights these accomplishments and plans for future RHIC running. 

{\bf Isobars at RHIC}
The 2018 RHIC run included isobar collisions of $^{96}_{44}Ru+^{96}_{44}Ru$ and $^{96}_{40}Zr+^{96}_{40}Zr$ to search for the chiral magnetic effect. The run should provide enough data to measure a significantly observable difference in the two collision species. To reduce the systematic uncertainties that arise from variations in run conditions a special run strategy was implemented which required synergy between the STAR collaboration and the RHIC Collider Accelerator Department. This strategy included alternating the isobar species between each beam store and maintaining a constant luminosity during a store and across isobar species. This program was successfully executed at RHIC and the necessary level of precision was achieved \cite{STAR:2021mii}. STAR also developed and executed a rigorous blind analysis method~\cite{STAR:2019bjg} when extracting these important results.

{\bf Beam Energy Scan II}

\begin{figure}[t]
\centering
\includegraphics[width=0.48\columnwidth]{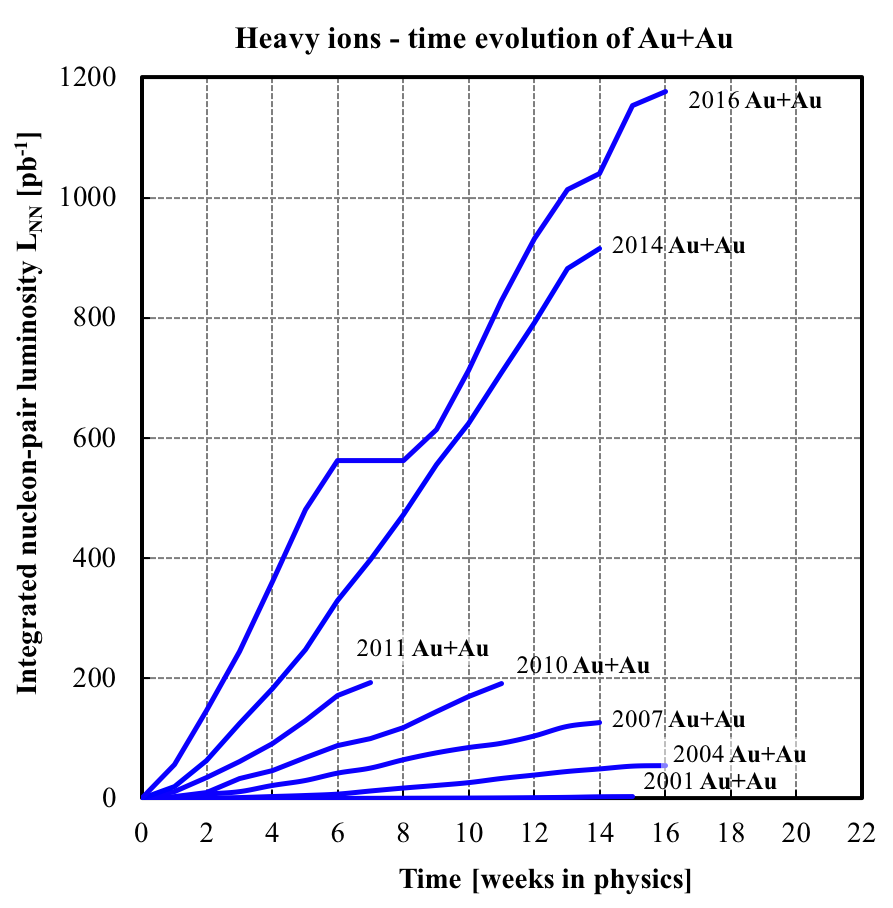}
\includegraphics[width=0.48\columnwidth]{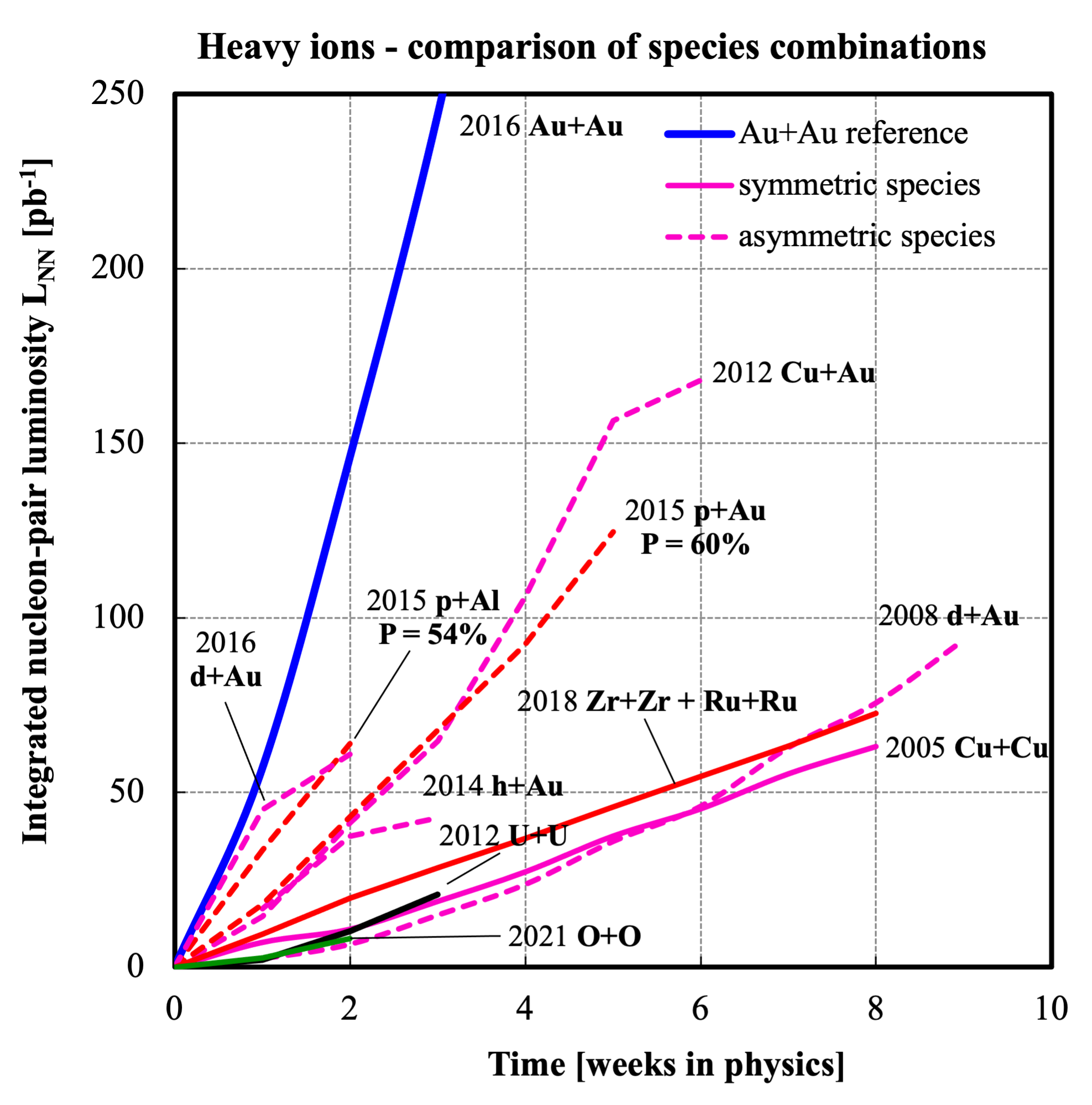}
\caption{Left panel: The integrated luminosity for all RHIC \AuAu{} Collisions. Right panel: The integrated luminosity comparing all heavy-ion collision species at RHIC. Symmetric collisions are denoted with a solid line while asymmetric collision systems are denoted with a dashed line.}
\label{fig:RHICLumi}
\end{figure}

In order to reach the desired luminosity, RHIC implemented an electron cooling upgrade, Low Energy RHIC electron Cooling (LEReC). The LEReC was successfully commissioned and brought into operation during the BES-II RHIC Runs in 2019--2021. To achieve energies below 7.7 GeV, a fixed-target program was engineered where a Au target was inserted in the STAR beam pipe~\cite{STAR:2020dav}. 
To maximize the physics output, the STAR detector at RHIC added several key subsystem upgrades: the inner TPC (iTPC)~\cite{Yang:2019bjr}, the Event Plane Detector (EPD)~\cite{Adams:2019fpo} and the endcap Time-of-Flight (eTOF)~\cite{STAR:2016gpu} Detector. These additions significantly improved the STAR detector tracking and particle identification capabilities, especially at large pseudo-rapidity and low transverse momentum regions. All of these upgrades were completed efficiently, and %
positively impacted the BES-II data.

{\bf STAR upgrades at forward rapidity}
Following the BES-II, STAR was equipped with the forward tracking system (FTS) and forward calorimetry system (FCS) enable measurements of neutral pions, photons, electrons, jets, and hadrons in the detectors that extend tracking and calorimeter coverage to $2.5< \eta <4.0$~\cite{Brandenburg:2021hrv}. The FTS consists of silicon mini-strip sensors as well as small-strip thin gap chambers (sTGCs). The FCS includes an electromagnetic calorimeter, which uses refurbished PHENIX lead-scintillator with SiPM-based readout electronics, as well as a hadronic calorimeter. In 2022, RHIC collided polarized protons at 500 GeV.
These upgrades and data are of particular interest to the cold QCD community and are useful in preparation of the future electron ion collider (EIC). However, starting with the 200~GeV \AuAu{} collisions in 2023, STAR will also utilize this extended rapidity range to address hot QCD topics such as measurements of long-range correlations to study early-time dyanamics of heavy-ion collisions.

{\bf sPHENIX}

\begin{figure}
    \centering
    \includegraphics[width=0.45\textwidth]{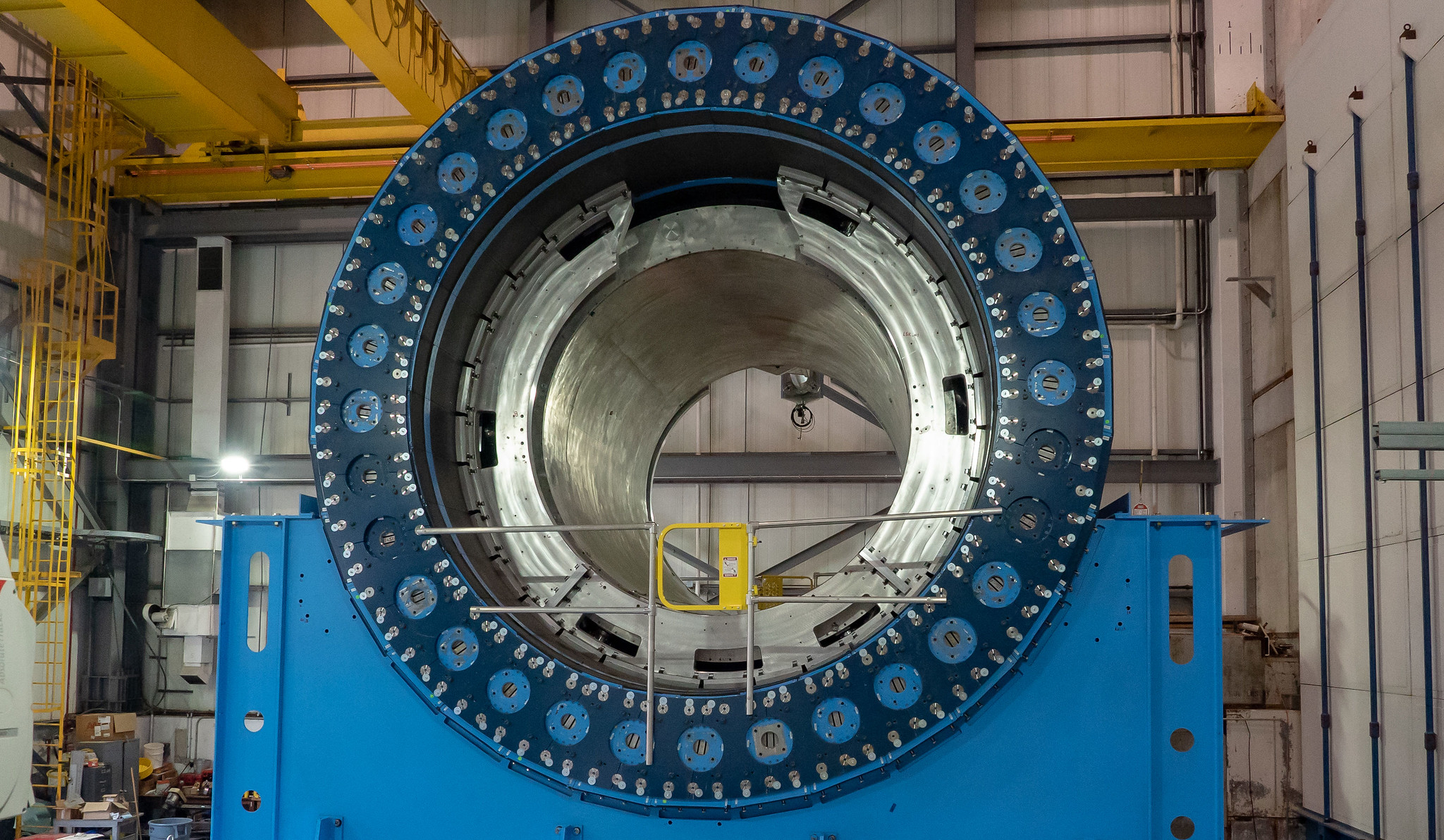}
    \includegraphics[width=0.42\textwidth]{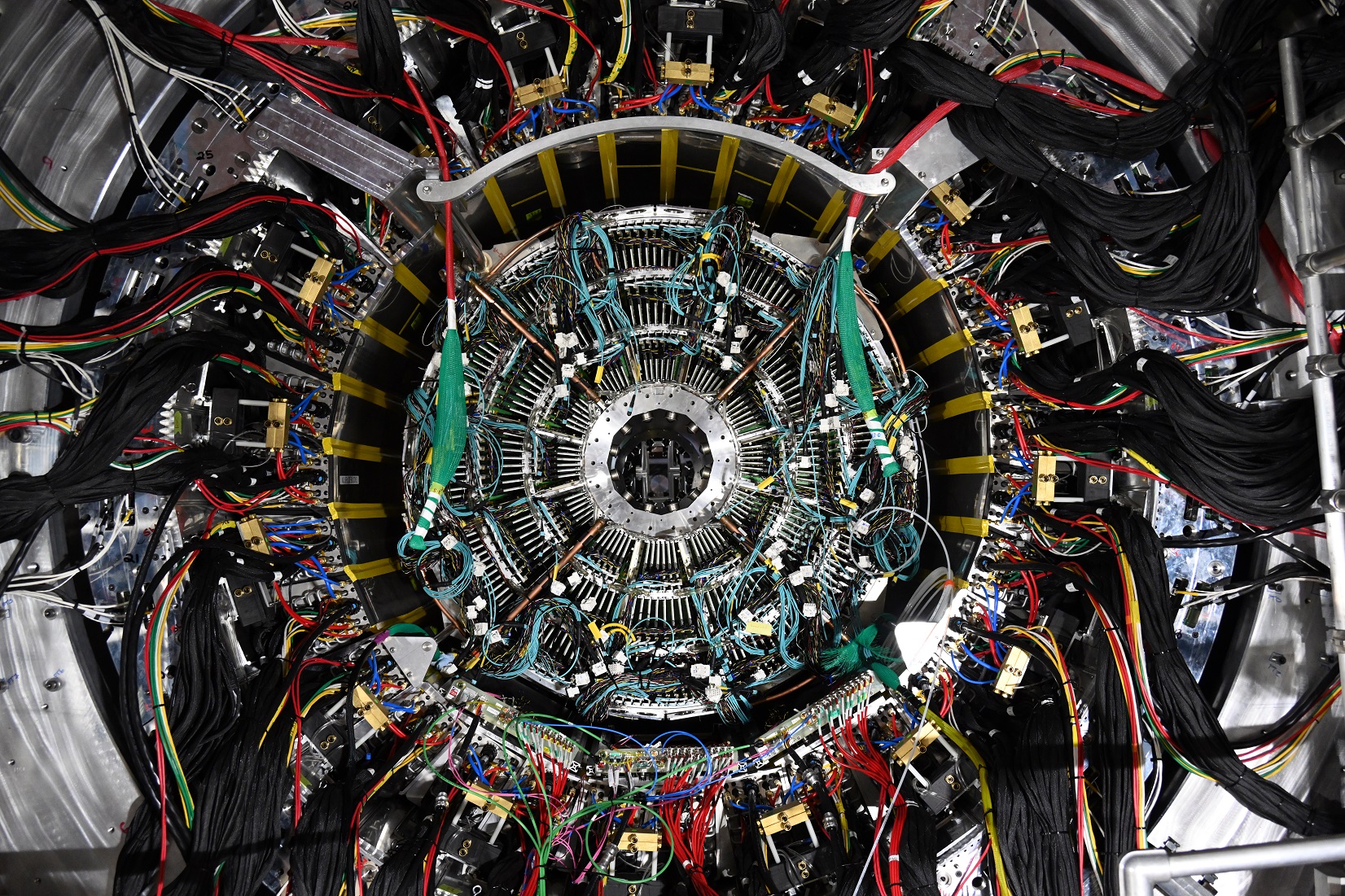}
    \caption{The sPHENIX outer hadronic calorimeter surrounding the solenoid magnet (left) and a close up image of the detectors inside of magnet prior to inserting the innermost detectors, INTT and MVTX (right).}
    \label{fig:sphenixphotos}
\end{figure}

In the period since the last \LRP{}, which recommended a "state-of-the-art jet detector at RHIC," the sPHENIX collaboration was formed and built the proposed experiment, which will start data taking in 2023. Photographs of the sPHENIX during assembly are shown in Fig. \ref{fig:sphenixphotos}. sPHENIX, the first new heavy-ion experiment at RHIC since the start of RHIC collisions 23 years ago, is designed specifically to measure fully reconstructed jets and upsilons in the quark-gluon plasma \cite{PHENIX:2015siv}. Such measurements required good hadronic and electromagnetic calorimeters as well as precise tracking. The tracking detectors, a Time Projection Chamber (TPC), a silicon tracker (INTT), a silicon pixel vertex detector (MVTX), and a Micromegas-based tracker (TPOT) enable detailed jet substructure measurements, as well as heavy flavor tagging. These detectors and a high rate streaming DAQ will enable precise measurements of rare probes such as heavy-flavor tagged and direct-photon tagged jets, and potentially observe the highly suppressed and elusive $\Upsilon$(3S) state in \AuAu{} collisions at 200~GeV.  The addition of the EPD will enable observables with respect to the reaction plane such as jet and open heavy flavor $v_n$ measurements.

RHIC’s run plans for the next three years include commissioning the sPHENIX detectors with \AuAu{} collisions in 2023, collecting high statistics datasets for \AuAu{} at 200 GeV in 2025 as well as \pp{} collisions and \pAu{} data in 2024. Precise \pp{} baseline measurements are essential as references for sPHENIX \AuAu{} measurements. Some of the projections for various observables, documented in \cite{sPHENIX:2022BUR} are highlighted in Sec.~\ref{sec:future:microscopic}.  It is clear there is interesting physics that RHIC, a unique facility, and its upgraded experiments could access beyond these three years. The beam use proposals for STAR and sPHENIX outline possibilities if the opportunity of additional RHIC running becomes available including exploring \OO{} collisions which would provide important insight into the transition from the smaller \pAu{} systems to larger \AuAu{} systems. The temperature dependence of any effect could also be compared to planned results from \OO{} collisions at a higher energy at the LHC. After operations cease, RHIC will begin transitioning into the EIC.

\begin{figure}[t]
\centering
\includegraphics[width=0.98\columnwidth]{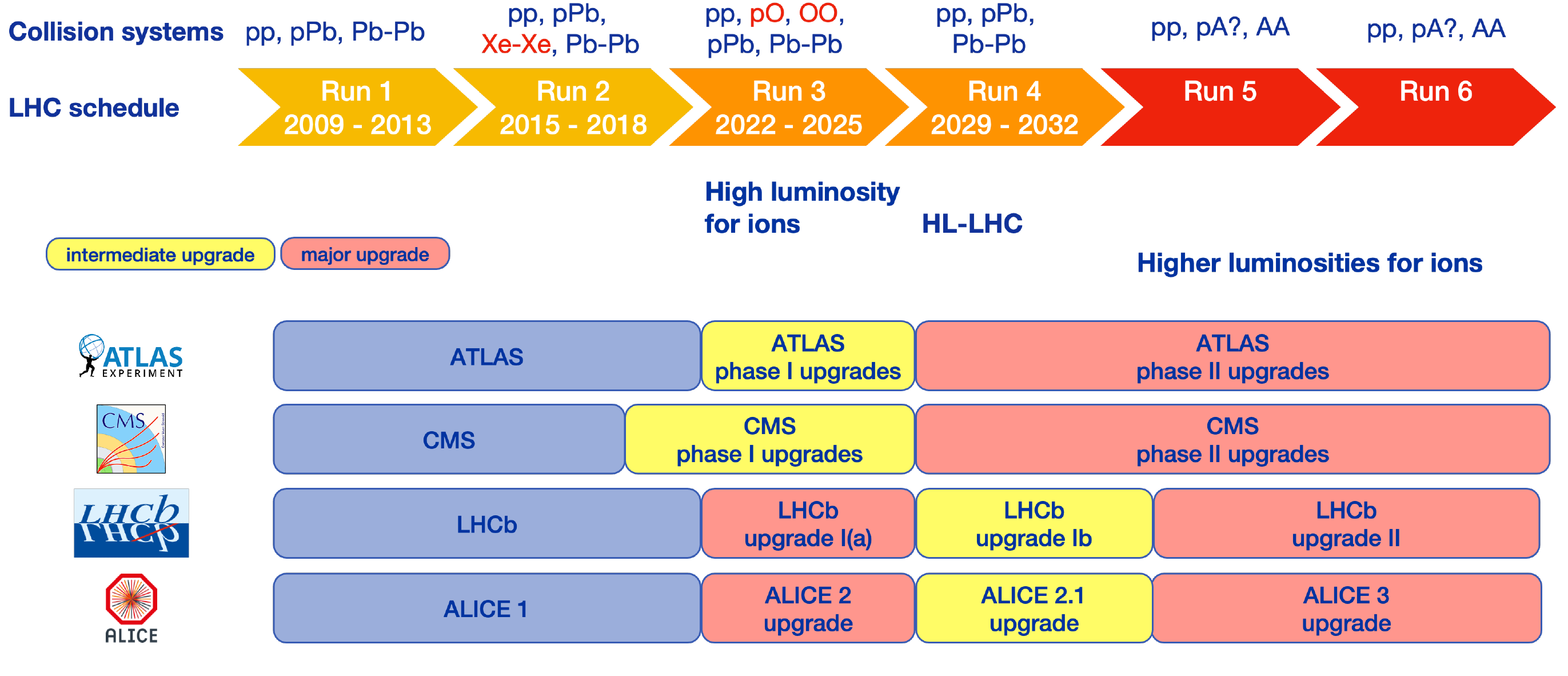}
\caption{Overview of upgrades for LHC experiments from 2022 and onward.}
\label{fig:LHCUpgrade}
\end{figure}

\subsection{LHC}

Figure~\ref{fig:LHCUpgrade} shows the timeline of upgrades for all four LHC experiments\footnote {From the plenary presentation ``The near- and mid-term future of the LHC" by Jochen Klein at Quark Matter 2022.}, in the context of the expected collision systems. The performance of the LHC is expected to lead to continuous improvements in provided luminosities for both proton and ion collisions from Runs 3-6, with the High Luminosity era (HL-LHC) beginning by Run 4. The LHC itself will be upgraded for HL-LHC with new magnets, cavities, and collimators etc~
\cite{Apollinari:2017lan}. By LHC Run 6, the provided integrated nucleon-nucleon luminosity for heavy-ion collisions is projected to be more than 100 times larger than what was provided for Run 2. Regarding the LHC experiments, prior to Run 3 (which started in 2022), ALICE underwent a major upgrade~\cite{ALICE:2023udb}. An increase of ALICE's recorded luminosities by two orders of magnitude has been enabled via the introduction of a continuous readout system in the TPC, and for several other detectors. The new ITS2 system offers a greatly improved tracking performance using new inner trackers, which have a reduction of factor three in radiation lengths in the inner most layers, and factor three improvement in hit resolution compared to the ITS system used in Runs 1\&2. ALICE has installed a new Forward Muon Tracker, with an ability (amongst others) to separate prompt and non-prompt $J/\Psi$ production. The physics possibilities for ALICE in this scenario, and other LHC experiments, are discussed elsewhere~\cite{Citron:2018lsq, ALICE-USA:2022glt}, and later in this paper. The major upgrade for the LHCb experiment also involved substantial gains in readout rates, comparable to those for ALICE. Tracking performance has been improved by the installation of a new vertex locator (VELO), a new upstream tracker (UT), and scintillating fibers (SciFi). Further details of LHCb upgrades for Run 3 can be found here~\cite{LHCb:2012doh} (Upgrade I(a)). The ATLAS upgrades for Run 3 involved increased statistics and an improved ZDC (with new reaction plane capabilities)~\cite{CERN-LHCC-2011-012}, while for CMS these have led to increased bandwidth and improved tracking performance~\cite{CMSTrackerGroup:2020edz} (Phase I). 

The LHC Long Shutdown 3 (LS3) period will involve substantial upgrades for ATLAS~\cite{Contardo:2015bmq} and CMS~\cite{ATLASCollaboration:2012ilu} (Phase II), in preparation for Run 4 (2029-2032), which marks the start of HL-LHC. Both ATLAS and CMS will achieve advances in their main tracking acceptance - roughly doubling $\eta$ acceptance to $|\eta| < 4$. CMS will install a MIP Timing Detector (MTD), which will provide charged hadron identification (PID) over $|\eta| < 3$~\cite{CMS:2667167}. A MIP timing detector will be installed in ATLAS over $2.4 < |\eta| < 4$ ~\cite{Casado:2811207}, which in addition to charged hadron identification, mitigates the effects of the pile-up in the forward region. ATLAS and CMS will also embark on a joint ZDC project~\cite{Longo:2825363}, with both experiments installing new ZDCs for Run 4, which are based on radiation-hard fused  (Cherenkov light producing) silica rods. An additional factor of two improvements of the pointing resolution in ALICE will be achieved by the installation of ITS3, via the introduction of wafer-scale ultra-thin silicon pixel sensors closer to the interaction point. These are also being pursued by the ePIC detector at the EIC. ALICE will also install a Forward Calorimeter (FoCAL), positioned at forward rapidities of $3.4 < \eta < 5.8$ with high granularity~\cite{ALICE:2020mso}. It is designed to probe the structure of nucleons and nuclei in the unexplored regime of Bjorken-$x$ down to $x\sim 10^{-6}$.

For LHC Runs 5\&6, a completely new detector has been proposed by the ALICE collaboration, named ALICE 3~\cite{ALICE:2022wwr}. It is a next-generation detector specifically designed for heavy-ion physics. Its main tracking system covers a large pseudorapidity range ($|\eta|<4$), and can reconstruct charged tracks down to $p_T \sim 50$ MeV/c. The lightweight and small radiation length design will greatly reduce the background and improve tracking resolution for electromagnetic and heavy-flavor probes. Compared to the ALICE setup in Runs 3\&4 (often referred to as `ALICE 2'), the pointing resolution at midrapidity is projected to be about three times better. This is in part achieved by placing a highly novel vertex detector 5mm from the beamline. The tracking system is placed in a superconducting solenoid with a field of up to $B= 2$~T, to obtain a momentum resolution of 1--2\% over a broad pseudorapidity and momentum range. This tracking is complemented by multiple sub-detector systems for particle identification; two TOF detectors and a RICH detector. These systems have the ability to identify leptons and photons in the thermal emission range of $0.05 < p_{\rm T} < 3$ GeV/c, which is inaccessible for other LHC experiments. Charged hadron identification on the $3\sigma$ level is possible up to $p_{\rm T} \sim 14$ GeV/c, and decay hadrons can be reconstructed much more efficiently and cleanly at low-\pt compared to ALICE 2. The fast readout systems will be able to record all of the expected heavy-ion luminosity provided by the LHC. The ALICE 3 program aims to collect an integrated luminosity of about 35~nb$^{-1}$ with \PbPb{} collisions and 18~fb$^{-1}$ with \pp{} collisions at top LHC energy. The potential to further increase the luminosity for ion running in the LHC by using smaller ions, e.g.\,$^{84}$Kr or $^{128}$Xe, as well as further runs with small collision systems, is being explored. The LHCb experiment will undergo a major upgrade for Runs 5\&6~\cite{LHCb:2018roe} (Upgrade II). It will have no centrality limitation for \AA{} collisions - its Run 4 upgrade will provide measurement capacities for \PbPb{} 30-100\% events. LHCb will also have enhanced PID capabilities.

\clearpage

\section{Future Prospects}

\label{sec:future}

The previous sections describe the substantial progress in understanding hot QCD matter since the last long-range plan. However, there are many open questions that we are well equipped to tackle in the next decade. This section will highlight some of those prospective areas of research from macroscopic to the microscopic nature of the medium produced in heavy-ion collisions.

\subsection{Macroscopic: collectivity, flow and thermal properties}
\label{sec:future:macroscopic}

\subsubsection{Open Questions and Future opportunities related to collectivity} %
With the advances made since the last long-range plan and the significant amount of data expected from the high luminosity era at the LHC (with additional upgraded detectors) as well as anticipated data from the Beam Energy Scan and sPHENIX, there are many open questions relevant to collectivity that are within our reach as well as new, exciting opportunities to explore:
\begin{itemize}[leftmargin=*]
    \item {\bf Identified particles, diffusion, and charge fluctuations} It is now possible to incorporate $g\rightarrow q\bar{q}$ interactions in the initial state such that baryon, strangeness, and electric charge fluctuations can be observed in the soft sector \cite{Carzon:2019qja}, opening the possibility of detailed charge correlation studies.  Coupling this with hydrodynamic simulations that allow for fluctuations of baryons, strangeness, and electric conserved charges provides an exciting prospect for studying a wide variety of observables related to charge fluctuations at the LHC and top RHIC energies where the system is still nearly boost-invariant and high luminosity measurements are possible. On the theoretical side, challenges include incorporating baryons, strangeness, and electric charge diffusion, 4-dimensional equations of state, and conservation of baryons, strangeness, and electric charges at particlization (see \cite{Lovato:2022vgq} for details).  However, this will have clear advantages because then one can constraint diffusion at low densities with high precision data such that it will provide an anchor for future work at large densities. Additionally, searches for the QCD critical point will require  the incorporation of critical fluctuations within realistic relativistic viscous fluid dynamics simulations and the ability to account for these fluctuations at particlization as well.  
    
    On the experimental side, many measurements exist that are sensitive to this type of physics (beyond the standard charge fluctuations measures, see Section~\ref{sec:progress:macroscopic:finite_muB}).   Two-particle correlations with net baryons can be used to explore these additional transport properties of the hydrodynamic evolution. The baryon diffusion constant $D_{B}$ is an example of such a parameter beyond $\eta/s$ and $\zeta/s$. It characterizes the mobility of baryon number and is predicted to be finite at the LHC despite the fact that $\mu_{B}\sim0$. A two-particle correlation function has been proposed to constrain $D_{B}$~\cite{Floerchinger:2015efa}. It explores correlations of net-baryon fluctuations as a function of azimuthal and rapidity separations. Such an analysis has yet to be carried out from LHC Run 1 or 2 data since it is statistically challenging. It will be greatly aided by the increase of two orders of magnitude in the \PbPb{} integrated luminosity foreseen for Runs 3\&4 for the ALICE detector, given its world-leading particle identification capabilities. The MIP detectors (enabling PID) installed in CMS and ATLAS from Run 4 will also provide an opportunity to refine and pursue such measurements.

    Charge balance functions \cite{Pratt:2017oyf,Pratt:2019pnd,Pratt:2022kvz,Pratt:2021xvg} are a useful method for constraint charge diffusion and will be possible to measure in a wider kinematic range with LHC upgrades. The recently published measurements of balance functions of identified charged hadrons $(\pi,\rm K,\rm p)$ in \PbPb\ collisions from Runs 1\&2 play an important role in constraining the charge diffusion coefficient $D_{e}$ for quarks~\cite{ALICE:2021hjb}. However, the central tracking acceptance of $-1< \eta < 1$ for Runs 1\&2 leads to large uncertainties in $D_{e}$ from ALICE balance functions data~\cite{Pratt:2021xvg}, with values between $0.5 <D_{e} (\emph{\rm{Lattice QCD})} < 4$ being permissible. These uncertainties will be significantly reduced when the same measurements are performed the ATLAS/CMS setups in Run 4, and ALICE 3 in Run 5, as the acceptance where identified charged hadron measurements can be made will increase to $-4< \eta < 4$. The reduction in $D_{e}$ uncertainties due to increases in the $\Delta \eta$ coverage are expected to be at least a factor 4.
    
    \item {\bf Hydrodynamics vs hadron transport at low beam energies} Due to the progress from the BES I, a number of relativistic viscous hydrodynamics codes with at least one conserved charged were able to compare to experimental data \cite{Denicol:2018wdp,Schafer:2021csj,Shen:2022oyg,Cimerman:2023hjw} (and one example of ideal hydrodynamics varying the equation of state \cite{Spieles:2020zaa}).  Generally, as the beam energy is lowered, hydrodynamics starts at lower temperatures and, therefore, the ratio of time spent in the QGP vs. hadron gas phase is tilted towards a longer hadron gas phase \cite{Auvinen:2013sba}.  
    It is unclear, however, at exactly what $\sqrt{s_{NN}}$ that the effect of the QGP phase becomes negligible.
    Hadronic transport models, benefiting from their ability to describe non-equilibrium evolution (including evolution through unstable regions of the phase diagram) as well as a natural inclusion of baryon number, strangeness, and charge transport, play an important role at these lower $\sqrt{s_{NN}}$ (see \cite{TMEP:2022xjg} for a comprehensive review of used models and \cite{Sorensen:2023zkk} for a deeper discussion on the theoretical developments needed in hadronic transport simulations). 
    Recent hydrodynamic studies estimate that hydrodynamics is applicable down to $\sqrt{s_{NN}}\sim 4-7.7$ GeV \cite{Schafer:2021csj,Shen:2022oyg}, although work remains to ensure full consistency of most hydrodynamic models with a first-order phase transition.
    Recently linear stability constraints were derived for a multi-component relativistic viscous fluid \cite{Almaalol:2022pjc}, which may aid in the determination of the applicability of hydrodynamics in this regime. However, much of this discussion also hinges on a solid understanding of both the equation of state and corresponding transport coefficients, which may display unique features if critical points or first-order phase transitions are present.   Thus, much work still remains.

    \item {\bf Second-order transport coefficients} From the previously discussed theoretical calculations of transport coefficients, there is an understanding that $\eta/s$ should have a minimum around the phase transition and a maximum in $\zeta/s$ (although not necessarily at the same location because they undergo a cross-over phase transition); there is some evidence of this from Bayesian analyses~\cite{Bernhard:2016tnd,Bernhard:2019bmu,JETSCAPE:2020mzn,Nijs:2020roc}.  However, depending on the hydrodynamic theory and assumptions made when deriving the hydrodynamic equation so motion, different second-order transport coefficients may exist that can even couple shear and bulk together \cite{Denicol:2010xn,Denicol:2012cn} or viscosity to diffusive currents \cite{Fotakis:2022usk,Almaalol:2022pjc,Sammet:2023bfo}. These second-order transport coefficients have been calculated in kinetic theory \cite{Denicol:2012cn,Denicol:2014vaa} and non-conformal holography \cite{Finazzo:2014cna}. A number of studies have been performed on the influence and necessity of these second-order transport coefficients in simplistic models 
    but significantly less work has been performed in 2+1 or 3+1 hydrodynamic simulations. The one exception is the Bayesian analysis in \cite{Nijs:2020roc} that confirmed the importance of these second-order transport coefficients in small systems. In addition, recent work has begun to explore the idea of third-order hydrodynamics \cite{Jaiswal:2013vta,Younus:2019rrt,deBrito:2023tgb} (although, it was shown that \cite{Jaiswal:2013vta} was acausal and unstable \cite{Brito:2021iqr}), but it is still to early to tell its influence yet.  Thus, now that tighter constraints exist on the first-order transport coefficients thanks to theoretical calculations and Bayesian analyses, more systematic studies of these second-order transport coefficients are needed. In particular, studies that can identify experimental observables sensitive to their values.  Most likely, their relevance is strongest in small systems, which emphasize the need for \OO{} collisions results from both the LHC and RHIC.

    \item {\bf Hydrodynamic development to better understand small systems} Small systems have challenged the hydrodynamic paradigm in many ways and a number of open questions still exist.  For instance, it is clear that a number of fluid cells have very large Knudsen and inverse Reynolds numbers in small systems (to the point where causality is violated).  One approach to overcome this issue and also account for strangeness enhancement is the core-corona approach \cite{Kanakubo:2019ogh,Kanakubo:2021qcw}. However, this has only been explored in ideal hydrodynamics and, therefore, it would be an important test in viscous codes that have become standard in the field.
    
    Jet-medium interactions may have a large impact in small systems but there may be numerical challenges when tackling such large gradients. To address this and other related challenges, it may be necessary to adopt practice from astrophysical fields, for example adaptive hydrodynamic schemes where the resolution varies depending on the local densities. Another approach may be anisotropic hydrodynamics \cite{Martinez:2010sc,Martinez:2012tu,Florkowski:2013lya,Bazow:2013ifa}, which is designed to resolve potentially large momentum-space anisotropies in the local rest frame. For further discussion of experimental observables and the phenomenology related to small systems, see Sections~\ref{sec:progress:mesoscopic} and \ref{sec:future:mesoscopic}.
    
    \item {\bf Rapidity dependence} Many of the opportunities that exist, for example, in small systems, for low beam energies, and for jet-medium interactions, require relativistic viscous hydrodynamic codes that are fully three dimensional (i.e. that the assumption of boost invariance is relaxed).  While a number of codes are now written in 3+1 (3 spatial, 1 time) dimensions, the third dimension significantly slows down computational time. Thus, computational advances with improved numerical algorithms are needed to overcome this computational cost.    Additionally, there are added concerns on the correct description of the initial state (see Sec.\ \ref{sec:progress:macroscopic:initial_cond} for further details).   Groups have already begun to explore the effects of rapidity dependence, exploring a number of questions that are only possible with fully three-dimensional simulations.  Is there a decorrelation or twisting of the event-plane angle as one varies rapidity \cite{Bozek:2017qir}? How do baryons, strangeness, and electric conserved charges propagate along the rapidity direction \cite{Fotakis:2019nbq}? How sensitive are fluctuations of conserved charges to kinematic cuts \cite{Vovchenko:2021kxx}? How are the initial conditions different for charm vs light quarks in a 3D picture \cite{Chatterjee:2017ahy}? What can be learned about the equation of state from the slope of directed flow \cite{FOPI:2003fyz,Nara:2016hbg}? Are there other levers offered by rapidity in the beam-energy scan~\cite{Du:2023gnv}?

\end{itemize}

\subsubsection{Narrowing the QCD critical point search}
The RHIC BES program has great potential for the discovery of the QCD critical point. Comprehensive and quantitative description of the dynamics of fluctuations, including non-Gaussian fluctuations, their freeze-out, and the effect of the critical point and the phase transition on the observable signatures need to be developed and implemented in simulations to realize that potential. 

Thermodynamically, the smooth crossover at small $\mu_B$, the critical point, and the first-order phase boundary at finite $\mu_B$, are all intrinsically connected and could exist consistently. Therefore, %
it is natural to study heavy-ion collisions at high baryon density in order to search for the first-order phase boundary. In addition, the QCD phase structure as well as the nuclear matter equation of state at high $\mu_B$ require detailed investigations. Baryonic interactions including nucleon-nucleon, hyperon-nucleon and hyperon-hyperon interactions are fundamental ingredients to understand QCD and the equation of state that governs the properties of nuclear matter and astrophysical objects such as neutron stars~\cite{Lonardoni:2014bwa,ALICE:2020mfd}. High-statistics measurements including observables for rare probes in low-energy nuclear collisions need to be carried on in the next phase of high-rate experiments. Opportunities for high $\mu_B$ measurements at FAIR are discussed in \cite{Almaalol:2022xwv}.

At higher densities, large collaborations such as MUSES \cite{MUSES} and NP3M \cite{NP3M} are working on providing an equation of state across the entire phase diagram and testing them in realistic merger simulations, where theory can be confronted against upcoming gravitational-wave data; these will be measured with unprecedented accuracy, starting already in May of 2023 \cite{KAGRA:2013rdx}. Better accuracy is also expected from upcoming nuclear and electromagnetic observational data. Together, these will allow us to determine what kind of phases dense matter presents and at what $\mu_B$ phase transitions take place. In combination with heavy-ion collisions data analysis from the BES, this will allow us to understand the degrees of freedom at play (hyperons, heavier resonances, deconfined quarks, etc.) and interactions that give shape to the QCD phase diagram at large density.

\subsubsection{Future opportunities to study thermal properties of the QGP}

Photons and dileptons are the only soft and penetrating probes of relativistic nuclear collisions. They will play a crucial role in probing the QCD phase structure with heavy-ion collisions at $\sqrt{s_{\rm NN}} < 50$\,GeV. The EM probes provide complementary information to hadronic measurements. To derive combined and robust experimental constraints on QGP properties, unified theoretical frameworks~\cite{Shen:2014vra, Putschke:2019yrg} with the Bayesian Inference method play a central role in future phenomenological studies~\cite{Bernhard:2019bmu, Paquet:2022wgu}.

The precise measurements of photon and dilepton spectra will also provide experimental constraints on QGP EM current-current correlation function at different $(T, \mu_B, \mu_S, \mu_Q)$ and the QGP electric conductivity~\cite{Kapusta:2006pm,Floerchinger:2021xhb}. They are valuable inputs for probing the inner workings of QGP, especially the effective degrees of freedom for electric charge carriers in the QGP. Systematic comparison with theory calculations from lattice QCD and other effective theories can elucidate how the strongly-coupled nature of QGP emerged from many-body interactions of quarks and gluons.

The STAR Beam Energy Scan (BES) program collected a vast amount of data in the $\sqrt{s_\mathrm{NN}}$ = 3-62 GeV region. With their upgraded inner TPC the STAR detector also improved the reconstruction of low-$p_{T}$ tracks. The improvement in the low-$p_{T}$ tracks would also enable better precision in the di-lepton and thermal photon reconstruction from the previous measurement~\cite{STAR:2018xaj}. As part of the Beam Energy Scan Phase II project, the STAR Collaboration collected over an order of magnitude more data than previously acquired in the energy range from 7.7 to 19.6 GeV, where the total baryon density changes substantially. These studies would allow us for a better understanding of the thermal radiation in the transition regions and also further clarify the connection between the chiral symmetry restoration and the $\rho$ meson mass broadening in the $M=1-1.5$\,GeV invariant mass region. In the future runs between 2023-2025, STAR will collect an order of magnitude or more statistics in comparison with the previous Au+Au datasets at 200 GeV. This will enable high precision measurements in both dilepton spectra and low-$p_{T}$ direct photons.

The ALICE detector also went under substantial upgrade in its tracking capabilities and data acquisition (ALICE-2). The ALICE-2 detector implemented a streaming readout with their tracking detector, allowing the collection of all the data at a much higher speed and without further bias from online triggers. At the same time, the conversion probability stays at a very similar level as in the previous runs, therefore, the total number of conversion photons in the upcoming proton-proton, proton-lead and lead-lead collisions will be in the order of magnitudes larger than in the previously published data. In Run 3 and 4, ALICE-2 will collect a sufficiently large data sample to perform a first measurement of thermal dilepton emission to determine the temperature of the early stage. With larger sampled luminosity, larger acceptance and excellent tracking of photon conversion products, ALICE-3 will further improve on these measurements of the temperature from the emission of photons and their elliptic flow. Future updates at ALICE should also improve measurements of yield fluctuations between neutral and charged mesons~\cite{ALICE:2021fpb}, which can provide complementary evidence of chiral symmetry restoration~\cite{Kapusta:2022ovq}. 

The precise double differential analysis of the dilepton production in both transverse momentum and mass is largely unexplored terrain. The most precise dilepton measurements are from SPS CERES~\cite{CERESNA45:1997tgc} and NA60~\cite{NA60:2008ctj}, while at these energies the Drell-Yan processes significantly contribute to smaller invariant masses. The measurements from RHIC at 200 GeV suffer less from the Drell-Yan background, however, the large correlated background from heavy flavor requires a high-precision vertex detector. The new ALICE-2 tracking would provide great improvement in the separation of the prompt and non-prompt dilepton production to isolate the thermal contribution from the background. ALICE-3 tracking capability would further eliminate the heavy flavor background contribution from intermediate then high mass regions compared to the ALICE-2 Run 3 and Run 4 data sets.

\begin{figure}[t]
\centering
\includegraphics[width=0.6\columnwidth]{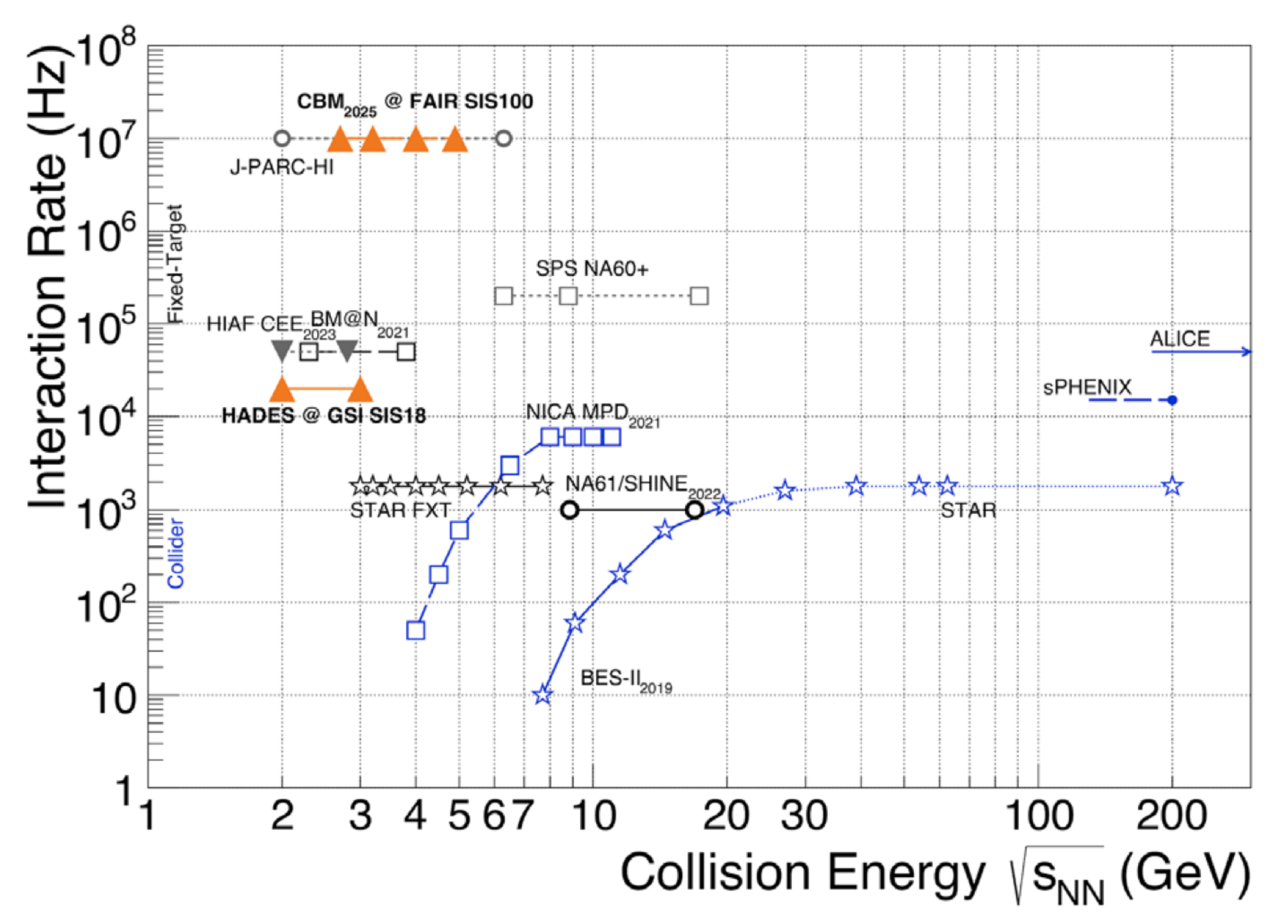}
\caption{Comparison of the interaction rate and collision energy of current and future experiments in the high-$\mu_{B}$ region.}
\label{fig:energyinteractionrate}
\end{figure}

The dilepton measurement at the high-$\mu_{B}$ region will be further explored with the new NA60+ upgrade at SPS in the 6-18 GeV collision energy range. The previously measured dilepton measurement~\cite{NA60:2008ctj} will be exceeded with the order of magnitude in statistics and the upgraded detector performance. The upcoming Facility for Antiproton and Ion Research (FAIR) will further explore the high-$\mu_{B}$ region, and the dedicated Compressed Baryonic Matter (CBM) experiment will be focused on the dilepton measurement in the 3-5 GeV collision region. Together with the STAR BES-II and STAR Fix target programs, the future detectors provide a high precision dilepton and thermal photon measurement lower energy regime around the expected critical point, see in Fig.~\ref{fig:energyinteractionrate}.

\subsubsection{Future opportunities for studying the initial conditions}

A window for collisions of new species will be opened at the LHC beyond 2028 and possibly before the shutdown of the RHIC upon successful completion of the sPHENIX program. More than 250 stable isotopes in the nuclide chart, 140 in isobar pairs or triplets, can be used~\cite{Giacalone:2021uhj}. Note that the nuclear structure and produced initial condition vary in a non-monotonic fashion with $N$ and $Z$, whereas the hydrodynamic response connecting the initial and final state varies smoothly and slowly with the mass number, $N+Z$. Hence, we can choose isobar or isobar-like systems with nearly identical hydrodynamic response but large structure differences to probe the initial condition. Clearly, continued effort is needed to identify species that maximize the scientific impact. For the moment, we point out four possible directions (see also ~\cite{Bally:2022vgo}):
\begin{itemize}
    \item  {\bf Initial condition in large systems}, whose knowledge is crucial for the precision extraction of QGP transport properties. We need to first calibrate the coefficients e.g. $c_{\mathcal{O},1}$--$c_{\mathcal{O},4}$ in Eq.~\eqref{eq:ns1} using systems that are close to those with well-known properties, such as doubly-magic $^{208}$Pb or $^{132}$Sn. These coefficients can then be used to make predictions for other species and check consistency with low-energy knowledge. 

    \item {\bf Initial condition from medium to small systems} to expose the role of sub-nucleon fluctuations, initial momentum anisotropy, and hydrodynamization process, by exploring isobar or isobar-like collisions in the region from $^{12}$C to $^{48}$Ca with different structures, e.g $^{16}$O+$^{16}$O vs $^{20}$Ne+$^{20}$Ne, which are nowadays accessible to cutting-edge {\it ab initio} calculations, with the aim of improving our understanding of the emergence of collectivity.

    \item {\bf Initial condition: $\snn$ dependence and longitudinal structure}. At high energy, the initial condition is controlled by the distribution of low-$x$ partons, which depends on the beam energy and the rapidity. Exploiting isobar ratios for bulk observables as a function of rapidity and $\snn$ may provide a new access route to the $x$-dependence of nPDF and gluon saturation, complementing thus the science goals of EIC.
    
    \item {\bf Initial condition for hard probes} is typically modeled by convoluting information from the Glauber model with the nPDF, which contributes to a large uncertainty in the relevant transport properties, such as $\hat{q}$~\cite{Apolinario:2022vzg,Xie:2022ght,JETSCAPE:2022ixz}. By constructing ratios of selected high-$p_T$ observables ($R_{AA}$ for example) at a fixed centrality, jet quenching effects are expected to largely cancel. Deviation of ratios from unity then provide access to flavor-dependent nPDF, tailored for each underlying hard-scattering process~\cite{Paukkunen:2015bwa,Helenius:2016dsk,Jonas:2021xju}. The determination of the impact parameter from bulk particles in combination with the observed hard processes will permit isobar ratios to access the differences in the transverse spatial distribution of partons, coming from structural differences in the collided nuclei.
    \end{itemize}

\subsubsection{Future opportunities to study chirality and vorticity in heavy-ion collisions}

The current status of the CME search demonstrates a situation of both challenges and opportunities. The challenges are nontrivial and will require considerable future resources to enable integrated efforts between theorists and experimentalists to perform dedicated modelings and comprehensive analyses that will quantify any residue uncertainties  such as non-flow corrections and conclusively extract any signal with high statistical significance. On the other hand, the opportunities of CME discovery, carrying magnificent implications for fundamental physics and substantial interests in other disciplines, are simply difficult to dismiss. Looking into the next decade, one anticipates a number of major thrusts for future developments.    

With the rich isobar data set available, post-blind analysis efforts are important in order to fully capitalize on the contrast power only available with such an isobar pair. Alternative analysis approaches (e.g. based on identical multiplicity selection or event-shape measurements between event-plane/spectator-plane) may also provide fresh insights. In particular, a quantitative understanding of the bulk property difference between the isobar systems appears feasible with further  theoretical studies. Once that is achieved, one can hope for an informed  baseline with careful calibration of nonflow effect for the isobar comparison and thus a convincing answer for the CME signal level in the isobar collisions.        

The \AuAu{} collision system is expected to have a higher CME signal fraction and holds strong future promise, given the aforementioned positive result from the event-plane and spectator-plane comparison analysis. The bottlenecks include the limited statistics and residue nonflow effect. The highly anticipated \AuAu{} collision runs at RHIC from 2023 to 2025 will provide events on the order of 20 billion that can help deliver a much-needed boost of statistics for the CME analysis. 
On the LHC side, the upcoming high-statistics measurements will soon provide an exciting opportunity for yet another scrutiny at extracting/constraining a potential CME signal at TeV energies.  
A close collaboration between theoretical and experimental efforts will further advance the simulation and analysis tools to substantially reduce various remaining uncertainties, thus paving the way for a decisive conclusion on the CME search.       

Last but not least, it is highly desirable to map out the beam energy dependence of the CME phenomenon. 
Past CME measurements in the low-energy region were based on Beam Energy Scan I data, while now there is an abundance of high-statistics data from the Beam Energy Scan (BES) II program. Understanding their implications will require the development of a CME  modeling framework  for low-energy collisions.  Furthermore, various new analysis methods and observables for the CME extraction have been mainly applied to top RHIC energy. Utilizing them for analyzing BES-II data will be very interesting. Ultimately, nailing down the beam energy dependence of CME will help reveal the QCD chiral symmetry restoration and contribute to determining the phase boundary via beam energy scan.

The global and local polarization measurements in heavy-ion collisions have opened a new direction in the study of the hottest and densest QCD matter, now under the fastest rotation~\cite{Becattini:2020ngo,Becattini:2022zvf}. 
As the spin polarization at the lowest order depends linearly on
the gradients of the hydrodynamic fields, it turns out to be a
very sensitive probe of the hydrodynamic evolution and thus can help quantitatively constrain bulk matter  properties. 
Despite the successful description of the average global polarization by various theoretical models, there still remain open questions. From the experimental point of view, more precise measurements are needed to constrain theoretical models and answer these questions. Analyses of high statistics BES-$\mathrm{I}\hspace{-1.2pt}\mathrm{I}$ data together with data from future experiments such as CBM~\cite{Almaalol:2022xwv} can reveal a complete picture of the energy dependence, especially at $\sqrt{s_{NN}}<10$ GeV, where there is little experimental data and theoretical models predict non-monotonic behavior. Quantifying the vorticity is important to better understand the QCD phase diagram as the vorticity also acts as the baryonic chemical potential~\cite{Jiang:2016wvv,Fujimoto:2021xix}. Also, the measurement of possible splitting in global polarization between $\Lambda$ and $\bar{\Lambda}$ is of great interest to constrain the lifetime of the magnetic field, which is also important for the study of chiral phenomena.  
In 2023 and 2025, high statistics data of \AuAu{} collisions at $\sqrt{s_{NN}}=200$ GeV will be collected by STAR with upgraded detectors in both mid and forward rapidities and by sPHENIX at RHIC. 
High statistics data at RHIC and the LHC will allow us to perform measurements sensitive to the electromagnetic field, such as charge-dependent directed flow including heavy-flavor hadrons.
The new data will also allow us to measure the polarization of multistrangeness ($\Xi$ and $\Omega$) more precisely which helps to shed light on possible spin and particle species dependence and to measure rapidity dependence of global polarization which is predicted differently by models, especially in the forward region. For the local polarization,  while considerable theoretical progress has been made with the newly found thermal shear contribution having potential for resolving the sign puzzle, more future studies are needed to understand the different implementation schemes as well as to achieve a quantitative explanation of data. A relevant direction is to extend the measurements to polarization induced by higher harmonic flow~\cite{Voloshin:2017kqp}, which is being studied by STAR. 
Generally speaking, the global and local spin polarization phenomenon could develop into an important probe of the hydrodynamic fields (such as their initial conditions and gradients evolutions) in heavy ion collisions.

Regarding the spin alignment measurements, interesting results of $\phi$ and $K^{\ast0}$ $\rho_{00}$ were observed both at RHIC and the LHC. However, the interpretation of the results is currently under discussion. Recently, the ALICE Collaboration observed $J/\psi$ transverse polarization relative to the event plane in the forward region ($2.5<y<4$) as $\lambda_\theta\sim 0.2$, corresponding to $\rho_{00}\sim 0.37$~\cite{ALICE:2022sli}. Heavy quarks are produced at a very early time by parton hard scattering, therefore there may be a large contribution from the initial magnetic field. It is of great interest to measure $J/\psi$ polarization at RHIC energy, where less regeneration in $J/\psi$ polarization, and helpful to understand how the different production mechanisms affect the spin polarization. The 2023+2025 runs at RHIC make measurements of $J/\psi$ polarization possible at midrapidity, which can be directly compared to the results of light vector mesons and helps to understand the origin of polarization of vector mesons.

Furthermore, there are new directions to explore vortical structure in heavy-ion collisions. 
Event-by-event density fluctuations superimposed on the collective flow field should produce vorticity ``hot spots" that may be probed by spin-spin correlations~\cite{Pang:2016igs} while jet with medium interactions should produce vortical toroids centered on the jet~\cite{Betz:2007kg,Tachibana:2012sa,Serenone:2021zef}.
Measurements of polarization induced by these phenomena would be experimentally difficult as the observable is sensitive to the experimental effects such as acceptance and efficiency. One possible direction will be a study in asymmetric collisions such as \AB{} or \pA{} collisions~\cite{Shi:2017wpk}, where a smaller projectile passes through the center of a larger target leading to a toroidal vortex~\cite{Voloshin:2017kqp,Lisa:2021zkj}. Data of $p$+Au collisions at RHIC taken in 2024 and $p$+Pb data at the LHC will have the potential to discover this unique vortical structure arising as a fluid behavior, which may also bring new insight into the collectivity in small systems.
Another interesting phenomenon that is possibly measurable is the generation of spin current called Spin Hall Effect~\cite{Liu:2020dxg,Fu:2022myl} whose signal is expected to be larger at lower collision energy and BES-$\mathrm{I}\hspace{-1.2pt}\mathrm{I}$ data will provide a good opportunity to search for the effect.

\subsection{Mesoscopic: emergence of the quark-gluon plasma and approach to equilibrium}
\label{sec:future:mesoscopic}
\subsubsection{Future Experimental Exploration of Collectivity in Small Systems}
\label{sec:future:mesoscopic:smallsystemexp}

There are several directions that can be explored experimentally in the next decade to improve our understanding of the emergence of collectivity in small systems. Additional collision systems with different initial geometries and structures,  as well as improved statistical precision for systems that have already been studied, may bring new insights. High-statistic data sets and improved detector capabilities can enable the exploration of new observables and new probes with different sensitivity to the underlying physical processes. Some examples of future measurements and directions are discussed below. 

Measurements in the smallest collision systems, such as photon-proton interactions accessed through ultra-peripheral \pPb{} collisions~\cite{ATLAS:2019pvn,CMS:2022doq} and $e^+e^-$ collisions~\cite{Badea:2019vey, Belle:2022fvl} are presently statistics limited. Additional \pPb{} data at the LHC will improve the multiplicity reach for $\gamma-$p collisions. A high statistics data sample on ultra-peripheral \AuAu{} collisions at $\sqrt{s_{_{\rm NN}}}=200$ will be collected during the upcoming RHIC Run 23 and 25. This along with the ultra-peripheral \dAu{} data from RHIC run 21 provide the opportunity to search for collectivity in $\gamma-$Au and $\gamma-$p collisions at $W_{\gamma N}\approx40$ with the extended pseudorapidity capability of the STAR detector~\cite{STAR:2022BUR}. There are also more archived data available from ALEPH at LEP2 for $e^+e^-$ collisions at energies higher than those presently studied.  The event multiplicity could reach more than 50 charged particles in these much simpler collisions that do not have complicated hadron structures or gluonic initial state radiation. Finding any evidence of collective flow in these events could improve our understanding of collectivity. Eventually, there will be new opportunities to study small systems at the Electron-Ion Collider~\cite{Shi:2020djm}.

In hadronic collisions, new measurements will be possible both at RHIC and at the LHC. Data from \dAu{} and \OO{} collisions at 200 GeV were already collected by the STAR experiment in 2021. These data sets include an upgraded STAR detector with extended coverage in rapidity (inner TPC), high-resolution event plane detector and forward-rapidity particle identification. A more direct comparison between the PHENIX and STAR measurements in \dAu{} collisions will improve the understanding of collective flow, nonflow correlations, and the three-dimensional system evolution in \dAu{} collisions. In addition, a high luminosity \pAu{} run is planned for 2024 with both sPHENIX and STAR taking data. The LHC planning also includes \pO{} and \OO{} runs in 2024. Correlation measurements in these collisions with different initial geometry and a smaller number of participating nucleons could provide new insights. 

Another avenue of exploration is to investigate new observables. Different physics mechanisms have been proposed as possible sources for particle azimuthal correlations in small systems. The relative contribution from each mechanism could be different for different particle species due to differing particle production mechanisms and in-medium interactions. One such example is the comparison of $v_2$ measured for light and heavy-flavor hadrons, including $J/\Psi$. Significant $v_2$ for charm has been observed in high-multiplicity \pPb{} collisions~\cite{ATLAS:2019xqc, ALICE:2017smo, CMS:2018loe, CMS:2020qul, CMS-PAS-HIN-21-001} and the mechanisms of these correlations, especially for $J/\Psi$, is not fully understood. Non-zero $v_2$ has also been observed for open charm hadrons in high-multiplicity \pp{} collisions, but not yet found for $J/\Psi$. Future measurements of long-range correlations of $J/\Psi$ in \pp{} collisions will give further handles to distinguish initial vs final state effects in the particle correlations. Correlations between $v_2$ and mean $p_T$ have also been proposed~\cite{Giacalone:2020byk} for distinguishing different origins of azimuthal correlations and are already pursued~\cite{ATLAS:2019pvn}. Further studies with multi-particle observables~\cite{Tuo:2022qm} will be needed for detailed understanding of the interplay between collective effects and nonflow correlations. Ultimately, comparisons to theoretical models that properly include the correlations between the bulk system and the hard-scattered partons will be needed. 

More broadly, the small-system evolution can be studied with event engineering, e.g. by selecting events with large and small $v_2$. The selection criteria could include both hard and soft observables, such as number of jets, soft particle multiplicity, mean $p_T$, and others. The correlation between $v_2$ and mean $p_T$ is just one of many possible observables. Recently, there is a proposal to search for collectivity in a QCD system where a single-parton is propagating in vacuum~\cite{Baty:2021ugw}. The measurement of $v_2$ using the particles inside the jet cone in the jet coordinate system is proposed. Such measurements are accessible with the high-energy jets produced at the LHC. Another interesting proposal is to search for vortex rings in small collision systems~\cite{Lisa:2021zkj}. Existence of the vortex toroids may provide new evidence for hydrodynamic collective flow in the smallest QGP droplets.

\subsubsection{The next steps for understanding small systems}
\label{sec:future:mesoscopic:smallsystemtheory}

Despite recent progress \cite{Schenke:2021mxx}, many outstanding questions remain concerning the description of small systems. The answer to those questions requires a number of new experimental and theoretical developments, as briefly discussed below.

Many more theoretical developments are needed in order to fully determine the applicability of hydrodynamics in small systems.
Most of what is understood about hydrodynamics comes from the gradient expansion. However, in rapidly expanding plasmas the gradient expansion can become divergent \cite{Heller:2013fn,Buchel:2016cbj,Denicol:2016bjh,Heller:2016rtz}, which motivated the idea of hydrodynamic attractors \cite{Heller:2015dha}. Though general properties of the hydrodynamic expansion have been worked out \cite{Grozdanov:2019kge,Grozdanov:2019uhi,Heller:2020uuy,Heller:2021oxl,Heller:2021yjh}, these developments have not yet been put to use in realistic phenomenological investigations, especially in the case of small systems. Furthermore, through anisotropic hydrodynamics \cite{Martinez:2010sc,Florkowski:2010cf,Alqahtani:2017mhy,Strickland:2017kux} it was understood that one may resum not only gradients but also the viscous stresses themselves, which significantly expands the regime of applicability of hydrodynamics. Further investigation is needed in this regard to determine the phenomenological impact of attractors and hydrodynamic resummation schemes. Additionally, even though hydrodynamic attractors are now starting to become part of the heavy-ion phenomenology toolbox \cite{Giacalone:2019ldn}, it remains unknown whether the QGP, realistically modeled as a rapidly expanding system in 3+1 dimensions, indeed approaches a hydrodynamic attractor at the relevant timescales needed in phenomenology. This question is even more pressing in small systems, due to their reduced size and lifetime. Despite some studies in this context \cite{Romatschke:2017acs,Denicol:2020eij,Heller:2020anv}, this question remains elusive due to the lack of a consistent framework capable of systematically investigating QCD in the far-from-equilibrium regime (at all values of coupling).

Additionally, in small systems where deviations from equilibrium can become very large, a systematic investigation of the nonlinear far-from-equilibrium properties of 2nd-order hydrodynamics, going beyond the initial results of \cite{Bemfica:2020xym}, is needed. Implementing causality constraints in simulations \cite{Chiu:2021muk,Plumberg:2021bme} of small systems can be important to avoid numerical instabilities and also to better constrain the properties of the pre-hydrodynamic phase  \cite{Kurkela:2018wud,Kurkela:2018vqr} and the importance of conformal invariance violation \cite{NunesdaSilva:2020bfs,Nijs:2020roc,daSilva:2022xwu}. 

The possibility of using different definitions of hydrodynamic variables (i.e., hydrodynamic frames) \cite{Bemfica:2017wps,Kovtun:2019hdm,Bemfica:2019knx,Hoult:2020eho,Bemfica:2020zjp,Noronha:2021syv} opens up a number of questions in the formulation of hydrodynamics that can be particularly useful in the description of small systems. For example, new terms and coefficients appear at 2nd order \cite{Noronha:2021syv} that have never been explored before in phenomenological simulations. Furthermore, because of the fluctuation-dissipation theorem, the choice of hydrodynamic frame should also influence how thermal stochastic fluctuations take place in the fluid. The simplest version of this statement is widely understood in the field already, given the  differences that appear when one considers fluctuations in the Eckart \cite{Eckart:1940te} or the Landau frame \cite{Landau1987-pq}. Further work is needed to systematically formulate 1st and 2nd order stochastic hydrodynamics in general hydrodynamic frames, going beyond existing work \cite{Akamatsu:2017zzl,Akamatsu:2017rdu,Martinez:2019bsn,An:2019osr}. Thermal fluctuations should become even more important with decreasing multiplicity, being especially relevant in small systems. Current phenomenological investigations 
\cite{Yan:2015lfa,Singh:2018dpk,Aasen:2022cid} must be generalized to systematically determine the size of such effects, and their experimental consequences, in small systems.

\subsubsection{Prospective studies of medium response to partonic excitation}
\label{sec:future:mesoscopic:medresponse}

 In recent years, groups are beginning to explore jets fully coupled to hydrodynamic simulations wherein energy is dumped into the fluid by using a source term in hydrodynamics \cite{Andrade:2014swa,Tachibana:2014lja,Okai:2017ofp} (see Section~\ref{sec:progress:microscopic} for the influence this has on jet observables). Additionally, breakthroughs in the theoretical description of jet-medium interactions by the inclusion of flow in energy loss calculations allow for an even better, self-consistent picture \cite{Rajagopal:2015roa,Sadofyev:2021ohn}.  An opportunity exists here to study the influence of this energy gain in the medium to study far-from-equilibrium phenomena. %
 The presence of a jet was shown to alter the $v_n(p_T)$ in the range of $p_T\sim 2-10$ GeV in \pPb{} collisions.  Thus, collective flow observables in this $p_T$ range in small systems provide an intriguing possibility to better understand jet-medium interactions. Another synergy has yet to be thoroughly exploited in that in initial studies \cite{CMS:2017xgk,Betz:2016ayq} multiparticle $v_n$ calculations and experimental data were explored at high $p_T>10$ GeV.  However, as we have discussed already in this section that there are an enormous number of collective flow observables such as symmetric cumulants, factorization breaking, $\rho(v_n,p_T)$ observables, etc that have never been calculated nor measured experimentally at high $p_T$.  Given the high luminosity era at the LHC and sPHENIX, this is a perfect opportunity to exploit these high statistics.

    To reveal the jet broadening effect from multiple soft scatterings and medium response experimentally, $Z$/photon-tag reduces ``survival bias'' narrowing observed in the inclusive jet shape.
    The increased data acquisition and acceptance of the sPHENIX will enable a dramatic increase in the number of photon-tagged jet observables previously studied at RHIC. At HL-LHC, the available statistics of $Z$/$\gamma$-tagged jet events will increase by a factor of 5-10. The large datasets will provide high-accuracy data on the $Z$/$\gamma$ tagged jet spectra, transverse momentum profile, fragmentation function, hadron angular distributions, and jet substructures.

    This large data set will also enable more differential measurements to separate the contribution from medium-induced momentum broadening via multiple soft scattering, medium-induced radiation, medium response, and medium recoil effects. Measurements such as $Z/\gamma$-hadron 2D angular correlation and $Z/\gamma$-tagged subjet multiplicity have been proposed. %
    Moreover, the large acceptance tracking detectors in ATLAS and CMS will enable us to study the long-range correlation between electroweak bosons, jets, and hadrons up to a $|\eta|<4$.    
   
\begin{figure}[htb]
\centering
\includegraphics[width=0.3\columnwidth]{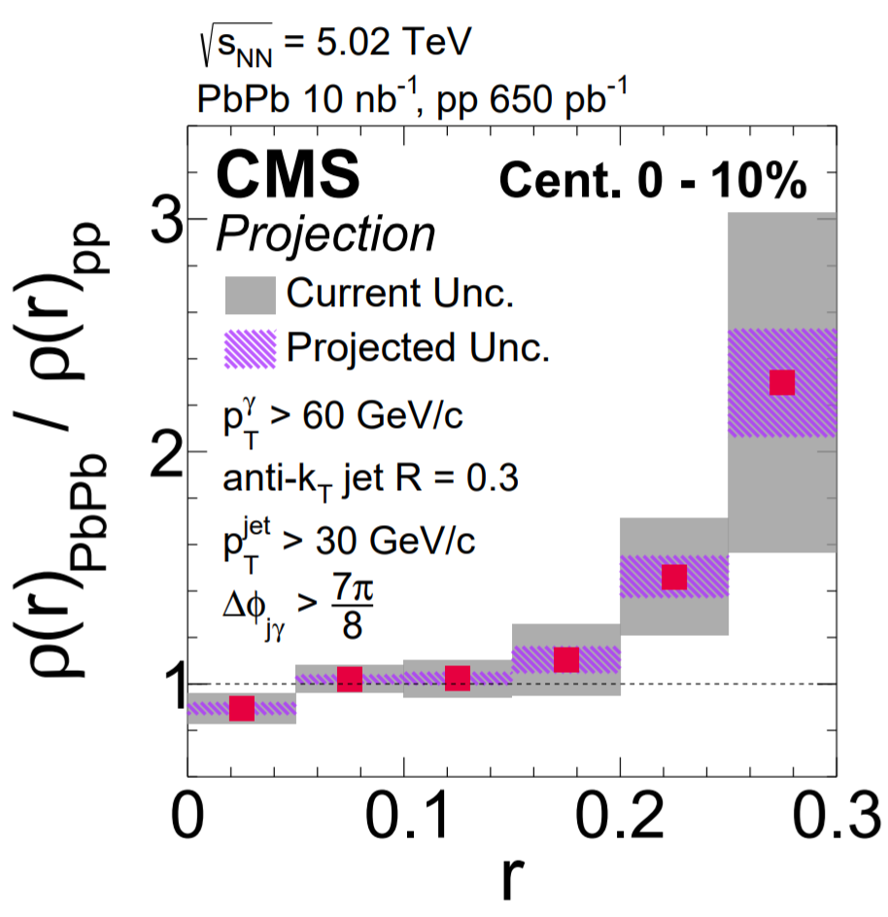}
\includegraphics[width=0.3\columnwidth]{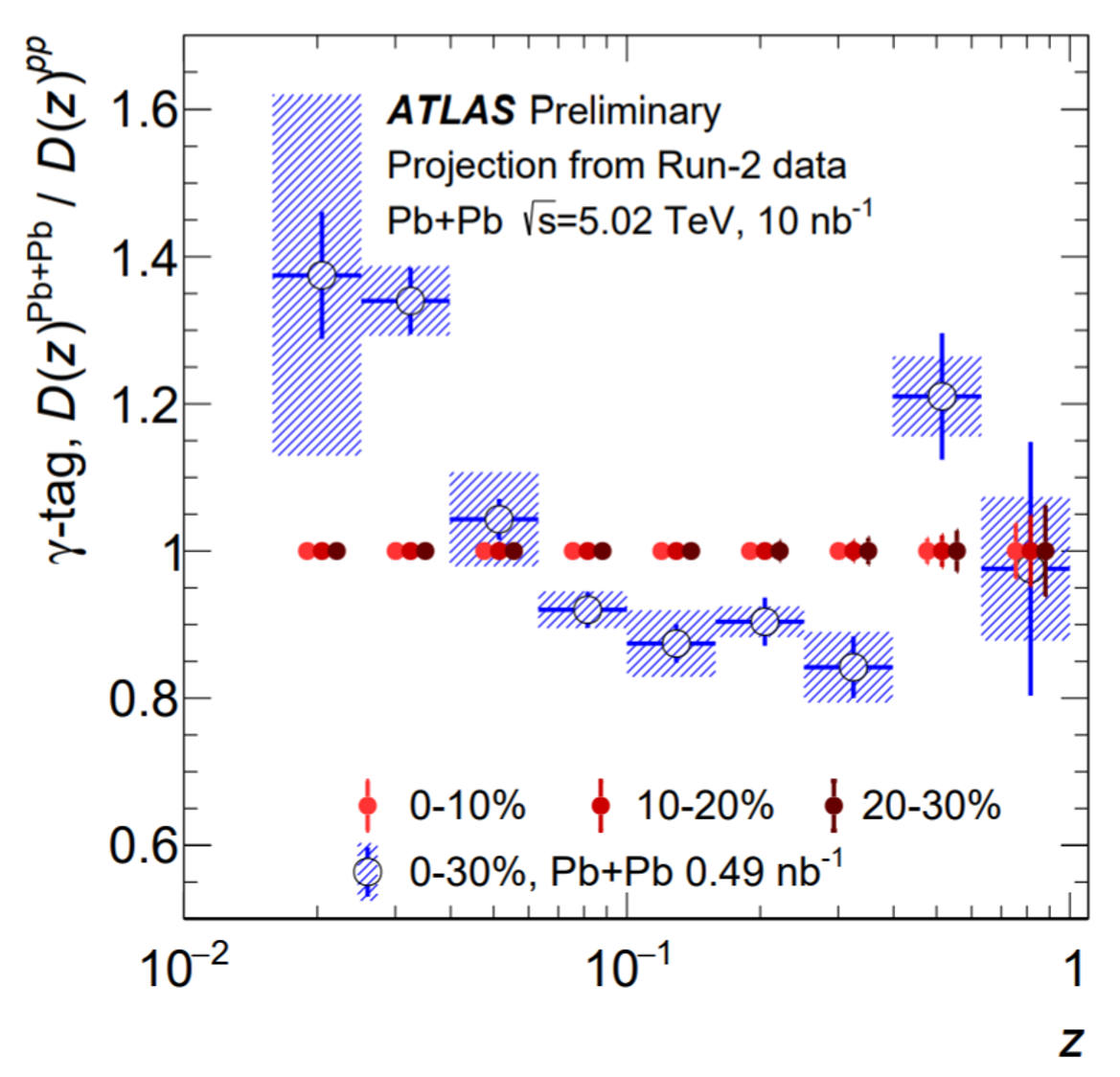}
\includegraphics[width=0.29\columnwidth]{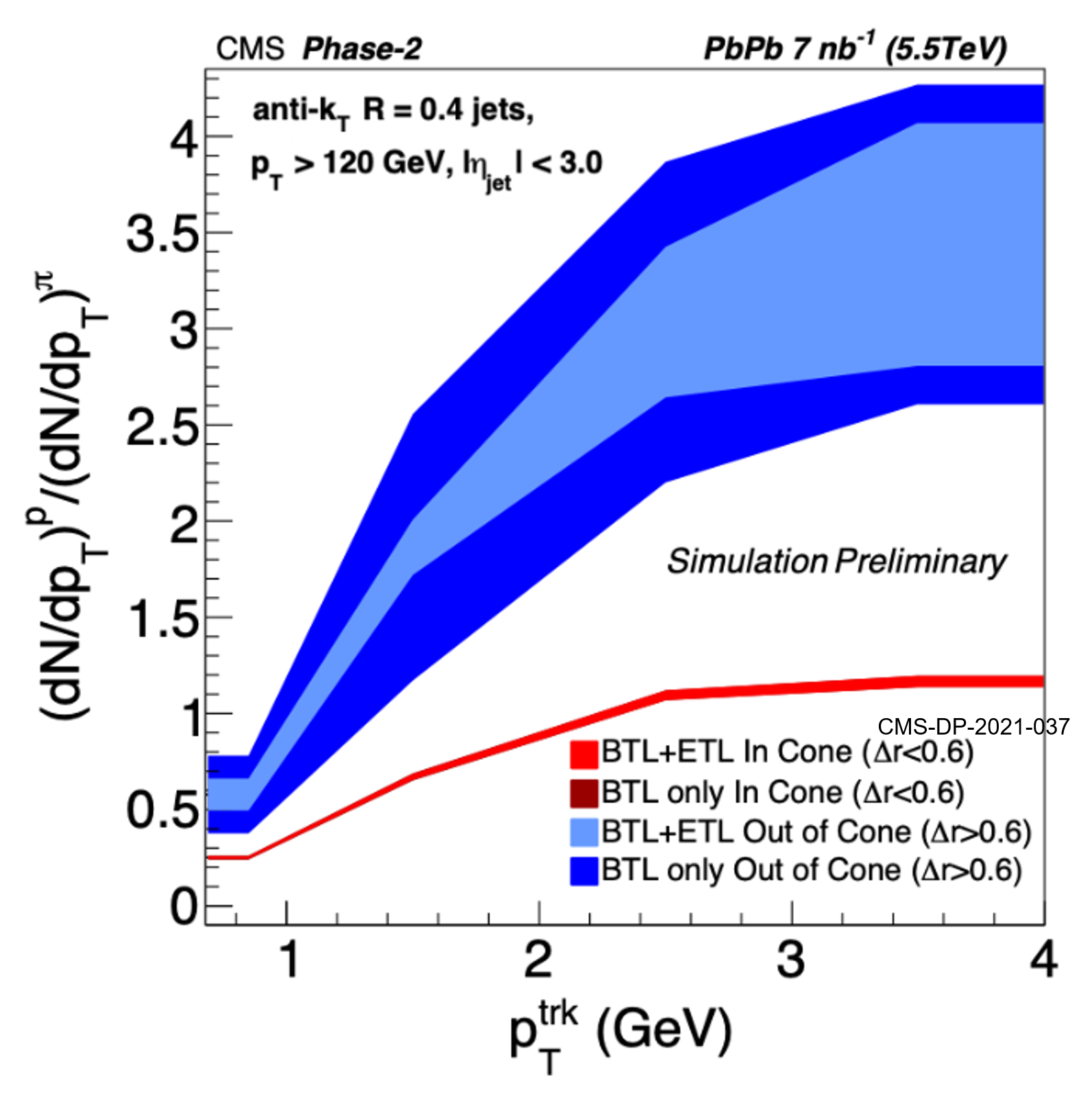}
\caption{Projections for LHC measurements related to medium response~\cite{Krintiras:2022ohr}.}
\label{fig:3.2.3:future}
\end{figure}

    The new capability of the CMS MTD will enable us to study the particle composition in the QGP wake. Example measurements include proton-to-pion ratios in and out of the jet cones as shown in Figure~\ref{fig:3.2.3:future}. The ALICE collaboration, who has studied particle ratios in jets for p+p and p+Pb collisions~\cite{ALICE:2021vxl,ALICE:2022ecr}, also expects to produce complementary measurements to these results in Pb+Pb collisions from Run~3 and Run~4 data.

\subsubsection{Further investigations of jet quenching in small systems}
\label{sec:future:mesoscopic:smallsystemjets}

As described in Sec. \ref{sec:progress:mesoscopic:small_size_limit_of_qgp} the lack of energy loss signatures in experimental measurements of jets in small system may not be so surprising due to the limited pathlength of the produced medium. How large does the collision system need to be, to be able to measure energy loss effects? New calculations at NLO for jets and leading hadrons indicate that the baseline spectrum in \pp{} collisions can now be calculated with high accuracy. This  will allow for even the smallest energy loss effects to become evident in intermediate systems such as those formed in light-ion collisions, e.g., \OO{}~\cite{Huss:2020dwe, Huss:2020whe}.

To explore this question experimentally, it has been proposed to quantify the suppression in collisions of light ions which would fill the gap between the existing data from \pA{} and heavy-ion \AA{} collisions. The LHC plans to record data from \OO{} collisions during Run 3 which will provide information on the system size necessary to transition from partons escaping the medium to jet quenching. sPHENIX is well equipped to further our understanding of the future LHC results as indicated in their beam use proposal if the opportunity for running light ions arises. Specifically they demonstrate the kinematic reach to measure jets and heavy-flavor observables in \OO{} and \ArAr{} collisions. RHIC measurements would provide temperature dependence for the physics observed in these light-ion collisions as well as probe lower momentum partons that may experience stronger quenching.

\subsection{Microscopic: Hard Probes}
\label{sec:future:microscopic}

The successful operation of sPHENIX and the complementary heavy-ion program at LHC as well as the subsequent analysis of all collected data is crucial for realizing the goals of exploring the QGP at smaller length scales as set forth in the previous \LRP. In particular, sPHENIX promises to deliver precision hard probes measurements to quantify properties of the QGP at the microscopic level, measurements that will be complementary to those planned at the LHC. This section will discuss planned measurements and opportunities for theoretical developments related to jets, open heavy flavor and quarkonia as well as electroweak and ultra-peripheral collision measurements related to nPDFs in the initial state of the collisions.

\subsubsection{Prospective jet measurements and theoretical advances}
\label{sec:future:microscopic:jets}
\begin{figure}
    \centering
    \includegraphics[width=0.9\textwidth]{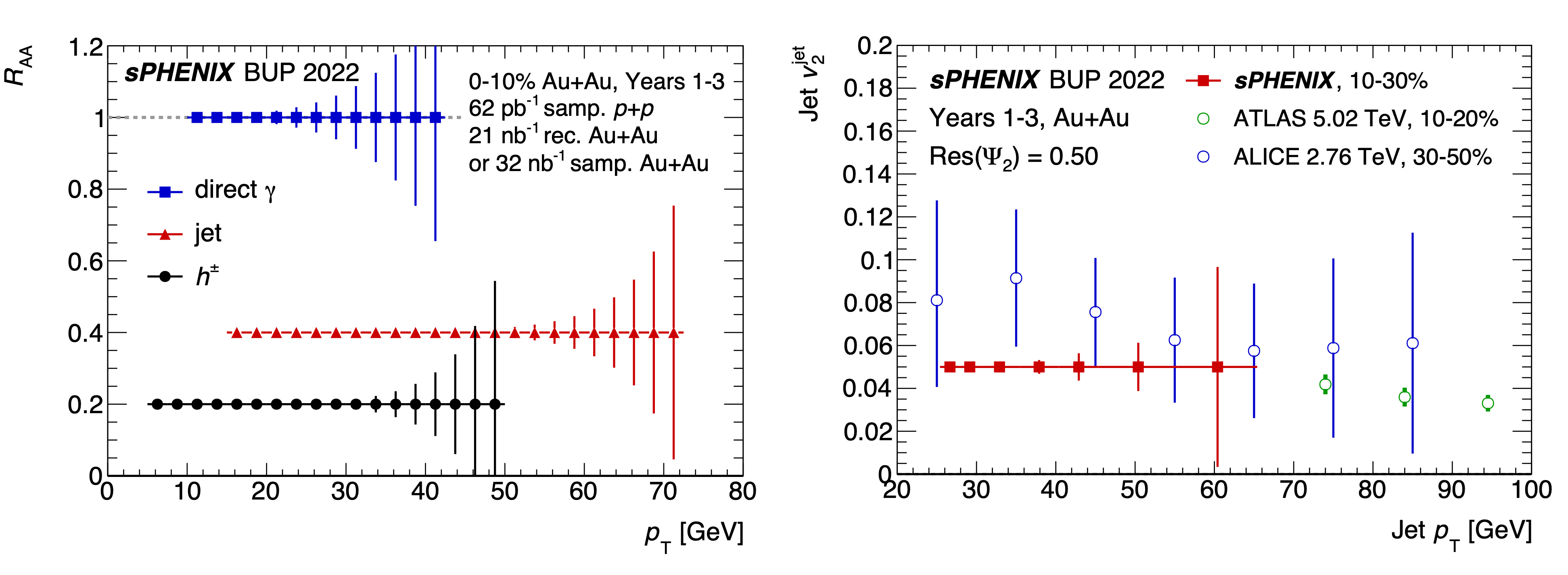}
    \caption{The anticipated statistical precision for sPHENIX to measure the $R_{AA}$ of jets, direct photons and hadrons (left) as well as the jet $v_2$ compared to existing measurements at the LHC~\cite{sPHENIX:2022BUR}}
    \label{fig:sphenixjetv2}
\end{figure}

The anticipated statistical precision for the $R_{AA}$ of jets expected from sPHENIX in 3 years of running is shown in the left panel of Figure~\ref{fig:sphenixjetv2} along with that of direct photons and hadrons which indicate a kinematic reach significantly beyond current RHIC measurements. The addition of the event plane detector (EPD) at STAR and sPHENIX will enable precise measurements of jet $v_n$ in \AuAu{} collisions at 200~GeV. Figure~\ref{fig:sphenixjetv2} shows the precision of the expected sPHENIX jet $v_2$ results. Jet $v_2$ provides insight to the pathlength dependence of the energy loss mechanism in the QGP. These new results at the lower temperatures at RHIC would be complementary to existing LHC measurements. In addition to studying energy loss at the jet level, the tracking system in sPHENIX will enable studies of the jet constituents to explore modification to the fragmentation functions as well jet substructure observables. 

{\bf Theoretical Advances}

Theoretical developments in the period covered by the next long range plan include developments in the underlying theory of energy loss, improvements in event generators, both in their theoretical content and in their computational abilities, and a deeper understanding of transport coefficients leading to further elucidation of the degrees of freedom of the QGP. 
Separate from all of these is the rise of machine-learning techniques, Bayesian analysis and uncertainty quantification that have changed the day-to-day mechanics of how theoretical developments are judged in comparison with experimental data. 
Continued developments in the aforementioned arenas imply the possibility of deep advances in our understanding of jet medium interactions and the nature and interaction of the prevalent degrees of freedom of a strongly interacting plasma at $T \sim (1 - 4 ) T_{\rm C}$.

In the period subsequent to the last \LRP, major theoretical developments have included improvements in Glauber enhanced Soft Collinear Effective Theory (SCET$_{\rm G}$)~\cite{Sievert:2019cwq, Huang:2013vaa}, developments in the overarching formalism of energy loss~\cite{Caucal:2018dla,Mehtar-Tani:2019tvy, Blaizot:2014bha,Casalderrey-Solana:2014bpa,Mehtar-Tani:2022zwf,Barata:2021wuf,Sirimanna:2021sqx,Arnold:2021mow,Arnold:2022fku}, as well as the incorporation of coherence effects~\cite{Casalderrey-Solana:2012evi,Kumar:2019uvu}. The stage is now set for the appearance of higher order calculations of energy loss, which should lead to precision calculations with realistic estimates of the theoretical uncertainty.

On the event generator side, improvements in individual simulators such as the development of jet modification with hydrodynamic response~\cite{Yang:2021qtl, Tachibana:2020mtb}, is coupled with the development of frameworks where a variety of different energy loss simulators can be modularly incorporated within an extensive end-to-end generator~\cite{Putschke:2019yrg}. In the upcoming interval of the current long range plan, we envision enhancements in the theoretical and phenomenological content of individual simulators, advancements in computational ability e.g., using machine learning routines to accelerate hydrodynamic simulations, 
and more elaborate Bayesian analysis combining hard and soft sectors in full 3+1D simulations. 

With the advent of theoretical advances, and improvements in simulation, along with the ability to conduct rigorous Bayesian comparisons between theory and data, it will become possible to finally constrain the underlying structure of the quark-gluon plasma. Current calculations of jet transport coefficients at higher order~\cite{Caron-Huot:2008zna}, will soon be enhanced with lattice calculations using 3D EQCD calculations~\cite{Panero:2013pla,Moore:2021jwe} and in full 4D lattice gauge theory~\cite{Majumder:2012sh,Kumar:2020wvb}. Comparing these calculations with extractions from Bayesian analysis~\cite{JETSCAPE:2020mzn,Ke:2020clc}, conducted using high statistics data from several colliding systems and energies, will allow us to reliably narrow down the quasi-particle structure of the plasma.

{\bf Jet substructure}

The next generation of substructure studies in heavy-ion collisions require higher theoretical accuracy and extending towards more differential measurements beyond inclusive ones. %
The emphasis here is on robust theoretical control in benchmarking a particular model of jet-medium interactions via calibrated Monte Carlo simulations. The novel experimental observables %
need to be sensitive to the physics of varying jet showers and topologies and simultaneously be robust to the heavy-ion background. While hard jet substructure is more immune to fluctuating heavy-ion background, its sensitivity to medium effects may be suppressed by the large-scale separation. On the other hand, soft jet substructure can be affected significantly by the medium.  A systematic approach from hard to soft jet substructure is essential to decouple physical effects at various energy scales. Observables based on the most energetic jet particles which are recoil-free as discussed in Sec.\ref{sec:progress:microscopic:jet_substructure} might be the starting point of this program. Also, in order to meaningfully and unambiguously extract information from soft jet substructure, combination of knowledge about local underlying event fluctuation beyond practical background subtraction schemes will be increasingly necessary. An initiation of the unification of hard and soft probes should be attempted in the 2023 \LRP{}.

With the upcoming measurements of several, different observables from the experiments at both RHIC and the LHC, one can eventually move towards a global analysis resulting in calibrated heavy-ion Monte Carlo models. In the coming years following the 2023 \LRP{}, enabled by the new sPHENIX detector designed with both hadronic and electromagnetic calorimetry, high precision charged particle tracking and secondary vertex upgrades along with streaming readout will result in large statistics to push into the next era of jet substructure measurements at RHIC. For inclusive jets, the high statistics dataset will result in a direct comparison of jets at RHIC and the LHC. Upgrades to the STAR detector at RHIC will provide a complementary handle on the studies of jet substructure in vacuum and in heavy-ion collisions. With Runs 3 and 4 at the LHC, the extension of the dimensions of substructure studies such as the dependence on the heavy quark masses, regions of the emission phase-space sensitive to early time dynamics, has the potential to extract information related to the early pre-equilibrium state of the plasma. 

{\bf Coincidence measurements with jets} 

One of the challenges in reconstructing jets at RHIC is the proximity to the size of the underlying event and upward fluctuations in the underlying event which give rise to so-called ``fake’’ jets. The contribution of these fake jets is greatly reduced when the jet is tagged by an opposing hard particle such as a direct photon or heavy flavor quark.  Photon tagged jets have long been referred to as the golden channel for studying energy loss in the QGP since the direct photon provides access to the kinematics of the initial hard scattering \cite{Dai:2013}. In addition, these measurements can be used to further our understanding of the medium response as discussed in the mesoscopic section (Section~\ref{sec:progress:mesoscopic:medium_response}). The b-tagging capabilities of sPHENIX and its high data acquisition rate will enable precise measurements of these rare probes.

{\bf Heavy-flavor tagged jets}

\begin{figure}
    \centering
    \includegraphics[width=0.7\textwidth]{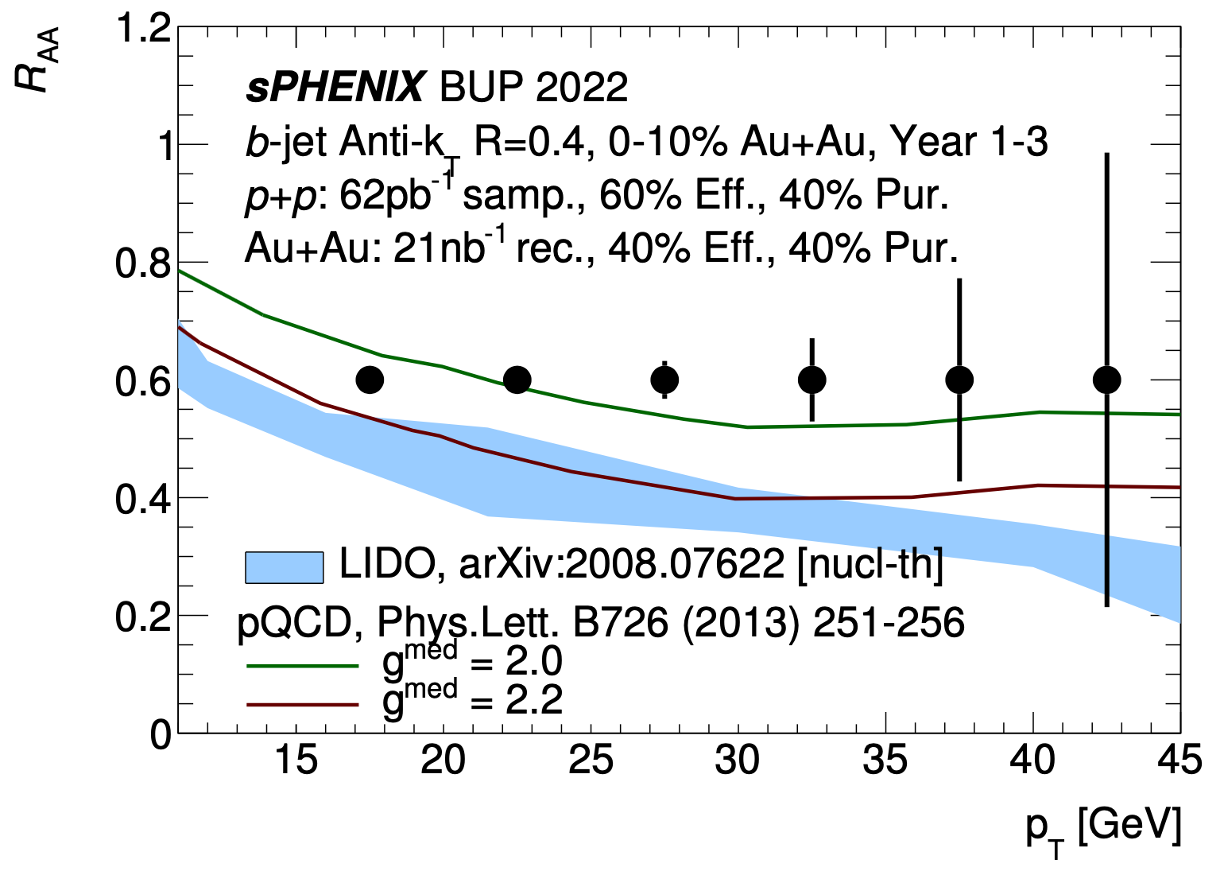}
    \caption{The anticipated statistical precision for measuring the $R_{AA}$ of b-tagged jets compared to curves from a pQCD calculations with two coupling parameters to the QGP medium, $g_{med}$ \cite{Huang:2013vaa}, and a blue band representing a calculation based on the LIDO transport model \cite{Ke:2020nsm}.}
    \label{fig:sphenixbjet}
\end{figure}

The new sPHENIX MVTX will enable the first b-jet measurements at RHIC. The sPHENIX b-jets measurements will extend the kinematic range of the existing LHC measurements to lower jet \pt where a larger heavy-quark mass effect is expected and further constrain heavy quark diffusion transport parameter in addition to providing insight to its temperature dependence. Figure~\ref{fig:sphenixbjet} show projections for the sPHENIX b-jet $R_{AA}$ compared to theoretical models \cite{Huang:2013vaa}, which it should be able to constrain. 

The LHC has published heavy-flavor tagged jet measurements as discussed in Section~\ref{sec:progress:microscopic:heavy_flavor_tagged_jets} and will continue to utilize this capability for more differential measurements with the additional data collected in future LHC runs. Extensions of previous measurements of the distribution of D mesons in jets~\cite{CMS:2019jis} to additional substructure measurements as well as comparisons to future RHIC studies probe the diffusion of charm quarks in the QGP and its dependence on temperature. Substructure measurements of b-jets compared to photon-jets (predominantly produced by light quarks) and inclusive jets (a significant fraction of which are produced from gluons) will further our understanding of the flavor dependent energy loss.

\subsubsection{Prospective heavy-flavor and quarkonia measurements and theoretical advances}
\label{sec:future:microscopic:heavyflavor}

{\bf Quarkonia} 

While tremendous progress has been made in the last decade in utilizing quarkonia to study the QGP in heavy-ion collisions, there is still a gap between a qualitative characterization of the QGP features as of now to a quantitative extraction of the QGP properties. To close the gap, advancements from both experimental and theoretical sides are urgently needed. The planned data-taking during 2023-2025 at RHIC \cite{STAR:2022BUR, sPHENIX:2022BUR} and LHC Run 3 (2022-2025) and Run 4 (2029-2032) with unprecedented statistics and detector upgrades opens the door for significantly improving the precision of the current quarkonium measurements and accessing to new quarkonium states in heavy-ion collisions \cite{Chapon:2020heu}. 

Quantification of the CNM effects is critically needed, especially at RHIC where no separate measurements of the three $\Upsilon$ states in \pAu{} collision are currently available. Many model calculations rely on experimental measurements to constrain the CNM effects, and the uncertainties in the experimental data are often the dominate ones \cite{Yao:2020xzw,Du:2017qkv}. Such a situation will be greatly improved with the 200 GeV \pAu{} dataset to be recorded in 2024 at RHIC. Another essential experimental input to model calculations is the total charm quark production cross section, which dictates the recombination contribution to the charmonia. The detector upgrades at both RHIC and the LHC will enable measurements of various charmed mesons and baryons down to zero \pt, from which the total charm quark yields can be extracted. See more details in Section~\ref{sec:progress:microscopic:experiment_open_heavy_flavor}.

\begin{figure}
    \centering
    \includegraphics[width=0.7\textwidth]{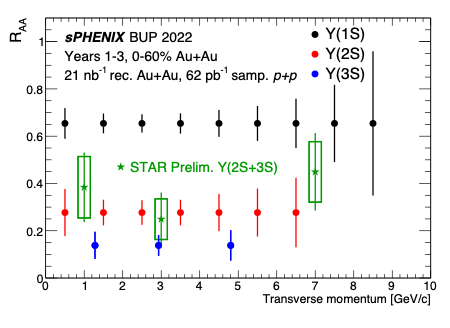}
    \caption{Projected statistical precision of the different upsilon states with 3 years of data collection with sPHENIX. The suppressed 3S state previously observed at LHC would be a new discovery at RHIC energies.}
    \label{fig:sphenixupsilon}
\end{figure}

At RHIC, the large sample of 200 GeV \AuAu{} collisions to be delivered in 2023 and 2025 and \pp{} collisions in 2024 will greatly improve the current measurements of the nuclear modification factors for $\Upsilon$(1S) and $\Upsilon$(2S) as a function of centrality and $\Upsilon$ \pt\ \cite{STAR:2022rpk}. Potentially, a real measurement of $\Upsilon$(3S) \raa\ could be achieved for the first time if a similar of level suppression as that observed at the LHC \cite{CMS:2022rna} is assumed as shown in the sPHENIX projections in Figure~\ref{fig:sphenixupsilon}. A precise measurement of the \jpsi\ flow at midrapidity in \AuAu\ collisions will also become available thanks to the increased statistics and detector upgrades at forward rapidity. The latter introduces a rapidity gap between \jpsi\ measurement and event plane reconstruction to reduce the non-flow effect, a major source of uncertainties in the previous measurement \cite{STAR:2012jzy}. Such a measurement will be instrumental in understanding the recombination contribution to \jpsi\ production and provide complementary information on the extent to which deconfined charm quarks are thermalized in the medium at RHIC. Furthermore, we will be able to access $\psi$(2S) mesons, the most loosely bounded quarkonium shown in Figure~\ref{fig:Quarkonia_Summary}, for the first time in 200 GeV \AuAu\ collisions, which will extend the lever arm with which quarkonia are used to probe the QGP. 

At the LHC, \PbPb\ collisions at $\sqrt{s_{\rm NN}}$ = 5.02 TeV with a total integrated luminosity of 13 nb$^{-1}$ are expected to be delivered in Run 3 and 4. With such a large data sample and enhanced detector capabilities, high-precision differential measurements in centrality, transverse momentum and rapidity for the three $\Upsilon$ states will become a reality \cite{CMS:2018bxx}. Elliptic flow of $\Upsilon$(1S) can also be measured with good precision, even though its constraining powering will be limited due to the small signal. Nevertheless, it will provide valuable insights into the $\Upsilon$ recombination and the degree of bottom quark thermalization in the QGP. Precise measurements of $\psi$(2S) production in these collisions will become feasible, which is expected to be decisive in distinguishing different models of charmonium recombination \cite{ALICE:2022wpn}. These measurements of the suppression levels for the excited quarkonium states are essential for interpreting the results of the ground states because of the feeddown contribution. Furthermore, the \pt\ reach of the \jpsi\ ($\Upsilon$) \raa\ will be extended to about 80 (50) GeV/$c$, compared to the 50 (30) GeV/$c$ with the current data \cite{CMS:2018bxx}. Such an increase in statistics and kinematic reach, together with $\psi$(2S) measurements, will greatly enhance our ability to study possible contributions of parton energy loss to the suppression of high-\pt\ \jpsi\ \cite{Zhang:2022rby}. In addition, the fixed-target program at the LHCb by injecting noble gases through the SMOG (System for Measuring Overlap with Gas) device has recently been upgraded and is now participating in the LHC Run 3 data-taking. It is anticipated to record about 15M \jpsi, 150k $\psi(2S)$ and 7k $\Upsilon$(1S) in \pAr{} collisions at $\sqrt{s_{\rm NN}}$ = 115 GeV, and similar statistics in \PbAr{} collisions at $\sqrt{s_{\rm NN}}$ = 72 GeV \cite{Bursche:2018orf}, bridging the gap between SPS and top RHIC energies. These data will prove valuable in studying the interplay of various contributions to quarkonium production in heavy-ion collision at varying collision energies. The P-wave charmonium states, $\chi_{c\rm{1}}$ and $\chi_{c\rm{2}}$ \cite{LHCb:2012af} whose binding energies are between those of \jpsi\ and $\psi(2S)$, will also become accessible in heavy-ion collisions by the LHCb experiment. Their measurements will provide additional tests of hot medium effect and constitute important inputs for understanding feeddown contributions to \jpsi.

A heavy-ion collision is a complex process that involves distinct phases with different dynamics. Its full understanding requires to piece together measurements of all the probes into a coherent picture, and there is no exception when it comes to quarkonia. With the wealth of high-precision quarkonium measurements to be carried out in the next decade, it is highly desirable to incorporate quarkonium physics into a common framework with other sub-fields in order to utilize the multi-messenger approach to study the properties of the QGP. Such an integration could pave the way for extracting the novel chromoelectric field correlator for quarkonium from the anticipated high-precision results based on the Bayesian analysis.

\textbf{Heavy Flavor measurements} 

The current heavy flavor data sets are already responsible to rule out transport models and constrain medium parameters, examples are the nuclear PDFs shown in Figure~\ref{fig:HF_bjorken_x} and the diffusion parameter in Figure~\ref{fig:diffusion_coefficient}. Detector upgrades and higher luminosity heavy-ion collisions will enable new channels and tools for the understanding of cold and hot medium. A better understanding of heavy quark hadronization and how it recombines in hot medium is needed. More precise measurements of $\Lambda_c$ and $D_s$ are expected at RHIC and LHC. Charm hadron spectroscopy, as explored by ALICE in \pp{} collisions \cite{ALICE:2021dhb}, is going to be extended to \pA{} and \AA{} collisions at LHC. The ALICE/ITS upgrade \cite{ALICE:2013nwm} is dedicated to heavy flavor measurements with the goal to access D-hadrons at rest. LHCb is already able to perform heavy flavor spectroscopy as part of its core flavor physics program. The current upgrade will enable LHCb to take more central \PbPb{} events and its new fixed target program will explore \pA{} and \AA{} collisions at $\sqrt{s_{NN}}\sim40-115$ GeV/$c$, expecting to measure 150M $D^0$s and 1.5M $\Lambda_c$s per year of operation \cite{Bursche:2018orf}. More exotic charm hadrons as expected to be seen for the first time in heavy-ion collisions. 

The b-quark hadronization is nearly unexplored experimentally. More abundant and precise B-meson measurements will be seen with the upgraded luminosities at LHC. The CMS experiment expects integrated luminosity of 5-10 nb$^{-1}$ in \PbPb{} during LHC Run3, a factor 3-5 increase from the current integrated luminosity. B-meson flow and top production in \PbPb{} can be explored at different \pt and event centralities. The upcoming high luminosity RHIC operation will also enable the exploration of b hadrons by STAR and sPHENIX. A silicon detector similar to the one used in ALICE has been installed in sPHENIX which will enable its b-hadron physics as one of its pillars \cite{Marshall:2022xug}.

The separation of initial and final state effects is still an experimental and theoretical challenge for the use of heavy flavor for QGP tomography. High statistics can also allow more measurements of heavy flavor correlation with hadrons, jets and direct photons in different event plane configurations. This kind of analysis are very sensitive to in-medium effects and pose a very good suppression of effects caused by nuclear medium before the hard scattering. 

Finally, the coming years will also be a preparation for the next decade heavy-ion program. Technical Design Reports and Letter of Intent have been prepared for LHC Run4,5 with specific proposals for heavy-ion physics. CMS plans for a MIP time of flight detector which will bring particle identification for CMS \cite{Butler:2019rpu} reducing the large combinatorial backgrounds CMS faces when dealing with HF decays. LHCb will continue pushing for a detector able to measure very central \PbPb{} collisions with better segmentation and timing capabilities and explore HF states with a difficult access to zero \pt, such as $D^*$, in its Upgrade2 program \cite{LHCb:2012doh}. The new-generation heavy-ion experiment proposed by the ALICE collaboration, ALICE 3, will be based on fast silicon trackers to enable the full luminosity that LHC can provide, full range particle identification and extend the acceptance for HF decays to $p_T~\sim0$ and rapidity coverage to $\eta\sim$4 \cite{ALICE:2022wwr}.

{\bf Heavy flavor and quarkonia theory developments}

 The theoretical progress in our understanding of heavy quark and quarkonium dynamics inside the QGP was discussed in Section~\ref{sec:progress:microscopic:experiment_open_heavy_flavor}. The key transport and thermodynamic properties of the QGP that can be probed by measuring heavy quark and quarkonium production in heavy-ion collisions are chromoelectric field correlators that are defined gauge-invariantly and nonperturbatively. The correlator for single heavy quark diffusion is different from that for quarkonium at low temperatures. Nonperturbative determination of the novel chromoelectric field correlator for quarkonium is an important theoretical question that should be addressed in the next \LRP{}, which has never been done. Both the Euclidean lattice QCD and AdS/CFT methods can be utilized to learn some aspects of this correlator at both zero and finite frequencies. Beyond these nonperturbative methods, one may investigate calculating this correlator by using real-time field theory simulation on a quantum computer in the next \LRP{}. Besides direct theoretical calculations, we can also use Bayesian analysis to extract this correlator from experimental data. The data at the RHIC energy is of particular importance in our extraction of the finite frequency dependence of the chromoelectric field correlator, since at the RHIC energy the low temperature regime of the QGP is more important. A good Bayesian extraction requires experimental data in a wide kinetic range with high precision, as well as solid theoretical calculations that have as little model dependence as possible. Current application of nonrelativistic effective field theories of QCD and the open quantum system framework is only applicable to quarkonium production at low $p_T$. How to generalize the framework to study quarkonium production at intermediate $p_T$ should be explored in the next \LRP{}. Quantum computing~\cite{DeJong:2020riy,deJong:2021wsd} and machine learning techniques should be investigated to speed up solving the Lindblad equation in the Brownian motion limit, which would be necessary if we want to study charmonium production in the open quantum system framework. The nPDF uncertainty is another big factor that can influence the Bayesian analysis, which we want to reduce as much as possible. Also, the nPDF uncertainty is small in observables like $R_{AA}$ ratios of different quarkonium states. The nonperturbative determination of the novel correlator for quarkonium in the next \LRP{} would be a joint effort between the theory, computation and experiment communities and it will involve not only the quarkonium community but also many other subfields of the heavy-ion and nuclear physics.

On the lattice side, the calculations of bottomonium properties and the complex potential have been performed on $N_{\tau}=12$ lattices. To better constrain the spectral functions one clearly needs calculations with larger $N_{\tau}$. As discussed above all calculations of the heavy quark diffusion coefficient have been performed in quenched approximation. The use of the recently developed gradient flow approach developed for this problem~\cite{Altenkort:2020fgs} will allow performing the calculations in QCD with physical quark masses. 
In addition to the heavy quark diffusion, which is encoded in the chromo-electric
correlator with the Wilson lines in the fundamental representation 
it is important to study the chromo-electric correlation function with the adjoint
Wilson line. As discussed above, this correlation function encodes the transport
properties of quarkonia. The lattice calculation of this correlation function
also needs lattices with a large temporal extent.
The calculations with larger $N_{\tau}$ and physical quark masses will require the use of exascale computing resources, which are allocated through ALCC and INCITE programs. To take advantage of the exascale resources existing lattice QCD codes have to be adapted to the ever-changing computational hardware. This will require funding from programs like SciDAC, which will support the workforce development of computational nuclear physicists and sustain this workforce needs in the long run. The lattice calculations with larger $N_{\tau}$ will not be limited to the calculations of bottomonium properties at non-zero temperature, the complex $Q\bar Q$ potential, and the heavy quark diffusion coefficient. With small additional computational investments and the use of extended meson operators, one would be able to access the in-medium properties of open heavy-flavor hadrons as well as of charmonia. The availability of lattices with larger $N_\tau$ will also benefit the study of charm fluctuations and charm baryon number correlations, allowing to fully control the discretization effects and provide more first principle QCD information on the charm production in \AA{} collisions. Furthermore, these lattices will be also used in the study of spatial bottomonium correlation functions and refine the estimates of the bottomonium melting temperatures. 

At high virtualities, the SCET formalism developed by \cite{Abir:2015hta} should be extended to obtain the heavy flavor pair production rate of $g\to Q+\bar{Q}$, while that SCET formalism should be combined with the $\hat{q}(Q^2)$ calculation in Ref.~\cite{Kumar:2019uvu} to provide a $\hat{q}(M,Q^2)$. At lower virtualities, an important extension to light flavor HTL-resummed AMY rates has been to consider finite-size effects. This has been achieved \cite{Caron-Huot:2010qjx} using Zakharov's light-cone path integral formalism \cite{Zakharov:1996fv,Zakharov:1997uu}, thus bridging the gap between infinite-sized medium result assumed by AMY and a finite-sized medium in the opacity expansion by Gyulassy-Levai-Vitev (GLV) \cite{Gyulassy:1999zd}. Finite-size effects of low-virtuality interactions between light flavor and the QGP are included within the Modular Algorithm for Relativistic Treatment of heavy IoN Interactions (MARTINI) Monte Carlo event generator \cite{Schenke:2009gb}. A natural continuation is to formally combine finite-size effects \cite{Caron-Huot:2010qjx} within the heavy flavor calculation of Ref.~\cite{Caron-Huot:2007rwy}. Some progress in that direction, combining elements of formal treatment with phenomenological input, is explored in Ref.~\cite{Ke:2019jbh}.

Drawing on the expertise of leading scientists in the U.S., the new nuclear theory topical collaboration, HEFTY will tackle challenges in describing heavy flavor quarks in QCD matter.  More specifically, the HEFTY collaboration aims to develop a comprehensive theoretical framework encompassing the evolution of heavy flavor from the initial production of heavy quarks in the early stages of the collisions, to their diffusion through the QGP as well as the hadronization process resulting in final-state heavy-flavor particles.

\textbf{Future Prospectives with Exotic Hadrons}

The only exotic hadron measured in heavy-ion collisions to date is the X(3872), which is generally considered to consist of a charm-anticharm pair and a quark-antiquark pair, $c\bar{c}q\bar{q}$.  Statistical hadronization models expect the recently discovered $T_{cc}^{+}$ hadron \cite{LHCb:2021vvq}, which is consistent with a tetraquark of the form $cc\bar{q}\bar{q}$, to be produced at nearly the same rate as X(3872), representing a huge enhancement due to recombination \cite{Hu:2021gdg}. Production of the "fully-charmed" $c\bar{c}c\bar{c}$ tetraquarks is also expected to be dramatically enhanced in \PbPb{} collisions \cite{Zhao:2020nwy}, and measurements would provide significant new constraints on models of charm quark hadronization.  There is also potential for measuring exotic states produced via interactions with photons in ultra-peripheral collisions \cite{Esposito:2021ptx}, which may be relevant for constraining projections of exotic hadron production at the forthcoming EIC.

The data on exotic hadron production in heavy-ion collisions remains limited by significant uncertainties.  However, all experiments at the LHC are pursuing relevant upgrades that will enable future measurements.  The recent upgrade to the ALICE TPC readout will enable significantly higher luminosity to be recorded, aiding measurements of relatively rare exotic states.  The CMS collaboration is pursuing the addition of particle ID with precision time-of-flight measurements and a large acceptance tracking system up to $|\eta|<4$. Those upgrades will help reduce combinatorial backgrounds and increase the acceptance when reconstructing exotic hadron decays. The LHCb experiment recently upgraded the entire tracking system, enabling measurements up to $\sim$30$\%$ centrality in \PbPb{} collisions.  A major upgrade to the LHCb fixed-target system will enable high-statistics \pA{} measurements at center-of-mass energies near $\sim100$ GeV, where the charm cross-section is relatively small and potential effects of $D$ meson recombination into hadronic molecules will be small.  In the further future, the ALICE3 and LHCb Upgrade II detectors will have full particle ID, precision vertexing, and fast DAQ systems, and will be well-suited for measurements of X(3872) and other exotic hadrons in central \PbPb{} collisions.   Future measurements of femtoscopic correlations between components of hadronic molecules may also yield new information on the size and masses of exotic molecular bound states.  There is also no theoretical consensus on describing exotic production in heavy-ion collisions.  However, as current research into exotic hadrons is largely driven by experiments, significant developments on the theoretical front can also be accomplished as more constraining data becomes available in the future.

\subsubsection{Prospective measurements and theoretical advances related to initial state nPDFs}
\label{sec:future:microscopic:IS}
{\bf Electroweak probes} 

All the processes described in Section~\ref{sec:progress:microscopic:hard_electroweak} are rare probes and thus directly benefit from increases in luminosity. In addition, the dominant uncertainties in some channels are evaluated using data-driven methods (e.g. photon purities), and thus larger luminosities also confer reduced systematic uncertainties as well. At the LHC~\cite{Citron:2018lsq}, approximately 13 nb$^{-1}$ of \PbPb{} and 1200 nb$^{-1}$ of $p$+Pb data may be collected by ATLAS and CMS in Runs 3 and 4, representing increases by a  factor of six to seven over that available in Runs 1 and 2. At the same time, measurements by ALICE and LHCb will provide measurements in a complementary kinematic range (such as at low photon-$p_\mathrm{T}$ or forward rapidity for heavy bosons). 

\begin{figure}[t]
    \centering
    \includegraphics[width=0.38\textwidth]{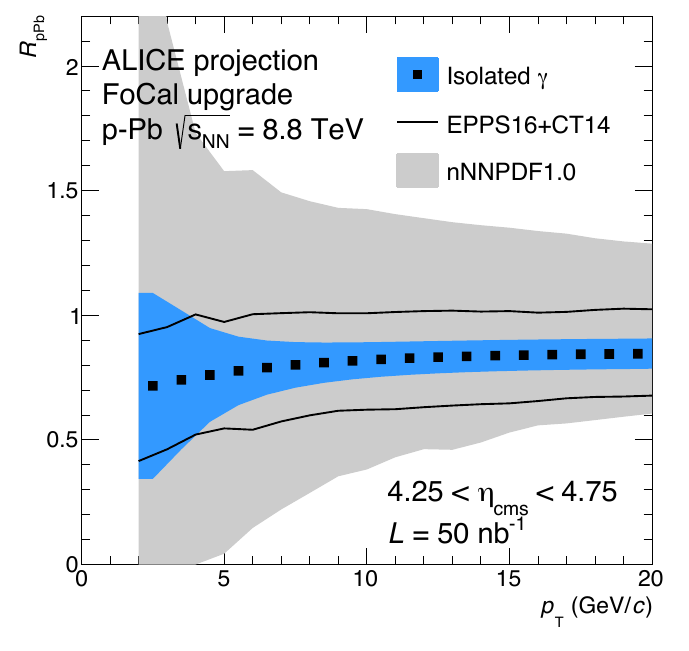}
    \includegraphics[width=0.54\textwidth]{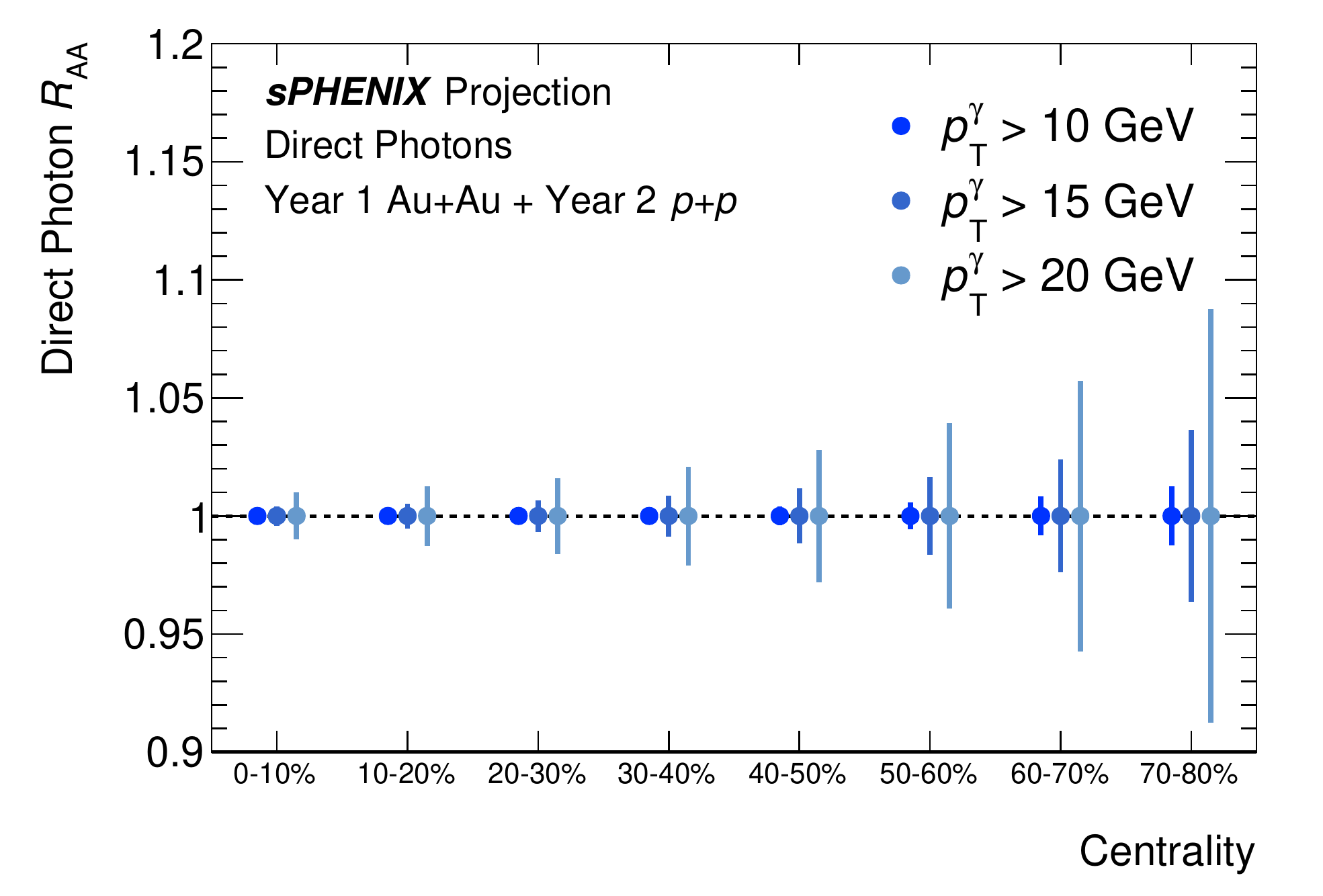}
    \caption{Left: Projected uncertainties for a measurement of the $R_{p\mathrm{Pb}}$ for forward photons with the ALICE FoCal, compared to the existing theoretical uncertainties~\cite{Eskola:2016oht,AbdulKhalek:2019mzd}. Right: Projected statistical uncertainties for a measurement of the $R_\mathrm{AA}$ for direct photons in sPHENIX as a function of centrality, using only Year-1 (2023) and Year-2 (2024) data.
    }
    \label{EW:future}
\end{figure}

The larger luminosity and improved detector capabilities may also allow access to new channels entirely, such as measurements of isolated di-photons in $p$+Pb collisions (produced dominantly via a quark box diagram and thus sensitive to gluon densities), forward photons with the ALICE FoCal~\cite{ALICE:2020mso} (probing very low nuclear-$x$, shown in the left panel of Figure~\ref{EW:future}), hadronic $W$ decays, and multi-differential measurements such as photons in coincidence with leading jets~\cite{ATLAS:2018xvi} (to better constrain the initial-state kinematics).

In the LHC Run 5, collisions of light ions with high luminosity would result in an enormous yield of electroweak probes for study. The smaller underlying event will also result in significantly improved systematic uncertainties, giving extraordinarily precise constraints on nPDFs in these systems. In fact, such constraints are expected to be necessary as the kinematic reach for jet probes will push far into the large-$x$ region, where the EMC effect is appreciable.

Finally, measurements of high-$p_\mathrm{T}$ direct photons are expected to be a flagship of the physics program of the sPHENIX experiment at RHIC, which will have its commissioning run in 2023. These are enabled by the large acceptance, high rate, effective triggering, and excellent energy resolution of the sPHENIX EMCal. sPHENIX will use direct photons for photon-tagged jet energy loss measurements, nPDF measurements, and as an early check of the centrality calibration (right panel of Figure~\ref{EW:future}).

{\bf Future studies with ultra-peripheral collisions}

{\em Resolving photon energy ambiguities based on ZDC neutrons} Due to the independent QED process of Coulomb excitation, different impact parameter of the two colliding nuclei would have a different probability of producing evaporated neutrons. By combining the ZDC neutron categories for a given rapidity of VM, the cross section of $\gamma^{\ast}+\rm A \rightarrow \rm{VM}+ A'$ can be obtained for both low- and high-energy photon configuration separately. This experimental technique can extend the kinematic reach down to $x\sim10^{-5}$ or up to $x\sim10^{-1}$. Based on Ref.~\cite{Guzey:2013jaa}, the photon-nucleus cross section corresponding to photon energy $k_1$ and $k_2$ can be related to the photon flux as follows, 
\begin{equation}
\label{eqa:upc-photon-resolve}
    d\sigma^{BnCn}/dy=\Phi^{BnCn}_{T.\gamma}(k_{1})\sigma_{\rm{\gamma^{\ast}+A\rightarrow J/\psi+A}}(k_{1}) 
 +\Phi^{BnCn}_{T.\gamma}(k_{2})\sigma_{\rm{\gamma^{\ast}+A\rightarrow J/\psi+A}}(k_{2})
\end{equation}
\noindent where $ \Phi^{BnCn}_{T,\gamma}$ is the average transversely polarized photon flux emitted from the A nucleus at the average rapidity in the selected $y$ interval; the superscript $BnCn$ corresponds to the neutron emission class such that $B$ and $C$ can be $0$ or $X$. Therefore, with three independent equations of $0n0n$, $0nXn$, and $XnXn$, the coherent cross section at photon energy $k_1$ and $k_2$ can be obtained (only two independent equations are needed to solve two unknowns, but three provides additional constraints on the result). Note that this method only applies to coherent \jpsi photoproduction. Without the photon energy ambiguities, the kinematic reach in Bjorken-$x$ can be extended to much lower and the energy $W_{\rm \gamma^{\ast}N}$ dependence can be directly obtained. This is an economic way to go to small-$x$ in UPCs.

{\em UPC $\phi$ photoproduction} Measuring $\phi$ photoproduction has been challenging in UPCs mostly due to its soft decay to two kaons: the daughter kaons have an average of $\sim 100~\rm{MeV/c}$ in transverse momentum, where tracking at such low \pt{} is already difficult and triggering UPC $\phi$ meson decay requires fast detector, e.g., calorimeter or Time-of-Flight system, that are sitting far away from the vertex. Therefore, the soft kaons do not have enough momentum to reach these fast detectors. This is the same for LHC experiments as well as for the RHIC STAR experiment. Recently, one possible solution has been proposed, given the larger cross section of $\phi$ than $J/\psi$ and the high luminosity of the accelerator machine, one can reconstruct $\phi$ from zero-bias or ZDC-based triggers only, without the requirement of fast detectors in the barrel. This new development in the trigger may provide a unique opportunity in studying photoproduction of $\phi$. In addition, although the EIC is the ultimate machine and experiment for exclusive production, photoproduction of $\phi$ can be equally if not more difficult
than that in UPCs. This measurement in UPCs at RHIC and at the LHC are truly complementary to $\phi$ measurements at the EIC in terms of the energy and virtuality coverage \cite{Arrington:2021yeb}.   

{\em Jet/high-\pt{} particle and VM photoproduction double ratio measurement} Inspired by the EIC White Paper~\cite{Accardi:2012qut}, one of the most important day-one measurements is to obtain the relative fraction of diffractive DIS over the inclusive DIS cross section in both $eA$ and $ep$ collisions. It was discovered at the $ep$ collider HERA that about $\sim12-15\%$ of the inclusive DIS cross section is diffractive. At the EIC, the most direct observation of gluon saturation predicted by the color glass condensate model is that this ratio will be enhanced in $eA$ than in $ep$; in other words, diffractive DIS cross section would be larger than $\sim12-15\%$ of the inclusive DIS cross section in $eA$ collisions. However, the other major paradigm, as discussed earlier in Section~\ref{subsubsec:vm}, is the nuclear shadowing model, which predicts qualitatively the opposite. In the Leading Twist Approximation, the diffractive DIS cross section is expected to be less in $eA$ than in $ep$. Therefore, this measurement becomes one of the best experimental tools to test the two models. 

This measurement requires high-energy $ep$ and $eA$ collisions with significant photon virtuality $Q^{2}$ at the EIC, which makes this impossible in UPCs. However, a similar observable might be possible in UPCs. Although the scale set by the photon virtuality in UPCs at the event level is small, the scale could be set by the mass or transverse momentum of the particle instead, e.g., the $J/\psi$ and/or a high-\pt{} particle. Exclusive $J/\psi$ photoproduction is a diffractive process, while photoproduction of single high-\pt{} particle or jet is an inclusive process. The ratio between them captures the key element of diffractive process over inclusive process, where the two models differ qualitatively. Therefore, a double ratio observable, $R_{R}$, in UPCs measurements can be roughly written as follows, 
\begin{equation}
    R_{R} = \rm \frac{\sigma^{exclusive}_{J/\psi} /  \sigma^{inclusive}_{jet,high-p_{T}}|_{A}} {\sigma^{exclusive}_{J/\psi}  /  \sigma^{inclusive}_{jet,high-p_{T}}|_{p} }.
\end{equation}
\noindent Here the $\sigma^{\rm exclusive}_{J/\psi}$ in heavy nuclei and proton have been measured by various experiments, while $\sigma^{\rm inclusive}_{\rm jet,high-p_{T}}$ has not yet been done. Theoretical predictions behind this observable in UPC events are being developed. This measurement might be the first rigorous experimental test to the saturation and nuclear shadowing model in UPCs. In addition, utilizing the wide kinematic reach from RHIC to the LHC could significantly advance our understanding of nuclear modification of the gluon densities in nuclei.

{\em Vector meson photoproduction using the FoCal detector at ALICE}
For Run 4 at the LHC (2029--2032), the ALICE experiment will have in operation the FoCal detector~\cite{ALICE:2020mso}. FoCal is a high-granularity, compact silicon-tungsten (\SiW{}) sampling electromagnetic calorimeter with longitudinal segmentation backed by a conventional high granularity metal and scintillating hadronic calorimeter. While FoCal has been designed for measuring direct photons at forward rapidity with high precision in \pp{} and \pPb{} collisions~\cite{ALICE:2020mso}, the study of UPCs is also an integral part of the FoCal physics program, as reported in~\cite{Bylinkin:2022wkm}. In particular, the future measurements of the energy dependence of $\sigma(\gamma \rm{p})$ and $\sigma (\gamma \rm{Pb})$ for the photoproduction of vector mesons using the FoCal detector should provide a clear signature of gluon saturation, probing nuclear gluon shadowing, and confronting several theoretical approaches. Here, we highlight two measurements in $\rm \gamma \rm{p}$ interactions discussed in Ref.~\cite{Bylinkin:2022wkm} (see this reference for other high-profile UPC measurements with FoCal).  
As mentioned above, results from HERA showed that the photoproduction cross section, $\sigma(\gamma~p~\rightarrow~J/\psi~p)$, increases with center-of-mass energy following a power law. The result from ALICE~\cite{ALICE:2014eof,ALICE:2018oyo}, extending down to $x \sim 10^{-5}$, showed no deviation from this behavior, although the uncertainties are large. The FoCal detector will provide access to an unexplored kinematic regime at small-$x$ where a different trend in the growth of the cross section might occur. At high energies, there are three non-linear QCD model predictions, namely, the Hot Spot model (CCT)~\cite{Cepila:2016uku}, the NLO BFKL~\cite{Bautista:2016xnp} and the CGC-based calculations~\cite{Armesto:2014sma}. To illustrate the prospects of observing saturation effects with FoCal, the projection of the energy dependence J/$\psi$ photoproduction cross section in the FoCal acceptance was obtained using the NLO BFKL model~\cite{Bautista:2016xnp}, as it is shown in Ref.~\cite{Bylinkin:2022wkm}. 
Similarly, if saturation occurs, the future FoCal UPC measurements would provide the first potential observation that the cross section of J/$\psi$ deviates from a power-law energy dependence at high energies.

In addition, as discussed earlier, the dissociative process has received recent interest~\cite{Mantysaari:2016ykx,Mantysaari:2016jaz,Cepila:2016uku}. The FoCal data are uniquely positioned to probe the fluctuations of the proton target configurations for gluonic saturated matter. The observation of a significant reduction of the measured cross section as energy increases will provide a clear signature of gluon saturation at high energies. Specifically, it will be clear to measure the ratio of dissociative-to-exclusive J/$\psi$ photoproduction as a function of $W_{\gamma \rm{p}}$, including the H1 data~\cite{H1:2013okq} and ALICE preliminary data, and comparisons to the BM and CCT predictions. See Ref.~\cite{Bylinkin:2022wkm} The FoCal projected points shown use the STARlight yield for the exclusive process, and the BM model prediction for the dissociative process.

{\em QED ($\gamma\gamma\rightarrow l^+l^-$)}
QED studies, e.g., with polarized photons, have been used and proposed as a tool to test and define the photon Wigner function~\cite{Zha:2018tlq,Wang:2021kxm,Klein:2020jom,Klusek-Gawenda:2020eja,CMS:2020skx,Sun:2020ygb,Zha:2021jhf}, to probe the properties of the Quark-Gluon Plasma~\cite{Brandenburg:2021lnj,STAR:2018ldd,ATLAS:2018pfw,Klein:2018fmp,Wang:2021oqq,An:2021wof,Klusek-Gawenda:2018zfz}, to measure nuclear charge and mass radii~\cite{Wang:2022ihj,STAR:2022wfe,Budker:2021fts,Brandenburg:2021lnj}, to study gluon structure inside nuclei~\cite{Hatta:2021jcd,Xing:2020hwh,Bor:2022fga} and to investigate new quantum effects~\cite{STAR:2022wfe,Zha:2018jin,Zha:2020cst,Xing:2020hwh,Dyndal:2020yen,Xu:2022qme}. Both RHIC and the LHC plan future data collection that will allow high precision multi-differential analysis of these $\gamma\gamma$ processes~\cite{SN0755STARBeam,ProspectsMeasurementsPhotonInduced}. Future measurements at STAR are expected to provide significantly higher precision measurements of the $e^+e^-$ transverse momentum spectra and the $\cos{4\phi}$ modulation. Additionally, multi-differential measurements, such as the $\cos{4\phi}$ modulation strength versus pair $p_T$, will be possible. The increased precision on the pair $p_T$ will provide additional constraining power to investigate the proposed final-state broadening effects. In addition to their effect on the $p_T$ spectra, final state interactions would wash out the $\cos{4\phi}$ modulation strength that results from the initial colliding photon polarization. The predicted future precision could be achieved for the $p_T$ and $\cos{4\phi}$ modulation measurements in future STAR analyses. The added precision in the $\cos{4\phi}$ modulation measurement is expected to allow experimental verification of impact parameter dependence predicted by the lowest order QED calculations (and therefore further test the $k_\perp$-factorization plus TMD treatment of the photon polarization~\cite{Li:2019yzy,Harland-Lang:2018iur}). The future data taking campaigns planned for the LHC experiments will also allow improved measurements from ALICE of the $\gamma\gamma \rightarrow e^+e^-$ process in a similar region of phase space as measured at RHIC, but in collisions with a much larger Lorentz-boost factor. Such measurements will provide further constraints on the treatment of the photon kinematic distributions over a range of photon energies. Similarly, future data taking and analyses by CMS and ATLAS~\cite{ATLAS:2022vbe,ProspectsMeasurementsPhotonInduced} will allow additional precision measurements of the $\gamma\gamma \rightarrow \mu^+\mu^-$ process in events with hadronic overlap, possibly shedding light on the presence (or lack) of medium induced modifications via differential measurements of the produced dilepton kinematics.

{\em The role of the ALICE 3 detector} Looking further ahead, by the mid/late 2030s, the proposed ALICE 3 detector \cite{ALICE:2022wwr} will have acceptance for both charged and neutral particles, over a very wide solid angle, with coverage expected for pseudorapidity $|\eta|<4$.  This will offer a very large increase in acceptance for more complex UPC final states.  Upgrades from other LHC experiments will also significantly benefit the UPC program.

\newpage 

\section{Summary}
\label{sec:summary}

This document highlights the tremendous progress made, since the 2015 \LRP{} in Nuclear Science, in understanding the properties of the quark-gluon plasma and how it develops in heavy-ion collisions from the microscopic to macroscopic level. The previous section on future prospects also illustrates the potential for future scientific discoveries and developments related to hot QCD, much of which is within our grasp in the coming decade. However, realization of these relies on support for the following:
\begin{itemize}
\item Successful operation of RHIC for the sPHENIX detector to fully achieve the scientific goals described in the previous \LRP{}. 
\item Continued leadership and upgrades for the LHC experiments to study high temperature QGP.
\item A thriving theory community to advance our description of QCD. 
\item Theory and experimental collaborations to develop tools such as Bayesian inference, which utilize the full suite of experimental data in extracting key quantities such as viscosity and the transport coefficients as well as furthering the interpretations of the experimental results.
\item The computing resources needed and novel approaches such as machine learning to overcome the difficulties in addressing the most challenging questions in the field.  
\item  A vibrant workforce to carry out the analysis of the wealth of experimental data collected including from BES-II.  \item Programs enhancing the participation of underrepresented minorities in nuclear physics, which will enhance our ability to recruit the workforce needed to accomplish these goals in the next decade.
\end{itemize}

With the start of sPHENIX operations concurrently with those of the upgraded STAR detector at RHIC and the increased luminosity at the LHC, there is much to explore experimentally in the next decade, while there are also many opportunities to advance our theoretical understanding of hot QCD. This document highlights the accomplishments as well as open questions in describing heavy-ion collisions from the initial state, to collectivity and thermalization to penetrating probes as well as the interplay between these. The next decade will further advance our understanding of hot QCD matter and propel us toward a more complete description of nuclear matter under extreme conditions as well as the evolution of heavy-ion collisions.  

\newpage
\bibliographystyle{hieeetr}
\bibliography{references_inspireshep,references_others}

\end{document}